\documentclass[a4paper, 12pt, twoside]{book} 

\usepackage[utf8]{inputenc}		
\usepackage[T1]{fontenc}		
\usepackage{lmodern}
\usepackage{ae,aecompl}
\usepackage[top=2.5cm, bottom=2cm, 
			left=3cm, right=2.5cm,
			headheight=15pt]{geometry}
\usepackage{graphicx}
\usepackage{eso-pic}	
\usepackage{array}

\usepackage[square,sort&compress,sectionbib]{natbib}		

\usepackage[french,main=english]{babel}
\usepackage{titlesec}
\usepackage{lastpage}
\usepackage[titletoc]{appendix}
\usepackage{caption}
\usepackage{subcaption}
\usepackage{cancel}
\usepackage{pdfpages}
\usepackage{sidecap}
\usepackage{amsmath}
\usepackage{amssymb}
\usepackage{amsfonts}
\usepackage{siunitx}
\usepackage{xcolor}
\usepackage{physics} 
\usepackage{empheq}
\usepackage{enumerate}
\usepackage{enumitem}
\usepackage[section]{placeins}	
\usepackage{epigraph}
\usepackage[font={small}]{caption}
\usepackage[english,nohints]{minitoc}		
\usepackage[notbib]{tocbibind}
\usepackage{import}
\usepackage{graphicx,calc}
\newlength\myheight
\newlength\mydepth
\settototalheight\myheight{Xygp}
\usepackage[colorlinks=true,citecolor=blue,urlcolor=red]{hyperref}
\usepackage[normalem]{ulem}
\DeclareMathAlphabet\mathbfcal{OMS}{cmsy}{b}{n}

\usepackage{fancyhdr}
\usepackage{natbib}
\renewcommand{\bibsection}{\section{References}}		

\newcommand{\er}[1]{Eq.~\eqref{#1}}

\newcommand*{\sumcirclearrowleft}{%
 \DOTSB
 \mathop{
  \mathchoice
   {\rlap{\kern.25em\rotatebox[origin=c]{-90}{$\circlearrowleft$}}{\sum}}
   {\vcenter{\rlap{\kern.2em\rotatebox[origin=c]{-90}{$\scriptscriptstyle\circlearrowleft$}}}{\sum}}
   {\sum}{\sum}
 }
}

\newcommand*{\sumcirclearrowright}{%
 \DOTSB
 \mathop{
  \mathchoice
   {\rlap{\kern.25em\rotatebox[origin=c]{90}{$\circlearrowright$}}{\sum}}
   {\vcenter{\rlap{\kern.2em\rotatebox[origin=c]{90}{$\scriptscriptstyle\circlearrowright$}}}{\sum}}
   {\sum}{\sum}
 }\slimits@
}

\usepackage{soul}

\let\oldpagenumbering\pagenumbering
\renewcommand{\pagenumbering}[1]{\fancyhead[L]{}\fancyhead[R]{}\cleardoublepage
	\oldpagenumbering{#1}
}

\usepackage{listings}
\definecolor{codegreen}{rgb}{0,0.6,0}
\definecolor{codegray}{rgb}{0.5,0.5,0.5}
\definecolor{codepurple}{rgb}{0.58,0,0.82}
\definecolor{backcolour}{rgb}{0.95,0.95,0.92}
\lstdefinestyle{pythoncodestyle}{
    backgroundcolor=\color{backcolour},   
    commentstyle=\color{codegreen},
    keywordstyle=\color{magenta},
    numberstyle=\tiny\color{codegray},
    stringstyle=\color{codepurple},
    basicstyle=\ttfamily\footnotesize,
    breakatwhitespace=false,         
    breaklines=true,                 
    captionpos=b,                    
    keepspaces=true,                 
    numbers=left,                    
    numbersep=5pt,                  
    showspaces=false,                
    showstringspaces=false,
    showtabs=false,                  
    tabsize=2
}

\setcitestyle{numbers,square,comp, unsrt} 
\usepackage{enumitem}
\usepackage{tikzpagenodes}

\definecolor{myblue}{rgb}{0,0.2,.8}

\makeatletter
\def\@ecole{école}
\newcommand{\ecole}[1]{
  \def\@ecole{#1}
}

\def\@specialite{Spécialité}
\newcommand{\specialite}[1]{
  \def\@specialite{#1}
}

\def\@directeur{directeur}
\newcommand{\directeur}[1]{
  \def\@directeur{#1}
}
\def\@subtitle{subtitle}
\newcommand{\subtitle}[1]{
  \def\@subtitle{#1}
}

\def\@encadrant{encadrant}
\newcommand{\encadrant}[1]{
  \def\@encadrant{#1}
}

\def\@jurya{}{}{}{}{}
\newcommand{\jurya}[5]{
  \def\@jurya{\scshape{#1},	& #2 & #3 & #4 & #5\\ }
}
\def\@juryb{}{}{}{}{}
\newcommand{\juryb}[5]{
  \def\@juryb{\scshape{#1},	& #2 & #3 & #4 & #5\\ }
}
\def\@juryc{}{}{}{}{}
\newcommand{\juryc}[5]{
  \def\@juryc{\scshape{#1},	& #2 & #3 & #4 & #5\\ }
}
\def\@juryd{}{}{}{}{}
\newcommand{\juryd}[5]{
  \def\@juryd{\scshape{#1},	& #2 & #3 & #4 & #5\\ }
}
\def\@jurye{}{}{}{}{}
\newcommand{\jurye}[5]{
  \def\@jurye{\scshape{#1},	& #2 & #3 & #4 & #5\\ }
}
\def\@juryf{}{}{}{}{}
\newcommand{\juryf}[5]{
  \def\@juryf{\scshape{#1},	& #2 & #3 & #4 & #5\\ }
}
\def\@juryg{}{}{}{}{}
\newcommand{\juryg}[5]{
  \def\@juryg{\scshape{#1},	& #2 & #3 & #4 & #5\\ }
}
\def\@juryh{}{}{}{}{}
\newcommand{\juryh}[5]{
  \def\@juryh{\scshape{#1},	& #2 & #3 & #4 & #5\\ }
}
\def\@juryi{}{}{}{}{}
\newcommand{\juryi}[5]{
  \def\@juryi{\scshape{#1},	& #2 & #3 & #4 & #5\\ }
}

\makeatletter
\newcommand{\pagedegarde}{
      \newgeometry{top=2cm, bottom=2cm, left=1cm, right=1cm}
        \begin{titlepage}
          \centering
          {\text{ }}
          \vspace{3cm}
          \centering
                {\Large {\bfseries PhD Thesis}}\\
                \vspace{2cm}
            
              {\huge  \bfseries{\@title}} 
                \vfill
                \vspace{2cm}
                  {\Large {Author: \bfseries \@author}} \\
                \vspace{2cm}
                  {\Large {PhD Supervisors: \bfseries \@directeur \ and \@encadrant}} \\
              \vspace{6cm}

                \vfill
            \centering
            \includegraphics[width=0.4\textwidth]{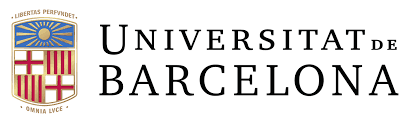}\\
              \vfill
        \end{titlepage}

      \restoregeometry  
}

\newcommand{\premierepage}{
      \newgeometry{top=2cm, bottom=2cm, left=1cm, right=1cm}
        \begin{titlepage}
        \centering
            {\text{ }}
              \vspace{7cm}
         
         {\huge  \bfseries{\@subtitle}} 
          \vfill
            {\Large {Programa de doctorat de F\'isica}} \\

          \vspace{2cm}
            {\Large {Autor: \bfseries \@author}} \\

            \vspace{2cm}
            {\Large {Directors: \bfseries \@directeur \ i \@encadrant}} \\

          \vspace{2cm}
            {\Large {Tutor: \bfseries Giancarlo Franzese}} \\
          \vfill

      \centering
      \includegraphics[width=0.4\textwidth]{other_images/logo_page_garde.png}\\
        \end{titlepage}
      \restoregeometry  
}

\newcommand{\dernierepage}{
      \newgeometry{top=2cm, bottom=2cm, left=1cm, right=1cm}
        \begin{titlepage}
        \centering
            \includegraphics[width=0.4\textwidth]{other_images/logo_page_garde.png}\\
        \vspace{1.5cm}
          {\Large {\bfseries PhD Thesis in Physics}}\\

          \vspace{2cm}
         
         {\huge  \bfseries{\@title}} 
          \vfill

          \vspace{0.5cm}
            {\Large {Author: \bfseries \@author}} \\

            \vspace{0.5cm}
            {\Large {PhD Supervisors: \bfseries \@directeur \ and \@encadrant}} \\

          \vfill

          \begin{center}
            \justifying
              We study topological defects in two-dimensional non-equilibrium systems, focusing on active extensions of the XY models, including activity, mobility and non-reciprocity. In a noisy Kuramoto lattice with short-range coupling, intrinsic frequency heterogeneity destroys quasi-long-range order and fragments the system into finite domains. Defects unbind at all temperatures and exhibit superdiffusive random walks, advected by evolving domain boundaries. By contrast, when oscillators are allowed to move in space, the system undergoes a Berezinskii–Kosterlitz–Thouless transition and regains quasi-long-range order, revealing the fundamental role of motility in sustaining coherence. We also analyse a non-reciprocal $O(2)$ model with vision-cone couplings and derive a continuum theory that captures the same physics. Non-reciprocity selects defect shapes, enriches the annihilation process, and reshapes patterns through advection. 
              Together, these results elucidate the fundamental role of activity and non-reciprocity in shaping topological defects and ordering in non-equilibrium systems.
          \end{center}
          \vfill

          \vspace{1.5cm}

      \centering
      \includegraphics[width=0.4\textwidth]{other_images/logo_page_garde.png}\\

        \end{titlepage}

      \restoregeometry  
}

\makeatother	
\author{Ylann Rouzaire}

\title{Topological defects in out-of-equilibrium systems}
\subtitle{Defectes topològics en sistemes fora de l’equilibri}
\specialite{Physics}
\directeur{Demian {Levis}}
\encadrant{Ignacio {Pagonabarraga}} 
\date{\today}

\jurya{Surname}{Name }{Title}{\textit{University}}{Role in the jury}
\juryb{Surname}{Name }{Title}{\textit{University}}{Role in the jury}
\juryc{Surname}{Name }{Title}{\textit{University}}{Role in the jury}

\hypersetup{
    pdfauthor={Ylann Rouzaire},
    pdfsubject={PhD Thesis Manuscript},
    pdftitle={Topological defects in out-of-equilibrium systems},
}

\begin{document}
\selectlanguage{english}
	\pagenumbering{arabic}
	\pagedegarde
	\premierepage
	\thispagestyle{empty}
	\selectlanguage{english}
	\cleardoublepage
	
\fancyhead[L]{} 
\fancyhead[R]{}

\section*{Acknowlegments/Remerciements}

This thesis would not have been possible without the support of a number of people whom I wish to thank here.
\newline 

First and foremost, Demian and Ignacio, thank you for making me the physicist I have become. Thank you for guiding me, teaching me, but also challenging and contradicting me throughout these five years. Thank you as well for your trust, your availability, and your constant enthusiasm. On the purely scientific side, I learned quite a lot from both of you, not only about physics but also about how to actually do physics. Thank you as well for giving me the freedom to explore my own ideas. And finally, thank you for introducing me to the very special world of scientific research, in particular through the many conferences I was lucky to attend.
\newline

My thanks also go to Fernando, Parisa, Dan, and Leticia, with whom I had the pleasure of collaborating during this PhD.

Elisabeth, thank you for your support, for our discussions about the academic world, and above all for the best advice: “You should contact Demian.”

Dan, thank you for welcoming me during my visit and for everything you taught me — scientifically, of course, but also about the realities of the research world and how it works. 
\newline 

Thank you to everyone at the University of Barcelona with whom I had the chance to talk science: the CUCA Lab, Javier, Lino, José, Gauthier, Ricard, Giacomo, Olga, Zaida, Gabriele, and many others. More formally, I acknowledge Demian and Ignacio for their careful reading of this manuscript, and Leticia and Gauthier for their constructive feedbacks.
\newline

Equally important, I want to acknowledge the various sources of funding that supported me over the years. First of all, the Spanish Ministerio de Ciencia e Innovación for the FPI scholarship that allowed me to complete this PhD free of mind. Thank you as well to CECAM in Lausanne for partially funding my first year as a visiting PhD student. Thanks also to the EPFL SEMP program for funding my Master’s project, and to the Fundació Montcelimar for funding a three-month visit to Dan’s group at the University of Geneva. Enfin, un grand merci à Christophe Pannatier et à Albert Grun de l'Ecole PrEP à Lausanne. Outre le financement de mon projet de Master et de mes 2 premières années de thèse, j'ai énormément appris en travaillant comme enseignant dans cette école unique, tant sur le plan pédagogique que sur le plan rhétorique.
\newline

Enfin, sur le plan personel, je tiens à remercier ma famille et mes amis pour leur soutien et leur curiosité au cours de ces années. J'espère que vous reconnaîtrez un peu de mes explications dans l'introduction de ce manuscrit. 

Núria, merci pour ton soutien malgré les sacrifices que ce PhD a demandés. 
Gràcies, Andreu, pel teu interès, els teus consells i per la teva relectura afectuosa del meu primer paper.

Paul, la référence aux empreintes digitales dans l'introduction est dédiée à ton enthousiame au sujet de ma thèse. 
Boris, merci pour ton aide et discussions sur le Machine Learning.
Et merci à vous deux pour l'accueil répété sur vos canapés ou lits respectifs lors de mes passages !

Gauthier, merci de m'avoir sorti d'une belle impasse au sujet du steep XY model. Je n'aurais sans doute pas eu le temps d'écrire le Chapitre~\ref{ch:steepXY} dans les temps sans toi.

Enfin, Salambô, merci pour ton soutien, tes conseils et ta vision du monde académique.

	\setcounter{tocdepth}{2}	
	\dominitoc						
	\tableofcontents

	 
 	 \chapter*{Abstract} 
	 \addcontentsline{toc}{chapter}{Abstract}  
	 
\section*{Abstract}
\begin{center}
\textbf{Topological defects in non-equilibrium systems}
\end{center}

We study topological defects in two-dimensional non-equilibrium systems, focusing on active extensions of the XY model, including activity, mobility and non-reciprocity. In a noisy Kuramoto lattice with short-range coupling, intrinsic frequency heterogeneity destroys quasi-long-range order and fragments the system into finite domains. Defects unbind at all temperatures and exhibit superdiffusive random walks, advected by evolving domain boundaries. By contrast, when oscillators are allowed to move in space, the system undergoes a Berezinskii–Kosterlitz–Thouless transition and regains quasi-long-range order, revealing the fundamental role of motility in sustaining coherence. We also analyse a non-reciprocal $O(2)$ model with vision-cone couplings and derive a continuum theory that captures the same large-scale physics. Non-reciprocity selects defect shapes, enriches the annihilation process, and reshapes patterns through advection. 
Together, these results elucidate the fundamental role of activity and non-reciprocity in shaping topological defects and ordering in non-equilibrium systems.
\newline

Keywords: Topological defects, XY model, Steep XY model, Kuramoto model, Non-reciprocal interactions, Active matter, Phase transitions, Berezinskii-Kosterlitz-Thouless transition, Non-equilibrium statistical mechanics.

\newpage

\section*{Resum}
\begin{center}
    \textbf{Defectes topològics en sistemes fora de l'equilibri}
\end{center}
En aquesta tesi hem ampliat el model XY, un model paradigmàtic de transicions de fase i defectes topològics, a situacions fora de l’equilibri, rellevants en particular per als sistemes biològics. Des de fenòmens de sincronització en oscil·ladors acoblats fins a sistemes no reciprocals, hem explorat com el desordre intrínsec, la mobilitat i la ruptura de la simetria acció–reacció afecten les propietats a gran escala dels sistemes model mitjançant una modificació del comportament dels defectes topològics en aquests sistemes.
\newline

Primer hem proporcionat una revisió completa del model XY en equilibri, dels defectes topològics i del seu paper en la transició de Berezinskii–Kosterlitz–Thouless i en la dinàmica de coarsening. Hem introduït un mètode numèric eficient per simular sistemes en una xarxa triangular utilitzant la mateixa estructura de dades que per a xarxes quadrades. També hem detallat el mètode numèric per calcular la funció de correlació espacial mitjançant transformades de Fourier, molt més eficient que el càlcul en espai directe.
Ambdós enfocaments han estat especialment útils per portar el nostre codi a GPUs i accelerar les simulacions.
\newline

Tot seguit hem considerat el model XY “steep” introduït per Domany \textit{et al.} en 1984. Aquesta extensió en equilibri del model XY, on la pendent del potencial és controlada per un paràmetre $p$, és coneguda per exhibir una transició de fase d’ordre primer per valors prou grans de $p$. Hem argumentat que el valor crític estimat $p_c \approx 15$ està d’acord amb un conjunt convergent d’observacions i analogies amb altres models. Hem determinat numèricament la temperatura crítica $T_c(p)$.
Hem establert que en aquest model hi ha dos tipus de defectes topològics, és a dir, defectes XY dividits i parets de domini, i hem introduït un algorisme per discriminar-los. Hem mostrat que els defectes puntuals de càrrega unitària del model XY es divideixen en dos defectes de càrrega semientera separats per una línia amb tensió no nul·la i longitud $\ell \sim p$.
\newline

Després ens hem centrat en el model de Kuramoto, un model paradigmàtic de fenòmens de sincronització on cada espín té la seva pròpia freqüència intrínseca. El model de Kuramoto ha estat tradicionalment molt estudiat en la comunitat de sistemes dinàmics, que se centra especialment en el caos, la teoria de bifurcacions i l’emergència de la sincronia. Tanmateix, les seves connexions amb la física estadística i els defectes topològics han estat en gran part ignorades, tot i que el simple acoblament del model de Kuramoto amb un bany tèrmic reforça la seva rellevància per a la mecànica estadística.
Així, hem creat un pont entre aquestes dues comunitats estudiant el model de Kuramoto des de la perspectiva de la mecànica estadística, centrant-nos en la transició de fase, els defectes topològics i la dinàmica d’ordenació.
Hem mostrat que la presència de desordre congelat destrueix l’ordre quasi a llarg abast del model XY i genera defectes no confinats i superdifusius.
L’analogia entre les trajectòries dels nostres defectes i les caminades aleatòries autoevitants no sols dona l’exponent correcte de superdifusió, $3/2$, sinó que també proporciona un argument simple i genèric per explicar el comportament superdifusiu en altres sistemes.
També hem argumentat que les conclusions del nostre treball són susceptibles d’aplicar-se a una àmplia gamma de sistemes, i suggerim que el simple model de Kuramoto en xarxa podria constituir un model mínim per als defectes superdifusius en diversos sistemes amb desordre congelat.
\newline

A continuació hem ampliat el model de Kuramoto a oscil·ladors mòbils en $2d$ i hem demostrat que una mobilitat suficient restaura de manera espectacular l’ordenació de tipus XY i la llei estàndard d’aniquilació de defectes. Aquest recorregut fora de l’equilibri també aporta una nova llum sobre el model XY en equilibri, tot revelant la robustesa del seu ordre quasi a llarg abast i de la seva dinàmica de defectes davant pertorbacions substancials. El procediment de suavització introduït en aquest context és general i es pot aplicar per coarse-grain sistemes de partícules fora de xarxa.
\newline

Finalment, hem analitzat un model XY no reciprocal on les interaccions direccionals trenquen el principi d’acció–reacció. Hem introduït un nucli de ponderació continu i no negatiu per ajustar la no reciprocitat de les interaccions de manera suau. Això ens ha permès obtenir una derivació molt senzilla per arribar a una teoria de camp continu a partir del model XY microscòpic. Hem posat de manifest l’emergència d’un terme de rotació actiu que no té contraparts en equilibri, fet que aclareix la naturalesa activa del model XY no reciprocal i la seva connexió amb la celebrada teoria de Toner–Tu per als estols.
La nostra derivació i els nostres resultats han enriquit la literatura sobre sistemes no reciprocals d’una sola espècie, especialment pel que fa al debat actual sobre quin és un bon model continu per a aquests sistemes o sobre la naturalesa de l’ordre a llarg abast.
A més, el terme rotacional explica la rica fenomenologia del model XY no reciprocal: el curl actiu remodela els defectes i obre la porta a nous escenaris d’aniquilació de defectes impossibles en equilibri.
Aquest terme actiu també advecta patrons, generant fenòmens de propagació unidireccional o fins i tot impulsant canvis en la topologia del sistema. Creiem que aquests resultats, potencialment vàlids més enllà del cas específic del model $\mathcal{O}(2)$, estimularan noves investigacions sobre sistemes no reciprocals generals, especialment en el context dels defectes topològics i la propagació de patrons.
\newline

Paraules clau: Defectes topològics, model XY, model de Kuramoto, interaccions no-reciproques, matèria activa, transicions de fase, transició de Berezinskii-Kosterlitz-Thouless, mecànica estadística fora de l’equilibri.

	\definecolor{darkblue}{rgb}{0.12,0.47,0.87}

 	\titleformat{\chapter}[display] {\fontsize{35pt}{35pt}\bfseries\centering}{\textcolor{myblue}{ Chapter \thechapter: \\ }}{40pt}{\Huge\centering}  		
	\titlespacing*{\chapter}{40pt}{40pt}{40pt} 
	
	\chapter{Introduction}

	\renewcommand{\bibsection}{\section{References}}		

	\fancyhead[L]{Chapter \thechapter} 
	\label{ch:intro}
\graphicspath{{figures/Ch1introduction}}

\section{Crowds and physics}
Humans crowds, swarm of fireflies, and cell tissues are very different objects and evolve at very different scales. Yet, they share common features, the most important being that they are all composed of a large number of individual units that interact locally with each other. They all exhibit emergent, large-scale collective behaviours, such as flocking or clustering, that arise from these local interactions.
As such, they can be studied using similar theoretical frameworks, looking for universal behaviours, generic and robust principles and simple models to describe and predict their collective dynamics. This is one of the goals of physics: understand the fundamental laws governing the behaviour of a system, regardless of its specific nature.

Because of their size, typically of the order of $10^2 - 10^{23}$ (from animal groups to gas molecules), these groups are often too complex to be understood using the approach of classical mechanics, which focuses on the behaviour and the evolution of individual particles. Also, the fate of a single individual is often irrelevant to understand the collective dynamics of the group, or for practical applications such as crowd management or disease spreading control.
The natural framework to study such systems is thus statistical physics, which provides a framework for understanding the collective behaviour of large ensembles of interacting particles. 
As its name suggests, statistical physics describes the macroscopic properties of a system using average values. 

\subsection{Phase transitions } \ \\
A central concept in statistical physics is that of a phase transition, which are changes in the macroscopic properties of a system that occur when a control parameter (such as temperature, pressure, or density) is varied. Those changes can be abrupt and discontinuous, meaning that small variations in the control parameter can lead to drastic changes in the system's state or behaviour.

In the context of thermodynamics, statistical physics relates the microscopic properties of individual particles (such as their position and velocity) to the macroscopic properties of the system (such as temperature, pressure, and volume). It explains how water can exist in different phases (solid, liquid, gas) depending on temperature and pressure, and describes phase transitions between these states.
In the context of magnetism, it relates the microscopic properties of individual spins (such as their orientation) composing a material to its macroscopic properties (such as magnetisation and susceptibility).
As they will be useful in the following, we briefly review two emblematic model of ferromagnetic transition: the Ising model and the Potts model.

The Ising model \cite{newell1953_review_Ising_model}, introduced in 1925 as a lattice model of ferromagnetism, has become a cornerstone of statistical mechanics. Individual units, called spins, lie on the sites of a regular lattice and can take two possible values, $S_i = \pm 1$, representing the two possible orientations of a magnetic moment (up or down). Neighbouring spins interact and tend, by aligning with each other, to minimise energy
\begin{equation}
  E = -J \sum_{\langle i,j \rangle} S_i S_j \ ,
  \label{eq:Ising_energy}
\end{equation}
where $J$ is the coupling strength ($J>0$ promotes alignment), and the sum runs over all pairs of nearest-neighbour spins $\langle i,j \rangle$.
The system is usually coupled to a thermal bath at temperature $T$, which introduces thermal fluctuations that tend to randomise the spin orientations.

This model exhibits a phase transition between an ordered phase at low temperature, where spins align over long distances and the material becomes magnetised, and a disordered phase at high temperature, where spins are randomly oriented and the material loses its magnetism. 
This behaviour is by essence a many-body emergent phenomenon, as it arises from the collective interactions of a large number of spins, and cannot be understood by studying a single spin in isolation.
The transition occurs at a critical temperature that depends on the dimensionality of the space. For two-dimensional ($2d$) systems, $T_c = 2J/(k_B \, \log (1 + \sqrt{2}))$~\cite{onsager1944crystal}. 
This phase transition is characterised by a sudden yet continuous change in the magnetisation of the material, defined as the average spin per site:
\begin{equation}
  P = \frac{1}{N} \left| \sum_{i=1}^{N} S_i \right| \,,
\end{equation}
where $N$ is the total number of spins in the system.
The magnetisation quantifies the degree of collective alignment and is an order parameter of this model, meaning that the physics of the system can be described by this single quantity.
As temperature is decreased across $T_c$, $P$ goes from zero (in the disordered phase $T>T_c$) to a non-zero value (in the ordered phase $T<T_c$). 
The phase transition is associated with a spontaneous symmetry breaking: above $T_c$, the system is symmetric under spin inversion (flipping all spins), while below $T_c$, the system spontaneously chooses one of the two possible magnetised states (all spins up or all spins down), breaking this inherent symmetry of the model Eq.~(\ref{eq:Ising_energy}).
Last, precisely at the critical temperature, the magnetisation, seen as a function of temperature, is continuous but its first derivative is discontinuous: this corresponds to a second-order phase transition.

The Potts model \cite{wu1982potts}, introduced in 1952 as a generalisation of the Ising model, extends the concept of spins to $q$ possible states instead of two. Each spin can take one of $q$ discrete values, $S_i = 1, 2, \ldots, q$. The interaction energy between neighbouring spins is given by
\begin{equation}
  E = -J \sum_{\langle i,j \rangle} \delta(S_i, S_j) \,,
\end{equation}
where $\delta(S_i, S_j)$ is the Kronecker delta, which is equal to 1 if $S_i = S_j$ and 0 otherwise.
The Potts model also exhibits a phase transition between an ordered phase at low temperature, where spins tend to align in the same state, and a disordered phase at high temperature, where spins are randomly oriented among the $q$ possible states. In $2d$, this phase transition occurs at the critical temperature $T_c = 2J/(k_B \, \log(1 + \sqrt{q}))$, recovering the Ising model result for $q=2$.
The nature of the phase transition depends on the value of $q$ and the dimensionality of the system. In $2d$, for $q \leq 4$, the transition is second-order, while for $q > 4$, it becomes first-order, characterised by a discontinuous jump in the order parameter.


\subsection{Synchronisation} \ \\

Although it does not fundamentally differ from the ordering phenomenon described above, the term \textit{synchronisation} is traditionally used to describe dynamical or out-of-equilibrium systems (among which active or living units) that spontaneously coordinate their behaviour in time or space, such as power plants \cite{filatrella2008analysis, dorfler2012synchronization}, metronomes \cite{pantaleone2002synchronization}, neurons \cite{sompolinsky1990global, grannan1993stimulus, sompolinsky1991cooperative}, or fireflies \cite{tyrrell2006fireflies}. 
Whether the tendency to synchronise is conscious or not does not matter from a physics point of view.
Examples include the synchronous flashing of fireflies, the coordinated beating of heart cells, the crowds clapping in unison, and the collective movement of bird flocks. Understanding how synchronisation emerges from local interactions is a key question in statistical physics and has important implications for various fields, including biology and social sciences.
Synchrony can refer to different aspects of individual units' behaviour. It can refer to phase synchrony, where oscillators adjust their phases to oscillate in unison. It can refer to motion synchrony, where individuals align their direction of motion with those of their neighbours, as in flocking or schooling behaviour. Finally, it can also refer to synchronisation of any generic status; individuals can align their opinions or their health status in the case of epidemics. 
In the following, we briefly review the first two types of synchronisation, focusing on celebrated models relevant for this thesis, namely the Kuramoto model of phase synchronisation and the Vicsek model of collective motion.

\paragraph{Phase synchronisation and the Kuramoto model \\}
Back in the 17$^\text{th}$ century, Huygens realised that two pendulum clocks sitting on the same board start, after some transient time, to beat at the same frequency in phase or in anti-phase \cite{huygens1895letters}. Since then, the study of synchronisation in large populations of oscillators has remained a recurrent problem across different sciences, ranging from physics to biology, computational social sciences or engineering \cite{strogatz2012sync, Pikovsky2003}.
Much progress in the understanding of synchronisation has been achieved through the detailed analysis of simplified model systems. In this context, the Kuramoto model of phase coupled oscillators has (and still does) played a central role \cite{Kuramoto,AcebronBonilla}.  
The original version of the model \cite{Kuramoto} considers a population of $N$ oscillators, each characterised by a phase $\theta_i(t)$ and a quenched intrinsic frequency $\omega_i$, drawn from a given distribution at initialisation. The oscillators are coupled through a sinusoidal interaction term that tends to synchronise their phases.
Ref. \cite{Kuramoto} considers all-to-all interactions and noiseless (hence deterministic) dynamics, such that the evolution equation for the phase of each oscillator reads:
\begin{equation}
  \frac{d\theta_i}{dt} = \omega_i + \frac{K}{N} \sum_{j=1}^{N} \sin(\theta_j - \theta_i) \,,
  \label{eq:kuramoto_original}   
\end{equation}
where $K$ is the coupling strength, and $N$ the total number of oscillators in the system.

In this all-to-all version, if the intrinsic frequencies are drawn from a unimodal distribution, the system exhibits a phase transition, involving a collective rotation (frequency locking) and phase synchronization ~\cite{KuramotoOriginal,AcebronBonilla, Sakaguchi1987}.  
This formally infinite dimension system exhibits a {second} order phase transition from order to disorder as the ratio between self-spinning and coupling strengths increases~\cite{KuramotoOriginal}. 

However, to describe real systems one often needs to go beyond the original formulation of the Kuramoto model. For instance, to describe oscillator systems at the micro-scale, such as genetic oscillators \cite{Mondragon2011}, noise should be taken into account, as well as shorter-range interactions, leading to finite-dimensional noisy extensions of the Kuramoto model \cite{ArenasRev}. 
It is known that if the oscillators lie on a regular lattice and their interactions are restricted to the nearest (lattice) neighbours, the emergent large-scale properties of the system strongly depend on the system dimensionality $d$~\cite{AcebronBonilla}. If intrinsic frequencies are sampled from a unimodal distribution, for $d\geq5$, the system fully synchronizes. For $d=3,4$, the system is disordered in phase, but exhibits frequency locking, i.e. all oscillators rotate at the same effective frequency \cite{Sakaguchi1987, hong2005collective, hong2007entrainment}.
In $2d$, the short-range Kuramoto model can only exhibit short-range order (SRO). The situation in this case, which is the object of a chapter of this thesis, is quite more involved than the original Kuramoto model and the corresponding literature far more scarce \cite{hong2015finite, lee2010vortices,Rouzaire_Kuramoto_PRL,rouzaire_Frontiers}.

\paragraph{Collective motion, the Vicsek model and the Toner-Tu equations\\}
Collective motion is another form of synchronisation, where individual units align their direction of motion with those of their neighbours, leading to coherent movement of the group as a whole. This phenomenon is observed in various biological systems, such as bird flocks, fish schools, and insect swarms \cite{couzin2005effective,attanasi2014collective, cavagna2023characterization}. 
The Vicsek model \cite{vicsek1995novel} is a seminal model that captures the essential features of collective motion. In this model, published in 1995, self-propelled particles move at a constant speed and update their direction of motion $\theta$ based on the average direction of their neighbours, with some added noise. The equations of motion read: 
\begin{equation}
\begin{aligned}
  \dot{\mathbf{r}}_i &= v_0\, \mathbf{e}(\theta_i) \,,\\
  \dot{\theta}_i &= J \langle \theta(t) \rangle + \eta/2\,\nu_i(t) \,,
\end{aligned}
\label{eq:vicsek}
\end{equation}
where $\langle \theta(t) \rangle$ denotes the average direction of the
velocities of particles (including particle $i$) being
within a circle surrounding the given particle, and $\nu$ is a random number chosen with a uniform probability from
the interval $[-1, +1]$ that emulates thermal noise \cite{vicsek1995novel}. 

As the noise amplitude or particle density is varied, the Vicsek model exhibits a phase transition from a disordered state, where particles move randomly like in a gas, to an ordered state, where particles move coherently in the same direction \cite{vicsek1995novel, chate2008modeling}. This ordered state is often referred to as flocking \cite{ginelli2016VicsekRev}. The interpretation of flocking in terms of active oscillators has been pointed out in a number of recent works \cite{chepizhko2010relation,chepizhko2021revisiting, levis2019activity, okeeffe2017oscillators}. 
\newline

That same year, Toner and Tu developed a continuum theory of flocking \cite{toner1995long}. The so-called Toner-Tu equations describe how a collection of self-propelled particles, such as birds in a flock or cells in a tissue, can spontaneously form coherent motion. They generalize the Navier-Stokes equations of fluid dynamics to systems that constantly consume energy at the microscopic scale to self-propel. In this framework, two continuous fields are introduced: the local density \( \rho(\mathbf{r}, t) \) and the velocity field \( \mathbf{v}(\mathbf{r}, t) \). In the incompressible limit, when the density variations are negligible and the velocity field satisfies \( \nabla \cdot \mathbf{v} = 0 \), the dynamics of \( \mathbf{v} \) follows  \cite{toner1995long, toner2005hydrodynamics, toner2012reanalysis}.
\begin{equation}
\partial_t \mathbf{v} + \lambda_1 (\mathbf{v}\cdot\nabla)\mathbf{v} + \lambda_2 (\nabla\cdot \mathbf{v})\mathbf{v} + \lambda_3 \nabla |\mathbf{v}|^2 
= K\Delta\mathbf{v} +  \alpha \mathbf{v} - \beta |\mathbf{v}|^2 \mathbf{v} 
+ \boldsymbol{\xi}\ ,
\label{eq:tonertu}
\end{equation}
where $K$ is a diffusion coefficient, and e noise term \( \boldsymbol{\xi} \) accounts for random fluctuations. 
The parameters \( \alpha \) and \( \beta \) control the onset of collective order and stem from the Landau like potential 
\begin{equation}
  V(\mathbf{v}) = -\frac{\alpha}{2} |\mathbf{v}|^2 + \frac{\beta}{4} |\mathbf{v}|^4 \,.
\end{equation} 
When \( \alpha, \beta > 0 \), the system spontaneously develops a macroscopic velocity $\sqrt{\alpha / \beta}$.
Finally, the terms $\lambda_{i}$ on the left hand-side describe nonlinear advection and alignment. More precisely, the $\lambda_1$ term represents the self-advection of the velocity field, accouting for self-propulsion in the direction of the orientation. The $\lambda_2$ term is called the divergence term and couples the velocity to its own local divergence. The $\lambda_3$ term, more difficult to intuitively understand, is sometimes referred to as the self-achoring term \cite{de2024supramolecular} and tends to reorient the velocity field towards regions of higher/lower speed. For an introduction to the Toner-Tu equations, see \cite{toner2018walking}.
\newline 

A striking prediction of these two models \cite{vicsek1995novel, toner1995long} is that, unlike equilibrium systems, flocks can sustain long-range orientational order even in two dimensions.
As such, 1995 is sometimes considered to mark the emergence of \textit{active matter} as a new subdiscipline of statistical physics. The field of active matter studies systems composed of individual units, interacting with their surroundings, that consume energy to generate motion or mechanical stresses \cite{vrugt2025exactly, ramaswamy2017active}. These units, often called ``agents" or ``particles", can be biological entities such as cells, bacteria, humans or animals, or synthetic objects like self-propelled colloids or robots. 
Since then, the field of active matter has grown rapidly, with many experimental and theoretical studies exploring the properties of various active matter systems, from bacterial colonies \cite{copenhagen2021topological, copenhagen2016self} to mosquito flocks \cite{attanasi2014collective, cavagna2023characterization} to human crowds \cite{bain2019dynamic, gu2025emergence}.
\newline

\section{Topological defects}
\begin{figure}[h!]
  \centering
  \includegraphics[width = 0.8\linewidth]{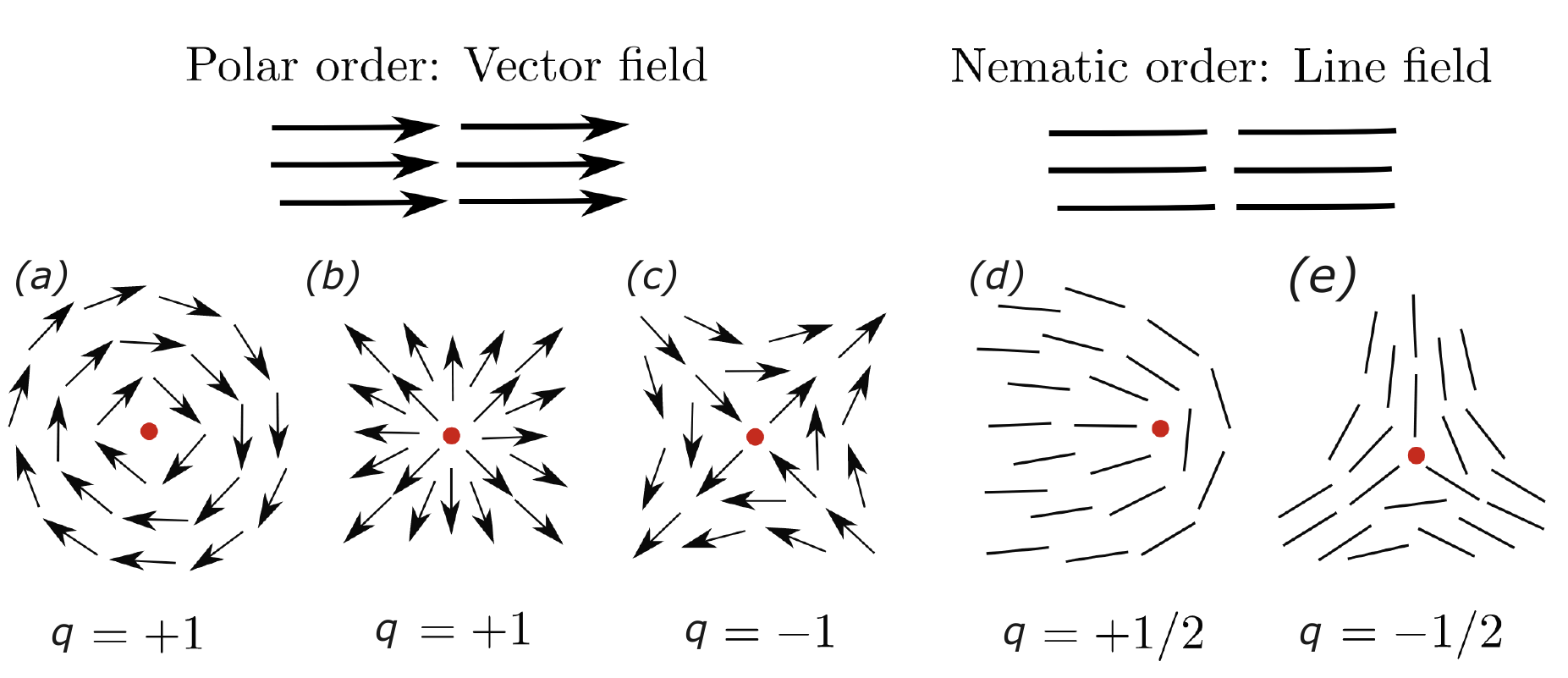}
  \caption{Sketches of different topological defects, of charge $q$, for polar order (a-c) and nematic (d,e) orientation fields. Image slightly adapted from \cite{shankar2022topological}.
  \textbf{(a)} Vortex, charge $q=+1$. The direction field (arrows) rotates by $+2\pi$ following a positive contour (in the counterclockwise direction) around the defect core, in red.
  \textbf{(b)} Source, charge $q=+1$. The direction field also picks a $+2\pi$ phase along a positive contour, but contrarily to the vortex configuration, the arrows point radially outwards from the core (no rotational, only divergence component). 
  \textbf{(c)} Charge $q=-1$ defect: the arrows rotates by $-2\pi$ following a positive contour around the defect core. 
  \textbf{(d)} Comet shaped $q=+1/2$ defect: the arrows rotates by $+\pi$ following a positive contour around the defect core. 
  \textbf{(e)} Triangle shaped $q=-1/2$ defect: the arrows rotates by $\pi$ following a positive contour around the defect core. 
  }
  \label{fig:sketch_defects}
\end{figure}

\begin{figure}[h!]
  \centering
  \includegraphics[width = 0.8\linewidth]{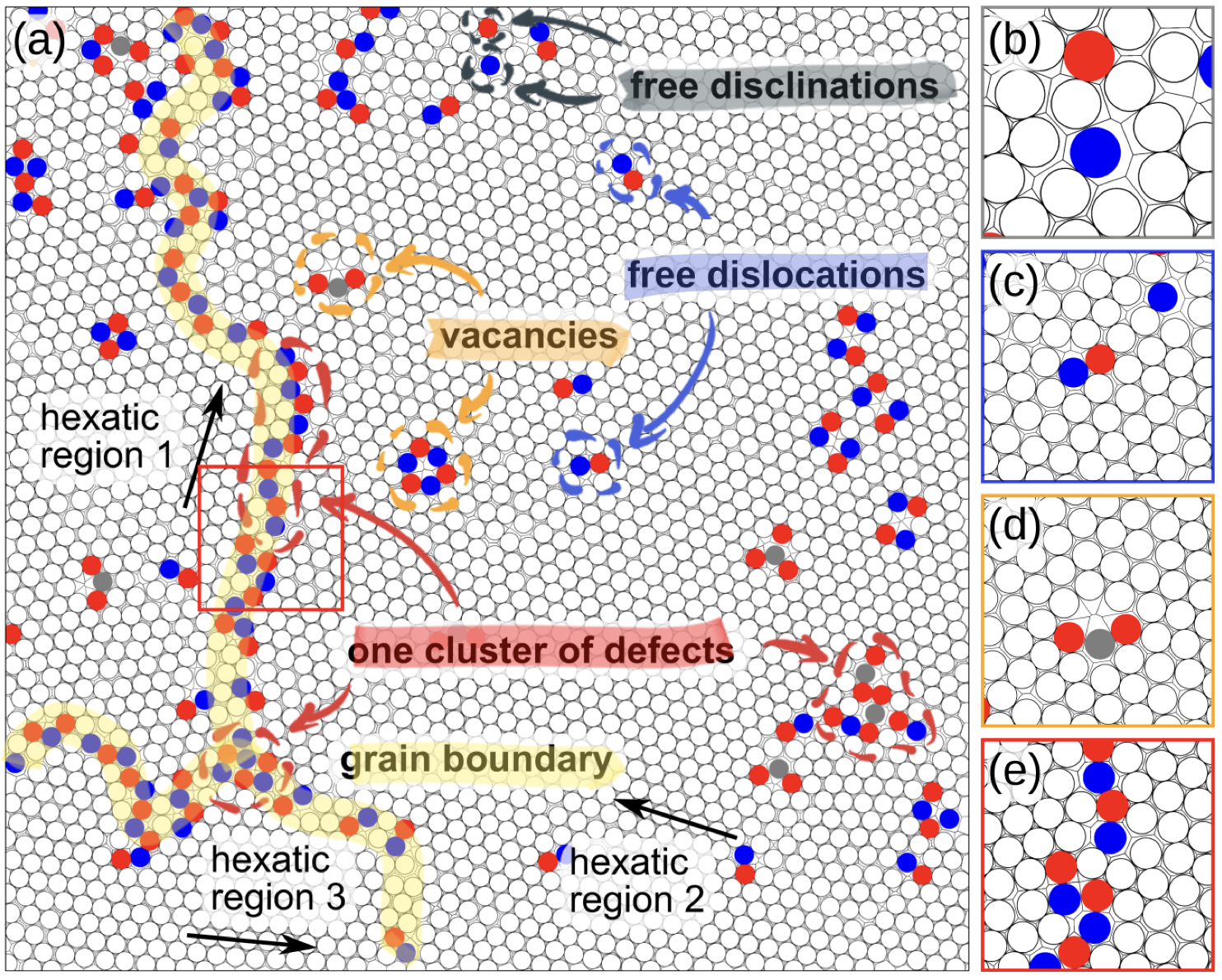}
  \caption{\textbf{(a)} Typical topological defects in a $2d$ crystal lattice of densely packed spherical particles. Image and caption from \cite{digregorio2022unified}. Particles in 5-fold cells are colored in red, 7-fold ones in blue, and all others in grey.
  Two disclinations are shown in \textbf{(b)}, a
  dislocation and a disclination in \textbf{(c)}, a vacancy in \textbf{(d)}, and a detailed view of two clusters of defects belonging to the same grain boundary (shown in yellow in (a) and delimiting regions with diﬀerent hexatic order) in \textbf{(e)}.
  }
  \label{fig:crystal_defects}
\end{figure}

In many interesting cases, syncronisation arises from local interactions promoting order/alignment. 
Yet, in every such systems, it is possible to observe local regions where the order is disrupted: this is what is called a \textit{topological defect}. 
More formally, defects are regions of space where the order parameter field is singular or undefined. We illustrate this concept with some sketches in Fig.~\ref{fig:sketch_defects} for $2d$ systems with polar or nematic order parameters and in Fig.~\ref{fig:crystal_defects} for $2d$ systems with positional order. 
Figure~\ref{fig:sketch_defects} is slightly adapted from Ref. \cite{shankar2022topological} (their Box 1).
Figure~\ref{fig:crystal_defects} is taken from Ref. \cite{digregorio2022unified} (their Fig. 2).

For instance, in a magnetic system where the order parameter is the local magnetisation vector, a defect can be a point or a line where the magnetisation vanishes or changes direction abruptly. A vortex in a fluid flow, where the velocity field circulates around a central point, creates a singularity at the center: the famous eye of the storm where the wind intensity vanishes, see Fig.~\ref{fig:sketch_defects}(a). Other examples, in crystalline lattices, are domain walls separating two regions with different directions (see yellow regions in Fig.~\ref{fig:crystal_defects}(a)), 
disclinations (see Fig.~\ref{fig:crystal_defects}(b)), or 
dislocations (see Fig.~\ref{fig:crystal_defects}(c)).

The term ``topological" refers to the fact that these defects are characterised by topological invariants, which are properties that remain unchanged under continuous deformations of the system. In layman words, one cannot remove a topological defect without cutting off or tearing some parts of the field configuration, or introducing other discontinuities.  
Defects are not physical objects per se, but rather a property of the field configuration. As such, they can be seen as collective structures of the system.

Defects arise in various physical systems; because they are topologically protected, they are robust to perturbations and can have significant effects on the system's properties, dynamics and function. Topology also implies conservation laws: defects can only be created or annihilated in pairs of opposite topological charge, and their total topological charge is conserved. In particular, on a closed surface, the total topological charge $Q$ is fixed by the topology and the boundary conditions of the surface: for a plane with periodic boundary conditions or a torus $Q=0$, for a sphere $Q=2$, etc.
\newline

\begin{figure}
  \centering
  \includegraphics[width = 0.75\linewidth]{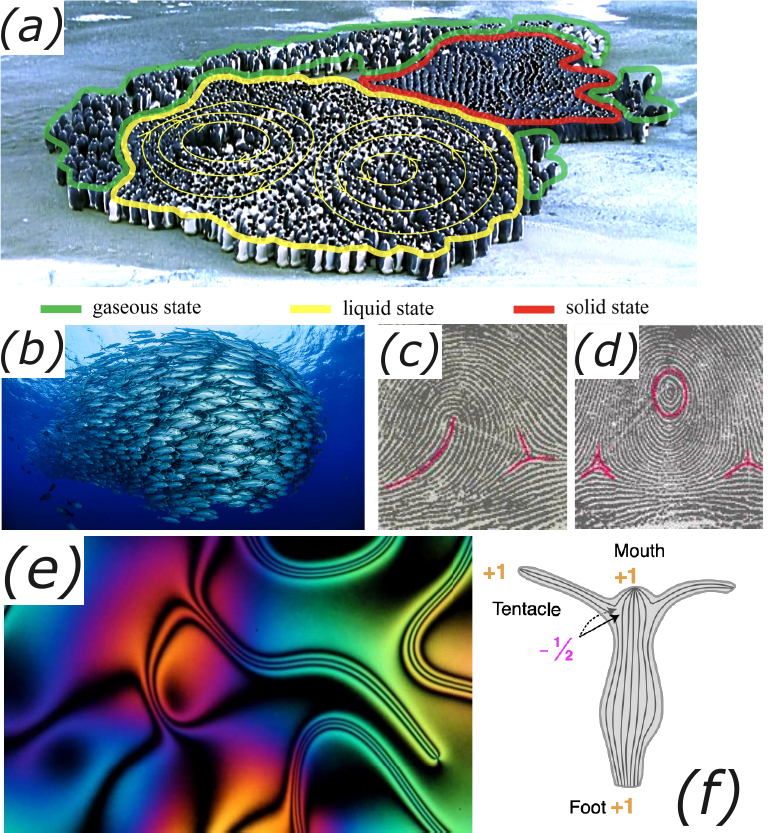}
  \caption{Examples of topological defects in various systems: 
  \textbf{(a)} Different states of matter in Emperor penguin populations (image from \cite{bratsun2025thermal}) 
  \textbf{(b)} A fish ball (image from \cite{fishball}) 
  \textbf{(c, d)} fingerprints, with topological defects highlighted in pink (image from \cite{Wikipedia_fingerprint})
  \textbf{(e)} nematic liquid crystal in an LCD screen (cover from \cite{chaikin1995principles})
  \textbf{(f)} Topological defects as organisation centers in \textit{Hydra} \cite{maroudas2021topological}. 
  }
  \label{fig:topdefs}
\end{figure}

To illustrate their ubiquity, we list various examples and illustrate some in Fig.~\ref{fig:topdefs}. 

Because they are visually more intuitive, let us start with defects in situations where synchronisation involves motion. Akin to the eye of the storm example mentioned above, it is known that Emperor penguins form vortices to protect the herd from cold and wind, see Fig.~\ref{fig:topdefs}(a) \cite{bratsun2025thermal}. The vortex centers are topological defects, where the direction of motion is undefined. This is reminiscent of the milling structures of certain malignant cancer cells to protect themselves from immunitary chemoattractant gradients \cite{copenhagen2018frustration}. Those also exist in higher dimensions, eg. in the collective organisation of a school of fish, called a fishball, creating a $3d$ vortex-like structure, see Fig.~\ref{fig:topdefs}(b).

Syncronisation can also take place in the phase of oscillators. 
The population of cilia on the surface of certain microorganisms is a classic example \cite{machemer1972ciliary, brumley2012hydrodynamic}. Those fixed cilia beat periodically to propel the organism in its environment, and they tend to synchronise their beating phases with their neighbours, leading to metachronal waves (think of Mexican waves in a stadium) that propulse the organism. The same mechanism is found in worms and centipedes. In this context, topological defects can be responsible for the abrupt loss of metasynchrony \cite{solovev2022synchronization2} when temperature is increased, hindering propulsion. 
The phase can also refer to the internal state of oscillators, such as the cell volume cycle phase in cardiac tissue. In this case, local synchronisation implies large-scale contractions of the heart \cite{karma1993spiral, christoph2018electromechanical}. Any perturbation of this (meta)synchrony impacts the heart function \cite{molavi2022spatiotemporal, rappel2022physics}, and can be due to topological defects in the phase field \cite{banerjee2025contraction, zhang2023pulsating, pineros2025biased}.  
\newline

In dense environments,
steric interactions (excluded volume effects) between units can lead to positional synchronisation, they arrange themselves in regular patterns, such as crystals or liquid crystals \cite{deGennesProst1993_LC}. 
In solids, topological defects can arise as dislocations or disclinations in the crystal lattice, disrupting the positional order \cite{anderson2017theory, digregorio2022unified}.
Those defects can impact the mechanical properties of the material, such as its strength and resistance to constraints, as structural defects tend to concentrate stresses and thus nucleate mechanical rupture \cite{hild1992statistical}.
After decades of research \cite{strandburg1988two, ryzhov2017berezinskii}, the melting process of two-dimensional solids is now understood as two successive phase transitions involving topological defects of different nature \cite{bernard2009event, kapfer2015two, thorneywork2017two, MarchettiRev, PRLino, Kosterlitz1973, Kosterlitz1974, berezinskii1971destruction}. Biological tissues can also exhibit positional order. In epithelial tissues like skin, topological defects can give rise to the well known patterns of fingerprints. Their existence, nature and location are one of the reasons why our fingerprints are unique. We highlight the structure imposed by defects with pink lines in Fig.~\ref{fig:topdefs}(c,d). Three kinds of defects are observed in fingerprints: vortices (charge $+1$, see Fig.~\ref{fig:sketch_defects}(a)), comets (charge $+1/2$, see Fig.~\ref{fig:sketch_defects}(d)) and triangle-shaped $-1/2$ defects, see Fig.~\ref{fig:sketch_defects}(e). 
Because the total charge has to be zero on a flat surface with periodic boundary conditions, a comet defect must be compensated by a triangle defect, see Fig.~\ref{fig:topdefs}(c), and a vortex by two triangle defects, usually located on both sides of the vortex, see Fig.~\ref{fig:topdefs}(d). 
Topological defects are thus essential to understand the dynamics and properties of solids and biological tissues.

While the solid phase has both translational and orientational order, and the isotropic liquid phase has neither, \textit{liquid crystal} is the broad class of soft materials that lie in between those two extremes~\cite{stephen1974physics, chaikin1995principles}. They can exhibit several phases (nematic, smectic, and cholesteric, among others), each defined by the symmetries they break. 
A \textit{nematic} is the simplest liquid-crystal phase: usually due to the units' elongated shape, molecules have orientational order (they tend to align along a common direction) but no translational order (their centers of mass are randomly distributed, as in a fluid).
Smectic phases have both orientational order and partial translational order. Cholesteric phases is a subcategory of nematics that break the parity symmetry, with a chirality given by a helical twist in orientation.
In these systems, topological defects arise as singularities in the orientation field. Due to the nematic symmetry of the molecules (head-tail symmetry), nematic defects can have half-integer charges, such as $+1/2$ and $-1/2$ defects, see Fig.~\ref{fig:sketch_defects}(d,e). In passive (equilibrium) nematics, the defects created during the quench from the isotropic to the nematic phase tend to annihilate in pairs of opposite charge over time: the system orders, the correlation length grows, and defects disappear leading to a defect-free nematic state at long times. This process is called phase ordering dynamics or coarsening dynamics. 
The phase transition from nematic to isotropic in two-dimensional passive liquid crystals is driven by the unbinding of pairs of $\pm 1/2$ defects,
following the famous Kosterlitz-Thouless (KT) mechanism \cite{Kosterlitz1973, Kosterlitz1974, berezinskii1971destruction}. 

The order parameter that describes the transition 
from fluid to superfluid helium in two-dimensions is a complex scalar field. Like the order parameter in nematics systems (the magnetisation modulo $\pi$), it is invariant under global rotations or addition of a global phase: we say it belongs to the $\mathcal{O}(2)$ symmetry group. All the systems described so far with an $\mathcal{O}(2)$ order parameter share the same universality class, meaning that the phase transition is ruled by the same KT mechanism. Finally, the two-dimensional XY model of magnetism, where the order parameter is also an $\mathcal{O}(2)$ planar vector, follows the same scenario; see Chapter \ref{ch:XY} for a complete review.

\textit{Active nematics} are an out-of-equilibrium version of nematic liquid crystals, where the constituent particles generate mechanical stresses \cite{doostmohammadi2018active}. In two-dimensions, this activity implies a bend instability \cite{thampi2014instabilities, thampi2016active}, which in turn spontaneously generates pairs of $\pm 1/2$ defects continuously, leading to a steady-state dynamics called active turbulence \cite{thampi2016active, alert2022active, alert2020universal}. In active nematics, the $+1/2$ defects are motile, propelling themselves along their head direction thanks to inner hydrodynamic flows, while $-1/2$ defects remain non-motile because of their 3-fold symmetric shape \cite{Giomi2013, giomi2014defect, giomi2015geometry, shankar2022topological}. 
In three dimensions the scenario is much more complex, with line defects called disclination lines, which can form loops and exhibit complex dynamics \cite{digregorio2024coexistence}.
\newline

Topological defects have important implications and applications in many fields ranging from material science to biology. 

In liquid crystal displays (LCD) screens, topological defects cause undesired visual artefacts \cite{chaikin1995principles}, see the texture in Fig.~\ref{fig:topdefs}(e). In that context, one seeks for conditions that prevent the spontaneous formation of defects. In some other situations, it has been shown that defects in active liquid crystals, considered as quasi-objects, exhibit a collective motion \cite{shankar2019hydrodynamics} that can be exploited to perform logic operations \cite{zhang2022logic}.  
Defects could also play a role in transport phenomena. Indeed, the depression at the center of a vortex-like defect in polar matter \cite{chardac2021topology} could be used as a trap, which can then be displaced and controlled by the motion of the defect. 

In biological systems, topological defects are usually detrimental to a correct functioning of the organism, as they disrupt synchrony by definition. However, the conditions imposed by topology can be exploited by living beings to their advantage. As stated earlier, on a closed surface that has the topology of a sphere, the total topological charge is equal to 2. 
The \textit{Hydra}, a unicellular organism, exploits this property in its regeneration process. If an arm were chopped off, a new one would grow
at the correct location to preserve the overall $+2$ defect charge of the organism, see Fig.~\ref{fig:topdefs}(f) \cite{maroudas2021topological}.
\newline

\section{Outline of the thesis}

We have seen that topological defects are ubiquitous in nature and arise in various physical systems, from fluids to solids to biological tissues. It is therefore crucial to understand their properties, their dynamics and the impact they have on the dynamics of the whole system. 
\newline 

\fbox{
  \centering
  \begin{minipage}{35em}
    \ \\ 
    In this thesis, we investigate the fundamental mechanisms governing topological defects in various systems, mostly out-of-equilibrium. In particular, we describe how the different ways we shall drive the model systems away from equilibrium impact the defects' creation, annihilation, interaction and dynamics. By understanding these mechanisms, we also shed light on the properties of the systems themselves. 
    \vspace{0.05cm}
  \end{minipage}
}
\vspace{1cm}
\newline 
This thesis is structured as follows. \\

We finish this introductory chapter with a discussion on the choice of the models used in this thesis, and with a presentation of notations and conventions used throughout this manuscript.
\newline

In Chapter \ref{ch:XY}, we introduce the XY model, a paradigmatic model in statistical physics, and we review its equilibrium properties, in particular the Berezinskii-Kosterlitz-Thouless phase transition driven by topological defects. We also study the out-of-equilibrium coarsening dynamics of the XY model following a quench from infinite temperature to below the critical temperature.
\newline

In Chapter \ref{ch:methods}, we present the methods used recurrently throughout this thesis, detail the numerical implementation of the model, and define the observables used to characterise and quantify collective behaviour.
\newline

In Chapter \ref{ch:steepXY}, we introduce the steep XY model, an equilibrium extension of the XY model where the interaction potential between spins can be made arbitrarily steep while keeping the continuous and polar properties of the interactions. We show how the steepness of the potential affects the topological defects, splitting them in half-integer charges that we characterize. 
\newline

In Chapter \ref{ch:kuramoto}, we turn to out-of-equilibrium systems and study the Kuramoto model of coupled oscillators with distributed intrinsic frequencies. We show that the presence of intrinsic frequencies destroys the quasi-long-range order of the XY model, and we characterise the topological defects dynamics. We show that the defects, far from being bounded as in the XY model, are genuinely free and superdiffuse. We rationalise this behaviour showing that defects share common features with self-avoiding random walks.  
\newline

In Chapter \ref{ch:kuramoto_mobile}, we extend the Kuramoto model to mobile oscillators, which move and interact with their nearest neighbours. We show that a sufficiently high mobility can restore the ordering dynamics and the topological defects annihilation law of the XY model.
\newline

In Chapter \ref{ch:nrxy_defects}, we introduce a non-reciprocal version of the XY model, where each spin interacts more with neighbours it looks at, breaking the action–reaction principle. We analyse how this non-reciprocity modifies defect dynamics, showing that defect shape and charge both control annihilation and coarsening. Finally, we show that the non-equilibrium term capturing the vision cone at the field theory level advects and reshapes patterns, alters the stability of winding states, and can even change the system’s topology.
\newline

This thesis is largely based on our 5 articles \cite{Rouzaire_Kuramoto_PRL,rouzaire_Frontiers,rouzaire2025activity,rouzaire2025nonreciprocal,rouzaire2025emergentdynamicsO2},
with some additional unpublished results, in particular in Chapter \ref{ch:steepXY}.

\section{Some comments on the choice of models}
We have seen that statistical physics studies simplified model systems to capture the essential features of more complex ones.
However, a same physical system can be modelled in different ways, depending on what one knows or looks for. 
In this section, to clarify the choices made in the rest of this work, we thus review different families of models, along with their specifities. 

\subsection{Continuum models, agent-based models and lattice models}
One generally distinguishes between \textit{continuum models, agent-based models}, and \textit{lattice models}.
Each of these approaches has its own advantages and disadvantages. 

Continuum models describe the physics at large scales, discarding microscopic details deemed irrelevant, but they require prior knowledge of the differential equations governing the studied system at the macroscopic scale. This is the case, for example, in fluid dynamics with the Navier–Stokes equations or in electromagnetism with Maxwell’s equations.

Agent-based models are often used to simulate systems whose particles exhibit complex behaviours and/or interactions. This is the case, for example, in molecular dynamics simulations, which model the interactions between the molecules of a system using the laws of classical mechanics. They are particularly well-suited for studying systems where microscopic details are essential, or systems whose large-scale laws are not (yet) known. 

Lattice models are, in the most common sense, the limiting case of agent-based models where the agents are restricted to live on a discrete lattice. These models are generally simplified to retain only the essential ingredients needed to highlight the collective phenomena of interest. They are simpler to program, faster to simulate than continuous space agent-based models and often better suited for analytical treatments.
\newline 

In this thesis, we will mostly work with lattice models. We will also use agent-based models in Chapter \ref{ch:kuramoto_mobile} to simulate systems where particle mobility in a continuous space is crucial. In Chapter \ref{ch:nrxy_defects}, building on the insights gained from lattice models, we will also construct continuum models to draw more general conclusions on the observed phenomena.

\subsection{Nature of the spins}
Spins can be discrete or continuous.
In the discrete case, spins can take a finite number of values, for example $+1$ or $-1$ in the Ising model \cite{newell1953_review_Ising_model}, or $1,2,3,\dots,q$ in the Potts model \cite{wu1982potts}. This can describe, for example, the two possible orientations of a magnetic moment in a ferromagnetic material (up or down), the $q$ possible states of an atom in a crystal lattice, the vote of an individual in a social network (for or against a proposal, or one of $q$ possible candidates) or the infection status (0=healthy, 1=infected).
In the continuous case, spins can take infinitely many values on continuous scale. This can describe for example the direction of motion of a self-propelled particle like in the Vicsek or Toner-Tu equations, see Eqs.~(\ref{eq:vicsek}) and (\ref{eq:tonertu}), or the opinion of an individual on a continuous scale (for example, from 0 to 1).
\newline

In this thesis, we will exclusively focus on continuous spin models, in particular the XY model and its variants, introduced in Chapter \ref{ch:XY}. 

\subsection{Nature of the interactions}
Spin interactions can take different forms: short-range, long-range, topological, or all-to-all.
In short-range models, each spin only interacts with its nearest neighbours, located at less than a given threshold distance. The exact value of this threshold is generally unimportant, as long as it is small enough for the interactions to be considered local relative to the system size. A simple rescaling argument allows this distance to be set to unity. Similarly, couplings that decay rapidly with distance (for example, exponentially) are considered short-range interactions.

In long-range models, each spin interacts with all the other spins of the system, with a force that typically decreases with the distance $r$. Classic examples are gravitational or Coulomb interactions: in $3d$ the force decreases as $1/r^2$, in $2d$, the force decreases as $1/r$.

The all-to-all model is a particular case of long-range interaction, where each spin interacts with all others with the same strength, regardless of distance.

Finally, topological interactions are those where each spin interacts with a fixed number of neighbours, independent of their distance. This is the case, for example, for interactions between cells in a biological tissue, where each cell interacts with its immediate neighbours but not with more distant ones. It is also the case for social interactions, often described by graphs, where each individual interacts with those in their immediate social environment, regardless of physical distance. Note that on a regular lattice, topological interactions with the $z$ nearest neighbours are equivalent to short-range interactions with a threshold distance equal to the lattice spacing.
\newline

In this thesis, we will exclusively focus on short-range interactions (with the $z$ nearest neighbours in lattice models and with particles located within a fixed threshold distance $R_0$ in the agent-based model of Chapter \ref{ch:kuramoto_mobile}).

\subsection{Symmetry of the interactions}
Spin interactions can also be classified according to their symmetry, for instance polar or nematic.
In polar interactions, spins have a defined head/tail axis, and two interacting spins tend to align their heads. This is the case, for example, for magnetic interactions between atoms in a ferromagnetic material, where spins tend to align in the same direction to minimize the energy of the system.
In nematic interactions, by contrast, spins have no defined head/tail axis, and two interacting spins tend to align without distinction between head and tail. This is the case, for example, for steric interactions between elongated bacteria, which tend to align parallel to each other without a preferential orientation.

In this thesis, we will exclusively work with polar interactions, and the global order of the system, when applicable, will therefore also be polar.
\newline

\subsection{Dimensionality of the system}
In physics, the spatial dimension of the system ($1d$, $2d$, $3d$) is a crucial parameter that greatly influences the system’s properties. This is for instance the case for the syncronisation scenario in the (short-range) Kuramoto model, as discussed earlier.

The Mermin–Wagner theorem \cite{mermin1979topological} (1966) states that $\mathcal{O(2)}$ symmetries cannot be spontaneously broken at finite temperature in systems with sufficiently short-range interactions in dimensions $d=1,2$. The intuition behind this theorem is that in these systems, long-range fluctuations can be created with no or little energy cost. Even if the energy cost is finite (provided that it does not scale with the system size), since they increase the entropy, they are favored and long-range order cannot be achieved. We will give more details on this theorem in Chapter \ref{ch:XY}, taking the XY model as an example. This theorem makes the study of $1d$ and $2d$ systems particularly interesting. 
The mechanisms allowing equilibrium systems to exhibit rich collective phenomena despite the absence of long-range order are indeed fascinating.
Out-of-equilibrium systems such as cell tissues, bacterial colonies, or fish in very shallow waters also evolve in two spatial dimensions and display a rich variety of collective phenomena. In particular, we will see that topological defects, which are at the core of our study, are of particular relevance in $2d$. 
\newline

In this thesis, we will focus exclusively on two-dimensional ($2d$) models.

\subsection{Lattice type}
The choice of the regular lattice on which the dynamics unfold (square, triangular, hexagonal, etc.) can affect the results quantitatively, but rarely qualitatively. In a regular lattice, each site has the same number of neighbours, called the coordination number, denoted $z$. For example, in a $2d$ hexagonal lattice $z=3$, in a $2d$ square lattice $z=4$, and in a $2d$ triangular lattice $z=6$. For higher coordination numbers, the discrete operators (gradient, Laplacian, etc) are smoother and the anisotropic ``errors" start at higher order: anisotropic errors start are reduced and the physics is expected to be closer to the continuum limit. 
Let us take the example of the continuous Laplacian operator. Expressed in Fourier space, it reads $\tilde{\nabla}^2 \propto k^2 + \alpha k^4 $, where $k = |\mathbf{k}|$ is the wave number and $\textbf{k} = (k_x, k_y) $ the wave vector. The $k^2$ term is isotropic, but the $k^4$ term can be anisotropic, depending on the lattice type.

On a square lattice with lattice spacing $a$, the discrete Laplacian reads 
\begin{equation}
  \tilde{\nabla}^2_{squ} \propto a^2k^2 - a^4(k_x^4 + k_y^4)/12 + \mathcal{O}(k^6) \,,
\end{equation}
which is anisotropic at order $k^4$. On a triangular lattice, it reads 

\begin{equation}
  \tilde{\nabla}^2_{tri} \propto a^2k^2 - 3a^4k^4/32 + \mathcal{O}(k^6) \,,
\end{equation}
which is isotropic at order $k^4$ and anisotropic only at order $k^6$.
\newline

Boundary conditions are also an important parameter. In this thesis, we will mainly use periodic boundary conditions (PBC) to minimize boundary effects and simulate an infinite system, giving it the topology of a torus (but not its geometry/curvature). Other types of boundary conditions exist, such as free boundary conditions (FBC), where spins at the edges have no neighbours outside the lattice, or fixed boundary conditions, where spins at the edges are set to a given value and are not updated.  Fixed and free boundary conditions are more realistic in some situations, for example when studying a finite system or a confined systems with solid walls. However, in addition to introducing possibly long-ranged edge effects, they have a strong impact on topological defects, as they modify the total topological charge of the system (fixed boundary conditions), or allow defects to enter/escape through the edges, thus breaking defect conservation (free boundary conditions).
\newline

In this thesis, since we want to focus on bulk properties and enforce topological charge conservation, we will use PBC unless otherwise stated. 
\vspace{1cm}

In summary, in this thesis we will focus on $2d$ lattice and agent-based models with continuous spins, short-range polar interactions, and periodic boundary conditions. We will see that even with these simple ingredients, a rich variety of collective phenomena can be observed, in particular regarding the behaviour of topological defects in out-of-equilibrium systems.

\section{Notations and abbreviations} \label{ch:notations}

\begin{table}[h]
  \centering
  \begin{tabular}{ |c | c |}
  \hline
    \textbf{Abbreviation} & \textbf{Meaning} \\  \hline
    PBC & Periodic boundary conditions \\ 
    KT & Kosterlitz-Thouless \\ 
    LRO & Long-range order \\ 
    QLRO & Quasi-long-range order \\ 
    SRO & Short-range order \\ 
    NR & Non-reciprocal \\ 
    NRXY & Non-reciprocal XY \\ 
    MSD & Mean square displacement \\ 
    $1d$ & One-dimensional \\ 
    $2d$ & Two-dimensional \\ 
    PDF & Probability density function \\ 
    SAW & Self-avoiding (random) walk \\ 
    TPP & Topologically protected patterns \\ 
    \hline
  \end{tabular}
  \caption{Abbreviations}
  \label{tab:abbreviations}
\end{table}

Note 1: we will use interchangeably the terms agents, particles, rotors, oscillators, units and spins to refer to the same objects.
In the same fashion, ``defects" and ``topological defects" are synonyms. \\
Bold quantities (e.g. $\mathbf{S}$) denote vectorial quantities, while non-bold quantities (e.g. $S$) denote scalar quantities.

\begin{table}[h]
  \centering
  \begin{tabular}{ |c | c |}
  \hline
    \textbf{Quantity} & \textbf{Symbol} \\  \hline
    Linear size of the system & $L$ \\ 
    Number of spins & $N=L^2$\\ 
    Coordination number (number of nearest neighbours) & $z = 4 \text{ or } 6$  \\ 
    Spin orientation (or phase) & $\theta$ \\ 
    Spin angular velocity & $\dot \theta$ \\ 
    Spin intrinisic frequency & $\omega$ \\ 
    Damping coefficient & $\gamma$ \\ 
    Coupling constant (to nearest neighbour) & $J=1$  \\ 
    Boltzmann constant & $k_B = 1$   \\ 
    Temperature & $T$  \\ 
    Kosterlitz-Thouless critical temperature & $T_\text{KT} \approx 0.89J\, (\square)\text{ or } 1.4J \, (\bigtriangleup)$  \\ 
    Angular distance (in Ch.\ref{ch:nrxy_defects}) & $\varphi_{ij} = \theta_i - u_{j}$  \\ 
    Kernel function & $g(\varphi)$  \\ 
    Sharp vision cone aperture & $\Theta$ \\ 
    Spatial correlation function & $C(r)$ \\ 
    Spatial correlation length & $\xi$  \\ 
    Number of defects & $n$  \\ 
    Polar order parameter & $P=|\frac{1}{N}\,\sum\limits_k \exp(i\theta_k)|$  \\ 
    Nematic order parameter & $Q=|\frac{1}{N}\,\sum\limits_k \exp(2i\theta_k)|$  \\ 
    Topological charge of a defect & $q$  \\ 
    Shape of a defect & $\mu \ (\text{mod} \ 2\pi)$  \\
    Energy & $E$  \\
    Free energy & $\mathcal{F}$  \\
    Hamiltonian & $\mathcal{H}$  \\
    Gaussian white noise or $C(r)\sim r^{-\eta}$ decay & $\eta$  \\
    Time & $t$  \\
    Time step & $dt$  \\
    Defect separation & $R$  \\
    Initial defect separation & $R_0$  \\
    Orientation field & $\mathbf{S}$ \\
    Unit length orientation field & $\hat{\mathbf{S}}$ \\
    Defect diffusion coefficient & $\mathcal{D}$ \\
    Particle velocity & $v_0$ \\

    \hline
  \end{tabular}
  \caption{Notations}
  \label{tab:notations}
\end{table}

Note 2: depending on the chapter, $\sigma$ will change meaning. 
In chapters \ref{ch:kuramoto} and \ref{ch:kuramoto_mobile}, it is the standard deviation of the intrinisic frequencies of the Kuramoto oscillators. In chapter \ref{ch:nrxy_defects}, it is the non-reciprocal parameter of the smooth vision cones. Historically, we have chosen the same symbol even if they are different quantities because they both represent the departure from the equilibrium XY case ($\sigma=0$) through distributions whose variance is related to $\sigma$.

	

	\chapter{The equilibrium XY model}
	\label{ch:XY}
	\graphicspath{{XY/figures/}}
\graphicspath{{figures/Ch2XY/}}


The XY model, also called the classical XY model or the planar $O(2)$ model, is a cornerstone of statistical physics as it is a paradigmatic model for the study of phase transitions and critical phenomena. It will be the reference equilibrium model of this thesis, from which we will depart to study specific out-of-equilibrium dynamics in the following chapters. 
It can be thought as a two-dimensional generalisation of the Ising model, where spins lie on a regular lattice and can point in any direction in the plane instead of being restricted to two states (up or down).
The only degree of freedom of each spin is its orientation $\theta_i$ (with respect to the some arbitrary axis), such that the spin vector reads 
\begin{equation}
  \hat{\mathbf{S}}_i = 
  \begin{pmatrix}
    \cos\theta_i \\
    \sin\theta_i
  \end{pmatrix} 
  \label{eq:spin_S}
\end{equation}

Spins interact with their $z$ nearest neighbours, and the Hamiltonian of the system is given by:
\begin{equation}
  \mathcal{H} = -J \sum_{\langle i,j \rangle} \hat{\mathbf{S}}_i \cdot \hat{\mathbf{S}}_j = -J \sum_{\langle i,j \rangle} \cos(\theta_i - \theta_j)
  \label{eq:XY_Hamiltonian}
\end{equation}
where $J$ is the coupling constant, and $\langle i,j \rangle$ denotes the sum over all nearest neighbours pairs. 
The energy of the system is minimized when all spins are aligned: in the ground state $\mathcal{H} = -JzN/2$, 
where $N$ is the number of spins in the system. Each spin in counted twice in each (reciprocal) link, hence the factor $1/2$ so that the sum runs over unique pairs of spins.



\section{Kosterlitz-Thouless (KT) phase transition}
The XY model features a competition between alignment and thermal noise. 
The alignment term stems from the minimisation of $\mathcal{H}$ and promotes order. The cosine non-linearity takes care of the periodicity of the angular variable $\theta$.
The system is coupled to a thermal bath of temperature $T$, which trivially promotes disorder.
We thus intuitively expect a phase transition between an ordered phase (where the spins are aligned) 
and a disordered phase (where the spins are randomly oriented) as the ratio $J/T$ increases.

In other celebrated models, such as the Ising model \cite{newell1953_review_Ising_model}, phase transitions are materialised as a change in the order parameter.  
In spin models, the scalar order parameter is the magnetisation $P$, defined in the case of XY model as the average of the spins orientations: 
\begin{equation}
  P = \frac{1}{N} \left|\,\sum\limits_{k=1}^N \exp(i\theta_k)\,\right| 
  \label{eq:XY_order_parameter}
\end{equation}
where $N$ is the number of spins in the system and $i$ is the imaginary unit.
Note that one could have used the alternative definition
\begin{equation}
  P =  \frac{1}{L} \sqrt{ (\sum\limits_{k=1}^N \cos \theta_k)^2 + (\sum\limits_{k=1}^N \sin \theta_k)^2}
  \label{eq:XY_order_parameter_alternative}
\end{equation}
and obtain the same qualitative results. 

In statistical mechanics, phase transition are traditionally defined as a change in the order parameter. 
If a system undergoes a first order phase transition, the order parameter displays a sudden jump.
For second order phase transition, only the first derivative of the order parameter with respect to the changing parameter is discontinuous. 
Following, in an infinite order transition, all the derivatives of the order parameter are continuous. 
\newline

However, in the XY model, the order parameter $P$ goes to zero in the thermodynamic limit: $P\sim1/L$ as $L\to\infty$. 
This is due to the so-called \textit{spinwave excitations}. 
Spinwaves are low-energy collective excitations around an ordered state, made possible by the continuous nature of the XY spins. 
In contrast with the discrete spins in the Ising \cite{newell1953_review_Ising_model}, Potts \cite{wu1982potts} or clock \cite{tobochnik1982properties, challa1986critical} models,
where the energy cost of having two neighbours in different states is finite, two XY spins can be arbitrarily close to each other and the associated energy cost arbitrarily small.  
Let us denote $\varepsilon$ the angular difference between two successive spins. Along each of the dimensions of the square lattice, 
the periodic boundary conditions impose $\varepsilon = 2\pi/L$, such that the energy cost of one bond is, in the $L\to\infty$ limit,
\begin{equation}
  -J\cos \varepsilon \approx -J (1-\varepsilon^2/2)
\end{equation}
Following, the total energy (summed over the $L^2$ bonds of the lattice) reads
\begin{equation}
\begin{aligned}
  E &= -JL^2 (1-\varepsilon^2/2)  \\ 
    &= E_0 + JL^2\varepsilon^2/2 \\
    &= E_0 + JL^2(2\pi/L)^2/2 \\
    &= E_0 + J 2\pi^2
\end{aligned}
\end{equation} 
where $E_0 = -JL^2$ is the energy of the ground state, where all spins are aligned.
The excess energy $2J\pi^2$ associated with the spinwave excitation is finite; 
for any finite temperature, the system will eventually excite this mode, which, 
by essence, implies that $P=0$ because the system contains spins of all the orientations. 

Therefore, if $P=0$ at all temperatures in the thermodynamic limit, regardless of the phase of the system, in what terms can one define a phase transition in the XY model ?
\newline 

The answer is that the XY model does undergo a phase transition, but not in the usual symmetry breaking sense.
Instead, the XY model undergoes a topological phase transition, known as the Kosterlitz-Thouless (KT) phase transition \cite{Kosterlitz1973,Kosterlitz1974}. 
The KT phase transition is an infinite order phase transition, meaning that the order parameter $P$ and all its derivatives are continuous at the critical temperature $T_{KT}$.
Instead, the phase transition is characterised by a sudden change in the asymptotic behaviour of the equal-time spatial correlation function $C(r, t)$, defined as:
\begin{equation}
  C(r, t) = \langle \hat{\mathbf{S}}_i(t)\cdot \hat{\mathbf{S}}_j(t) \rangle_{|\mathbf{r}_i - \mathbf{r}_j|=r} = \langle \cos(\theta_i(t) - \theta_j(t)) \rangle_{|\mathbf{r}_i - \mathbf{r}_j|=r}
  \label{eq:XY_spatial_correlation_function}
\end{equation}
where $\langle \cdot \rangle$ denotes an average over all pairs of spins $(i,j)$ separated by a distance $r$.
In the steady state, $C(r,t)$ becomes independent of time and is simply denoted $C(r)$.
For finite temperatures below the critical temperature $T_{KT}$, the spatial correlation function decays as a power-law:
\begin{equation}
  C(r) \sim r^{-\eta(T)} \qquad \text{for } T < T_{KT}
  \label{eq:XY_Cr_power_law}
\end{equation}
This is a signature of quasi-long-range order (QLRO), meaning that the correlation length is infinite while the order parameter $P$ is zero.

The spinwave approximation, which is valid at low temperatures because it assumes that the phase differences between neighbouring spins are small, predicts $\eta(T) = T/(2\pi)$ \cite{Kosterlitz1974}. 
Precisely at $T_{KT}$, the spatial correlation function decays as a power-law with a universal exponent $\eta(T_{KT}) = 1/4$ \cite{Kosterlitz1974}.

Above $T_{KT}$, the spatial correlation function decays exponentially to zero:
\begin{equation}
  C(r) \sim \exp(-r/\xi) \qquad \text{for } T > T_{KT}
  \label{eq:XY_Cr_exponential}
\end{equation}
where $\xi$ is the correlation length. This is a signature of short-range order (SRO), meaning that the correlation length $\xi$ is finite and the order parameter $P$ is zero.

\begin{figure}[h!]
  \centering
  \includegraphics[width = 0.99\linewidth]{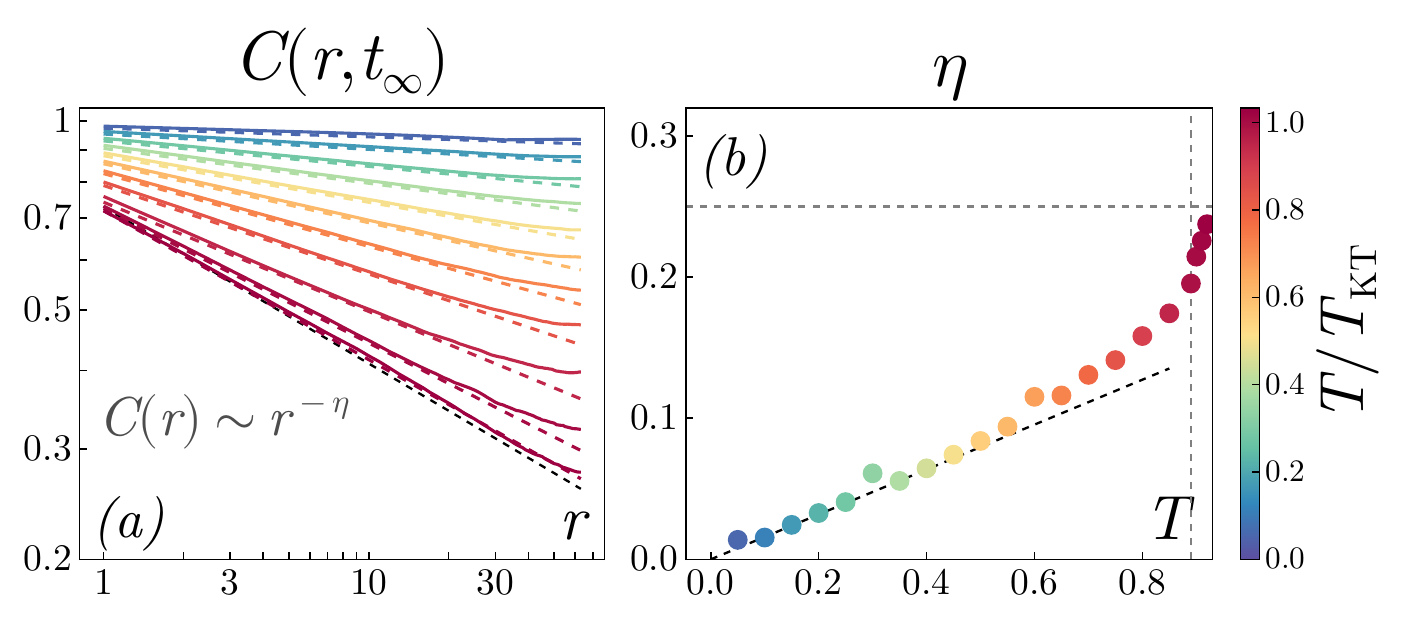}
  \caption{
  \textbf{(a)} The correlation function $C(r)$. The coloured dashed lines are the best fits $\sim\,r^{-\eta}$. The black dash line is $0.71r^{-1/4}$.
  \textbf{(b)} The correlation length $\eta$,  extracted from $C(r) \sim r^{-\eta}$. The dashed line is $T/(2\pi)$, the dotted horizontal line is $\eta = 1/4$, the vertical dotted line is $T=T_{KT} = 0.89$. 
  These measurements were performed for $L=128$ because they are numerically more demanding (average over more realizations and longer equilibration times are needed), in particular close to the critical temperature, explaining the shift of the critical temperature value from the large $L$ limit $T_{KT} \approx 0.89$.
  }
  \label{fig:C_eta_SS}
\end{figure}

The change from a power-law decay to an exponential decay of the spatial correlation function $C(r)$ at $T_{KT}$ is the hallmark of the Kosterlitz-Thouless phase transition.
We illustrate this change with our own results in Fig.~\ref{fig:C_eta_SS}. In panel (a), we plot $C(r)$ for different finite temperatures below $T_{KT}$, together with the corresponding exponent $\eta(T)$ in panel (b). We indeed observe a power-law decay of $C(r)$ for $T<T_{KT}$. For $T>T_{KT}$ (not shown), we observe an exponential decay of $C(r)$, as expected. 
For temperatures below $T=0.55$, we find that $\eta(T)$ is very well approximated by the spinwave prediction $\eta(T) = T/(2\pi)$, shown as a dashed line in panel (b). However, the spinwave approximation does not take into account the presence of peculiar collective structures, called topological defects. Those become more and more numerous as the temperature increases, explaining why the spinwave approximation fails to predict $\eta(T)$ close to $T_{KT}$. We represent the spinwave prediction as a dashed line in Fig.~\ref{fig:C_eta_SS}(b). We also plot the Kosterlitz-Thouless prediction $\eta(T_{KT}) = 1/4$, with a horizontal dotted line. This prediction arises from the combination of the spinwave regime result with renormalisation group calculations of the vortex fugacity at the critical temperature~\cite{Kosterlitz1974}.

\section{Topological defects in the XY model}
This change in $C(r)$ is due to a change in the behaviour of topological defects. 
In the $2d$ XY model, the topological defects are point-like excitations where the phase $\theta$ is undefined.
They are local minima of the energy and cannot be suppressed by a continuous deformation of the spin field. 
Topological defects are characterised by a winding number $q$, also called the defect \textit{charge}, defined by the sum of the phase differences along the bonds of the dual plaquette $\sumcirclearrowleft \Delta \theta_{i,j}=2\pi q$.
More precisely, for the XY model, the orientation field imposed by an ideal and isolated single of charge $q$ 
stems directly from the double integration of the Poisson equation $\Delta\theta(x,y) = 0$
and is therefore given by 
\begin{equation}
  \theta(x,y) = q\arctan(y/x) + \mu
  \label{eq:arctan_defect}
\end{equation}




\subsection{Charge of the topological defects}
In our framework, because the interactions between spins are polar, the topological charge $q$ is an integer (in contrast with a nematic alignment giving rise to half-integer charges).
Defects with a topological charge $|q|\geq 2$, although not prohibited \textit{per se} in the XY model, are not stable and very rapidly decay to $|q|=1$ defects. 
Indeed, the integration of the field \er{eq:arctan_defect} in a square box of size $L$ and lattice spacing $a$
shows that the energy of a defect of charge $q$ is given by $-J\pi q^2 \log (L/a)$. 
Because of the $q^2$ term, the energy of a defect increases with the square of its charge, and therefore,
defects of charge $|q|>1$ are energetically unfavourable: they are unstable to thermal fluctuations and decay into unit charges $|q|=1$.
This claim will remain true in the rest of this thesis, even when we will consider out-of-equilibrium dynamics, even though there is a priori no argument to support it. 
Thus, throughout this manuscript, defects will always be understood to be of unit charge, $|q|=1$.

\subsection{Shape of the topological defects}

\begin{figure}[h!]
    \centering
    \includegraphics[width=0.85\linewidth]{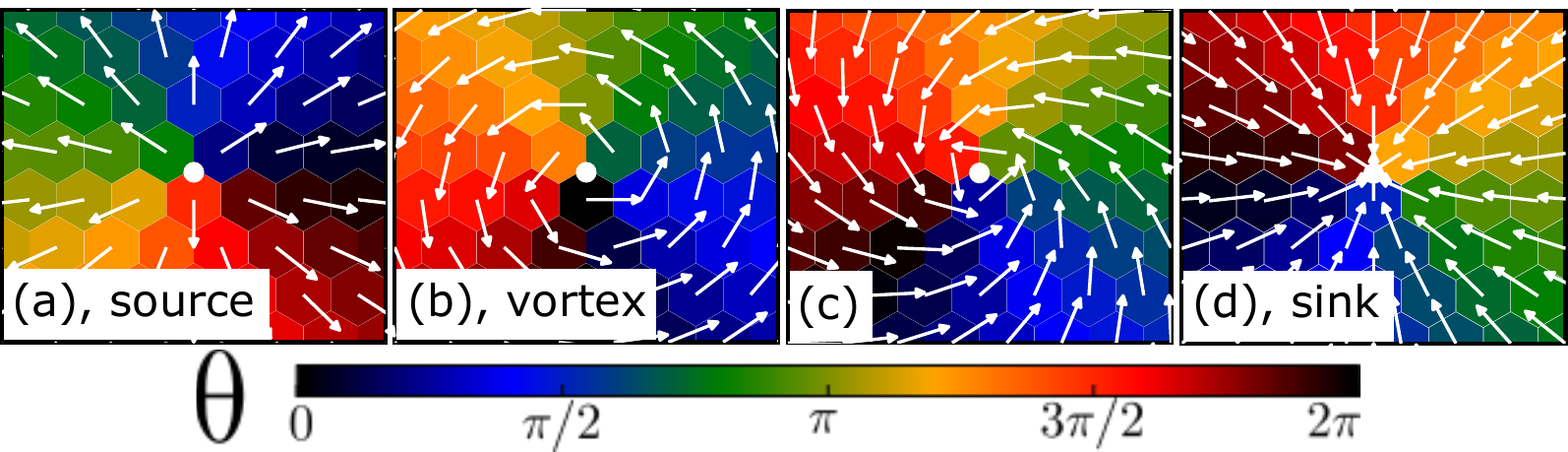}
    \caption{Four $+1$ defects with different shapes \textbf{(a)}  $\mu_+ = 0 $
    \textbf{(b)}  $\mu_+ = \pi/2 $
    \textbf{(c)}  $\mu_+ = 3\pi/4 $
    \textbf{(d)}  $\mu_+ = \pi $. 
     White dots show  the defect core, on the dual (hexagonal) lattice. The colour code displays the phase $\theta$ of each spin.}
    \label{fig:defects_shape_snapshots}
\end{figure}

\begin{figure}[h!]
    \centering
    \includegraphics[width=0.85\linewidth]{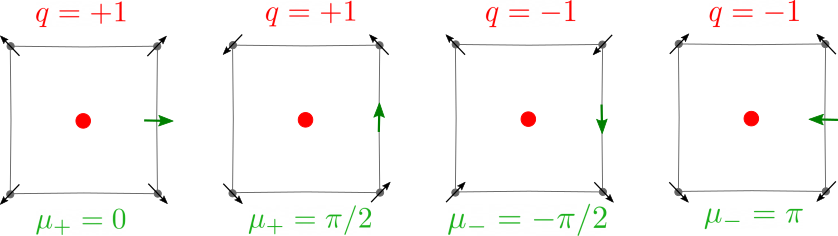}
    \caption{The defects (red) live at the center of the dual plaquettes. The spins (black) live on the lattice, i.e. on the corners of the plaquette. An easy way to visually identify $\mu$ is to imagine the orientation an imaginary spin (green) would have if  located at the right edge of the defect ($y=0$ and thus $\arctan (y/x) = 0$).}
    \label{fig:square_plaquette_spin_right}
\end{figure}

The integration constant $ \mu_\pm$ is a pseudo-scalar (modulo $2\pi$). In contrast with the defect charge, it is not a topological invariant, meaning that it can be changed by a continuous rotation of the spin field. For $-1$ defects, $\mu_-$ corresponds to a global rotation of the defect by $\mu_-/2$. For $+1$ defects, $\mu_+$ physically represents the \emph{shape} of the defect, giving in particular \emph{sources} ($\mu_+ = 0$), \emph{sinks} ($\mu_+ = \pm\pi$) and \emph{vortices} ($\mu_+ = \pm \pi/2$). 
We illustrate various defect shapes in Fig.~\ref{fig:defects_shape_snapshots}. The colour of the spins corresponds to their orientation $\theta_i$, 
on a cyclic colour scale from black ($\theta = 0$), to blue ($\theta = \pi/2$) to green ($\theta = \pi$) to red ($\theta = 3\pi/2$) to black again ($\theta = 2\pi$).
This colourscale will be used throughout this thesis to represent the orientation of the spins. Topological defects of unit charge are the points where all the colours meet.
Note that $\mu$ can be thought of as the value of the orientation field $\theta$ at the immediate right side of the defect, as sketched in Fig.~\ref{fig:square_plaquette_spin_right} with the imaginary green arrows.

In the equilibrium XY model, since the equation of motion is invariant under a global rotation of the spins $\theta \to \theta + \alpha$, the shape is irrelevant for any physical phenomena. If this still holds for the Kuramoto model, it is not the case for the non-reciprocal XY model, where the actual shape of the defect plays a crucial role in their dynamics.    

\subsection{Defect density at infinite temperature}
At initialisation, the system is at $T=\infty$: the spins are randomly oriented, the polarisation $P$ is close to zero (in fact, $P\sim N^{-1/2} = L^{-1}$ due to the central limit theorem), the correlation length $\xi=0$ and the correlation function $C(r)$ is a Kronecker function.
It is known that the defect density at infinite temperature is equal to $1/3$ \cite{jelic2011quench}, but this value is often mentionned without proof. In this paragraph, we provide our own derivation to prove it.

At infinite temperature, the spins are randomly oriented. Let us work on a square lattice for simplicity. The spins on the four corners of a plaquette, labelled $1,2,3,4$, have independent orientations, which we take in $[-\pi, \pi[$ instead of $[0,2\pi[$ for mathematical convenience (see the integrals later). 
Thanks to the rotation invariance of the system, we can set $\theta_1 = 0$. Let us call $\theta_2 =x, \theta_3 = y$ and $\theta_4 = z$.

To handle the angular periodicity analytically, we write the modulo function as 
\begin{equation}
  f(t) = t - 2\pi\,g(t)
\end{equation}
where $g(t)$ is the step function defined as
\begin{equation}
  g(t) = 
\begin{cases}
  1,  & t > \pi, \\[6pt]
  -1,  & t < -\pi \\[6pt]
  0,  & \text{else, ie. when  } -\pi \leq t \leq \pi
\end{cases}
\end{equation}

The winding number $q$ of the plaquette is given by 
\begin{equation}
\begin{aligned}
  2\pi\,q &= f(x) + f(y-x) + f(z-y) + f(-z)  \\
        &= x + (y-x) + (z-y) - z - 2\pi \left( g(x) + g(y-x) + g(z-y) + g(-z) \right) \\
        &= - 2\pi \left( g(x) + g(y-x) + g(z-y) + g(-z) \right) \\
        &= - 2\pi \left( g(y-x) + g(z-y) \right) 
\end{aligned}
\end{equation}
where the last equality stems from the fact that $x$ and $z$ are in $[-\pi,\pi[$, so that $g(x)=g(-z)=0$.

The probability of $g(y-x) = 1$ is
\begin{equation}
  \begin{aligned}
    \mathbb{P}\Big(g(y-x) = 1\Big) &= \mathbb{P}(y-x > \pi) \\
    &= \mathbb{P}(x < y-\pi) \\
    &=\frac{y_+}{2\pi}
  \end{aligned}
\end{equation}
where $y_+ = \max(y, 0)$.
Similarly, 
\begin{equation}
  \begin{aligned}
    \mathbb{P}\Big(g(y-x) = -1\Big) &= \frac{(-y)_+}{2\pi} \\
    \mathbb{P}\Big(g(z-y) = +1\Big) &= \frac{(-y)_+}{2\pi} \\
    \mathbb{P}\Big(g(z-y) = -1\Big) &= \frac{y_+}{2\pi} \\
    \end{aligned}
\end{equation}
The events $g = 0$ are computed by complement. For instance,  
\begin{equation}
  \begin{aligned}
    \mathbb{P}\Big( g(y-x) = 0 \Big)  &= 1 - \mathbb{P}\Big(g(y-x) = +1\Big) - \mathbb{P}\Big(g(y-x) = -1\Big)\\
    &= 1 - \frac{y_+}{2\pi} - \frac{(-y)_+}{2\pi} \\
    &= 1 - \frac{|y|}{2\pi} \\
    \end{aligned}
\end{equation}
The same goes for $g(z-y)$.

To obtain a defect of charge $q=+1$, we thus need $g(y-x) = 1$ and $g(z-y) = 0$ or vice-versa. The probability of this event is
\begin{equation}
  \begin{aligned}
    \mathbb{P}\Big(q=+1\Big) &= \mathbb{P}\Big(g(y-x) = 1 \text{ and } g(z-y) = 0\Big) + \mathbb{P}\Big(g(y-x) = 0\text{ and }  g(z-y) = 1\Big) \\
    &= \mathbb{P}\Big(g(y-x) = 1\Big)\cdot\mathbb{P}\Big(g(z-y) = 0\Big) + \mathbb{P}\Big(g(y-x) = 0\Big)\cdot\mathbb{P}\Big(g(z-y) = 1\Big) \\
    &= \left( 1-\frac{|y|}{2\pi} \right)\left( \frac{(y)_+}{2\pi} + \frac{(-y)_+}{2\pi} \right) \\
    &= \left( 1-\frac{|y|}{2\pi} \right) \frac{|y|}{2\pi}\\
  \end{aligned}
\end{equation}

Finally, since $y$ is independent and uniformly distributed in $[-\pi,\pi[$, the probability of having a defect of charge $q=+1$ is
\begin{equation}
  \begin{aligned}
    \mathbb{P}\Big(q=+1\Big) &= \frac{1}{2\pi}\int_{-\pi}^{\pi} \left( 1-\frac{|y|}{2\pi} \right)\frac{|y|}{2\pi} dy \\
    &= \frac{1}{\pi}\int_{0}^{\pi} \left( 1-\frac{y}{2\pi} \right)\frac{y}{2\pi} dy \\
    &= \int_{0}^{1} \left( 1-\frac{\tilde y}{2} \right)\frac{\tilde y}{2} d\tilde y \\
    &= \frac{1}{6}
  \end{aligned}
\end{equation}

We have computed the probability of having a defect of charge $q=+1$. By symmetry, the probability of having a defect of charge $q=-1$ is also $1/6$. 
The probability of having one defect ($|q|=1$) on a plaquette (in other words, the defect density), is thus $1/3$.

\subsection{Peierls-like argument and the Kosterlitz-Thouless phase transition}
A simple argument, known as the Peierls argument, can be used to understand the existence of a phase transition in the XY model.
The idea is to consider the free energy of the single isolated defect of charge $q$ in a system of size $L$. 
We have seen that the associated energy is $J\pi q^2 \log (L/a)$. The defect can be located anywhere in the system, at each of the $L^2$ sites of the lattice.
The entropy associated with the defect position is $S = k_B \log(L^2/a^2)$, where $a$ is the defect core size, which is equal to the lattice spacing in the XY model.
The free energy of the defect is then given by:
\begin{equation}
  \mathcal{F} = E - TS = J\pi q^2 \log (L/a) - k_B T \log(L^2/a^2) = (J\pi q^2 - 2k_B T) \log(L/a)
  \label{eq:XY_defect_free_energy}
\end{equation}
Because both the energy and the entropy scale logarithmically with the system size $L$, both terms are of the same order of magnitude
and the free energy can be positive or negative depending on the temperature $T$. 
At high temperatures, the free energy is negative, and the defects proliferate in the system and perturb it: the system is in a disordered phase.

At low temperatures, the free energy of a single defect is positive, isolated defects are thus suppressed. 
However, the energy of a pair of defects of opposite charges $+q$ and $-q$ separated by a distance $R$ scales as $2J\pi q^2 \log (R/a)$, independently of the system size $L$. Indeed, such a \textit{bounded pair} does not perturb the system at long distances, as the far fields induced by each on the two defects cancel each other out. More precisely, the perturbation induced by a pair of defects is limited to a distance of the order of the separation $R$ between them, a posteriori justifying the scaling of the energy with $L$ for isolated defects and with $R$ for bounded defect pairs.
Therefore, at low temperature, pairs of bounded defects are energetically and entropically favourable. 

Between the high and the low temperature phases, there exists a critical temperature $T_{KT}$ : from the simple argument above, one can deduce that
$T_{KT} = \frac{J\pi q^2}{2k_B} = \pi/2 \approx 1.57$ for unit charge defects $|q|=1$ and in a unit system where $J=k_B=1$. While the order of manitude is correct, this argument does not provide an accurate estimate of $T_{KT}$.
With numerical simulations of the XY model, one gets $T_{KT} \approx 0.89$ for the square lattice and $T_{KT} \approx 1.4$ for the triangular lattice; we shall come back to this point in the next paragraph.  
\newline

Nevertheless, the Peierls argument still provides a good intuition of the existence of such a topological phase transition driven by the proliferation of topological defects.
A Kosterlitz-Thouless phase transition should occur whenever the energy and the entropy of a single isolated defect scale in the same way with the system size $L$. 
It also justifies why the the spatial dimension $d=2$ is special for the XY model. On one hand, the only possible defects are point-like: they can therefore be located anywhere in the $2d$ system, and the associated entropy scales as $S \sim \log(L^2)$. This is for instance not the case in $d=3$, where the defects can be either singular point or line defects, which modifies the scaling of the entropy with $L$. 
On the other hand, the energy of a point-like defect in $d$ dimensions can be obtained by integrating the energy density of the far field induced by the defect over a $d$-dimensional sphere of radius $L$.
\begin{align}
    \frac{1}{2}\,(\nabla{\theta})^2 &= \frac{1}{2}\,\left(\nabla{q\arctan(\frac{y}{x})}\right)^2\\
    & = \frac{q^2}{2}\,
    \left(
  \frac{1}{r^2}
    \begin{pmatrix}
      -y \\
      x
    \end{pmatrix} 
    \right)^2 \\
    & = \frac{q^2}{2}\,\frac{r^2}{r^4} \\
    & = \frac{q^2}{2}\,\frac{1}{r^2}
\end{align} 

Because the energy density scales as $r^{-2}$, the energy of an isolated defects in a $d$-dimensional system of linear size $L$ scales as:
\begin{equation}
  E \sim \int_a^L r^{d-1} r^{-2}
  \sim \int_a^L r^{d-3}
  \sim
  \begin{cases}
    L^{d-2}  & d \neq 2 \\[6pt]
     \log L & d = 2    
  \end{cases} 
\end{equation}
We thus see that the energy of a point-like defect scales as $\log L$ only in two dimensions.
\newline

\subsection{Determination of the Kosterlitz-Thouless critical temperature}
On the square lattice, the value of the critical temperature has been deeply investigated and debated over 3 decades \cite{olsson1995monte, Hasenbusch2005, weber1988monte, hasenbusch1997computing, schultka1994finite}, finally coming down to the value $T_{KT} = 0.89294(8)$ \cite{Hasenbusch2005}.

However, the literature is surprinsingly scarce regarding the triangular lattice case. Butera presented in 1994 a high-temperature expansion and concluded $T_{KT}=1.47$ \cite{butera1994high}. They improved it to $T_{KT}=1.46$ in 2008 \cite{butera2008high} and confirmed this value $T_{KT}=1.46$ in 2021 \cite{butera2008further}.
A study of the helicity modulus but for modest lattice sizes (up to $L=80$) found $T_{KT}=1.48$ \cite{sun2022numerical}.

\begin{figure}[h!]
    \centering
    \includegraphics[width=0.65\linewidth]{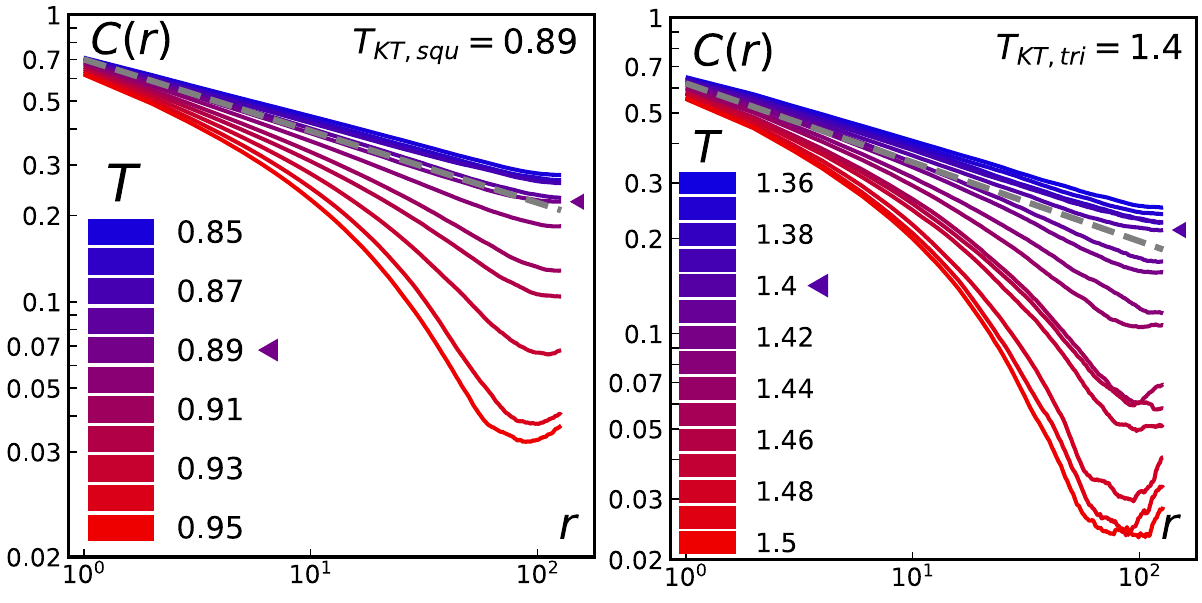}
    \caption{Spatial correlation function $C(r)$ computed for a $L=256$ system. \textbf{(left)} Square lattice, we obtain $T_{KT}\approx 0.89$ \textbf{(right)} Triangular lattice, we obtain $T_{KT}\approx 1.4$.}
    \label{fig:TKT}
\end{figure}

So far, the values reported were found in the literature. Here, we directly measure the spatial correlation function $C(r)$ for different temperatures in $L=256$ systems. We define the critical temperature $T_{KT}$ as the largest temperature for which $C(r)$ is well described by a power-law decay. We present our results in Fig.~\ref{fig:TKT} for the square lattice (left panel) and for the triangular lattice (right panel). For the square lattice, we obtain $T_{KT}=0.89\pm 0.01$, in agreement with the literature. For the triangular lattice, we obtain $T_{KT}=1.40 \pm 0.02 $, relatively close (4\%) to the literature values cited above. 

Let us denote by $T_4$ (resp. $T_6$) the critical temperature for the square (resp. triangular) lattice.
Interestingly, the ratio $\frac{T_6}{T_4} = \frac{1.4}{0.89 } \geq 1.57 \neq \frac{6}{4}$, meaning that there is more going on than a simple rescaling due to the number of neighbours. One explanation might lie in the possibly different mobility of the defects in both lattices. 
\newline

Finally, note that the critical temperature $T_{KT}$ is size dependent, and that the value $T_{KT} = 0.89$ is only valid in the thermodynamic limit $L\to\infty$.
In practice, the critical temperature $T_{KT}(L)$ follows the finite size scaling law \cite{chung1999essential, bramwell1994magnetization, tomita2002probability}:
\begin{equation}
  T_{KT}(L) = T_{KT} + \frac{c}{(\log L)^2} + \mathcal{O}\left(\frac{1}{(\log L)^3}\right)
  \label{eq:TKT_finite_size_scaling}
\end{equation}
where $c$ is a non-universal constant that appears to be small, at most of the order of $0.3$ (given that with $L=256$ we obtain the correct value with an asbolute error of $0.01=c (\log L)^{-2}$). 

This scaling law \er{eq:TKT_finite_size_scaling} is a direct consequence of the divergence of the correlation length approaching the critical temperature $T_{KT}$ from above: 
\begin{equation}
  \xi(T) \sim \exp\left(b\sqrt{\frac{T_{KT}}{T-T_{KT}}}\, \right)
  \label{eq:correlation_length_finite_size_scaling}
\end{equation}
where $b$ is a constant. 
For finite systems, the correlation length $\xi$ is limited by the system size $L$. One thus has $\xi \sim L$, which translates into $T-T_{KT}\sim (\log L)^{-2}$.
\newline

In this thesis, we will mostly work with $L = 256$ systems, which is large enough to consider the critical temperature $T_{KT}$ as the thermodynamic limit value. 

\section{Dynamics of the XY model}

The equilibrium properties of the XY model are independent of the choice of the dynamics, as long as the dynamics satisfies the detailed balance condition. However, the time evolution of the system, in particular the coarsening process (=relaxation process from a disordered initial configuration to an equilibrium state, over which the order grows progressively), 
naturally depends on the choice of the dynamics. 
One can choose among various family. If one wants to rely on the concept of energy, one can opt for Monte Carlo dynamics \cite{harrison2010introduction} or Glauber dynamics \cite{marro2005nonequilibrium, martinelli2004lectures}. If instead one prefers to work at the force level, one can choose the molecular dynamics \cite{hansson2002molecular, hollingsworth2018molecular} framework, based on the Newton equations, or Langevin dynamics \cite{bussi2007accurate}, based on the Langevin equation.

In many out-of-equilibrium systems, the dynamics does not derive solely from the minimisation of an energy\footnote{One could actually use a force deriving from a well defined, equilibrium energy, and drive the system away from 
equilibrium with coloured noise \cite{paoluzzi2018effective} for instance. }. 
Approaches based on the force are particularly well suited to study active matter systems, as they allow to easily break the detailed balance condition and drive the system out of equilibrium.
In this thesis, we will exclusively work with the Langevin dynamics.

\subsection{Langevin dynamics for the XY model}
 
The force is given by the derivative of the Hamiltonian with respect to the angle. For the XY model, it gives:
\begin{equation}
  \text{force on spin } i = -\frac{\partial \mathcal{H}}{\partial \theta_i} = J\sum_{j \in \partial_i} \sin(\theta_j - \theta_i)
  \label{eq:XY_force}
\end{equation}
where $\partial_i$ denotes the set of nearest neighbours of spin $i$.

In the most general form the dynamics of the spins is given by the underdamped Langevin equation:
\begin{equation}
  m \ddot{\theta}_i + \gamma \dot{\theta}_i = J\sum_{j \in \partial_i} \sin(\theta_j - \theta_i) + \sqrt{2k_B T \gamma} \ \eta_i(t)
  \label{eq:XY_Langevin_underdamped}
\end{equation}
where $\ddot{\theta} = d^2\theta/dt^2$ is the angular acceleration, $\dot{\theta} = d\theta/dt$ is the angular velocity,
$m$ is the mass of the spins, $\gamma$ is the friction coefficient, and $\eta_i(t)$ is a Gaussian white noise term with zero mean and unit variance:
\begin{equation}
  \langle \eta_i(t) \rangle = 0\ , \quad \langle \eta_i(t)\, \eta_j(t') \rangle = \delta_{ij}\, \delta(t - t')
  \label{eq:XY_noise}
\end{equation}

However, in active matter, the inertial effects are usually negligible compared to the friction, and the inertial term can be neglected.
One can therefore work in the limit $m/\gamma\to0$ and obtain the overdamped equation of motion of the XY model \er{eq:XY_Langevin}:
\begin{equation}
  \gamma \dot{\theta}_i = J\sum_{j \in \partial_i} \sin(\theta_j - \theta_i) + \sqrt{2k_B \gamma T} \ \eta_i(t) 
  \label{eq:XY_Langevin}
\end{equation}

In the following, we show that one can redefine time to absorb all the parameters of the equation of motion, except for the temperature $T$.


Let us divide the equation \er{eq:XY_Langevin} by the coupling coefficient $J$. One obtains:
\begin{equation}
  \frac{\gamma}{J} \frac{d\theta_i}{dt}=\sum_{j \in \partial_i} \sin \left(\theta_j-\theta_i\right)+\sqrt{\frac{2 k_B T \gamma}{J^2}}\  \eta_i(t)   
  \label{eq:XY_Langevin_adim1}
\end{equation}

We then normalise time $\tilde{t}=t J / \gamma$ and define $\tilde{T}=k_B T / J$. One thus obtains

\begin{equation}
  \frac{d\theta_i}{d\tilde t}=\sum_{j \in \partial_i} \sin \left(\theta_j-\theta_i\right)+\sqrt{\frac{2 \tilde{T} \gamma}{J}} \ \eta_i(\tilde{t} \gamma / J)
  \label{eq:XY_Langevin_adim2}
\end{equation}

Finally, since the delta function in the variance of the white noise $\eta_i$ satisfies $\delta(a x)=\frac{\delta(x)}{|a|}$, one obtains the dimensionless equation of motion:

\begin{equation}
  \frac{d{\theta}_i}{d\tilde{t}}=\sum_{j \in \partial_i} \sin \left(\theta_j-\theta_i\right)+\sqrt{2 \tilde{T}} \ \eta_i(\tilde{t})
  \label{eq:XY_Langevin_adim3}
\end{equation}
We then can drop the tildes and understand the variable $\theta$ as implicitly depending on $t$. 
We thus obtain the following equation:
\begin{equation}
  \dot{\theta}_i =\sum_{j \in \partial_i} \sin \left(\theta_j-\theta_i\right)+\sqrt{2 {T}} \ \eta_i({t})
  \label{eq:XY_Langevin_adim}
\end{equation}

This motivates the choice of working in a system of units where $J=k_B=\gamma=1$, which we will do in the rest of this thesis. We will thus express times in units of $\gamma/J$ and temperatures in units of $J/k_B$.
In the rest of this thesis, this final equation \er{eq:XY_Langevin_adim} will be the reference equation of motion for the XY model, 
and it will serve as a basis we will depart from to study out-of-equilibrium versions of the XY model.

\subsection{Interaction between topological defects in the XY model} \label{sec:XY_defect_interaction}
Now that we are equipped with a dynamics to simulate the XY model, we can study the dynamics of the topological defects, for $T<T_{KT}$ (recall that for $T>T_{KT}$, the system is disordered and defects proliferate, making their individual defect dynamics hardly trackable).

The fact that these are also called topological \textit{charges} is not a coincidence: they interact with each other, and the interaction is long-ranged.
Indeed, a first computation similar to the one done above for the energy of a single defect shows that the interaction energy between two defects of charges $q_1$ and $q_2$ separated by a distance $R$ in a $2d$ space is given by:
\begin{equation}
  E_{12} = -2J\pi q_1 q_2 \log(R/a)
  \label{eq:XY_defect_interaction_energy}
\end{equation}
where $a$ is the lattice spacing. The interaction is attractive for opposite charges and repulsive for like charges and the resulting force decays like $1/R$. \footnote{Note that in $3d$, the interaction between two electrostatic charges would decay like $1/R^2$ and the energy like $1/R$.} 
This is similar to the interaction between two point-like charges in $2d$ electrostatics; this is why the term \textit{charge} is used and why the interaction is called \textit{Coulomb-like}; cf. \cite{minnhagen1987coulombgas} for a review on $2d$ Coulomb gases and the Kosterlitz-Thouless phase transition.

However, while the static attraction between two defects is indeed logarithmic, the defect dynamics is not purely Coulombian. Indeed, the mobility of a defect is not a simple constant, as they experience a friction that depends logarithmically on the size of the region they influence, due to the rearrangement of the spins around them as the defect moves \cite{pleiner1988dynamics}. In the case of a pair of defects, that influenced region is limited to the distance $R$ separating them. 
The friction force thus scales as $\dot R \,\log R $, where $\dot R$ is the velocity of each defect, see \cite{YurkeHuse1993} and references therein. 
Because the dynamics is overdamped, the friction and attractive forces have to balance each other, giving: 
\begin{equation}
  \log(R) \dot R \sim 1/R
  \label{eq:XY_defect_dynamics}
\end{equation}
One can integrate \er{eq:XY_defect_dynamics} by part to obtain \cite{YurkeHuse1993}: 
\begin{equation}
  R^2 \left( \log(R) - \frac{1}{2} \right) \sim t^* - t \equiv \tilde t \ .
  \label{eq:XY_defect_dynamics_integrated}
\end{equation}
In \er{eq:XY_defect_dynamics_integrated}, $t^*$ is the time (timestamp) of the annihilation of the defect pair and $\tilde t$ is the \textit{time to annihilation} (the time left before the time of annihilation). In other words, $\tilde t$ is the timespan between the current time $t$ and the moment of the annihilation $t^*$. Note that the time of annihilation depends on the initial distance $R_0$ between the two defects, on the temperature and in particular on the noise realisation. $t^*$ is thus a random variable centered on a deterministic mean, we will come back to this point below. For now, consider $t^*$ as this deterministic mean value. 
We complete the analysis of Ref.\cite{YurkeHuse1993} by providing the solution for $R$:
\begin{equation}
R(\tilde{t}) = \exp\!\left( W\!\left(\frac{2\pi}{\nu}\tilde{t}/2 \right) \right),
\label{eq:XY_defect_dynamics_solved}
\end{equation}
where $W(x)$ is the Lambert $W$ function (also known as the Product Log function) and $\nu$ is proportional to the friction coefficient $\gamma$ in the Langevin equation \er{eq:XY_Langevin_adim}. 

By comparison, if one does not take into account the space-dependence of the motility, one would obtain $\dot R \sim 1/R$, resulting in $R\sim \sqrt{\tilde t}$. 
\newline

\begin{figure}
  \centering
  \includegraphics[width = 0.99\linewidth]{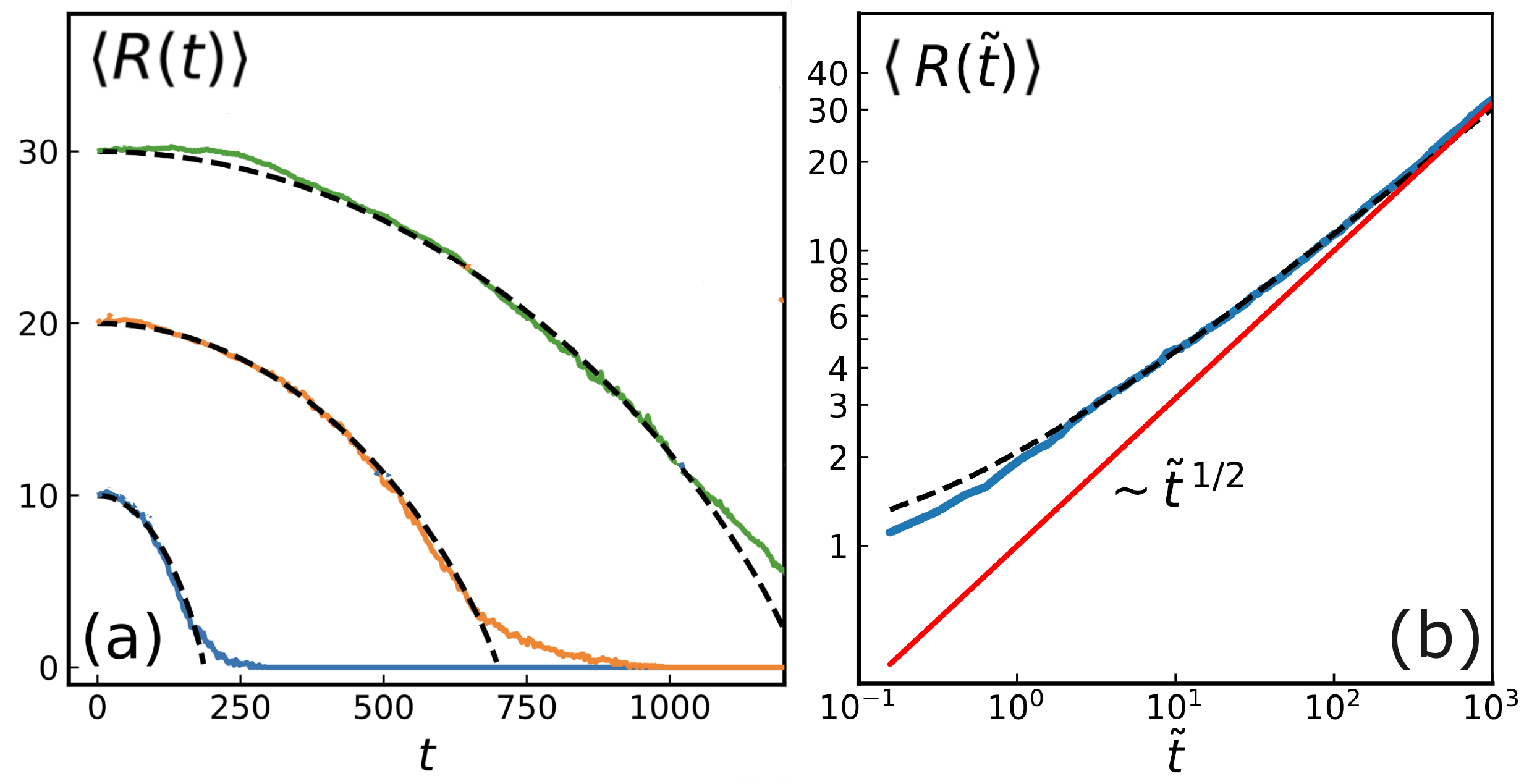}
  \caption{
  \textbf{(a)}
  The distance $R(t)$ is averaged over 120 independent runs, for three different initial separation distances $R_0$. (blue: $R_0 =10$, orange: $R_0 = 20$ and green: $R_0 = 30$). Black dashed lines are the predictions for an overdamped dynamics with a Coulomb interaction potential.
  \textbf{(b)}
  We recover the results of Ref. \cite{YurkeHuse1993} (see their Fig.\,6). The data (blue) is fitted by expression \er{eq:XY_defect_dynamics_solved} with $\nu = 1/2$ (dash) and is notably distinct from a simple diffusive trajectory (red), highlighting the importance of the space dependency of the motility.} 
  \label{fig:Rt_XY}
\end{figure}

We illustrate this annihilation phenomena reporting our own data. We manually create a pair of defects in a system at a temperature $T < T_{KT}$. We then measure the distance $\langle R(t) \rangle$ (averaged over 120 realisations) between the two defects as a function of time and report the results in Fig.~\ref{fig:Rt_XY}(a), for three different initial distances $R_0 = 10, 20, 30$ in a $L=64$ system. The black dashed line are the predictions for an overdamped dynamics with a Coulomb interaction potential \er{eq:XY_defect_dynamics_solved}. At short times, the distance $R(t)$ decreases as expected, but at longer times, the curves flatten out and depart from the expected behaviour. 
This is not a failure of the model, but rather an artefact arising from the fact that different realizations have different annihilation times $t^*$ that we explain in the next paragraph.

One has two choices of representations. \\
The first possibility is to perform the average $\langle R(t) \rangle$ over realizations not yet annihilated, neglecting the realizations already annihilated (for which $t > t^*$). Individual curves all have the shape of the fit but with different endtimes $t^*$. 
The second option is to perform the average $\langle R(t) \rangle$ over all realizations, defining $R(t \geq t^*) = 0$ for each trajectory. 
This method has the advantage to preserve the continuity of the average (and therefore the clarity of the figure), at the cost of the flattening (as more and more zeros enter the average when time increases) of the curve for small $R$. We chose the second method in Fig.~\ref{fig:Rt_XY}(a).
\newline

Since the problem arises from the distribution of $t^*$, let us reverse time and consider instead the \textit{time to annihilation} $\tilde{t}$. 
In this equivalent description, $\tilde{t}=0$ is the moment of annihilation: all trajectories start at $\tilde{t} = 0$ ($R(\tilde{t}\,=\,0) = 0$)  and defects move away from each other with increasing $\tilde{t}$, see Fig.~\ref{fig:Rt_XY}(b) for the same data as in Fig.~\ref{fig:Rt_XY}(a) for $R_0 = 30$. 
The interest of $\tilde{t}$ is that now the crucial part of the dynamics, when the $1/R$ force is relevant because $R$ is small, can be analysed with averages that are meaningful and more precise \cite{YurkeHuse1993, Rouzaire_Kuramoto_PRL, rouzaire_Frontiers}.  

Figure~\ref{fig:Rt_XY}(b) reproduces the results of~\cite{YurkeHuse1993}, proving that the discrepancy between the dotted line and the data in Fig.~\ref{fig:Rt_XY}(a) is a consequence of the data representation choice.

\subsection{Dynamics following a quench} \label{sec:dynamics_quench}

\begin{figure}[h!]
  \centering
  \includegraphics[width = 0.99\linewidth]{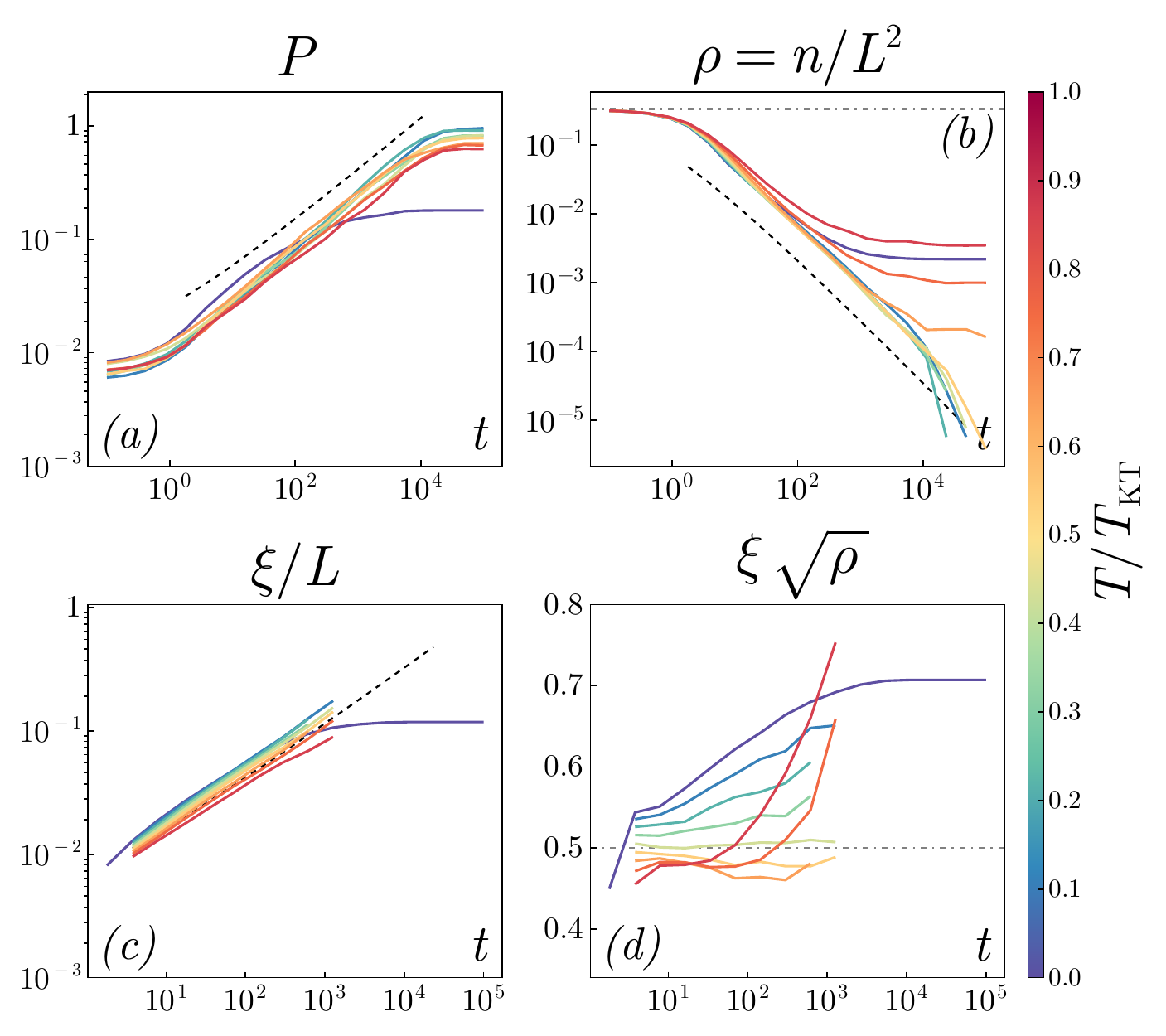}
  \caption{Dynamics of the XY model following a quench from $T=\infty$ to different $T<T_{\text{KT}}$ (see colourbar). Time evolution of 
  \textbf{(a)} the polarisation $P$ The dash line is $0.04\sqrt{t / \log(10t)}$.
  \textbf{(b)} the defect density $\rho = n/L^2$.  The dash line is $0.08 \log t / t$. 
  \textbf{(c)} the correlation length $\xi$, extracted from $C(\xi) = 1/e$. The dash line is $0.011\sqrt{t / \log(10t)}$.
  \textbf{(d)} the quantity $\xi \sqrt{\rho}$. The dotted horizontal line is at $1/2$. All panels are for $L=256$ systems. 
  }
  \label{fig:quench_XY}
\end{figure}

In this section, we provide a brief overview of the dynamics of the XY model following a quench from $T=\infty$ to $T<T_{KT}$. 
We focus on the main observables of interest: the polarisation $P$, the correlation length $\xi$, the correlation function $C(r)$ and the defect density $\rho = n/L^2$, where $n$ is the total number of topological defects in the system.
We have computed the time evolution of these observables in Fig.~\ref{fig:quench_XY} for different temperatures $T<T_{KT}$ (see colourbar). 

For $0<T<T_{KT}$, the defect density decreases as $\rho\sim \log t /t$, while the polar order parameter $P$ and the correlation length $\xi$, defined as $C(\xi) = 1/e$, both grow over time as $\sqrt{t/\log t\,}$ \cite{YurkeHuse1993, jelic2011quench}, see Fig.~\ref{fig:quench_XY}(a,b,c). We represent these XY predictions as black dashed lines in the whole figure. The logarithmic corrections stem from the logarithmic dependence of the defect mobility on the distance between defects explained in section~\ref{sec:XY_defect_interaction} \cite{YurkeHuse1993, bray2000breakdown}.
At $t=0$, we recover the expected defect density $\rho(t=0) = 1/3$ at infinite temperature, as seen in Fig.~\ref{fig:quench_XY}(b) (the grey dashdotted line is $1/3$).

The correlation length $\xi$ is to first order given by the typical distance between defects, considered uniformly distributed in a homogeneous space. One thus has $\xi \sim 1/\sqrt{\rho}$, or, equivalently $\xi \sqrt{\rho} \sim$ constant. We indeed observe in Fig.~\ref{fig:quench_XY}(d) that $\xi \sqrt{\rho}$ is approximately equal to $1/2$ for all temperatures. Unfortunately, we do not have a clear explanation for why the average distance between defects is twice the correlation length. This numerical value $1/2$ might also depend on the definition of the correlation length $\xi$ (here defined as $C(\xi) = 1/e$). 
A clear departure from this scenario is a signature that a different physics is at play and introduces another relevant lengthscale, as we shall see in the next chapters.

\begin{figure}[h!]
  \centering
  \includegraphics[width = 0.99\linewidth]{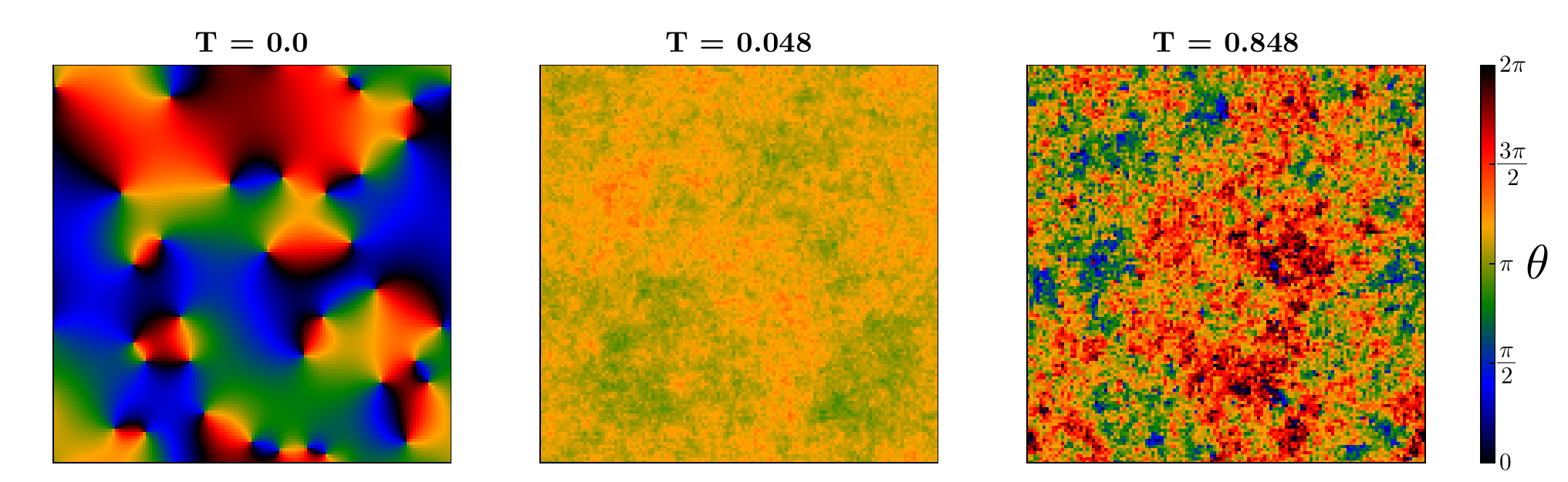}
  \caption{Snapshots of $L=128$ systems (square lattice, so $T_{\text{KT}}=0.89$) in the steady state ($t=10^5$) following a quench from $T=\infty$ to 
  \textbf{(a)} $T/T_{\text{KT}} = 0$
  \textbf{(b)} $T/T_{\text{KT}} = 0.054$
  \textbf{(c)} $T/T_{\text{KT}} = 0.952$. 
  }
  \label{fig:quench_XY_snapshots}
\end{figure}

Finally, we illustrate in Fig.~\ref{fig:quench_XY_snapshots} steady state snapshots at three different temperatures. In finite systems, at low temperatures (and even for $L=512$ systems), the steady state is effectively defectless, see Fig.~\ref{fig:quench_XY_snapshots}(b). Close to the critical temperature, the fluctuations in the orientation field are large, see Fig.~\ref{fig:quench_XY_snapshots}(c) and pairs of defects are continuously created and annihilated immediately. At zero temperature, the system is frozen in a metastable state with many defects remaining, see Fig.~\ref{fig:quench_XY_snapshots}(a). At zero temperature, the dynamics is deterministic, the system can get trapped in metastable states and defects have small or zero mobility. In the coarsening dynamics, this results in a  steady state with a finite density of defects, displaying short-range order in the (exponentially decaying) correlation function, see the blue/violet curve in each panel of Fig.~\ref{fig:quench_XY}.

This is why in the rest of this thesis, when we want to study the dynamics of defects, we will always work at small but finite temperatures (typically $T>0.05 T_{\text{KT}}$), to avoid these metastable states and ensure that defects can freely move.

	\chapter{Methods}
	\label{ch:methods} 
	\graphicspath{{methods/figures/}}
\graphicspath{{figures/Ch3methods/}}


In this chapter, we summarise the methods and measurements used recurrently in the following chapters. 
The methods only used at a particular point in the thesis will be introduced in the corresponding chapter.

\section{Time integration of the Langevin equation}
We have seen in Chapter \ref{ch:XY}, \er{eq:XY_Langevin_adim} that the overdamped Langevin formulation of the XY model reads:
\begin{equation}
  \dot{\theta}_i = J\sum_{j \in \partial_i} \sin(\theta_j - \theta_i) + \sqrt{2k_B \gamma T} \ \eta_i(t) 
  \label{eq:XY_Langevin3}
\end{equation}

We numerically integrate this equation in time with the standard explicit Euler-Maruyama method:

\begin{equation}
  \theta(t + dt) = \theta(t) + dt \cdot J\sum_{j \in \partial_i} \sin(\theta_j(t) - \theta_i(t)) + \sqrt{2\,dt\,k_B T} \ \eta 
  \label{eq:Euler-Maruyama}
\end{equation}
where $\eta$ is a random number drawn from a unit Gaussian at every timestep:   
\begin{equation}
  \eta = \text{randn()} \sim \mathcal{N}(0,1)
\end{equation}
One then has to choose an appropriate timestep $dt$. If $dt$ is too small, the simulation will be longer to run, for the same final result. If $dt$ is too large, the numerical scheme is unstable and/or the fastest dynamics/oscillations are not well resolved. For the XY model, one has to resolve the alignment term and the thermal noise. We impose an angle difference of at most $\delta =\pi/10$ (arbitrary value) between two successive time steps, i.e. $\Delta\theta = |\dot\theta|\cdot dt < \delta$.
In the worst case scenario, when each one of the four neighbours produces a unit coupling, the alignment term is of amplitude $4J=4$, giving $dt = \delta =/4$.
Because spins are locally aligned, the situation where the four neighbours of a spin $i$ are perpendicular to $i$ never happens in actual simulations, even close to topological defects. We thus relax this constraint from 1/4 to 1/3 of the maximum value, giving $dt = \delta/3 = 0.1$, so that this alignment term is not pointlessly over-constraining the timestep value.

We now turn to the thermal noise, of the order of $\sqrt{2k_B T}$, which is typically of the order of $0.1 - 1$ in this thesis. To ensure that typical thermal fluctuations are well resolved, we compute the expectation value of the absolute value of the thermal noise:
\begin{equation}
\sqrt{2T\, dt} \, \mathbb{E}|\eta(t)| = \sqrt{2T\, dt} \sqrt{2/\pi} = \sqrt{4T\, dt / \pi} < \delta ,
\end{equation}
leading to $dt < \delta^2 \pi / (4T)$. For typical values of $T=0.1-1$, this gives $dt < 0.08 - 0.8$.
Note that all spins are updated sequentially at every time step.

The final timestep $dt$ is thus the minimum of the two constraints above, i.e.
\begin{equation}
  dt = \min\left(\frac{\pi}{30}, \frac{\delta^2 \pi}{4T}\right) \ .
\end{equation}
Any additional term in the equation of motion, such as the non-reciprocal term in the non-reciprocal XY model, or the intrinsic frequencies in the Kuramoto model, will further constrain $dt$ to ensure that we correctly resolve all the physical phenomena at play. We will add specific details in each chapter.
In a nutshell, throughout this thesis, we use timesteps of the order of $dt \sim 0.01 - 0.1$.

Note that the code, fully written in Julia language, is accessible publicly on GitHub at \href{https://github.com/yrouzaire}{github.com/yrouzaire}. 
Julia is a high-level programming language, similar to Matlab/Python, but usually faster because it is compiled just-in-time (such that the user experience is the one of an interpreted language). Visit \href{www.julialang.org}{julialang.org} to install it.

\section{Numerical implementation of the square and triangular lattices}
In this section, we explain how all the values of the orientation field $\theta_i$ are stored in a square matrix for numerical efficiency, even when simulating a system on a physical triangular lattice. This allows to efficiency access the memory and easily port the code to GPUs.

\begin{figure}[h!]
    \centering
    \includegraphics[width=0.6\linewidth]{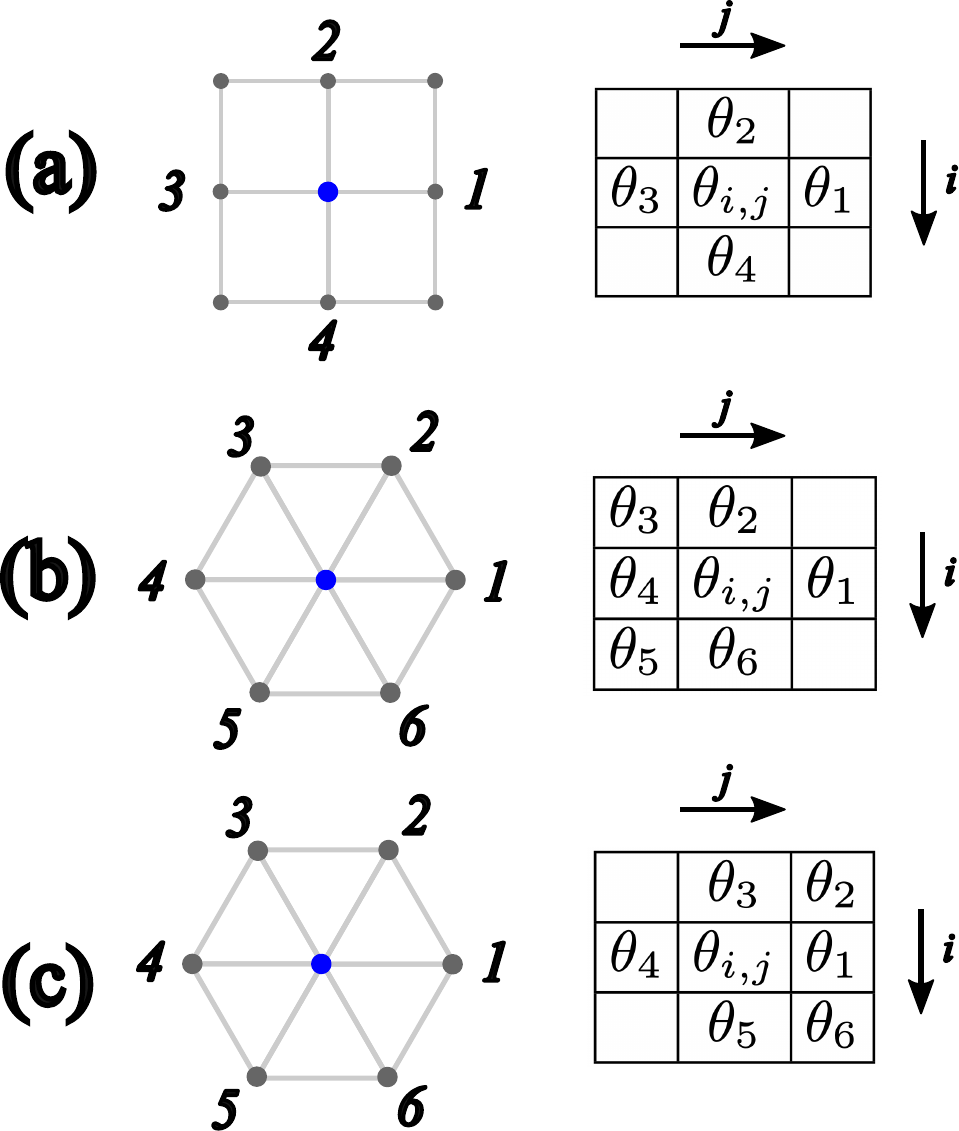}
    \caption{Nearest neighbours structure for \\
    \textbf{(a)} square lattice, for all $i,j$.\\
    \textbf{(b)} triangular lattice, for $i$ even and all $j$.\\
    \textbf{(c)} triangular lattice, for $i$ odd and all $j$.\\
    A triangular lattice can be thought as a square lattice with all odd rows laterally displaced by one half of the lattice spacing.}
    \label{fig:matrix_triangular}
\end{figure}

To do so, we encode the topology of the network in the definition of the nearest neighbours, as shown in Fig.~\ref{fig:matrix_triangular}. The indices $i=1, ..., L$ correspond to the rows and run from top to bottom. The indices $j=1, ..., L$ correspond to the columns and run from left to right. 

\section{Detection of topological defects}

A large part of this thesis relies on the tracking over time of the position (and shape) of the defects, living in the dual space.  
The dual lattice of a square lattice is also square. The dual lattice of a triangular lattice (6 nearest neighbours) is a hexagonal lattice (3 nearest neighbours). We call plaquette the smallest building block of the dual lattice.  

A topological defect is plaquette with a charge (or winding number) $q \neq 0 $, defined by the sum of the phase differences along its four/six bonds 
\begin{equation}
    \sumcirclearrowleft \Delta \theta_{i,j} = 2\pi q 
\end{equation}. 

\begin{figure}[h!]
    \centering
    \includegraphics[width=0.8\linewidth]{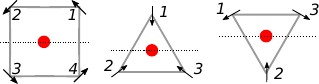}
    \caption{Defects (red) lie at the centre of the dual plaquettes while the spins live on the lattice nodes (grey). Left panel: the unique plaquette of the square lattice. Centre and right panels: the two plaquettes (resp. up and down) of the triangular lattice. Dotted lines are horizontal and pass through the defects; the angles $\theta_j$ and $u_j$ are expressed with respect to that axis in the counter-clockwise direction. }
    \label{fig:plaquettes}
\end{figure}

Once a $q=\pm1$ defect is located, i.e. one knows $(x_{core},y_{core},q)$; we then naturally define $\mu_\pm = \text{Arg}\left\{\sum_j \exp(i(\theta_j - q\,u_j))\right\}$, where the sum runs over the spins $j$ surrounding the plaquette where the defect lies, $i$ is the imaginary unit and Arg is the function that returns the phase of a complex number: Arg$(re^{i\theta}) = \theta$. 
Each neighbour $j$ in that sum makes a certain angle $u_j= \arctan(y_j/x_j)$ with respect to the horizontal axis. Irrespective of its position, we note its orientation $\theta_j$. We sketch this in Fig.~\ref{fig:plaquettes}, where the red dots are the defects cores. For example, for the square lattice (left panel), $u_j = j\pi/4, \ j=1,2,3,4$. For the upward (inner) triangular plaquette (central panel),  $u_{1,2,3} = \pi/2,-5\pi/6,-\pi/6$.
If the defect is located on a downward triangular plaquette (right panel), one has $u_{1,2,3} = 5\pi/6,-\pi/2,\pi/6$.

\section{Creation of a configuration with one or two defects}
To investigate some phenomena, such as the diffusion of a single defect or the annihilation mechanism of a defect pair, it is always simpler to study one or two manually created defects, instead of considering an "\emph{in-vivo}" system where the many defect pairs can influence each other or annihilate, making the tracking over sufficiently long time scales complicated. 
In this section, we describe how to manually create one and two defects in a controlled manner, in particular regarding the desired shape of the defects (useful in the non-reciprocal XY model).  

Creating a configuration with a single defect of charge $q$ and shape $\mu$ at position $(x_0, y_0)$ simply follows from Eq.(\ref{eq:arctan_defect}) : 
\begin{equation}
    \theta(x,y) = q \,\text{arctan}\left(\frac{y-y_0}{x-x_0}\right) + \mu
\end{equation}
Since the total charge in this case is $q\neq0$, it is not possible to enforce periodic boundary conditions to evolve this system. 

\begin{figure}[h!]
    \centering
    \includegraphics[width=0.5\linewidth]{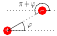}   
    \caption{ 
Sketch of a pair of defects and definition of the polar angle $\varphi$ that the negative defect makes with respect to the positive defect. 
    }
    \label{fig:creation_pair}
\end{figure}

To create a pair of defects, one must be able to create 2 defects with controlled shapes $\mu_+$ and $\mu_-$. For simplicity, we work in polar coordinates, where $\varphi=\arctan(y/x)$ (depicted in Fig.~\ref{fig:creation_pair}), such that the field contribution from a defect reads $\theta = q\varphi + \mu$. 

It follows that if one creates a positive defect with $\mu_1$ and a negative defect with $\mu_2$, both fields will add up. 
At the core of the negative defect, the sum of the two contributions is (each contribution is put between brackets for clarity) 
\begin{equation}
    \mu_- = \left[\mu_1 + (+1)\varphi \right] + \left[\mu_2\right]  = \mu_1 + \mu_2 + \varphi \ .
    \label{eq_app:mu-}
\end{equation}
At the core of the positive defect, the sum of the two contributions is 
\begin{equation}
    \mu_+ =  \left[\mu_1 \right] + \left[(-1)(\varphi+\pi) +\mu_2\right] = \mu_1 + \mu_2 - \pi -\varphi \ .
    \label{eq:mu-}
\end{equation}
Those are the concrete equations we use to create defects with controlled $\mu_+,\mu_-,\varphi$ (only 2 of these 3 parameters are independent), tuning $\mu_1 + \mu_2$ to obtain the desired $\mu_+,\mu_-$.  \\
Also, one can get rid of the artificial $\mu_1,\mu_2$ by combining these equations to obtain: 
\begin{equation}
    \mu_- - \mu_+ =  \pi + 2\varphi \ .
    \label{eq_app:geom_adequation}
\end{equation}
{Since the shapes are equivalent modulo $2\pi$, Eq.(\ref{eq_app:geom_adequation}) can also be written }
\begin{equation}
    \mu_+ - \mu_- =  \pi - 2\varphi \ .
    \label{eq_app:geom_adequation2}
\end{equation}
As it creates the smoothest possible field, this equation represents the state of least distortion. It also highlights that only two of these three parameters are independent.
In this thesis, and without loss of generality, we work with $\varphi=0$: the defects are horizontally aligned and the positive defect is on the left side.

\section{Computation of the spatial correlation function}
We define the spatial correlation function
\begin{equation}
    C(r) = \langle \boldsymbol{S}_i \cdot \boldsymbol{S}_j \rangle 
    = \langle \cos (\theta_i - \theta_j) \rangle 
\end{equation}
where spins $i$ and $j$ are separated by a Euclidian distance $r=|i-j|$. The brackets $\langle \cdot\rangle$ denote an ensemble average over realisations of the thermal noise and over all spins. 

A naive and straighforward implementation implies three nested for loops: two for $1 \le a,b \le L$ scanning all the spins ($L$ is the linear size of the system), and another one to scan the distance $1 \le r \le L/2$. Since it relies on computations in real space, we also refer to it as the "real space method".

Computing directly such a quantity system scales as $\mathcal{O}(L^3)$, which is more than $\mathcal{O}(L^2)$ cost of evolving the system in time. For large systems, it can even become prohibitive if this measurement has to be done many time along the dynamics, for instance to compute the correlation length as a function of time during the coarsening. We will come back to orders of magnitude below in the discussion about Fig.~\ref{fig:XY_times_C_eta_L}. For numerical efficiency, one can compute $C(r)$ in the Fourier space: this reduces the numerical cost to $\mathcal{O}(L^2\,\log L)$. 
First note that, by trigonometric identities, 
\begin{equation}
    \cos (\theta_i - \theta_j) = \cos (\theta_i)\cos (\theta_j) + \sin (\theta_i)\sin(\theta_j)
\end{equation}
The term $\cos (\theta_i)\cos (\theta_j)$ is a convolution and can be easily handled in Fourier space. 

We note the Fourier transform operator $\mathcal{F}$ and its inverse $\mathcal{F}^{-1}$. 
We note $\odot$ the Hadamard element-wise multiplication: if $x$ is a vector of components $x_i$,  then $x\odot y$ is a vector of the same length of components $x_i\cdot y_i$.

With this notation in mind, the $2d$ spatial correlation function can be expressed as 
\begin{equation}
    C_{2d} = C(r_x, r_y) = \left| \mathcal{F}^{-1}\left(\left|\mathcal{F}[\cos \theta] \odot \mathcal{F}[\cos \theta]  + \mathcal{F}[\sin \theta] \odot \mathcal{F}[\sin \theta]\right|  \right) \right|
\end{equation}
In a continuum simulation, $ C_{2d} $ would be a radial function. Here, because of the discrete nature of the underlying lattice, $ C_{2d} $ is slightly anisotropic. To obtain the usual unidimensional $C(r)$, we take the mean of the two principal components (along the real axes $(r_x, 0), (0,r_y)$) and normalise to get $C(r=0)=1$. \\

\begin{figure}[h!]
    \centering
    \includegraphics[width=\linewidth]{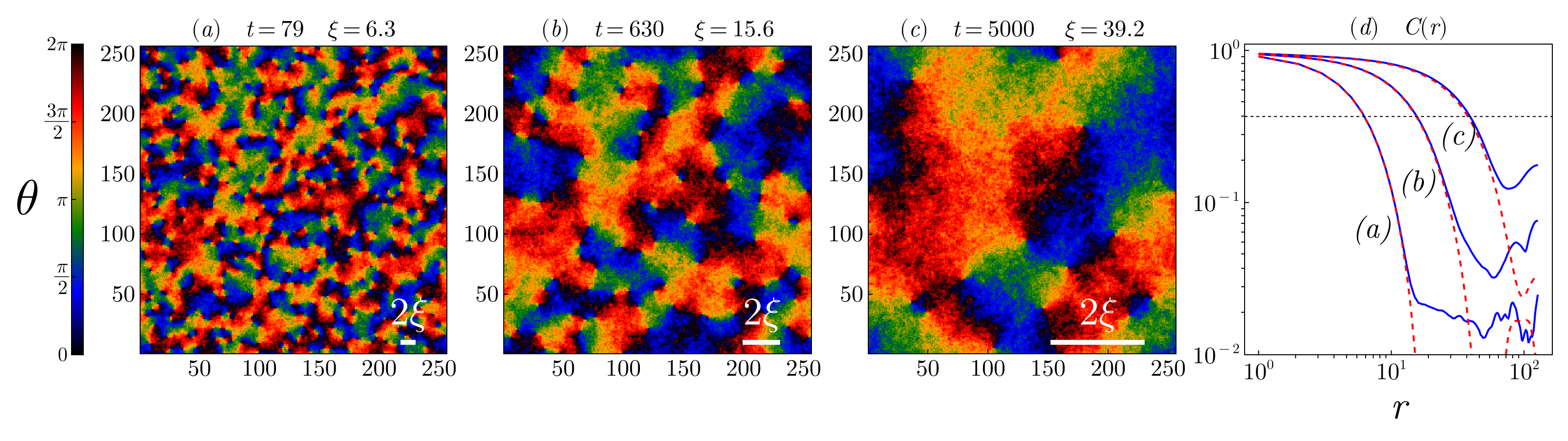}   
    \caption{ \textbf{(a-c)}  Three snapshots of $L=256$ system during the coarsening process, at different times $t$ and spin-spin correlation lengths $\xi$ indicated above. The white bars at the bottom right of each panel represent the length $2\xi$ (with a factor 2 for the sake of readability of panel (a)) \textbf{(d)} Spatial correlation function $C(r)$ for three different times corresponding to the panels (a-c): Naive implementation in red dash and Fourier implementation in solid blue. The horizontal line is $C=1/e$, used to define $\xi$.}
    \label{fig:comparison_2methods_Cr}
\end{figure}

We plot the results of both methods in Fig.~\ref{fig:comparison_2methods_Cr} (naive implementation in red dash and Fourier implementation in solid blue).  
Inspired by the typically exponential decay of the spatial correlation function of the disordered phase, we define the spin-spin correlation length from $C(r)$ as $C(r=\xi) = 1/e$, independently of the method chosen to compute $C$.

The computations in real space and in Fourier space give indistinguishable results for small distances, in particular giving the exact same value of $\xi$. On the other hand, at large distances, the Fourier method is noisier for a reason we do not fully understand yet. This is a problem if one wants to study the large-distance behaviour of $C(r)$, but not if one only wants to extract $\xi$. We now compare the computational time of both methods on CPU, against the runtime for a system of size $L=256$ and $R=32$ independent realisations (a usual number). Those simulations have been performed on a laptop with an Apple M1 Pro chip. For CPU computations and for these system sizes, the memory specs are not relevant.

We present the results in Fig.~\ref{fig:XY_times_C_eta_L}. 
The fact that a single Langevin update of the $R=32$ systems (panel (a), blue crosses) takes as much time as computing $C(r)$ with the Fourier method (green diamonds) is a coincidence. the Figure~\ref{fig:XY_times_C_eta_L} also shows that the Fourier method is very efficient and can be considered as a negligible overhead to the time integration of the system. It also shows that the real space method (orange circles), scaling as $L^3$, is prohibitively slow for large systems.
The ratio of the two times, plotted in panel (b), shows that the Fourier method is much faster than the naive method, by a factor $300$ for $L=256$.
All this discussion is for CPU computations. In contrast with the Fourier method, the naive method is not easily ported to GPU. This makes the Fourier approach even more advantageous. We do not report GPU timings here, since they depend a lot on the specifications of the GPU, in particular its memory caches. One can expect a speedup of typically a factor $10-50$ compared to CPU computations (ie. factor 20 for the spins update). 
Finally, note that copying data from the CPU to the GPU to perform CPU measurements and back is a significant overhead, calling for on-the-fly GPU measurements instead of storing the data and post-processing them, when possible.

\begin{figure}[h!]
    \centering
    \includegraphics[width=\linewidth]{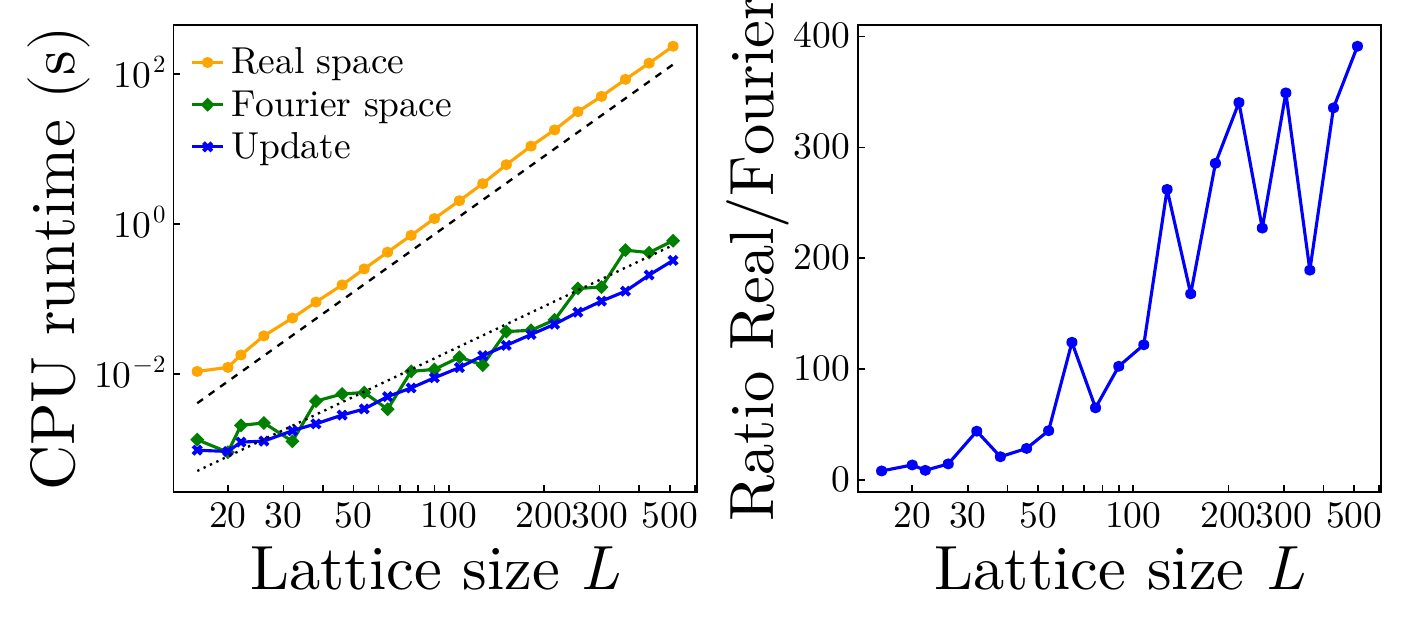}   
    \caption{ \textbf{(a)} Computational time of a single Langevin update of $R=32$ systems (blue crosses), of the computation of $C(r)$ with the Fourier method (green diamonds) and with the real space method (orange circles), as a function of the system size $L$. 
    The dotted line is $2\cdot 10^{-6} L^2$ and the dashed line is $10^{-6} L^3$.
    \textbf{(b)} Ratio of the computational times of the real space and Fourier space methods.
    }
    \label{fig:XY_times_C_eta_L}
\end{figure}



		
\chapter{The steep XY model} \label{ch:steepXY} 
\graphicspath{{steepXY/figures/}}
\graphicspath{{figures/Ch4steepXY/}}

\section{Abstract}

We consider the steep XY model introduced by Domany \textit{et al.} \cite{Domany}. This equilibrium model is an extension of the usual XY model, where the steepness of the potential is controlled by a parameter $p$. It is known that for a steep enough potential, this model exhibits a first order phase transition. However, little is known regarding the physical mechanisms responsible for this change as $p$ increases, nor on the role of topological defects. The critical value $p_c$ at which this change occurs is not known either. In this work, we first give the $p$ dependence of the critical temperature $T_c(p)$ and we show that the steady-state of the low temperature phase, for all $p$, is well explained by the spinwave theory. We then characterise the two kinds of topological defects, namely split XY defects and boundary defects, and introduce an algorithm to discriminate between them. We show that the pointlike, unit-charge defects of the XY model split in two half-integer defects separated by a line with non-zero tension of length $\ell \sim p$.  
Finally, we argue that the estimated critical value $p_c \approx 15$ is in good agreement with a converging body of observations and analogies with other models. 
\newline 

This chapter covers unpublished results (as of today \today). 

\section{Introduction}

Common lore in the theory of phase transitions is that they are classified in universality classes according to  the dimension of space, the dimension of the order parameter, the underlying symmetries of the energy function and whether the range of the interactions is short or long ranged. In two dimensions ($2d$), a model with short-range interactions, a planar order parameter and global rotational symmetry would then belong to the 
Berenzinskii-Kosterlitz-Thouless class, with a continuous phase transition of infinite order driven by the collective behaviour of topological defects \cite{Kosterlitz1973, Kosterlitz1974, berezinskii1971destruction}. 
In particular, the melting of $2d$ solids \cite{strandburg1988two, digregorio2022unified} was believed to fall in the same class of universality and to feature two successive BKT phase transitions. The solid-hexatic transition involves dislocations defects, while the hexatic-liquid transition features disclinations defects. 
This scenario, described by the Kosterlitz-Thouless-Halperin-Nelson-Young (KTHNY) theory
\cite{halperin1978theory, nelson1979dislocation, young1979melting}    
is in sharp contrast with the first order transition of $3d$ solids.
Yet, a large body of molecular dynamics simulations \cite{van1980melting, toxvaerd1981computer, toxvaerd1980phase, abraham1980melting} and experiments \cite{leemann1990percolative}, prior to 1984, indicated a $2d$ melting scenario rather compatible with a first order transition for sufficiently hard interaction potentials. 

This phenomenology reminds of the one of the $q-$Potts model  \cite{wu1982potts}, briefly introduced in Chapter \ref{ch:intro}, which exhibits a change from second to first order phase transition as the number $q$ of different spin species is increased above $q_c=4$.
\newline

\begin{figure}[h!]
 \centering
     \includegraphics[width=\linewidth]{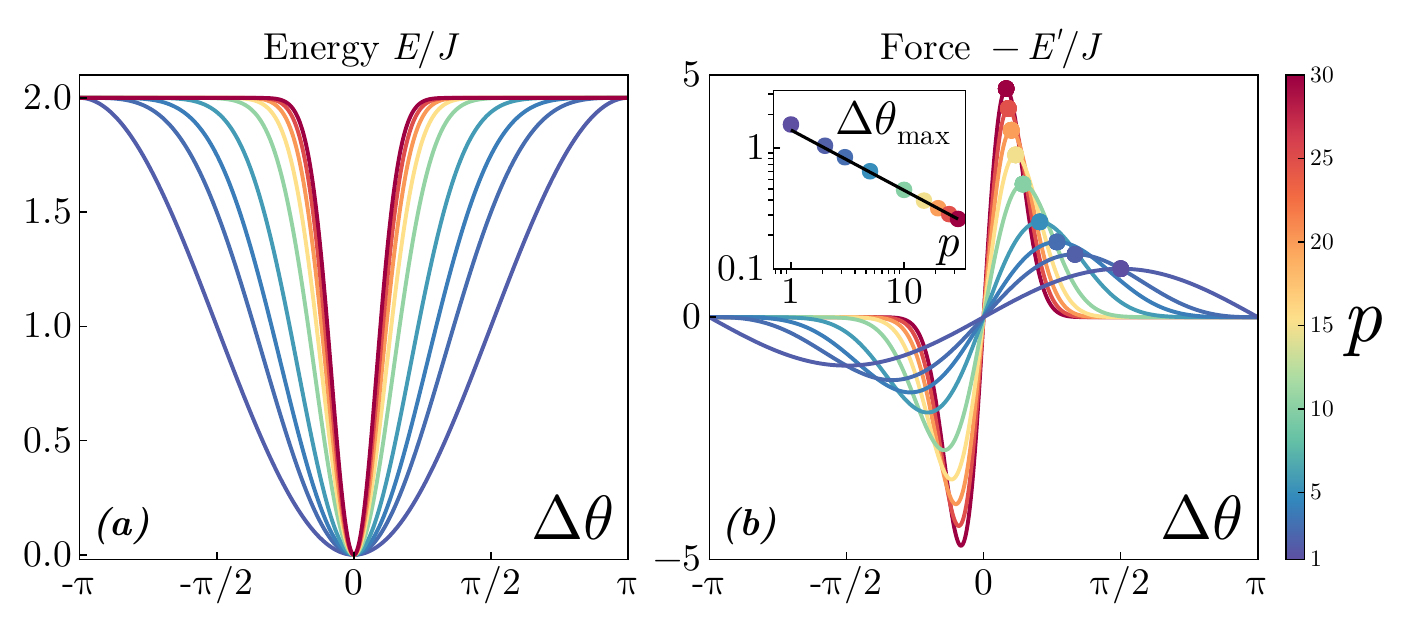}
     \caption{\textbf{(a)} Energy $E/J$ of the steep XY model, for increasing $p$ from blue to red (see colorscale to the right).  
     \textbf{(b)} Force $- \frac{1}{J} \,dE/dx$. The circles represent the maximum of the force profile $f_\text{max} = f(\Delta \theta_\text{max}) $ for each $p$. \underline{Inset:} $\Delta \theta_\text{max}(p)$. The black line is $\sqrt{2/p}$, see main text. 
     }
     \label{fig:pot_force}
 \end{figure}

Domany, Schick and Swendsen suggested in 1984 that the {strong universality} described above does not necessarily hold. Inspired by the Potts model, they introduced a modified version of the standard XY model, retaining all the features of the XY universality class while adding an integer parameter $p$, controlling the steepness of the potential \cite{Domany}.
In this model, which we call the steep XY model, the energy of a bond reads
\begin{equation}
    E(x) =2J \left[1- \left(\cos \frac{x}{2}\right)^{2 p}\right] 
    \label{eq:domany_potential}
\end{equation}
where $J$ is the coupling constant, $x = \Delta \theta = \theta_j-\theta_i $ is the angular difference between 2 neighbouring spins $i$ and $j$. One recovers the standard XY model for  $p=1$, and the potential becomes steeper with $p$, see Fig.~\ref{fig:pot_force}(a). We have changed $p^2\to p$ compared to the original formulation of Domany \textit{et al.} \cite{Domany} to simplify the notation. We convert values into our own convention when referring to articles using Domany's one.

The potential Eq.~(\ref{eq:domany_potential}) exhibits interesting properties as $p$ varies. Let us first examine the derivatives of $E$. 
The sum of the forces imposed by the neighbours $j$ on spin $i$ reads 
\begin{equation}
    \begin{aligned}
-\frac{\partial E}{\partial \theta_i} & =2J \sum_{j} \frac{\partial}{\partial \theta_i}\left(\cos \frac{\theta_j-\theta_i}{2}\right)^{2 p} \\
& =4J p \sum_{j}\left(\cos \frac{x }{2}\right)^{2 p-1} \left(-\sin \frac{x }{2}\right)\cdot\left(-\frac{1}{2}\right)  \\
& =2Jp \sum_{j}\left(\cos \frac{x }{2}\right)^{2 p-1} \sin \frac{x }{2} \ \ \ \ (*)\\
& =2Jp \sum_{j}\left(\cos \frac{x }{2}\right)^{2 p-2}\left(\cos \frac{x }{2}\right) \sin \frac{x }{2} \\
& =Jp \sum_{j}\left(\cos \frac{x }{2}\right)^{2 p-2} \sin (x)  \ . \ \ \ (**)
\end{aligned}
\label{eq:force}
\end{equation}
Importantly enough, the coupling constant $J$ of the classical XY model becomes $Jp$ in the steep XY model. In Eqs.~(\ref{eq:force}), we have derived two equivalent equations, marked with $(*)$ and $(**)$. 
While Eq.~$(**)$ makes a clearer analogy with the usual $\sin x$ alignment term of the XY model, Eq.~$(*)$ is 10\% faster to numerically evaluate, because the same variable $x/2$ appears twice, allowing for optimization. 
We plot the force profile in Fig.~\ref{fig:pot_force} (b). 
\newline 

We now compute the second derivative
\begin{equation}
\frac{\partial^2 E}{\partial x^2}= Jp \cos\left(\frac{x}{2}\right)^{2p-2} 
\big(  
1-p + p \cos x
\big)
\label{eq:second_derivative}
\end{equation}
and solve $\partial^2E/\partial x^2  =0$ to obtain the location of the maximum force:  
\begin{equation}
    \Delta \theta_{\text{max}} = \cos^{-1}\left( 1 -p^{-1}\right) \sim \sqrt{\frac{2}{p}} + \mathcal{O}(p^{-3/2}) \ .
    \label{eq:theta_c}
\end{equation}
The scaling $\Delta \theta_{\text{max}}\sim \sqrt{2/p}$ comes from the Puiseux series. It also naturally appears in the small angle expansion of $E$, using $\cos x \approx 1 - x^2/2$ and $\log (1+x) \approx x$: 
\begin{equation}
  \cos^{2p} \left(\frac{x}{2}\right) 
\approx \left(1 - \frac{x^2}{8}\right)^{2p}   
\approx  e^{2p \, \log\left(1 - \frac{x^2}{8}\right)}  
\approx e^{-\frac{1}{2}\left(\frac{x}{\sqrt{2/p}}\right)^2 }  \ .
\end{equation}
In the XY model, the force is non-zero for all $0<|\Delta \theta|<\pi$, which means that even with a large angle difference, two neighbouring spins do interact. On the contrary, in the steep XY model for large $p$, 
two neighbouring spins with $|\Delta \theta| > 2\Delta \theta_\text{max}$ will effectively not interact, see Fig.~\ref{fig:pot_force}(b),
like two different spin species in the $q-$Potts model. 
Following this analogy, we consider that one species covers $4\Delta \theta_\text{max}$.  The number of ``spin species" $q$ therefore reads
\begin{equation}
   q =  \frac{2\pi}{4\Delta \theta_\text{max}} = \frac{\pi}{2\cos^{-1}\left( 1 -p^{-1}\right)}  \ .
   \label{eq:potts_q}
\end{equation}
Solving for $p$, one obtains
\begin{equation}
    p =\frac{1}{1 - \cos(\pi/ 2q)}  \ .
   \label{eq:potts_p}
\end{equation}
In the $q-$Potts model, the order-disorder phase transition is second order for $q\le q_c = 4$  and first-order for $q> q_c$. \er{eq:potts_p} translates it into $ p> 13.2$. Because $p$ is an integer, one has $ p_c = 14$. 
We will come back to this analogy later. 
\newline


At low temperatures, we expect angle differences $x$ between two neighbours to be small. In this limit, one can expand the cosine and use $(1+x)^a = 1+ax$ to show that $E$ becomes quadratic
\begin{equation}
    \begin{aligned}
E = 2J\left[ 1 - \cos^{2p} \left(\frac{x}{2}\right) \right] 
&\approx 2J\left[ 1 - \left(1 - \frac{x^2}{8}\right)^{2p} \right]  \\
&\approx 2J\left[ 1 - \left(1 - 2p\frac{x^2}{8}\right) \right]  \\
&= \frac{1}{2}\,Jp\,x^2  \\
\end{aligned}
\end{equation}
highlighting the harmonic nature of the potential close to 0. Importantly, the curvature in $x=0$ is $Jp$, as one could have derived from Eq.~(\ref{eq:second_derivative}) . 
\newline

The numerical results of Domany \textit{et al.} \cite{Domany} show that the transition becomes first order for large $p$. 
For $p=50$, the steep XY model exhibits a clear hysteresis for temperatures $1 < T/J < 1.005$. In this very narrow range, they observe a clearly bimodal energy probability distribution, implying a non-zero latent heat, a hallmark of phase coexistence. This phase coexistence, itself a signature of a first order transition, has been observed in previous numerical simulations \cite{toxvaerd1980phase} and rigorously proven by van Enter and Shlosman in 2002~\cite{vanEnter1,vanEnter2} using the method of Reflection Positivity~\cite{Shlosman} and Chessboard Estimates on square lattices. Concretely, they proved the existence of a temperature at which two different Gibbs states exist and have the same free energy. However, the mathematical proof does not provide any information about the nature of the low temperature phase, nor it gives the specific value $p_c$ at which the order of the phase transition changes. 
In systems of Burgers vectors, numerical simulations suggested, somehow counter-intuitively, that the transition could change from infinite to first order as the core energy\footnote{The core energy of a defect is the energy necessary to create it, regarless of the energy needed to sustain the distortion it imposes around itself.} is lowered~\cite{saito1982_PRL, saito1982_PRB}. It also appear that, if the potential is made steeper, loops of dislocations appear and populate the system like grain boundaries, a mechanism suggestive of first order transitions~\cite{saito1982_PRL}. 
\newline

Renewed interest in this problem arose recently, as it is now known that the solid-hexatic transition indeed falls in the BKT universality class while first order hexatic-liquid transitions have been found with massive numerical simulations for passive~\cite{BernardKrauth} and weakly active~\cite{PRLino} hard sphere systems, see also the discussion in the introduction of \cite{digregorio2022unified} and the references [22-29] therein. 

To the best of our knowledge, the physical mechanisms responsible for this drastic change in the order of the phase transition of the steep XY model remain elusive. 
In particular, the literature on the fate of the topological defects \textit{à la} XY, key structures for the XY model to fall in the BKT universality class, is very scarce \cite{mila1993first, SinhaKumar}. 
In this work, we first give the $p$ dependence of the critical temperature $T_c(p)$ and we show that the steady-state of the low temperature phase, for all $p$, is well described by the usual spinwave theory. We then characterise the two kinds of topological defects in this model, namely split XY defects and boundary defects, and introduce an algorithm to discriminate between them. We show that the pointlike, unit-charge defects of the XY model split in two half-integer defects separated by a line with non-zero tension of length $\ell \sim p$.  
Finally, we argue that the estimated critical value $p_c \approx 15$ is in good agreement with a converging body of observations and analogies with other models. 
\newline

To study the steep XY model numerically, we need to prescribe ourselves a dynamics.
As for all the chapters of this thesis, we will use the Langevin dynamics framework. The equation of motion thus reads 
\begin{equation}
\begin{aligned}
\gamma \frac{d \theta_i}{d \tilde t}
&=Jp \sum_{j}\left(\cos \frac{\theta_j-\theta_i }{2}\right)^{2 p-2} \sin (\theta_j-\theta_i)  +\sqrt{2\gamma \,k_B\tilde T\,}\ \eta({\tilde t}) \\
\end{aligned}
\label{eq:langevin}
 \end{equation}
 where $\gamma$ is the friction coefficient, $k_B$ is the Boltzmann constant, $\tilde T$ is the temperature of the thermal bath and $\eta$ is a Gaussian white noise, with zero mean and unit delta correlated variance 
 \begin{equation}
     \langle \eta_i(t)\ \eta_j(t')\rangle = \delta_{ij}\,\delta(t-t')
 \end{equation}

The angle $\theta$ has no dimensions.   
The white noise $\eta$ has units of $1/\sqrt{\text{time}}$, because its variance is delta correlated and the delta function obeys $\delta(ax) = \delta(x)/|a|$ . 
 
The dimension equation thus reads   
  \begin{equation}
\frac{\gamma}{\text{time}} =Jp +\sqrt{\frac{\gamma \,k_B\tilde T\,}{\text{time}}}\ .
 \end{equation}
One can extract two time scales: $\gamma/k_B\tilde T$ is associated to the diffusion time scale and
\begin{equation}
    \tau = \gamma/Jp 
\end{equation}
is the time scale associated to synchronisation between neighbours.
One can divide Eq.~(\ref{eq:langevin}) by $\tau$ :  
\begin{equation}
\frac{d \theta_i}{d t}
=\sum_{j}\left(\cos \frac{\theta_j-\theta_i }{2}\right)^{2 p-2} \sin (\theta_j-\theta_i)  +\sqrt{\frac{2 T}{p}\,}\ \eta({t})
\label{eq:langevin_divided_by_tau}
 \end{equation}
 where $t = \tilde t/\tau$ and $T = k_B\tilde T/J$. We thus express time in units of $\tau$, and temperatures $T$ in units of $J/k_B$, and energies in units of $J$ as in the rest of this thesis. 
Replacing the deterministic part of the force Eq.~(\ref{eq:force}$**)$ by Eq.~(\ref{eq:force}$*)$, one obtains the equation of motion we actually integrate in time with the Euler-Maruyama method, with a time step $dt = 0.1$~: 

\begin{equation}
\frac{d \theta_i}{d  t}
=2\sum_{j}\left(\cos \frac{\theta_j-\theta_i }{2}\right)^{2 p-1} \sin \left(\frac{\theta_j-\theta_i }{2}\right)  +\sqrt{\frac{2 T}{p}\,}\ \eta({ t}) \ .
\label{eq:langevin_in_code}
 \end{equation}
\newline


\section{Order-disorder phase transition}
\begin{figure}
    \centering
    \includegraphics[width=0.95\linewidth]{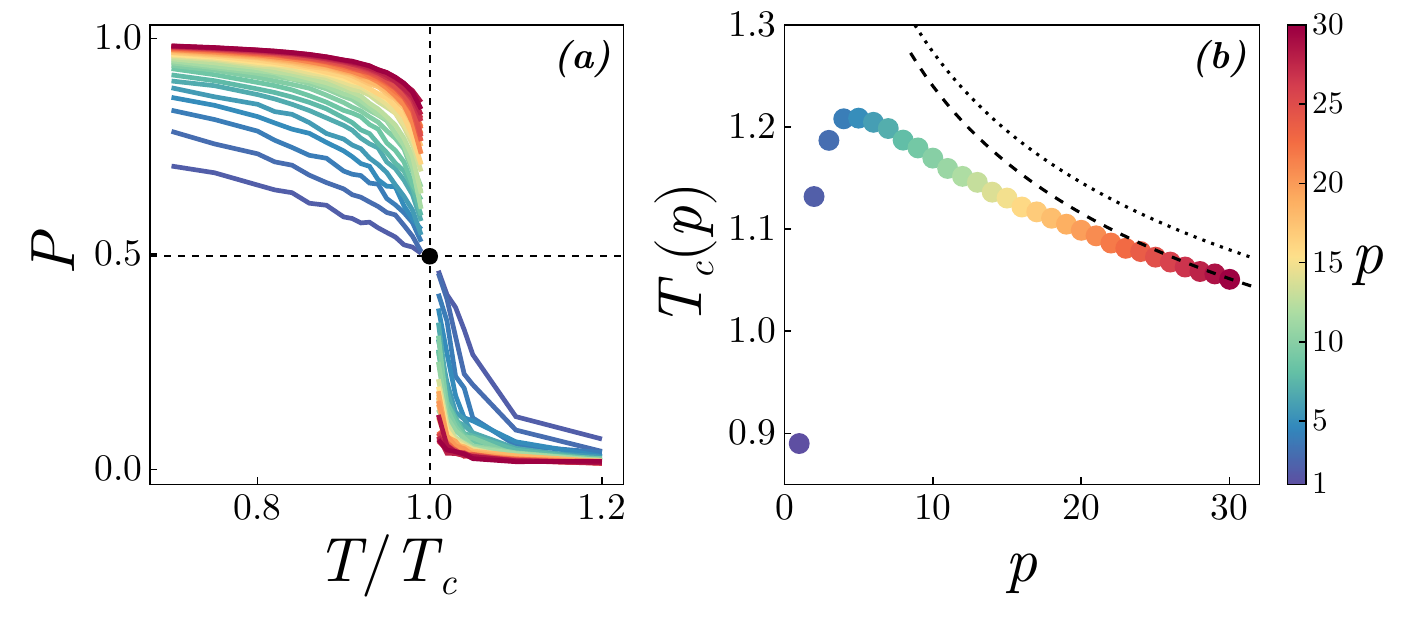}       
    \caption{\textbf{(a)} $L=128$, $P$ is averaged over $32$ independent realizations of the thermal noise. No hysteresis is observed and random initial configurations give noisier results, so we start from an ordered configuration. The central point $T=T_c\,,\, P = 0.495 $ is common to all curves, by definition of $T_c$. We have removed on purpose the lines connecting to the central point to highlight the increasingly sharp jump between the low and high temperature phases as $p$ increases. \textbf{(b)} The critical temperature, estimated from $P(T_c) = 0.495$, as a function of $p$. The dash line is $0.653\,T_{c,\text{Potts}}$, the dotted line is $0.6705\,T_{c,\text{Potts}}$, see main text and Eq.~(\ref{eq:Potts_Tc_q}). 
}
    \label{fig:P_T_Tc}
\end{figure}

For all $p$, the system undergoes an order-disorder phase transition at a critical temperature $T_c(p)$. We first define it using the polar order parameter $P$.  
For the XY model ($p=1$), we know that $T_c = T_\text{KT}$ ($T_\text{KT}=0.89$ on the square lattice). 
Due to the spin wave excitations, the polarization vanishes at all temperatures in the thermodynamic limit: 
$\lim \limits_{L\to\infty} P = 0$ . 
In reality, finite size effects are strong and one observes finite values at low temperatures~\cite{bramwell2001magnetic, palma2002finite}. 
As seen in Fig.~\ref{fig:P_T_Tc}(a), in the low temperature regime, the polarization $P$ increases with $p$ whereas in the high temperature regime, it decreases with $p$. This stems directly from the change of the force between two neighbours, cf. Fig.~\ref{fig:pot_force}. Indeed, for large $p$, two spins with a small angle difference are tightly coupled together, and the noise is comparatively less important than the force term. On the other hand, if $|\Delta \theta| \gtrapprox 2\Delta \theta_\text{max}$, the alignment force is almost zero and the dynamics is only ruled by white noise.  These two trends are amplified with increasing $p$, such that the jump in polarization between the two phases increases, suggestive of a first order transition.  

For small $p$, the polarisation, as a function of temperature, seems continuous.
In particular, for $L=128$ and $p=1$, one finds $P(T_\text{KT}) =  0.495$. For $p\ge1$, we define, by linear interpolation,  $T_c(p)$ as the temperature for which a $L=128$ system has an average polarisation $P(T_c(p)) =  0.495$. We report the results in Fig.~\ref{fig:P_T_Tc}(b). For small $p$, the critical temperature increases with $p$, consistent with the transformation of the coupling constant $J\to Jp$ from the XY model to the steep XY model. Around $p=5$, $T_c$ reaches its maximum before slowly decreasing. 
Domany \textit{et al.} reported that $T_c(p=50) = 1.01$  and $T_c(p=100) = 0.91$. One can now compare our results with the prediction based on the analogy with the $2d$ Potts model, for which the critical temperature is 
\begin{equation}
T_{c, \text{Potts}}(q) = \frac{2}{\log \Big(1 + \sqrt{q}\Big)} \ . 
    \label{eq:Potts_Tc_q}
\end{equation}
Using Eq.~(\ref{eq:potts_q}) to express $q$ as a function of $p$, one obtains
\begin{equation}
T_{c, \text{Potts}}(p) = \frac{2}{\log \Big(1 + \sqrt{\frac{\pi}{2\cos^{-1}(1 - 1/p)}}\Big)} \ . 
    \label{eq:Potts_Tc_p}
\end{equation}
The numerical values given by Eq.~(\ref{eq:Potts_Tc_p}) are much larger than the ones we obtained numerically or reported by \cite{Domany}: for instance, $T_{c, \text{Potts}}(30) = 1.61$ and $T_{c, \text{Potts}}(50) = 1.50$. We thus introduce a multiplicative factor to fit the data at large $p$, see the dashed line plotting $0.6705\,T_{c, \text{Potts}}(p)$
in Fig.~\ref{fig:P_T_Tc}(b). 
The decreasing trend is similar but not identical, neither quantitatively, nor qualitatively. It would appear that the critical temperature cannot be extrapolated from one model to the other, even though the steep XY model was originally designed to interpolate between the usual XY model and the Potts model as $p$ is increased \cite{Domany}.

\begin{figure}
    \centering
    \includegraphics[width=0.99\linewidth]{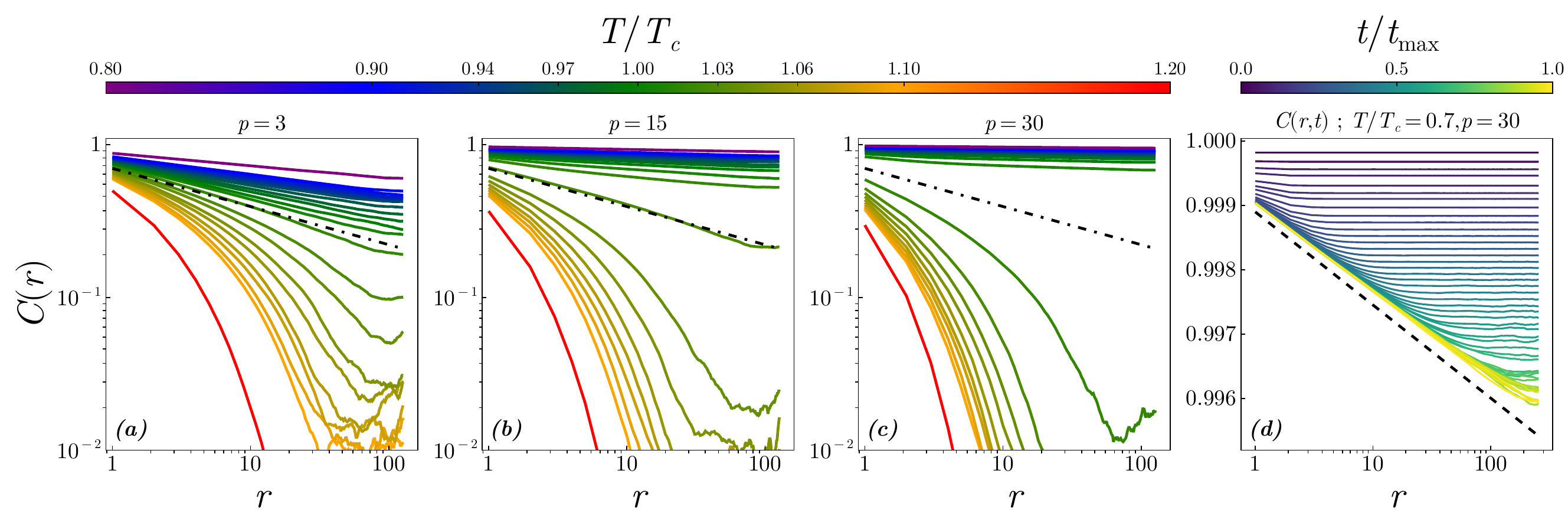}       
    \caption{Steady-state spatial correlation functions $C(r)$ for $L=256$, averaged over 16 realisations (ordered initial configuration) and \textbf{(a)} $p=3$, \textbf{(b)} $p=15$,  \textbf{(c)} $p=30$. Dashdotted line is $r^{-1/4}$, the power law decay of the XY model at $T=T_c$. Different colours are different temperatures, see color bar in the middle. \textbf{(d)} In log-log scale,  $C(r, t)$ for $L=512$, averaged over 16 realisations  (ordered initial configuration) and $T/T_c = 0.1, p=30$. Different colours represent different times, from $t=0$ to $t_\text{max} = 10^6$. The dashed line represents a $r^{-0.00063}$ decay. The exponent predicted by the spinwave approximation is  $\eta = T/(2\pi p) = 0.00053$.
    }
    \label{fig:Cr_across_Tc} 
\end{figure}

In the XY model, the KT transition is characterised by a drastic change in the spatial correlation functions $C(r)$, from algebraic to exponential.
Therefore, we now study $C(r)$ to confirm the results above based on $P$.  The XY scenario is recovered for $p=3, 15, 30$, see Fig.~\ref{fig:Cr_across_Tc}(a-c). We used the same ratios $T/T_c$ for all $p$, coloured from purple (low $T$) to red (high $T$).  The separation between the correlation functions in the low $T$  and the high $T$ phase increases with $p$, a behaviour consistent with the separation of the polarisation $P$ in both phases, as seen in Fig.~\ref{fig:P_T_Tc}. Beyond this qualitative difference, there seems to be no abrupt change in behaviour that could indicate a critical $p_c$ between first order and BKT transition.  

\begin{figure}
    \centering
    \includegraphics[width=0.85\linewidth]{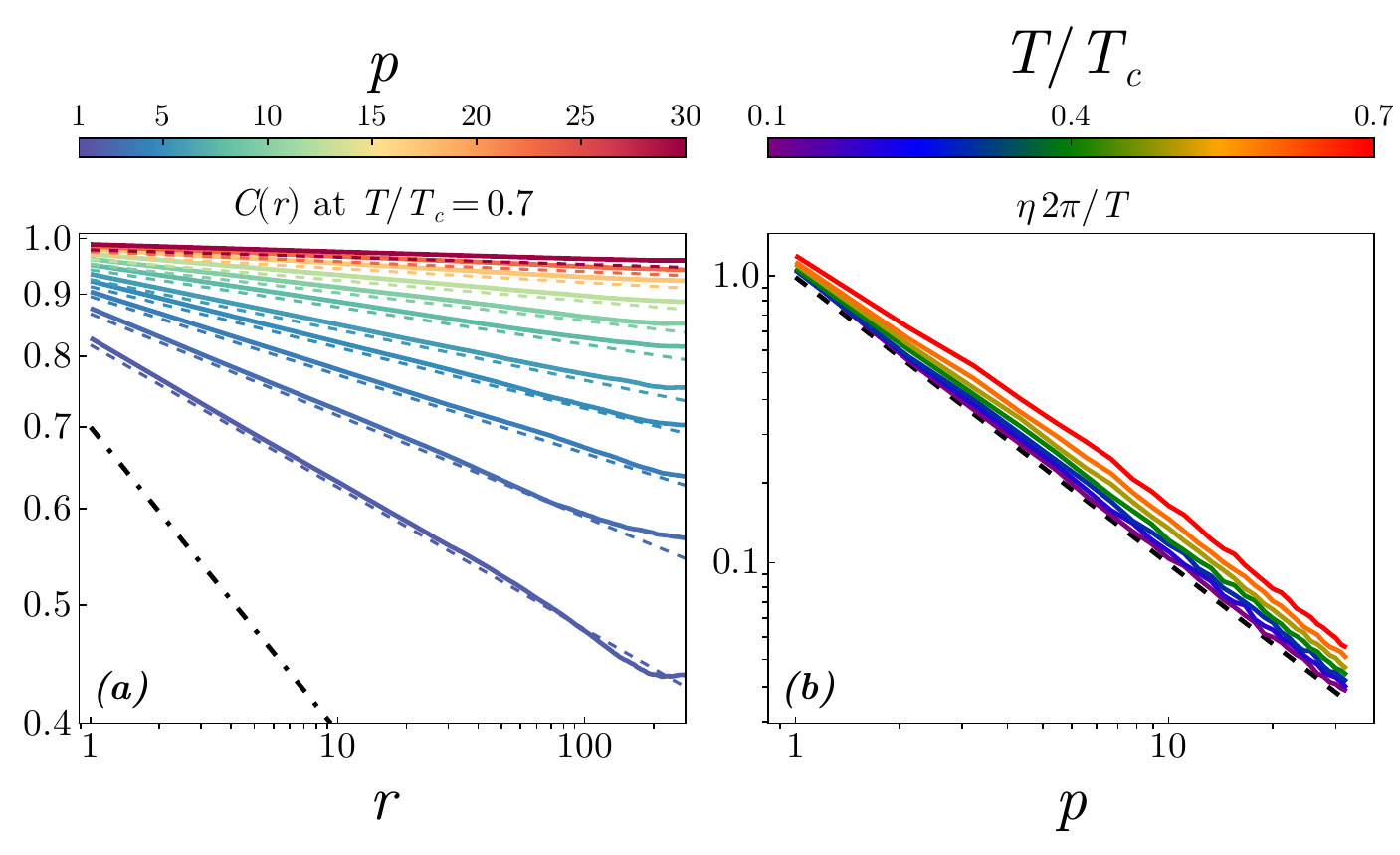}       
    \caption{\small $L=512, R = 16$ (ordered initial configuration). \textbf{(a)} $C(r)$ at $T/T_c = 0.7$.
    Dashed lines are power law of the form $a\,r^{-\eta}$, where $a$ and $\eta$ are fitted to the curves for $1\le r \le 100$ (the straight part only). The fit curves are shifted down by 1\% on purpose for better visualization: we plot $0.99a\,r^{-\eta}$.  \textbf{(b)} The value of the exponent $\eta$ fitted in the left panel. The dashed line is $1/p$. 
    }
    \label{fig:Cr_low_temp}
\end{figure}

The data confirm the criticality of the low temperature phase for $p=3, 15, 30$. We show in Fig.~\ref{fig:Cr_across_Tc}(d) the time evolution $C(r, t)$ for $T/T_c = 0.7, p=30$, from flat (ordered initial condition) to a power law with a very small exponent (dashed line is $r^{-\eta}$, with $\eta = 6.3\cdot 10^{-4}$), showing that the system has indeed reached its steady-state. 
We now place ourselves in the low temperature regime $0.1 \le T/T_c \le 0.7$ and study more carefully the decay of the steady state correlation functions to determine the nature of this phase and discriminate between a long-range order (LRO) and a quasi-long-range order (QLRO).   
In Fig.~\ref{fig:Cr_low_temp}(a), the $C(r)$ clearly appear to be power laws $r^{-\eta}$. We fit them to extract the decay exponent $\eta(p, T)$. In the XY model, the spinwave theory predicts $\eta = T/(2\pi J)$ and relies on the quadratic part of the potential for small angles. In the steep XY model, the potential remains quadratic at all $p$ the effective coupling (=the curvature) is $Jp$, such that we expect $\eta = T/(2\pi Jp)$.
We test this prediction by plotting $2\pi\,\eta /T$ against $p$ (with $T$ measure in units of $k_B/J$) in Fig.~\ref{fig:Cr_low_temp}(b). The data clearly follow $1/p$ (dashed lines) for all temperatures, indicating that the spinwave theory extends to $p>1$, with, once again, no qualitative difference between small and large $p$.  

\section{Topological defects}
In the classical XY model, the order-disorder phase transition is understood in terms of the collective behaviour of topological defects. As in the XY case, the invariance of Eq.~(\ref{eq:langevin}) under parity $\theta_i \to -\theta_i$ and global rotation $\theta_i \to \theta_i + \theta_0$ ensures that the defects dynamics is independent of their charge and shape.  
In the XY model, defects have a topological charge $q = \pm 1$. Indeed, defects with $|q|>1$ are not stable and rapidly decay to multiple $|q|=1$ charges. Fractional charges $|q|<1$ (such as the half-integer charges $|q|=1/2$ of the nematic liquid crystals) never appear as they impose a jump in the close orientation field, which is not energetically favourable. 

\begin{figure}
    \begin{center}  
    \includegraphics[width=\linewidth]{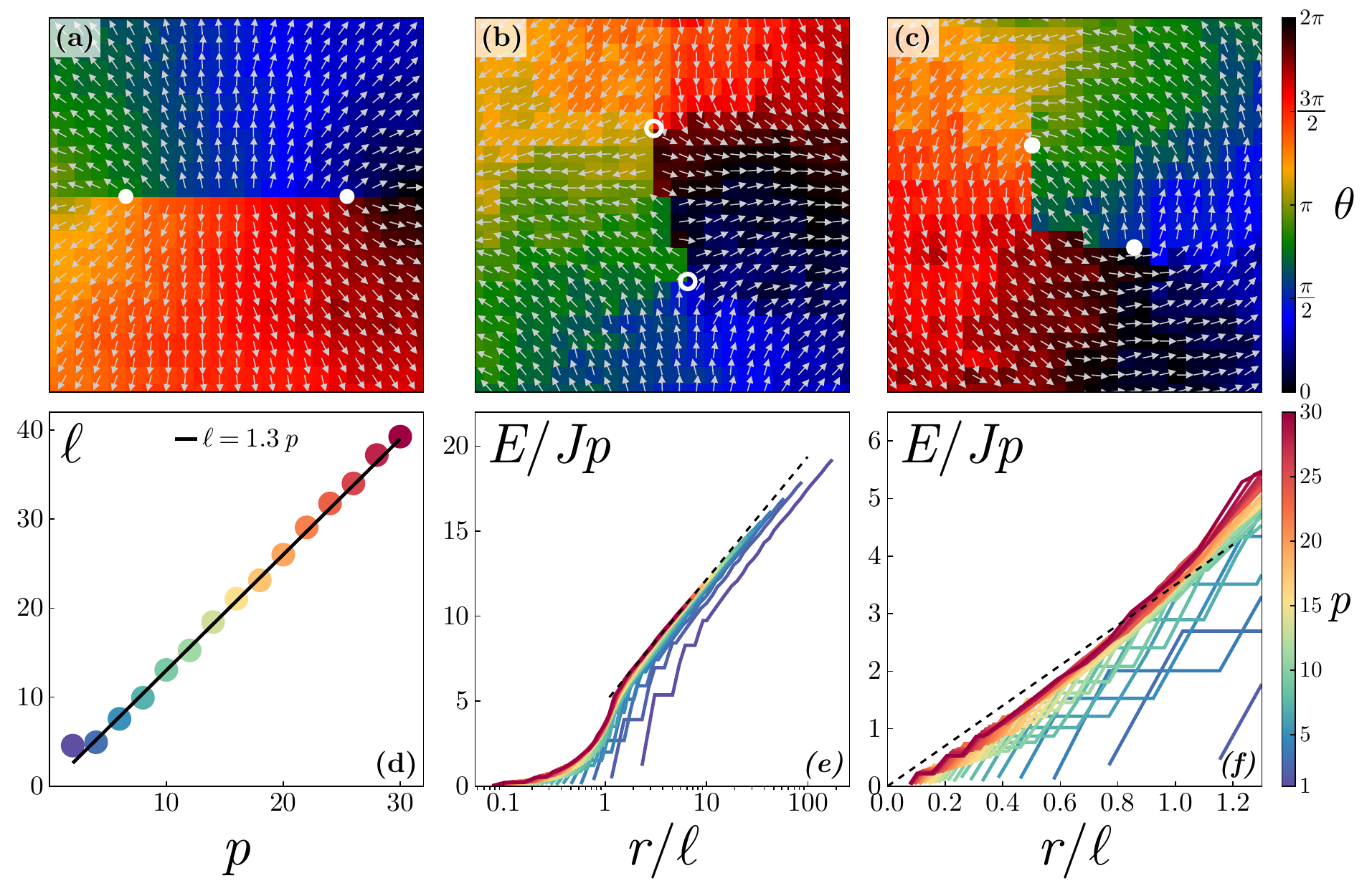}
    \caption{ 
      \textbf{(a-c)} Orientation field around defect pairs for $p=10$ (zooms on regions of a larger system). The $+1/2$ (resp.  $-1/2$) defects are represented with filled (resp. empty) circles. $\ell$ is the line length in units of the lattice spacing. The color scale on the right corresponds to the orientation of the spins, also indicated by the arrows. (a) total charge $+1$, $T/T_c = 0.1, \ell = 13$ (b) total charge $-1$, $T/T_c = 0.4, \ell = 11$  (c) total charge $+1$, $T/T_c = 0.4, \ell = 13$. 
    \textbf{(d)} Average line length against $p$. The black line is $1.3p$.    
    \textbf{(e-f)} The energy measured in the simulations, as a function of the distance $r$ from the center of the two defects, for different $p$ (see colorbar on the bottom right). (e) Note the log x-axis. The dash line is $\pi\log(r/\ell) + 6$. (f) zoom on the small $r/\ell$ regime. Note the linear x-axis. The dash line is $3.5(r/\ell)$.
    }
    \label{fig:illustration_XYlike_defect}
    \end{center}
\end{figure}

The picture completely changes for $p>1$, even though the non-linearity $\sin \Delta\theta$ in the main Eq.~(\ref{eq:langevin}) remains polar (as opposed to a nematic alignment $\sin (2\Delta\theta)$). 
Instead of being point-like, topological defects in the steep XY model (for all $p$) split into two equal half-integer charges, connected by a line of length $\ell$, as illustrated in Fig.~\ref{fig:illustration_XYlike_defect}(a). At both ends of this line, the orientation field points in opposite directions: two neighbouring spins along the line thus have an angular difference $\varepsilon = \pi/\ell $. 
Two simple arguments predict that $\ell \sim \sqrt{p}$.  For large enough $\ell$, $\varepsilon$ is small and the potential can be linearized for small angles: $E = \frac{1}{2}Jp \,\varepsilon^2 \sim Jp /\ell^{2}$. In this limit,  the equipartition theorem applies and the average energy per bond is $k_BT$, a constant independent of $p$. Thus, $Jp /\ell^{2} \sim \text{cst}$, leading to $\ell \sim \sqrt{p}$. This prediction is confirmed by another argument: the spins along the line have to interact with their neighbours for the line to be a unique, coherent structure. This implies $\varepsilon= \pi/\ell  < \Delta\theta_{\text{max}} \sim 1/\sqrt{p}$, and thus  $\ell \sim \sqrt{p}$.

We measure the length $\ell$ of this line as follows. First, we manually create a single isolated XY defect of charge $q$ at the center of a large box of linear size $L=300$. Because the model is invariant under parity and global rotation, we can focus on $q=+1$ defects with a phase $\theta_0 =0$. We let the system relax at low temperature $T/T_c(p) = 0.1 $, with free boundary conditions. The $+1$ charge spontaneously splits into two $+1/2$ charges, that progressively move apart from each other. We wait for the steady state (at this low temperature, the composite object remains close to the center and does not approach the boundaries). 
We then count the number $m$ of spins that have at least one neighbour $j$ with $|\theta_i -\theta_j| > \Delta\theta_\text{max}$. Those specific spins surround the line between the two defects, such that $m\approx2\ell$. This procedure is robust to changes of the chosen threshold: one obtains the exact same results using $\Delta\theta_\text{max}(p)$, $2\Delta\theta_\text{max}$ or a fixed value (eg. $\pi/8$). Altogether, the measurement of $\ell$ indicates that 
\begin{equation}
    \ell \sim 1.3\,p \ ,
\end{equation}
see Fig.~\ref{fig:illustration_XYlike_defect}(d). 
This discrepancy between data ($\ell \sim p$) and simple scaling arguments ($\ell \sim \sqrt{p}$) calls for a more thorough analysis.
We thus carefully list the different contributions to the energy $E(\ell)$ of the structure illustrated in Fig.~\ref{fig:illustration_XYlike_defect}(a).  
The previous scaling arguments only considered the (horizontal) interaction between spins  with angular difference $\varepsilon$. The associated cost, stemming from the linearized region of the potential integrated over both sides of the line of length $\ell$, reads 
\begin{equation}
    E_{elastic} = 2\ell \, \frac{1}{2}Jp \,\varepsilon^2 = \frac{\pi^2\, Jp}{\ell} \ .
\end{equation}
$E_{elastic}$ is minimized for $\ell \to \infty$, and thus does not correctly describe our object. 
The line connecting the two defects separates two distinct regions (blue and red in Fig.~\ref{fig:illustration_XYlike_defect}(a)). The bottom blue row interacts with the top red row.  In contrast with the first interaction between spins along one side of the boundary that lead to $E_{elastic}$, those angular differences are large and lie beyond the critical angle $\Delta\theta_{\text{max}}$, such that the energy per bond is of the order of the energy plateau value $2J$, see Fig.~\ref{fig:pot_force}(a):
\begin{equation}
    E_{plateau} = 2J\ell \ .
\end{equation}
Imposing 
\begin{equation}
    \frac{\partial\ (E_{elastic} + E_{plateau})}{\partial \ell} = 0
\end{equation}
to find the length $\ell$ of the line minimizing its energy once again implies $\ell \sim \sqrt{p}$, like the simple scaling arguments, in contrast with the numerical measurement. The missing ingredient is the interaction energy between the two topological defects. Because they have the same charge, they repel each other and thereby tend to increase $\ell$. If one assumes the usual logarithmic interaction of the XY model, but now with a coupling constant $Jp$, one has 
\begin{equation}
    E_{defect} = 2 E_{core} - \frac{\pi\,Jp}{4} \log \left(\frac{\ell}{a}\right)  
\end{equation}
where $a$ is the lattice spacing, and $\frac{1}{2}$ is the charge of each defect and $E_{core}$ its core energy. The minus sign of the second term reflects the repulsion between same sign charges. We define 
\begin{equation}
\begin{aligned}
    E_{0}  &=  E_{plateau}  + E_{defect} + E_{elastic} \\ 
                  &= 2J \ell  \left[ - \frac{\pi\,Jp}{4} \log \left(\frac{\ell}{a}\right)  + 2 E_{core} \right] + \frac{\pi^2\, Jp}{\ell} \\ 
                  &\equiv A\,\ell -B\,\log \ell + \frac{C}{\ell} + \text{ cst} \\ 
\end{aligned}
\end{equation}
Imposing $\partial E_{0}/\partial \ell = 0$ and multiplying by $\ell^2$, one finds a unique positive solution 
\begin{equation}
    \ell = \frac{\pi\,p}{16}\left( 1 + \sqrt{1+\frac{64}{p}}\right)
    \label{eq:ell}
\end{equation}
which, for large $p$, indeed predicts the correct scaling $\ell \sim p$ and implies, at leading order in $p$, that $E_0(\ell) \sim p$. To obtain the scaling  $\ell \sim p$, one can neglect $E_{elastic} $ but $E_{defect}$ is essential. 

Each elbow in the line, see Fig.~\ref{fig:illustration_XYlike_defect}(b), costs an extra energy $2J$. Those configurations are less probable and in fact rarely observed at low temperatures, such that the line connecting two defects preferentially follows the lattice axes. It provides a preferred direction of motion of the defect pair. Indeed, moving perpendicular to the line connecting the two defects would imply a drastic rearrangement of $\mathcal O(\ell)$ spins, which makes it very unlikely in a dissipative medium. Instead, moving in the direction of the line only requires $\mathcal O(1)$ spin flips. 

Each elbow does cost energy
but contributes to the entropy $S$ because it increases the number of possible configurations $W$.
For instance, there exist
\begin{equation}
    W = \dbinom \ell 1 = \ell
\end{equation}
configurations with a unique elbow along the line of length $\ell$. 
Working in terms of the free energy $F = E - TS= E - T\log W$ thus rationalises why, at higher temperatures, the line connecting the two defects often contains elbows. Those effectively bend the axis and orient it in any direction, decoupling it from the underlying lattice symmetry, see Fig.~\ref{fig:illustration_XYlike_defect}(c). 

The energy $E_0$ follows from a description of the orientation field close to the defects and the line connecting them. Far from the defect pair, for $r\gg \ell$, the two half-integer defects behave like a $|q|=1$ charge and the far field contribution to the energy of the whole structure is the same as a genuine unit charge in an XY model with coupling constant $Jp$. For $r>\ell$, we define this energy as
\begin{equation}
    E(r) = E_0 + \pi\,Jp\log\left( \frac{r}{\ell} \right)
\end{equation}
and $E_0(\ell)$ plays the role of a core energy, describing the contribution of scales $r<\ell $, which corresponds to the size of the split defects. 
\newline

To confirm the theory developed above, we have to measure the energy $E(r)$ inside a square of linear size $r$, centered on the center of mass of the two $1/2$ defects. To do so, we repeat the procedure used to measure the line length $\ell$, using a (large) simulation box of size $L = 300$. However, one has to deal with an additional subtlety. If one evolves the system at low but finite temperature $T>0$, the measured energy will scale like $k_BT\,r^2$, overriding the expected linear or logarithmic scalings. On the other hand, if one lets the system relax at $T=0$, the initial $+1$ defect does not correctly split into two fractional charges and one line connecting them. 
Instead of spontaneously breaking the isotropic symmetry around the defect core, at zero temperature the axis often develops in a cross structure, which in turn might affect the scaling of the energy with the system size. This is why we resort, for this specific measurement of $E(r)$, to an annealing protocol. We start from a low but finite temperature (we typically use $T_0 = 0.1\,T_c(p)$) and let the system evolve for $t_{\max}/2$. 
We typically use $t_{\max}= 2\cdot 10^5$ so that after a time $t_{\max}/2$, the axis has reached its steady-state length for all $p\le 30$.
The second part of the dynamics is the annealing part, where the temperature is linearly decreased over time, from $T_0$ at $t_{\max}/2$ to $0$ at $t_{\max}$. A too rapid annealing protocol would freeze in space the thermal fluctuations of the field, impacting the scaling of $E(r)$. 

\begin{figure}[h!]
    \centering
    \includegraphics[width=0.7\linewidth]{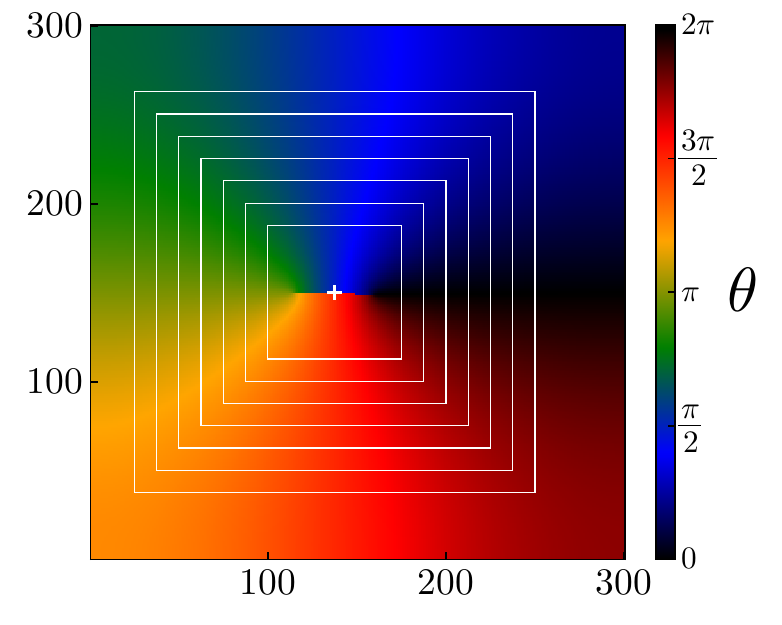}
    \caption{\small The nested subsystems of linear size $r$ used to compute the energy of a single defect configuration. We have not plotted the smallest square sizes to leave the $\theta$ field close to the defect visible. The largest square is $0.75L$. Here, $L=300, p=30, T = 0, t=t_{\max}= 2\cdot 10^5$. 
    }
    \label{fig:sketch_protocol_energy_single_defect}
\end{figure}

At $t=t_\text{max}$, we locate the two $+1/2$ defects, we compute they center of mass, and we draw nested squares of various sizes $2 \le r<0.75 L $ around it, see Fig.~\ref{fig:sketch_protocol_energy_single_defect}. All squares are centered on the defects' center of mass to comply with the isotropic symmetry of the $\arctan (y/x)$ functional form of an XY defect, but the center of mass itself is not necessarily at the center of the large $L$ simulation box due to thermal diffusion. Finally, we compute the energy of each square to obtain $E(r)$, and report the results in  Fig.~\ref{fig:illustration_XYlike_defect}(e, f). 
Because $E_0\sim p$, we expect $E\sim p$ for such a split defect and we plot $E/p$ against $r/\ell$. 

At large distances compared to the length of the composite defect, $E$ follows the expected XY-like $\pi\log r$ scaling, see dashed lines in Fig.~\ref{fig:illustration_XYlike_defect}(e). At distances $r <  \ell$,  when the box does not contain the $q = 1/2$ defects, the leading term is the line tension, explaining why $E$ scales linearly with $r$, see Fig.~\ref{fig:illustration_XYlike_defect}(f). 
\newline

The same linear relation was obtained by Mila in a simplified model designed to mimic the Domany model~\cite{mila1993first}. Instead of a continuous function like the Domany potential, Mila opted for a piecewise function: quadratic for small angle differences $|x|<x_c$, with fixed, unit curvature and constant otherwise. Concretely, the potential reads 
\begin{equation}
    \frac{E_\text{Mila}(x)}{J} = 
    \begin{cases}
      \frac{1}{2}{x}^2 \quad \text{if } |x|<x_c\\
       \frac{1}{2}{x_c^2} \quad \text{if } \pi>|x|>x_c\\
     \end{cases} \ .
     \label{eq:potential_mila}
\end{equation}
The potential Eq.~(\ref{eq:potential_mila}) shares essential features with the Domany model and is thus expected to exhibit the characteristic phenomenology of the steep XY models. 
In these models, a physically relevant ratio is 
\begin{equation}
    r = \frac{ \text{curvature in zero} }{ \text{plateau value} }
\end{equation}

\begin{figure}[h!]
 \centering
     \includegraphics[width=\linewidth]{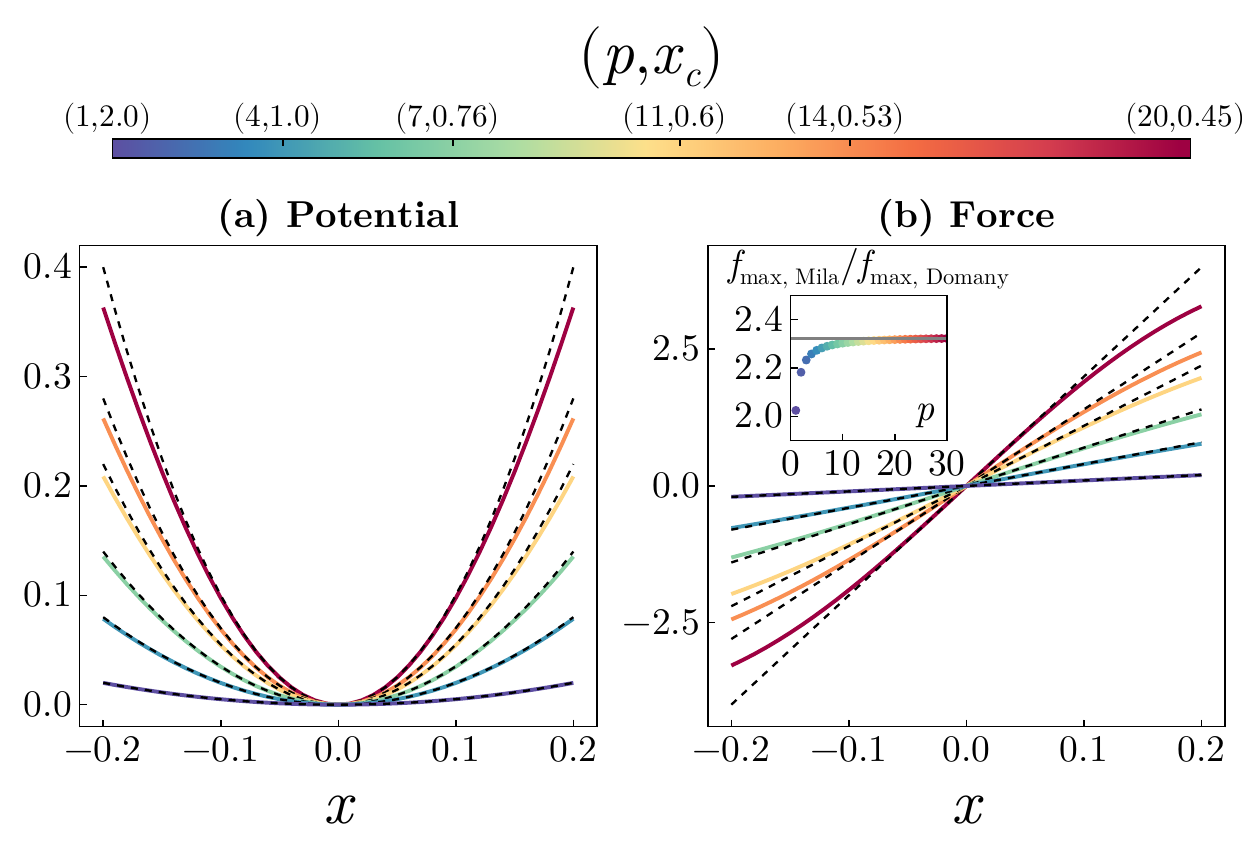}
     \caption{\small For both the Domany (solid colored lines) and modified Mila (dashed black lines) models, we plot \textbf{(a)} the potential  \textbf{(b)} the force. The colorbar translates the equivalence $p = 4x_c^{-2}$ between $x_c$ (Mila) and $p$ (Domany), see main text. \underline{Inset}: the ratio between the maximum force for both models against $p$. One can compute that $f_{\text{max, Mila}}(p)= 2\sqrt{p}$ and $f_{\text{max, Domany}}(p) = 2p \cos^{2p-1}(1 / \sqrt{2p}) \sin(1 / \sqrt{2p})$. 
    The grey line is $y=2.32$. }
     \label{fig:pot_force_mila}
 \end{figure}

For the Domany model, $r=Jp/(2J)= p/2$. For the Mila potential Eq.~(\ref{eq:potential_mila}), $r=2/x_c^2$. Equating the two provides the equivalence relation 
\begin{equation}
    p = 4/x_c^2 \ .
    \label{eq:equivalence_domany_mila}
\end{equation}
Multiplying Eq.~(\ref{eq:potential_mila}) by $4/x_c^2$  bridges with the original Domany model: 
\begin{equation}
    E_\text{Mila, modified}(x) = \frac{4}{x_c^2} \ E_\text{Mila} =
    \begin{cases}
     2J \,x^2/x_c^2 \\
       2J \\
     \end{cases}
     = \ 
     \begin{cases}
      \frac{1}{2}Jp x^2 \quad \text{if } |x|<x_c\\
       2J \qquad \ \text{ if } \pi>|x|>x_c\\
     \end{cases} 
     \ .
\end{equation}
We plot $E_\text{Mila, modified}$ and the corresponding force $f_\text{Mila, modified}$ in Fig.~\ref{fig:pot_force_mila} (solid colored lines for the Domany model and black dashed for the modified Mila model). 

All the results obtained with Eq.~(\ref{eq:potential_mila}) remain valid upon the transformation $E\to pE$. In particular, the energy of a $+1/-1$ defect pair configuration is measured via Monte Carlo simulations. Concretely, the positions of the $|q|=1$ defects (which we believe correspond to the centers of mass of the half-integer split defects) are fixed to specific locations separated by a distance $r_c$. Mila states that the interaction ``can be approximated by"
\begin{equation}
\frac{E_\text{defect, Mila modified}(r)}{Jp} = 
\begin{cases}
x_c^2 + \lambda \,x_c^2\,(L/a) & \text{if } r \leq r_c \\
E_0(p) + 2\pi \log(L/a) & \text{if } r >r_c
\end{cases}
\end{equation}
with $\lambda = (1+\sqrt{2})/2$ \cite{mila1993first}.
As understood in the discussion of Fig.~\ref{fig:illustration_XYlike_defect}, for $r>r_c$ the far field of split defects is that of unit charges, leading to the expected logarithmic interaction energy. The extra factor of 2 in front of the $\log r$ might be explained by the maximum force in the Mila model being approximately twice the maximum force in the Domany model, cf. inset in Fig.~\ref{fig:pot_force_mila}(b). 
Below $r_c$, the energy scales linearly with the distance between the centers of mass of the two split defects, and it is found that $r_c \sim p$. Those results, together with the orientation field illustrated in Fig.~1(b) of \cite{mila1993first}, are consistent with our understanding in terms of two half-integer split defects. 

\section{Domain boundaries}
\begin{figure}
    \begin{center}  
    \includegraphics[width=0.75\linewidth]{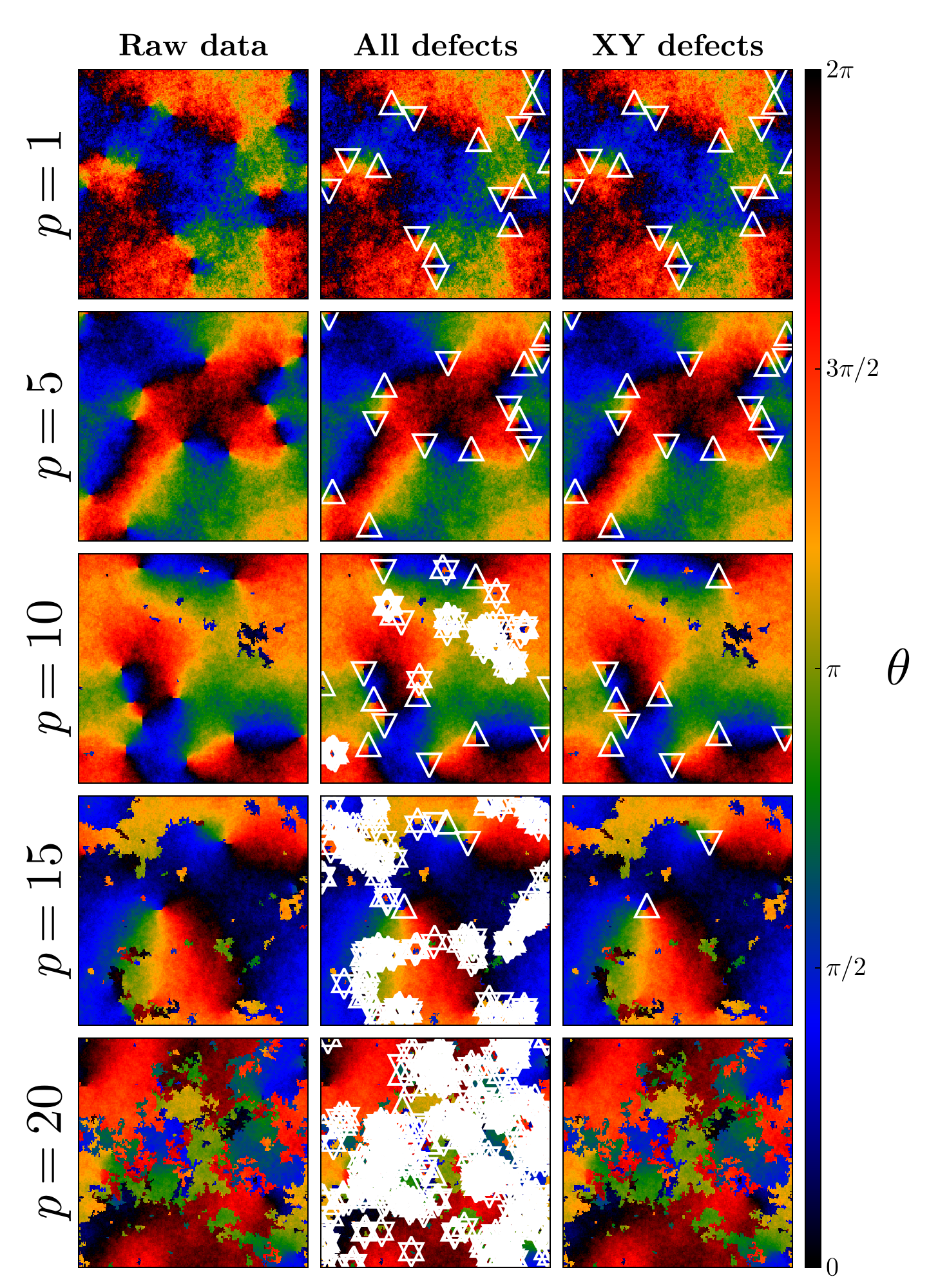}
    \caption{Snapshots of 5 systems during the coarsening dynamics. For all the systems, $L=256, T/T_c = 0.2$. Each row corresponds to a given $p$ : $p=1,5,10,15,20$. The time at which the snapshots are taken is $t=5000$ for $p=1,5,10$ and  $t=10\,000$ for $p=15, 20$, simply because the coarsening dynamics slows down with increasing $p$.
    For a given $p$, the three columns represent the orientation of the same system, with the usual color scale to the right. The left column displays the raw system, the central column highlights all the topological defects (as identified by the standard algorithm) and the right column highlights only the XY-like defects, as identified by the standard algorithm and then filtered by the protocol explained in the main text. 
    }
    \label{fig:illustration_defects}
    \end{center}
\end{figure}

We have characterized the structure of isolated topological defects in the steep XY model. We now turn to the coarsening dynamics after a quench from a disordered initial configuration to low temperature $T<T_c$.
We provide in Fig.~\ref{fig:illustration_defects} representative snapshots of 5 systems during the coarsening at $T/T_c = 0.2$. Each row corresponds to a different $p$. We first focus on the first column, where we plot the raw spins configurations with the usual cyclic color code. 

For the usual XY model ($p=1$), one recover the typical XY picture, with pointlike defects, bounded in pairs.
For moderate values of $p\sim 2-9$, each $\pm1$ defect exhibits a small core axis, corresponding to the line connecting the two half-integer defects. At large distance however, each pair of half-integer defects can be thought of as a unique integer defect: the overall picture remains qualitatively the same.
For $p=10$, the core axis of the composite defects is longer, thus more easily spotted. However, composite defects are not the only objects at play anymore: multiple clusters of different orientations appear across the system.  In contrast with the smooth Ising domain boundaries, here the clusters are very rough; their stability is a consequence of the almost inexistent interaction between neighboring spins with an angular difference larger than $\Delta \theta_\text{max} \sim 1/\sqrt{p}$. 
Further increasing $p$, those clusters populate the system and split defects become scarcer. For $p\ge20$ (bottom left panel of Fig.~\ref{fig:illustration_defects}), it seems that defects are completely absent. 

To confirm this visual impression, we run for all $p$ the standard procedure to identify topological defects in the usual XY model and superimpose the locations of the defects (up triangles for $+1$, down triangles for $-1$) onto the (same) orientation fields in the second column. 
But because the clusters boundaries can be thought of as lines of $\pm 1$ defects due to translational invariance along the boundary, the standard algorithm is unable to discriminate between true (split) defects and domain boundaries, as displayed in the central column of Fig.~\ref{fig:illustration_defects}. 
The visual inspection of the $p=10$ case displays a typical situation where a human eye would see the difference between the split defect structures where all the colors meet on one hand, and the dark blue clusters immersed in the ordered green/yellow area on the other hand, whereas the standard algorithm is incapable to do so.  

\newpage

We thus need a different identification protocol of the split defects. Because we know that half-integer defects come in pairs of the same charge at a distance $\ell \sim p$, we can design an algorithm to detect unit charge defects. Our algorithm consists in a filter that takes all the defects detected by the standard procedure and classifies them in two categories: split defect or domain boundary.

\begin{figure}[h!]
    \centering
    \includegraphics[width=0.99\linewidth]{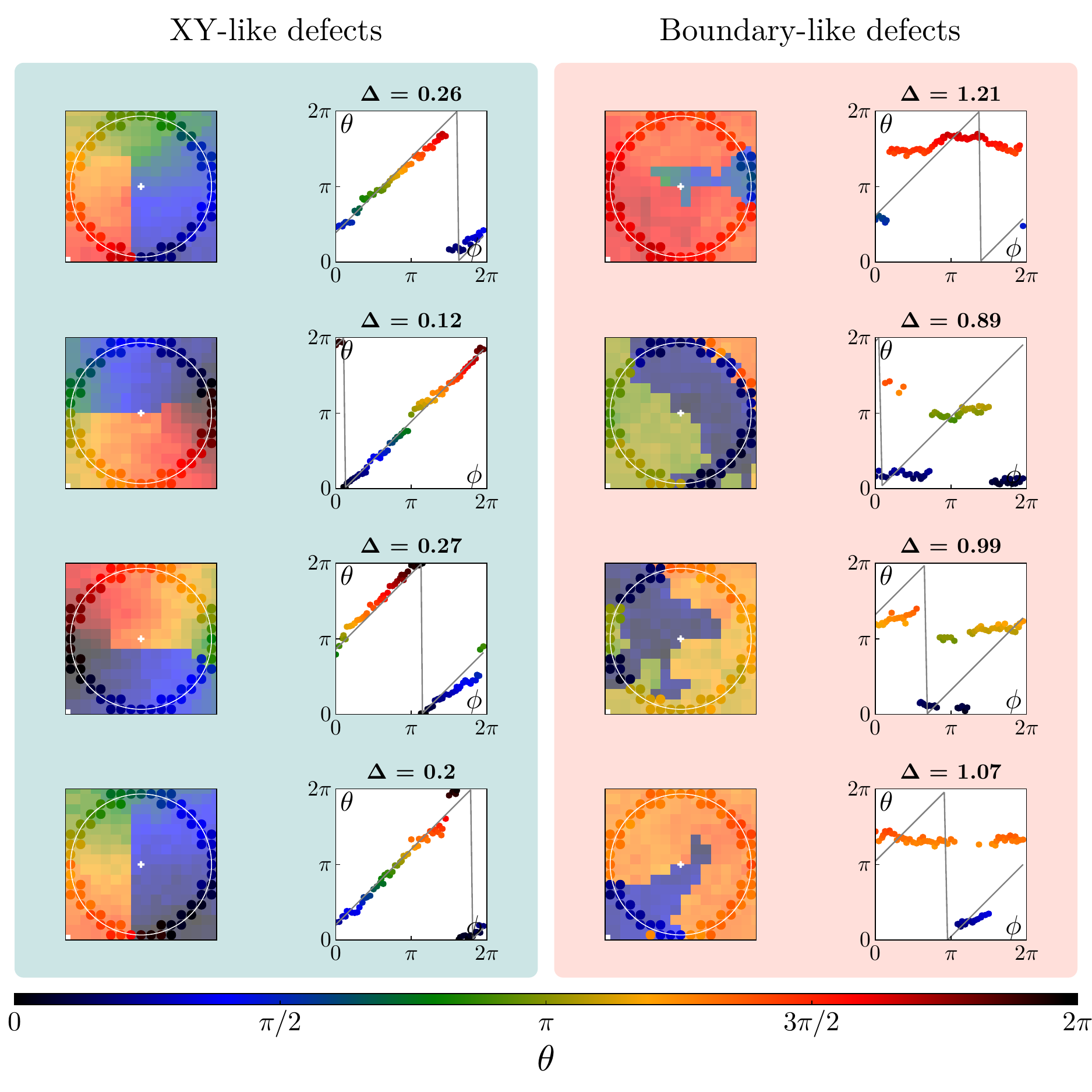}
    \caption{\small 
    Illustration of the protocol to discriminate between split defects (left block, light green) and boundary defects (right block, light red), for $p=10$. In each block, each row represents a defect randomly chosen in the system. In each block, the left column is a zoom on the orientation field. The white circle is a contour of diameter of $1.1\,\ell$, centered on the defect core (white cross). The colored dots highlight the spins along the contour used in the right column to fit $\theta = q\,\phi + \theta_0$. The best fit is plotted in thin grey line. The distance $\Delta$ (Euclidian, accounting for $2\pi$ periodicity) between the data and the best fit is reported above each plot in the second column of each block. Split defects are defined as the ones for which $\Delta<0.75$. Boundary defects are defined  as the ones for which $\Delta > 0.75$.
    }
     \label{fig:sketch_protocol}
\end{figure}

The protocol we use is illustrated in Fig.~\ref{fig:sketch_protocol}. 
Recall that a charge $q$ at position $(x_0,y_0)$ imposes a field $\theta = q\arctan (\frac{y-y_0}{x-x_0}) + \theta_0 $ around itself. 
For each candidate plaquette of charge $|q|>0$ and location $(x_0, y_0)$ identified by the standard algorithm, we draw a circular contour of diameter $d$ parametrized by the angular coordinate $\phi = \arctan (y/x)$. We take $d = 1.10 \,\ell = 1.43\,p$ (taking a 10\% margin to ensure that both half-integer defects and the connecting line are included in case of a split defect). 
We then collect the orientation field $\theta$ along this contour (colored circles in Fig.~\ref{fig:sketch_protocol}) and fit it with the expression $q\,\phi + \theta_0$ (see grey line, $\theta_0$ is the only fitting parameter). 
We then compute the goodness of fit $\Delta = |\theta - (q\,\phi + \theta_0) |$, accounting for the $2\pi$ periodicity of the variable $\theta$.  If $\Delta < 0.75$ (arbitrary robust threshold), the $\theta$ profile is well fitted by an ideal XY defect far field, and the defect is classified as a split defect, see left block of Fig.~\ref{fig:sketch_protocol} (in light green). The cases $\Delta > 0.75$ correspond to configurations where the ideal XY defect far field is a very bad fit because the orientation field along the contour exhibits a clear discontinuity, consistent with domain boundaries, see right block of Fig.~\ref{fig:sketch_protocol} (in light red).

This simple method works remarkably well. In the third column of Fig.~\ref{fig:illustration_defects}, we only superimpose the locations of the defects classified as split defects by our algorithm. The comparison between the last two columns of Fig.~\ref{fig:illustration_defects} shows that this modified algorithm gives intuitive results and successfully excludes all the spurious boundary defects. 
It confirms our first visual impression, namely that, even though split defects are stable and exist for all $p$ when created manually (we have measured their axis length $\ell$ for $p\in [1, 30]$), these XY like defects do not appear -- or in a very scarce manner -- in actual realizations of the systems for $p\gtrsim 15$. 


\begin{figure}[h!]
 \centering
     \includegraphics[width=\linewidth]{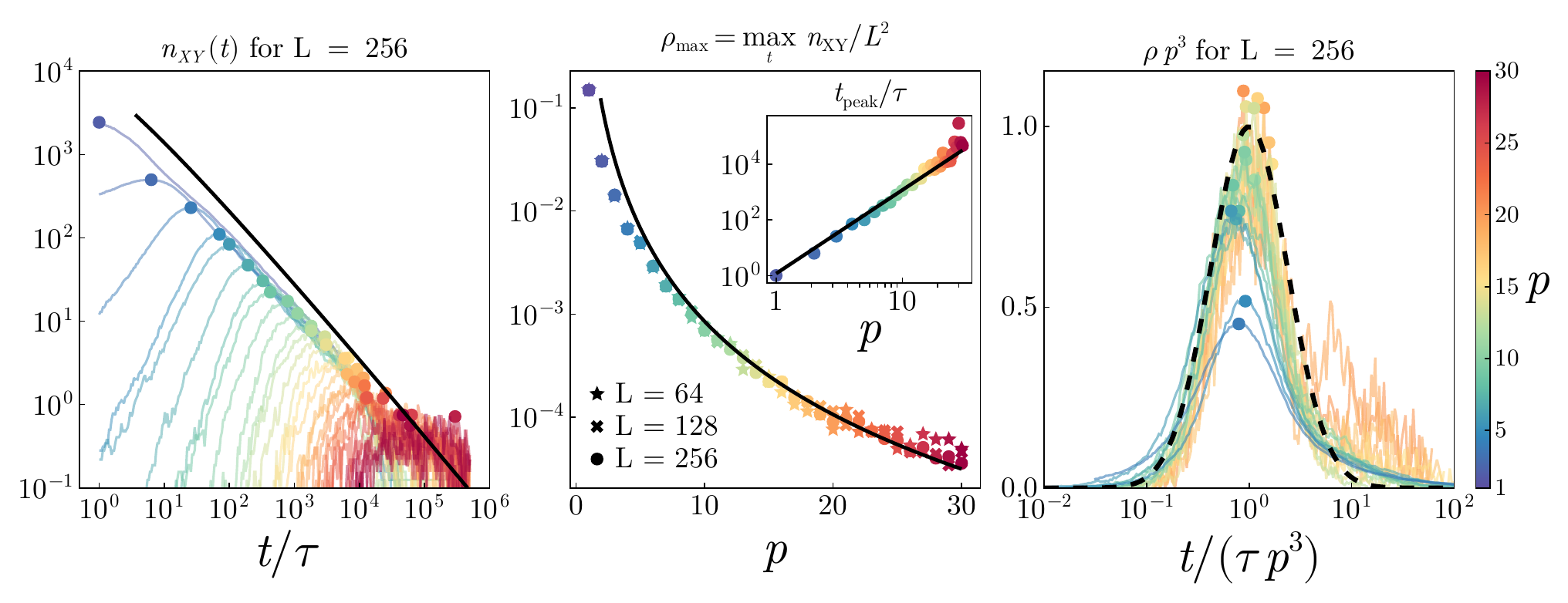}
     \caption{\small 
     \textbf{(a)} Number of XY defects $n_{XY}$ over time, for a $L=256$ system. The black line is $(\log t) / t$, the usual XY coarsening scaling. For all the panels, $T/T_c = 0.2$ . 
     \textbf{(b)} $n_{XY}$ against $p$ for different system sizes $L$ (different symbols). The black line is $0.85 / p^3$. 
     \underline{Inset:} $t_\text{peak}$, the time at which the XY defect density reaches a maximum.  The black line is $(0.85 / p^3)^{-1}$.  
     \textbf{(c)} $\rho\,p^3$ as a function of $t/p^3$. 
    The black dashed line is $\exp(-\frac{3}{2}\log^2 x)$. 
     }
     \label{fig:nXY}
 \end{figure}

We can now investigate the time evolution of the defect density $\rho = n/L^2$ for different $p$. We present the results for $L=256$ and $T/T_{KT} = 0.2$ in Fig.~\ref{fig:nXY}(a). 
For the $p=1$ case (dark blue), we recover the expected XY coarsening dynamics $\rho\sim (\log t)/t$ (black line), with $\rho(t=0) = 1/3$ as expected. For $p>1$, the number of defects first increases in time, until it reaches a maximum (colored circles) and then decays following the XY scaling. 
The curves for different $p$ have similar shapes; the maximum value $\rho_\text{max} = \max\limits_t \rho(t)$ decreases with increasing $p$ and is attained at longer times $t_\text{peak}$. Figure~\ref{fig:nXY}(b) shows that $\rho_\text{max} \sim p^{-3}$ independently of the system size $L$ (different symbols correspond to different $L$), and $t_\text{peak}\sim p^3 $ (inset), \textit{a posteriori} validating the shape invariance of $\rho$ as $p$ varies. 
Interestingly, it is not possible to extract a typical $p_c$ from the polynomials $p^3$, such that neither $t_\text{peak}$ nor $\rho_\text{max}$ indicate a critical $p_c$ separating the KT transition regime from the first order transition regime.  

\section{Rationalization of the critical $p_c$}
It has been shown that in the steep XY model, the transition becomes first order for large enough $p$ \cite{Domany, vanEnter1, vanEnter2}. However, the critical value $p_c$ at which the change in the order of the transition occurs remains unprecise. Domany \textit{et al.} \cite{Domany} report that $1 \le p_c \le 50$.  The claim $p_c=10$ of \cite{vanHimbergen84}, based on the specific heat, seems to be contradicted by the unimodal nature of the distribution of the energy for that $p$.  The same reference, along with \cite{SinhaKumar}, reports that $p\approx 16$ is the largest $p$ for which there is no jump in the defects density. It is not clear, however, if the defects considered in these studies are only the split defects involved in the BKT transition or if the boundary defects are also counted as defects. Those articles do not mention the existence of defects of different nature. 

We are still currently working on drawing a clearer picture regarding the emergence of a critical $p_c$. Preliminary results indicate that $p_c\approx 15$. Unfortunately, those results are not definitive enough to be reported in this thesis. Yet, we argue that this value is consistent with a converging body of observations and analogies with similar models. 
\newline 

The BKT transition mechanism entirely relies on the collective behaviour of topological defects and, more precisely, on their ability to attract each other, move and annihilate in pairs. We have seen that in the steep XY model, those unit charge defects split into two tightly bind half-integer defects. At large distances however, those interesting structures behave like unit charges and follow the BKT annihilation mechanism. Those structures, once formed, are stable at all $p$. For $p\gtrsim 10$, clusters appear, together with the defects on their boundaries. For $10 \lesssim  p\lesssim 15$, split defects and boundary defects coexist, but the latter do not seem to hinder the annihilation mechanism of the split defects (ongoing investigation), such that the BKT mechanism remains valid.
Above $p\approx 15$, boundary defects clearly dominate over XY-like split defects. Those are very rarely seen in actual systems, see Fig.~\ref{fig:illustration_defects} for snapshots or Fig.~\ref{fig:nXY} for the count of split defects. This suggests that the BKT mechanism is not the governing one anymore, changing the nature, and thus the order, of the phase transition.

\begin{figure}[h!]
 \centering
     \includegraphics[width=\linewidth]{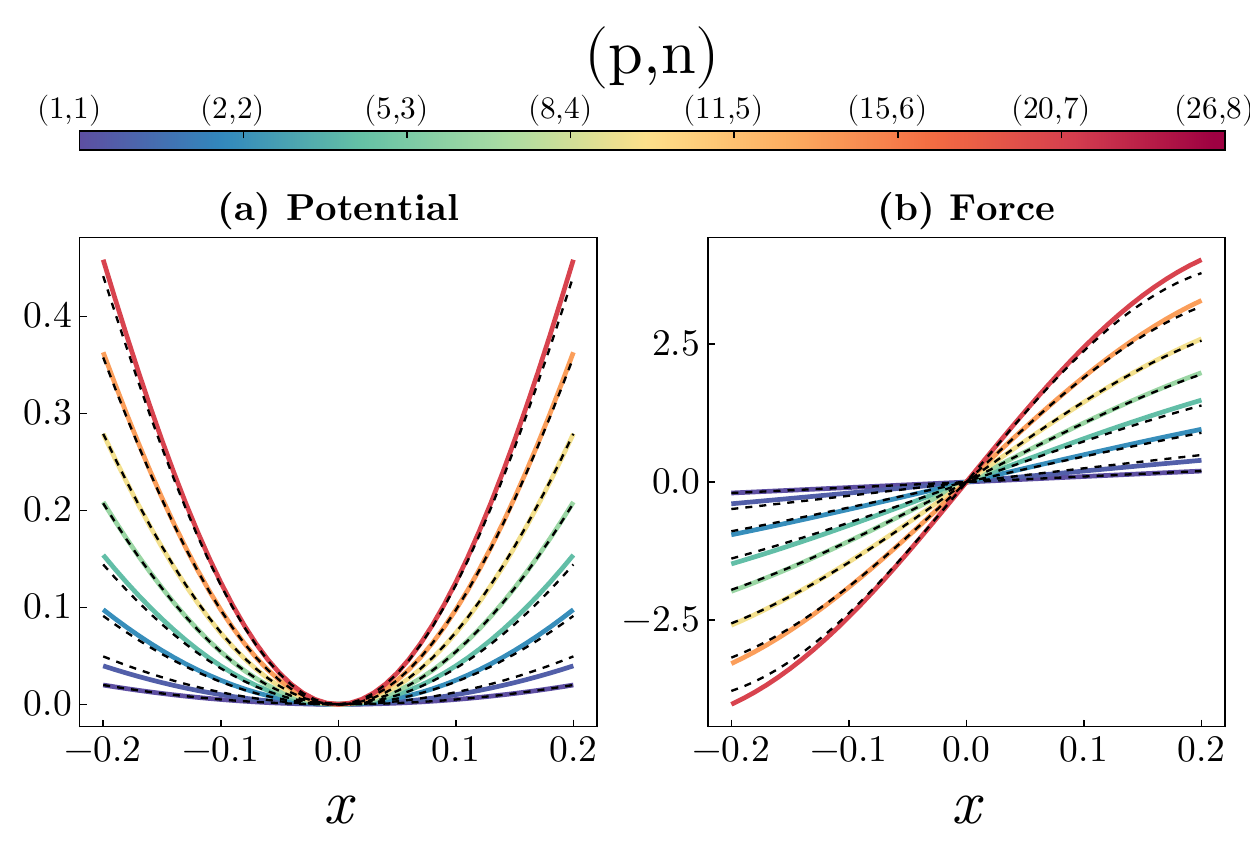}
     \caption{\small For both the Domany (solid colored lines) and Zukovic (dashed black lines) models, we plot \textbf{(a)} the potential  \textbf{(b)} the force. The colorbar translates the equivalence $p(n) =(n+1)(2n+1)/6$ between $n$ (\v{Z}ukovi\v{c}) and $p$ (Domany) see main text. 
     The small discrepancy between the dash and solid lines is due to the necessary rounding to nearest integer of $p$. }
     \label{fig:pot_force_np}
 \end{figure}

The claim $p_c\approx 15$ is supported by the study of a steep Heisenberg model. Ref.~\cite{blote2002phase} reports that a similar steep $\mathcal{O}(3)$ model in $2d$ also exhibits a (weak) first order transition for  $p>p_c=16$ and an ``enhanced first character" for $p\ge 20$. 
Higher order XY models also exhibit the same change in transition. In a recent article~\cite{zukovic2025crossover}, \v{Z}ukovi\v{c}  studies the potential
\begin{equation}
    E_n(x) = -\frac{1}{n}\, \sum\limits_{k=1}^n \cos(kx) =-\frac{\cos(x) + \cos(2x) + \ ...  \ + \cos(nx)}{n}  
    \label{eq:hamiltonian_zukovic}
\end{equation}
where $x = \Delta \theta = \theta_j - \theta_i$.  For small $x$, the potential Eq.~(\ref{eq:hamiltonian_zukovic}) becomes steeper as $n$ increases, and plateaus (with oscillations around 1) for large $x$ and thus shares features with the steep XY models of Domany \cite{Domany} and Mila \cite{mila1993first}. 
To be precise, the original model does not contain the $1/n$ rescaling; we have added it so that the energy does not scale with $n$ and to make the connection easier with the Domany model. Since it is only a change in the energy scale, the conclusions of  \cite{zukovic2025crossover} remain valid, the only consequence would be a similar rescaling of the critical temperature.  
The main result of \cite{zukovic2025crossover} is that for $n<6$, the order-disorder transition is of KT type while for $n\ge 6$, it becomes first order. \v{Z}ukovi\v{c} proves the discontinuous nature of the transition by  looking at the Binder cumulant of the polarization and conducting a finite-size analysis, but does not provide any physical explanation. 
To apply this result to the Domany potential, we need an equivalence between $n$ and $p$. We suggest to equate the curvature in $x=0$ in both cases. 
In the \v{Z}ukovi\v{c} model, it reads 
\begin{equation}
   \left. \frac{\partial^2E_n}{\partial x^2}\right|_{x=0} = \frac{1}{n}\sum\limits_{k=1}^n k^2= \frac{(n+1)(2n+1)}{6}
\end{equation}
We thus obtain the relation $p(n) =(n+1)(2n+1)/6$, rounded to the nearest integer. 
For small angles, the agreement for the potential and the force between the two models is very good. In Fig.~\ref{fig:pot_force_np}, we plot the \v{Z}ukovi\v{c} potential for $n=1,2, 3,...,8$ in dash, and the Domany potential for $p(n)=1,2, 5,...,26$ in solid coloured lines. 
As such, the critical power $p_c$ for the Domany model is given by the critical value of $n_c$ for the \v{Z}ukovi\v{c} model, which is $n_c=6$. This corresponds to a critical $p_c = p(n_c) = 15$. 

Finally, we come back to the original motivation of Domany and co-authors, namely the similarity with the $2d$ $q-$Potts model. We have seen that the Potts critical temperature Eq.~(\ref{eq:Potts_Tc_q}) does not match that of the steep XY model. Relying on this equivalence to obtain $q$ as a function of $p$ gives incorrect predictions, like the claim $q=5.5\sqrt{p}$ at the very end of Ref.~\cite{Domany} .
Indeed, we know that for above $q_c=4$, the phase transition of the Potts is first order, and $q=5.5\sqrt{p}$ would imply that already $p=1$ gives $q>q_c$. 
If instead one relies on the analogy between Potts discrete species and the critical angle difference above which two spins will not interact, one obtains the equivalence Eq.~(\ref{eq:potts_p}). Solving $q(p_c) = q_c$ gives $p_c = 14$, in good agreement with the observations.

\newpage


\chapter{The short-range lattice Kuramoto model} \label{ch:kuramoto} 
\graphicspath{{figures/Ch5kuramoto/}}
\section{Abstract}


We consider a non-equilibrium extension of the two-dimensional (2D) XY model, equivalent to the noisy Kuramoto model of synchronization with short-range coupling, where rotors sitting on a square lattice are self-driven by random intrinsic frequencies. We study the static and dynamic properties of topological defects, investigate the ordering dynamics following a quench and establish how self-spinning affects the Berezenskii-Kosterlitz-Thouless phase transition scenario.
The non-equilibrium drive breaks the quasi-long-range ordered phase of the 2D XY model into a mosaic of ordered domains of controllable finite size, scaling as the inverse of the standard deviation of the distribution of intrinsic frequencies. Defects generically unbind at any temperature and perform a random walk reminiscent of the self-avoiding random walk, advected by the dynamic network of boundaries between synchronised domains, featuring long-time superdiffusion, with the same anomalous exponent $3/2$ in the mean-square displacement, although with different scaling of the full distribution of displacements.
\newline

This chapter covers the results published in 
\begin{center}
  Rouzaire and Levis, Physical Review Letters (2021) \cite{Rouzaire_Kuramoto_PRL} \\
  Rouzaire and Levis, Frontiers in Physics (2022) \cite{rouzaire_Frontiers}. 
\end{center}
  
  \section{Introduction}
Much progress in the understanding of synchronisation has been achieved through the detailed analysis of simplified model systems. In this context, the Kuramoto model of phase coupled oscillators has (and still does) played a central role \cite{Kuramoto,AcebronBonilla}.  
In the original version of the model \cite{Kuramoto}, considering only all-to-all interactions and noiseless (hence deterministic) dynamics, if the intrinsic frequencies are drawn from a unimodal distribution, the system exhibits a phase transition, involving a collective rotation (frequency locking) and phase synchronization ~\cite{KuramotoOriginal,AcebronBonilla, Sakaguchi1987}.  
This formally infinite dimension system exhibits a {second} order phase transition from order to disorder as the self-spinning strength increases~\cite{KuramotoOriginal}. 

Self-spinning obviously implies a constant energy supply, which at the microscopic scale is dissipated into the environment, which also provides a source of noise.  
Thus, to describe oscillator systems at the micro-scale, such as genetic oscillators \cite{Mondragon2011}, noise should be taken into account, as well as shorter-range interactions, leading to finite-dimensional noisy extensions of the Kuramoto model \cite{ArenasRev}. 
It is known that if the oscillators lie on a regular lattice and their interactions are restricted to the nearest (lattice) neighbours as in the XY model, and $T>0$,  the emergent, large-scale properties of the system strongly depend on the system dimensionality $d$~\cite{AcebronBonilla}. If intrinsic frequencies are sampled from a unimodal distribution, for $d\geq5$, the system fully synchronizes. For $d=3,4$, the system is disordered in phase, but exhibits frequency locking, i.e. all oscillators rotate at the same effective frequency \cite{Sakaguchi1987, hong2005collective, hong2007entrainment}.
In $d=2$, the system can only exhibit short-range order (SRO). 
The situation in this case, which is the object of this chapter, is quite more involved than the original Kuramoto model and the corresponding literature far more scarce \cite{hong2015finite, lee2010vortices,Rouzaire_Kuramoto_PRL,rouzaire_Frontiers}. 

The problem of synchronisation in the physical sciences has more recently experienced a resurge of interest in the context of active matter. Active matter stands for a class of soft matter systems, composed of coupled interacting agents that convert energy from their surrounding into some kind of persistent motion \cite{shankar2022topological}, such as biological units oscillating at a given rate. Such injection of energy at the level of each single constituent, in the presence of dissipation, drives active systems out-of-equilibrium. At the collective level, interactions between active agents result in the emergence of a variety of coherent states. A broad class of active systems spontaneously self-organise into synchronised states characterised by the coherent motion of self-propelled individuals, a phenomenon called flocking (a term borrowed from the spectacular example of the murmuration of starling birds) \cite{ginelli2016VicsekRev}. The interpretation of flocking in terms of active oscillators has been pointed out in a number of recent works \cite{chepizhko2010relation,chepizhko2021revisiting, levis2019activity}. Indeed, the most salient model of flocking, namely, the Vicsek model, can be mapped to the 2D noisy Kuramoto model in the limit where the oscillators are not self-propelled along the direction defined by its phase, and all of them have the same intrinsic frequency.  

Closer to the standard synchronisation setup of oscillators lying on a static substrate, synchronised states have been studied in numerous active matter systems in the absence of self-propulsion. A salient example are active filament carpets, such as cilia or flagella attached on a surface. Cilia, for instance, perform a beating cycle (usually  modelled as a phase oscillator), which generates a net hydrodynamic flow (at  low Reynolds number) that then affects the motion of their neighbours \cite{golestanian2011hydrodynamic, solovev2022synchronization}.  Such filaments, modelled as coupled oscillators, can then synchronise, to optimise a biological function such as propelling microorganisms. 
Chiral colloidal fluids composed of spinning colloidal magnets constitute another promising novel venue to investigate synchronisation at the micro-scale \cite{soni2019odd,massana2021arrested}. In these systems, an external oscillatory field drives the colloidal magnets, making them rotate around their body axis at a given frequency imposed by the field.   

Here we investigate the collective dynamics of Kuramoto oscillators, with short range coupling, in two dimensions (2D), in contact with a thermal bath and with a Gaussian distribution of intrinsic frequencies. In the absence of driving,  or intrinsic oscillations,  the system is equivalent to the 2D XY model, which exhibits a topological Berezinskii-Kosterlitz-Thouless (BKT) transition driven by the unbinding of topological defects \cite{berezinskii1971destruction, Kosterlitz1973,Kosterlitz1974}. As such, it provides a natural playground to study the role played by topological defects in systems of coupled oscillators. The dynamics of topological defects in out-of-equilibrium systems has been extensively investigated over the last decade in soft active systems \cite{shankar2019hydrodynamics, LinoDefects},  where it has been found that, in many instances, defects self-propel, or superdiffuse \cite{GiomiPearce, Rouzaire_Kuramoto_PRL, Bililign2021, bowick2022symmetry}. 
A recent study of this model shows that defects become free upon self-spinning, although two regimes remain clearly distinct, a vortex-rich and a vortex-poor one \cite{Rouzaire_Kuramoto_PRL}.
Here we focus our attention on the dynamics of the system following different kind of quenches, from a random initial state, where many defects proliferate, to the low temperature (noise) regime where defects are scarce. 
The coarsening dynamics following an infinitely rapid quench has been extensively studied in model statistical physics systems \cite{bray1994phase_ordering_kinetics}, but has received little attention in the context of synchronisation \cite{LevisPRX}. Of particular interest for us, the study of the coarsening dynamics of the 2D XY model has shown that topological defects, here vortices, diffuse, interact and annihilate, in a way that sheds light on the mechanisms underlying the non-equilibrium relaxation of the system \cite{YurkeHuse1993,rojas1999dynamical, bray2000breakdown, berthier2001nonequilibrium, jelic2011quench}. 
In this chapter, we aim at characterising the dynamics of the noisy Kuramoto model in 2D through the analysis of the density of vortices and the characteristic growing length. We then examine the random motion of vortices and discuss the fundamental differences displayed by the system as compared to its equilibrium limit. To do so, we use as a key control parameter the variance of the distribution of frequencies.

\section{The model}

We consider a set of rotors (or spins) arranged on a $L\times L$ square lattice with periodic boundary conditions (PBC). Building on the adimensional overdamped Langevin equation of the XY model, the evolution of the phase $\theta_i$  of the Kuramoto rotors reads
\begin{equation} \label{eq:model} 
 \dot{\theta}_i= \sigma \omega_i + \sum_{ j\in\partial_i} \sin(\theta_j - \theta_i) + \sqrt{2T\,}\eta_i(t) 
 \end{equation}
Like in the XY model framework, the sum runs over the (4) nearest-neighbours of $i$ and $\eta_i$ is a Gaussian white noise of unit variance. The specificities of the Kuramoto model are the presence of a random intrinsic frequency $\omega_i$ for each rotor. Those are quenched, meaning that they do not evolve in time, and are drawn at initialisation from a Gaussian distribution of zero mean and unit variance, ie. $\omega_i\sim \mathcal{N}(0,1)$. 
Since Eq.~(\ref{eq:model}) is invariant under the transformation $\theta\to \theta - \Omega t$, one can choose a distribution of frequencies with zero-mean without loss of generality. 
It follows that $\sigma \omega_i\sim \mathcal{N}(0,\sigma^2)$, such that the parameter $\sigma$ represents the population heterogeneity between the rotors, or, in other words, quantifies the dispersion of the intrinsic frequencies.
We recall that, like in the XY adimensional framework, time is expressed in units of $J^{-1}$, where $J=1$ is the coupling strength between nearest-neighbour rotors. 
\newline

We numerically integrated Eq.~(\ref{eq:model}) using the Euler-Maruyama scheme, typically using $L=200-500$. Intrinsic frequencies impose an additional constrain on the timestep $dt$ one can use to integrate the equation of motion. Indeed, one has to resolve the fastest rotors, which are those with the largest intrinsic frequency. This imposes a timestep $dt\leq \frac{\pi}{10}\,/\,\max\limits_i |\omega_i|$, where $\pi/10$ is a arbitrary factor to ensure that the phase difference between two successive time steps for the fastest rotor is small enough, i.e. $\Delta\theta = \omega_{\max} \, dt < \pi/10$.

We use the bound $\mathbb{E}\max\limits_{i} |\sigma\,\omega_i| < \sigma\sqrt{2\log N}$, inspired from \cite{ref_bound_dt_omega} and summarized hereafter.
If one defines $Z = \max_{i} \omega_i$, by Jensen's inequality one obtains
\begin{align}
        \exp \{t \mathbb{E}[Z]\} &\leq \mathbb{E} \exp \{t Z\} \\
        &=\mathbb{E} \max _i \exp \left\{t \omega_i\right\} \\
        &\leq\sum_{i=1}^N \mathbb{E}\left[\exp \left\{t \omega_i\right\}\right]\\
        &=N \exp \left\{t^2 \sigma^2 / 2\right\}
        \label{eq:bound_dt_omega}
\end{align}
where the last equality follows from the definition of the Gaussian moment generating function.
Rewriting this, $\mathbb{E}[Z] \leq \frac{\log N}{t} + \frac{t \sigma^2}{2}$
Now, set $t = \frac{\sqrt{2 \log N}}{\sigma}$ to obtain $\mathbb{E}\, Z= \mathbb{E}\max\limits_{i} |\sigma\,\omega_i| < \sigma\sqrt{2\log N}$. \\
We thus have $dt =  \pi/10 \, (2\sigma^2\log N)^{-1/2}$, which is of the order of $0.05 - 0.1$ for typical values of $\sigma$ and $N = L^2$.
\newline

The distribution of natural frequencies introduces quench disorder in the XY model. The way it is introduced though, is fundamentally different to what is typically done in disordered systems, namely, adding disorder in the interactions or an external random field \cite{cardy1982random,le1995replica, agrawal2021domain}. The distribution of intrinsic frequencies drives the system out-of-equilibrium. Adding a term 
\begin{equation}
\sigma\sum_i\theta_i\omega_i
\end{equation}
to the XY Hamiltonian 
\begin{equation}
\mathcal{H}\to \mathcal{H}'=\mathcal{H}+\sigma\sum_i\theta_i\omega_i
\end{equation}
would provide the same equation of motion Eq.~(\ref{eq:model}) when writting: 
\begin{equation}
\dot{\theta}_i = -\frac{\partial \mathcal{H}'}{\partial \theta_i} + \sqrt{2T}\,\eta_i
\end{equation}
However, $\mathcal{H}'$ is now unbounded, as a result of the constant injection of energy into the system needed in order to sustain the intrinsic oscillations. The quench disorder introduced by the intrinsic frequencies drives the system out-of-equilibrium and cannot be mapped to random field or random coupling XY models. 

\section{First snapshots}
We first study in this section the dynamics of the model following a infinitely rapid quench from an initially disordered state where all the phases are picked from a homogeneous distribution between 0 and $2\pi$.

\begin{figure}[h!]
  \centering
		\includegraphics[width=0.9\linewidth]{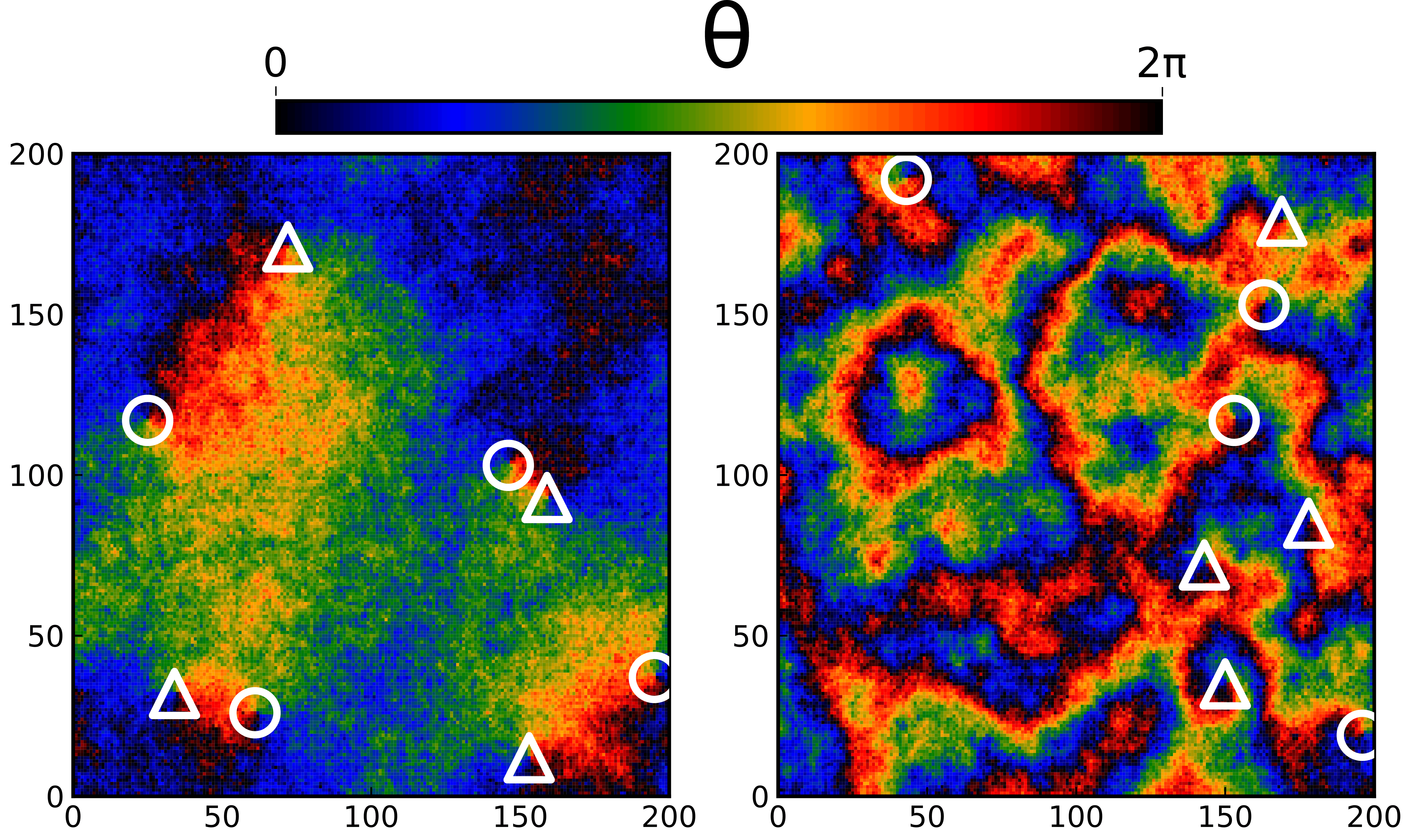}
		\caption{
			Snapshots for $T =0.2$, \textbf{(a)} $\sigma^2=0$ and \textbf{(b)} $\sigma^2=0.1$. The phase of each oscillator is represented by a colour scale. Vortices (antivortices) are represented by circles (triangles).
		}
    \label{fig:snapshots_kuramoto}
	\end{figure}

In order to illustrate the nature of the low temperature state, we show in Fig.~\ref{fig:snapshots_kuramoto} representative steady state snapshots for (a) $T=0.2$ and $\sigma^2=0$ (b) $T=0.2$ and $\sigma^2=0.1$. Both snapshots contain the same number of defects: four vortices (circles) and four antivortices (triangles), visually identified as the points where black, red, yellow, green and blue regions meet. This correspond to plaquettes with a winding number $q=\pm 1$, defined by the sum of the phase differences along its four bonds $\sumcirclearrowleft \Delta \theta_{i,j}=2\pi q$.

At equilibrium, when spins are not forced, one recovers the classical XY model picture, with large regions of spins sharing the same colour, ending at topological defects of opposite charge.  As the Kosterlitz-Thouless theory \cite{Kosterlitz1973,Kosterlitz1974} predicts, those defects visually come in pairs.  
From the snapshots one already observes a clear connexion between the typical size of correlated domains of mostly parallel spins, and the typical distance between defects. We will go back to this point later on. 

On the contrary, upon self-spinning, the first impression is qualitatively different, even though the number of defects is identical in both panels (here 8). The phase field pattern emanating from the vortices  does not extend over large distances, and the typical size of oriented domains seem decoupled from the defects' locations, as seen in Fig.~\ref{fig:snapshots_kuramoto}(b). Ordered regions are in this case rather localised, and one can no longer pair vortices by visual inspection of the snapshots.

\section{Spatial correlation function}

\begin{figure}[h!]
  \centering
		\includegraphics[width=0.99\linewidth]{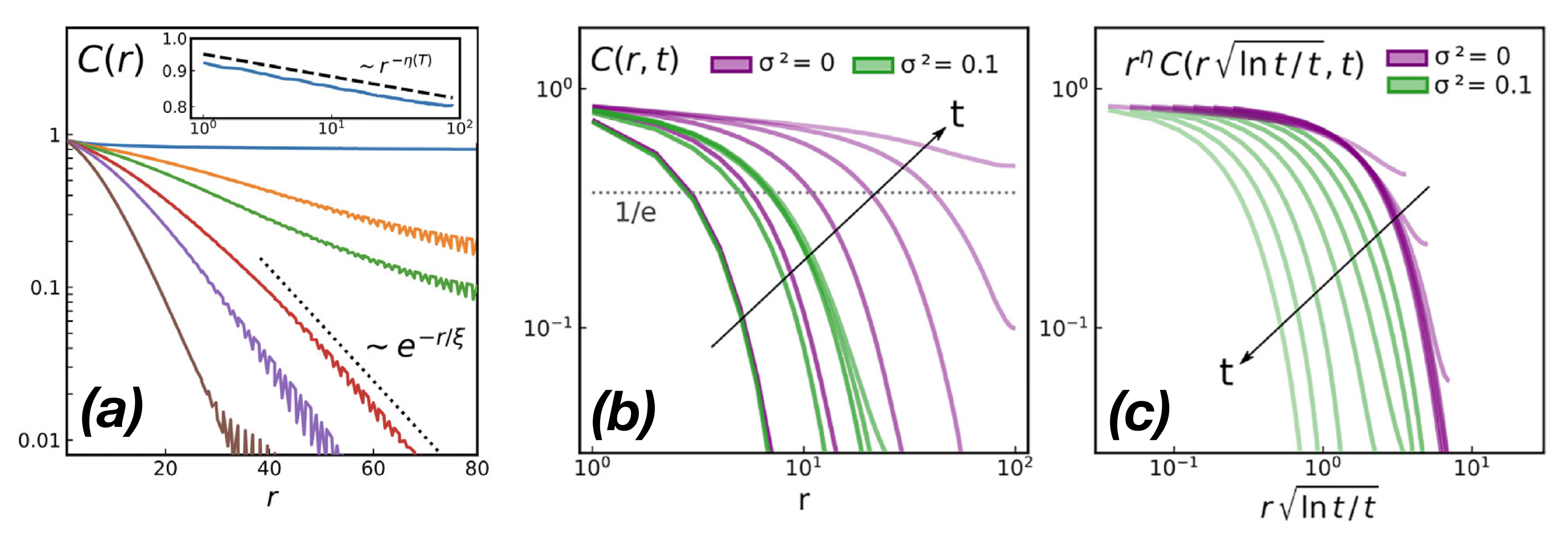}
		\caption{
      \textbf{(a)} $C(r)$ (in log-log) at $T=0.2$ and $\sigma^2 = 0,0.005,0.01,0.02,0.03,0.06$ (from blue/top to brown/bottom). \underline{Inset} : the $\sigma^2=0$ case, showing $C(r)\sim r^{-\eta}$, with $\eta(T)=T/2\pi$.
      \textbf{(b)} Time evolution of the space correlation function $C(r,t)$ at different times $t\approx 4,20,100,400,2000,10000$ 
      Greater times are represented by lighter colours, as the arrows indicate.
      \textbf{(c)} Time evolution of  the rescaled correlation function ($\eta = T/2\pi$ from the spin-wave theory) against the rescaled space. In panels (b) and (c), the purple curves correspond to the XY model at $T=0.4$ and the green curves correspond to a forced system at $T=0.4$ and $\sigma^2=0.1$. 
		}
    \label{fig:kuramoto_Cr}
	\end{figure}
The computation of the steady state spin-spin correlation function $C(r)$ supports this picture. As reported in Fig.~\ref{fig:kuramoto_Cr}(a), for $\sigma=0$, correlations in the low-$T$ phase decay algebraically $C(r)\sim r^{-\eta(T)}$, signature of quasi-long-range order, with $\eta(T)=T/2\pi$ as predicted by the spinwave theory, see Chapter \ref{ch:XY}. On the other hand, $C(r)$ decays exponentially for $\sigma>0$ at any finite temperature, signalling the absence of quasi-long range order in the system and defining a correlation length $\xi$. 

We show in panel (b) the spatial decay of $C(r,t)$ at different times following a quench to $T=0.4$, for $\sigma^2=0$ (in purple) and for $\sigma^2=0.1$ (in green). The data is shown in a log-log scale to easily discriminate between algebraic and geometric decays. 
Panel (c) confirms that the dynamic scaling hypothesis \cite{bray1994phase_ordering_kinetics} is consistently fulfilled for $\sigma=0$ as all the curves $C(r,t)$ collapse into a single master curve in the scaling regime once the space variable has been rescaled by the characteristic growing length. This is however not the case for $\sigma>0$, where the curves do not collapse onto a single master curve, but rather show a clear dependence on time.

\section{Characteristic length scale}

\begin{figure}[h!]
  \centering
		\includegraphics[width=0.99\linewidth]{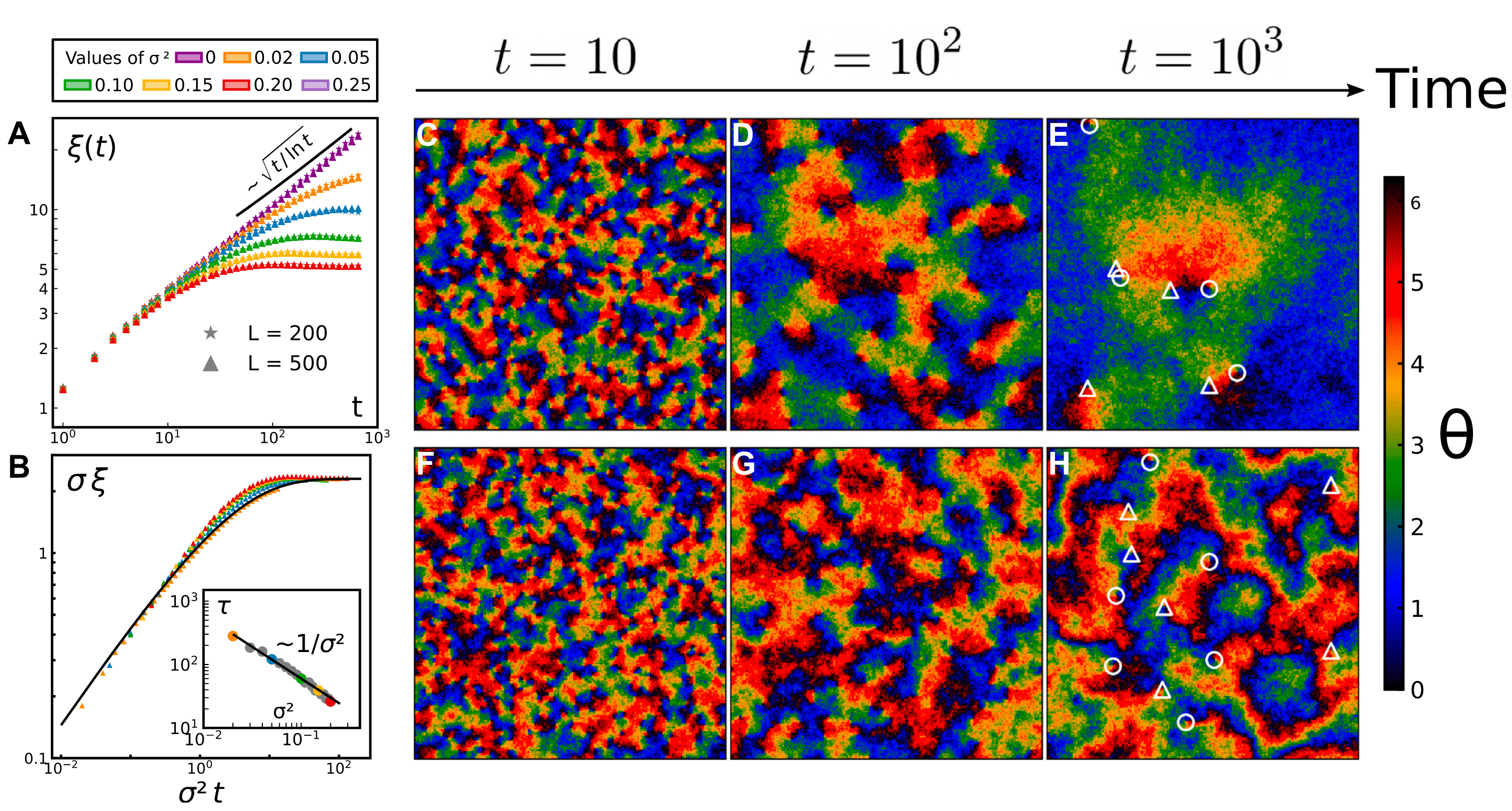}
		\caption{  Time evolution of the characteristic length scale (a-b) illustrated by typical snapshots of a $200 \times 200$ system (c-h).  \textbf{(a)} Growth of the typical length scale $\xi(t)$ for $T = 0.4$ and several $\sigma$, as indicated in the legend above. \textbf{(b)} the same data where $\xi$ and $t$ have been rescaled by $\sigma$ and $\sigma^2$, respectively. All curves collapse onto a single master curve $f(x)=a(1-\exp(-b\sqrt{x}))$ with $a = 2.3$ and $b = 0.65$. \underline{Inset} : variation of the transients time $\tau$ defined in Eq.(\ref{eq:tau_def}) against the self-spinning intensity. \textbf{(c, d, e)} Representative configurations at different times showing the evolution of the system after a quench (for $T = 0.2$ and $\sigma^2 = 0$). The phase of each oscillator is represented by a cyclic colour scale. In the last panel only, vortices (anti-vortices) are represented by circles (triangles). \textbf{(f, g, h)} Representative configurations at different times now for $T = 0.2$ and $\sigma^2 = 0.1$.}
    \label{fig:kuramoto_xi}
	\end{figure}

To study the coarsening, or phase ordering, dynamics, we define a correlation length $\xi(t)$, extracted from $C(\xi(t),t) = 1/e$ and plot its time evolution in Fig.~\ref{fig:kuramoto_xi}(a) for different values of $\sigma$. 
For $\sigma=0$, we recover without surprise the well-known scaling $\xi(t) \sim \sqrt{t/\log t}$ \cite{jelic2011quench,rojas1999dynamical,YurkeHuse1993}. The logarithmic correction takes root in the logarithmic dependence of a defect mobility on its size; note that it is only relevant for large $t$. Such logarithmic correction to the usual $\sim \sqrt{t}$ growth law in systems with non-conserved order parameter dynamics also appears in all related quantities such as the evolution of the number of vortices $n$ or the mean square displacement (MSD) of a single  vortex, as we shall discuss in more detail later on. 

For $\sigma>0$, the results qualitatively differ: independently of the system size $L$, the growing length $\xi(t)$ saturates at a finite value at long times. Such steady value decreases as $\sigma$ increases. This is consistent with the limit case $\sigma\to \infty$, corresponding to a system of decoupled oscillatorst. Interestingly, Fig.~\ref{fig:kuramoto_xi}(b) shows that all curves for $\sigma>0$ collapse into a single master curve if both distances and time are rescaled. This master curve follows 
$f(x) = a(1-\exp(-b\sqrt{x}))$
from which we conclude that 
\begin{equation}
  \label{xieq}
  \xi(t) = \frac{a}{\sigma}\, \left( 1 - e^{-b\sqrt{\sigma^2\,t}}\right) \ ,
  \end{equation}
with $a =  2.3$, a constant related to the long time limit of $\xi$, and $b=0.65$. 

Equation (\ref{xieq}) contains information about both short time and long time dynamics. At short times, one recovers the short-time dynamics of the XY model. Indeed, by expanding $\xi(t)$ to first order in $t$, one gets 
\begin{equation}
  \sigma \, \xi(t)\approx ab \, \sqrt{t}
\end{equation}
$ab =1.5$ thus represents the slope of the initial $\sim \sqrt{t}$ coarsening. 

At intermediate times, the system crosses over from its short time dynamics, following the passive XY scaling, to its steady state. We define this transient time $\tau(\sigma)$ as the time for which the equilibrium correlation length is equal to the steady state out-of-equilibrium one: 

\begin{equation}
\xi(\sigma = 0, \tau) = \xi(\sigma, t \to \infty)\ .
\label{eq:tau_def}
\end{equation}

We plot it in the inset of Fig.~\ref{fig:kuramoto_xi} and obtain $\tau \sim 1/\sigma^2$: the wider the frequency distribution is, the faster is the relaxation process. This intuitive tendency is easily captured by a trivial scaling argument: the logarithmic correction is a long term effect, so we drop it and assume for the sake of simplicity that $\xi$ first follows an equilibrium growth $\sqrt{t}$. As it eventually saturates at a steady state value $\xi_\infty \sim 1/\sigma$, the crossover between both regimes thus occurs when

\begin{equation}
\sqrt{\tau} \sim \xi_\infty \sim 1/\sigma \Rightarrow \tau \sim 1/\sigma^2,
\end{equation}

confirming the scaling naturally contained in Eq.~(\ref{xieq}). 

At long times, in the steady state, the characteristic length boils down to
\begin{equation}
\lim_{t\to\infty} \xi(t) = \xi_\infty \sim \frac{1}{\sigma}
\end{equation}
One can understand this scaling in two different ways. 
\paragraph{Argument 1 : intrinsic frequencies sustain a non-zero angular difference between nearest-neighbours at mechanical equilibrium\\}


We simplify the problem and consider a 1D chain of spins and, for simplicity, we work at $T = 0$. First, we consider two spins $i$ and $k$ isolated from the rest of the lattice. From their respective equations of motion, one obtains
\begin{equation}
\frac{d}{dt} \Delta \theta = \Delta \omega - 2 \sin \Delta \theta,
\end{equation}
with $\Delta \theta = \theta_i - \theta_k$ and $\Delta \omega = \omega_i - \omega_k$. In the steady state, this yields $\Delta \theta_0 = \sin^{-1}(\Delta \omega / 2)$.

Let’s now consider the problem of a 1D chain of $N$ spins. The frequency difference between each pair of oscillators $(j, j+1)$ is denoted $\Delta \omega_j$, and its steady phase difference $\Delta \theta_j$. The phase difference accumulated along $n$ links reads
\begin{equation}
\delta_n = \sum_{j=i}^{i+n} |\Delta \theta_j| .
\end{equation}

\begin{figure}[h!]
  \centering
		\includegraphics[width=0.55\linewidth]{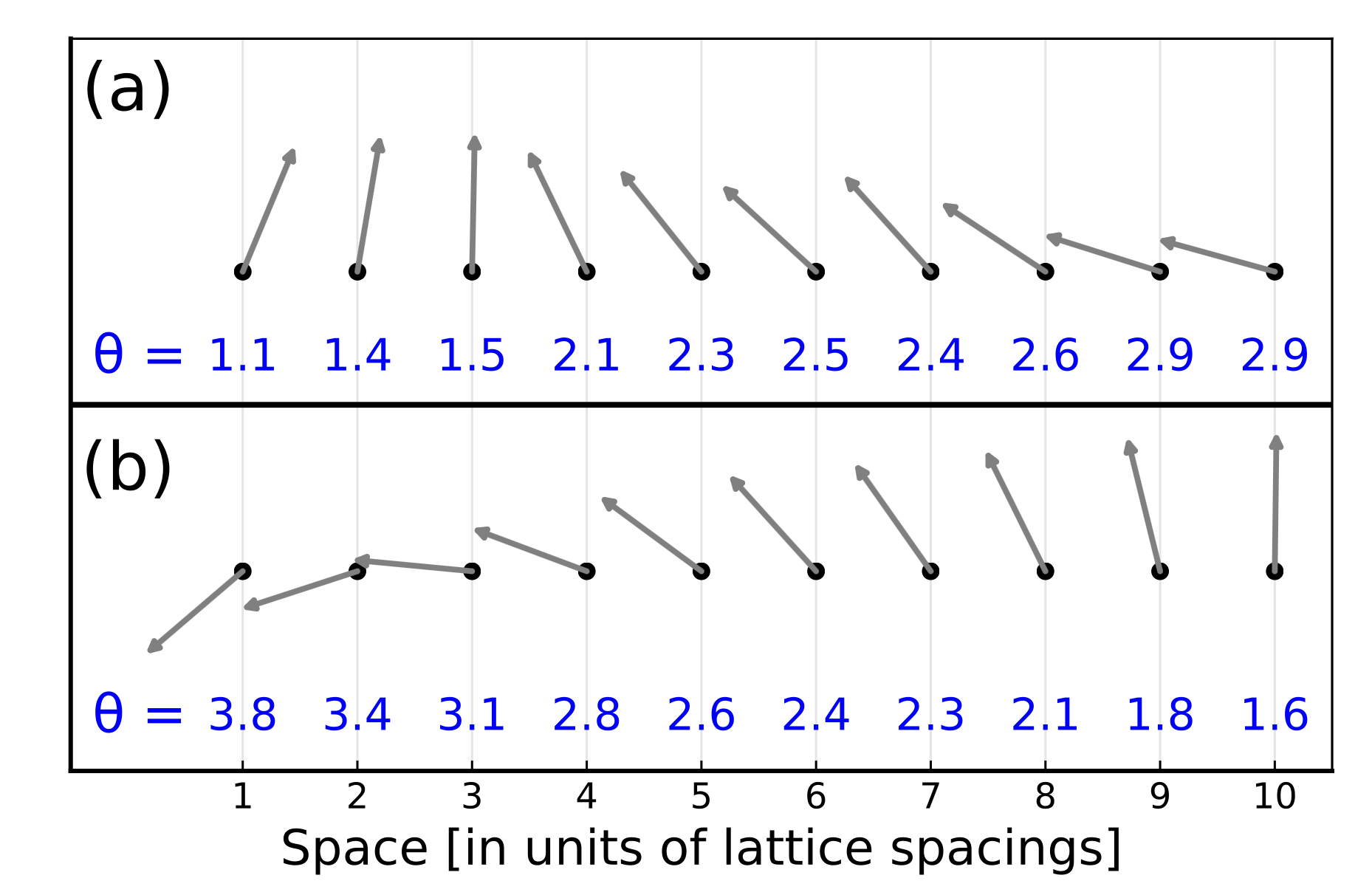}
    \includegraphics[width=0.44\linewidth]{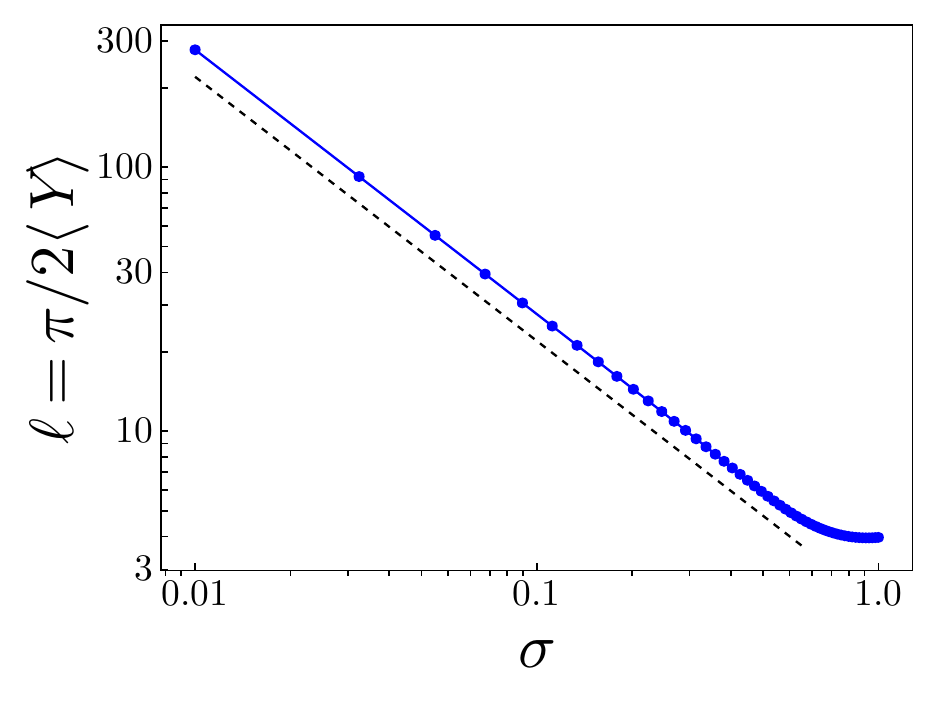}
    \caption{
    The left panels illustrate two typical independent situations selected from the steady-state of a $L = 200$ system at $\sigma^2 = 0.1$ and $T = 0$. We see that locally, spins tilt in one direction.
    \textbf{(a)} In this particular configuration, the characteristic length $\ell = 7$ because spins \#1 and \#8 are almost perpendicular.
    \textbf{(b)} Here also, $\ell = 7$ since spins \#3 and \#10 are almost perpendicular.
    The right panel \textbf{(c)} shows the characteristic length $\ell$ as a function of $\sigma$ (in log-log scale). The dashed line us $\ell=2.2/\sigma$, which is in good agreement with the numerical data. A perfect fit is given by $\ell=2.8/\sigma$. 
    } 
    \label{fig:1dchain_argument}
	\end{figure}

In actual system’s configurations, successive $\Delta \theta_j$ do not compensate; on the contrary, they accumulate in one direction (hence the absolute value above). To illustrate this point we show in Fig.~\ref{fig:1dchain_argument}(a,b) a detailed view of two typical configurations of the 2D system along one direction (at $T=0$, for clarity).

Then, we define a characteristic length $\ell$ as the typical distance between two spins that have an accumulated phase difference equal to $\pi/2$, i.e.
\begin{equation}
\delta_\ell = \sum_{j=i}^{i+\ell} |\Delta \theta_j| = \pi/2
\end{equation}

We then approximate the latter expression by $\delta_\ell \approx \ell \langle |\Delta \theta_0| \rangle$, where $\langle |\Delta \theta_0| \rangle$ is the expectation value of $|\Delta \theta_0|$, thus
\begin{equation}
\ell = \frac{\pi}{2 \langle |\Delta \theta_0| \rangle}
\end{equation}
The rest of the argument is purely technical and aims at proving that $\langle |\Delta \theta_0| \rangle \sim \sigma$. 
To carry out the derivation explicitly, we simplify notations by defining
\begin{equation}
X = \frac{\Delta \omega}{\sqrt{2} \sigma}, \qquad Y = \Delta \theta_0 = \sin^{-1}(\Delta \omega / 2) = \sin^{-1}(\sigma X / \sqrt{2})\ .
\end{equation}

The goal is now to express the probability density function $f_Y(y)$ of the random variable $Y$ in a closed-form. Eventually one wants to compute $\langle |Y| \rangle = \langle |\Delta \theta_0| \rangle$.

Since $\omega_i \sim \mathcal{N}(0, \sigma^2)$, one has $\Delta \omega \sim \mathcal{N}(0, 2\sigma^2)$, leading to $X \sim \mathcal{N}(0,1)$. Note that the case $|X| > \sqrt{2}/\sigma$, where the $\sin^{-1}$ is no longer defined, is rare enough to be neglected. The probability to encounter such a case is given by $1 - \text{erf}(1/\sigma)$, which, for the most unfavored value of $\sigma^2$ investigated, is of the order of $10^{-3}$.

The cumulative density function of $Y$ is given, for $-\pi/2 \leq y \leq \pi/2$, by
\begin{equation}
\begin{aligned}
F_Y(y) &= \mathbb{P}(Y \leq y) \\
  &= \mathbb{P}\left( \sin^{-1}(\alpha X / \sqrt{2}) \leq y \right) \\
  &= \mathbb{P}\left(X \leq \frac{\sqrt{2}}{\sigma} \sin y \right) \\
  &= \Phi \left( \frac{\sqrt{2}}{\sigma} \sin y \right) \\
  &= \frac{1}{2} \left[ 1 + \text{erf}\left( \frac{\sin y}{\sigma} \right) \right]
\end{aligned}
\end{equation}

Naturally, $F_Y(y < -\pi/2) = 0$ and $F_Y(y > \pi/2) = 1$. From $f_Y(y) = \dv{F_Y(y)}{y}$ it follows that
\begin{equation}
f_Y(y) = \frac{\cos y}{\sqrt{\pi} \sigma} \exp\left( - \left( \frac{\sin y}{\sigma} \right)^2 \right).
\end{equation}

Naturally, $f_Y(y < -\pi/2) = f_Y(y > \pi/2) = 0$.

Let’s compute the expectation value of $|Y|$:
\begin{equation}
\begin{aligned}
\langle |Y| \rangle 
  &= \int_{-\pi/2}^{+\pi/2} |y| f_Y(y) \, dy \\
  &= 2 \int_0^{\pi/2} y f_Y(y) \, dy \\
  &= \frac{2}{\sigma \sqrt{\pi}} \int_0^{\pi/2} y \cos y \exp\left( -\frac{\sin^2 y}{\sigma^2} \right) dy \\
  &= \frac{2}{\sqrt{\pi}} \int_0^{1/\sigma} \sin^{-1}(\sigma z) \, e^{-z^2} \, dz,
\end{aligned}
\end{equation}
where $z = \sin y / \sigma$.

We carry out this integration numerically to compute $\langle |Y| \rangle = \langle |\Delta \theta_0| \rangle$, which nicely reproduces the $1/\sigma$ scaling of the characteristic length $\xi$ extracted from the simulations, see Fig.~\ref{fig:1dchain_argument}(c).

\paragraph{Argument 2 : intrinsic frequencies screen the influence of a distant topological defect\\}
An easier and more straighforward argument to understand the scaling of the characteristic length $\xi$ is to consider the influence of a distant topological defect on a given rotor. In the absence of intrinsic frequencies, a distant defect generates a field disturbance that decays algebraically with distance as $1/r$. 

In the presence of intrinsic spinning, the field disturbances generated by the defects only expand over relatively short length scales, at any temperature. Consider Eq.~(\ref{eq:model}) at $T=0$ for a test spin at distance $r$ from a topological defect. 

Distance $r$ from a topological defect. The defect generates a field $\theta(x, y) = \arctan(y/x)$ around itself. As a first approximation, we replace the influence of the spin’s neighbourhood ($\partial$) by the Coulomb force generated by the defect, namely:
\begin{equation}
\sum_{\partial} \sin(\Delta \theta) \approx \sum_{\partial} |\nabla \theta(x, y)| = \frac{c}{r}.
\end{equation}

The constant $c$, homogeneous to a length, depends on various parameters, in particular the defect’s charge. One thus gets
\begin{equation}
\dot{\theta}_i = \sigma \omega_i + \frac{c}{r}.
\end{equation}
It now becomes clear that for $r > c/(\sigma \omega)$, $\dot{\theta}$ is dominated by the forcing term. As $\omega$ is a random variable with zero mean and unit variance, averaging over the whole system proves that beyond a typical distance $\xi \sim 1/\sigma$, the spins’ dynamics is no longer primarily governed by the topological defects’ influence but rather by the intrinsic driving frequency. 

Importantly, this second argument explains the same scaling without resorting to the crude 1D approximation of the first argument. It however relies on the assumption that the force generated by a defect still decays as $1/r$ even at short distances in the active case. 
All in all, both arguments lead to the same conclusion, namely that the characteristic length $\xi$ scales as $1/\sigma$ in the steady state, a signature of the breakdown of quasi-long-range order in the system, together with the spatial limitation of the $1/r$ influence of topological defects, key prerequisite to belong to the BKT universality class (recall the Peierls argument introduced in Chapter \ref{ch:XY}).

\section{Defect density thoughout the coarsening}
\begin{figure}[h!]
  \centering
    \includegraphics[width=0.85\linewidth]{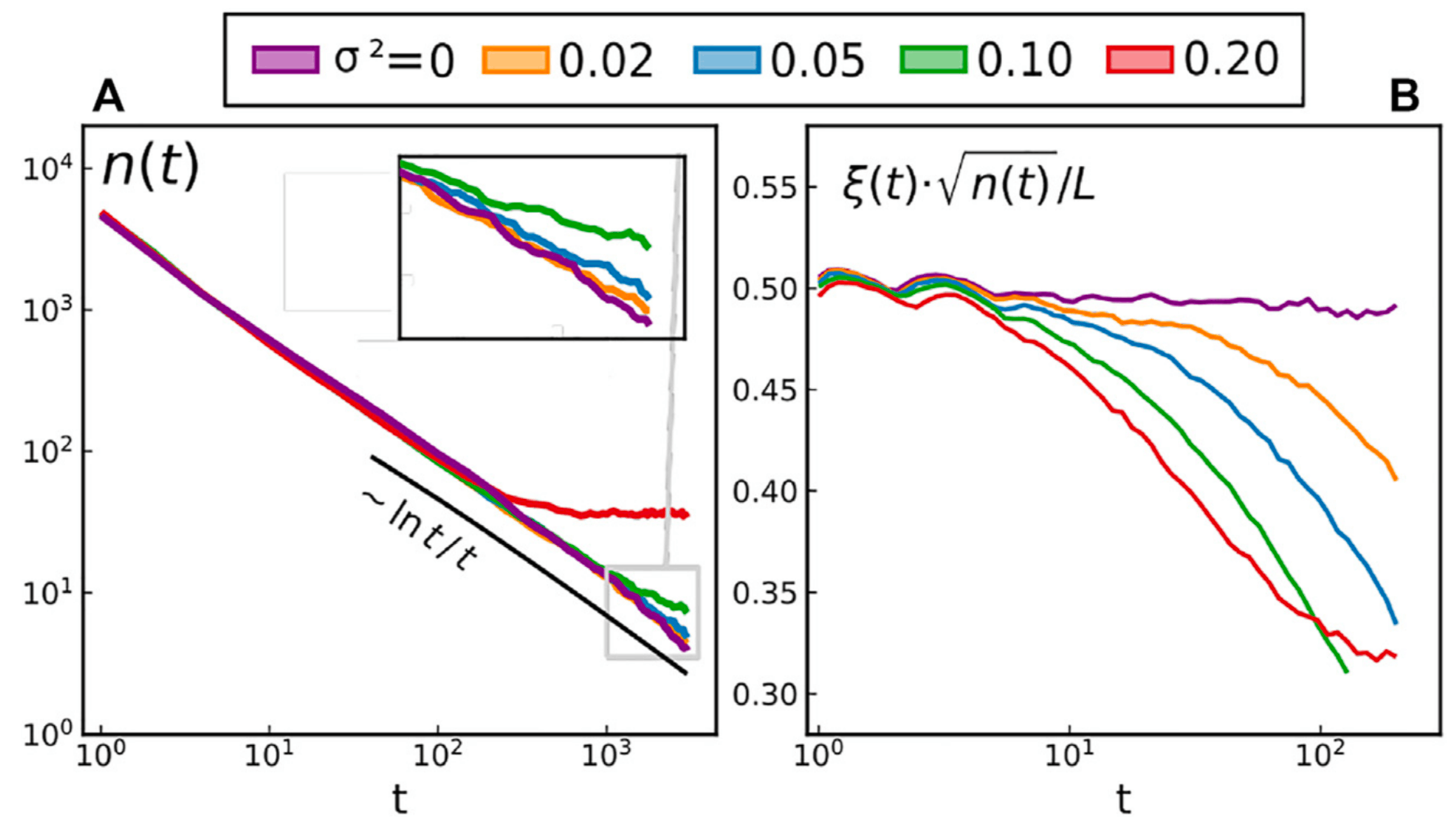}
    \caption{Time evolution of \textbf{(a)} the total number of defects $n$ for different forcing intensities and $T=0.4$ \textbf{(b)} the quantity $\xi\,\sqrt{\rho}$ where $\rho = n/L^2$ is the defect density. Note that because the defects are (statistically) homogeneously distributed, $\sqrt{\rho}$ is the average distance between two of them. }
    \label{fig:kuramoto_nt}
  \end{figure}

We complete the analysis of a system following an infinitely rapid quench from an initially disordered state to a low temperature  
by reporting the total number of defects $n(t)$ for different driving strengths $\sigma$ in Fig.~\ref{fig:kuramoto_nt}(a).

For the equilibrium XY model (purple), one obtains $n(t) \sim \log t\,/\,t$. 
The similarity of the equilibrium long-term scaling for $\xi(t)$ and for $n(t)$ is not a coincidence. The correlation length $\xi$ is intimately related to the total number of defects $n$ in the equilibrium XY model. Indeed, as a first approximation, $\xi$ is given by the average distance between defects; the system being homogeneous, it follows that $\xi \sim 1/\sqrt{\rho\,} =  L/\sqrt{n}$. We indeed recover that at equilibrium, $\xi\sqrt{n}$ is constant over time, cf. Fig.~\ref{fig:kuramoto_nt}(b). For $\sigma>0$, $n(t)$ eventually saturates at a finite steady state value, as the number of defects at a given temperature increases with $\sigma$. This departure from the  XY equilibrium behaviour is also clearly reflected by the fact that  $\xi$ is no longer given by the typical distance between defects, as shown in Fig.~\ref{fig:kuramoto_nt}(b). The steady state of the Kuramoto model is characterised by both a finite number of defects and a finite correlation length which are, \emph{a priori}, unrelated. 

\begin{figure}[h!]
  \centering
  \includegraphics[width=0.99\linewidth]{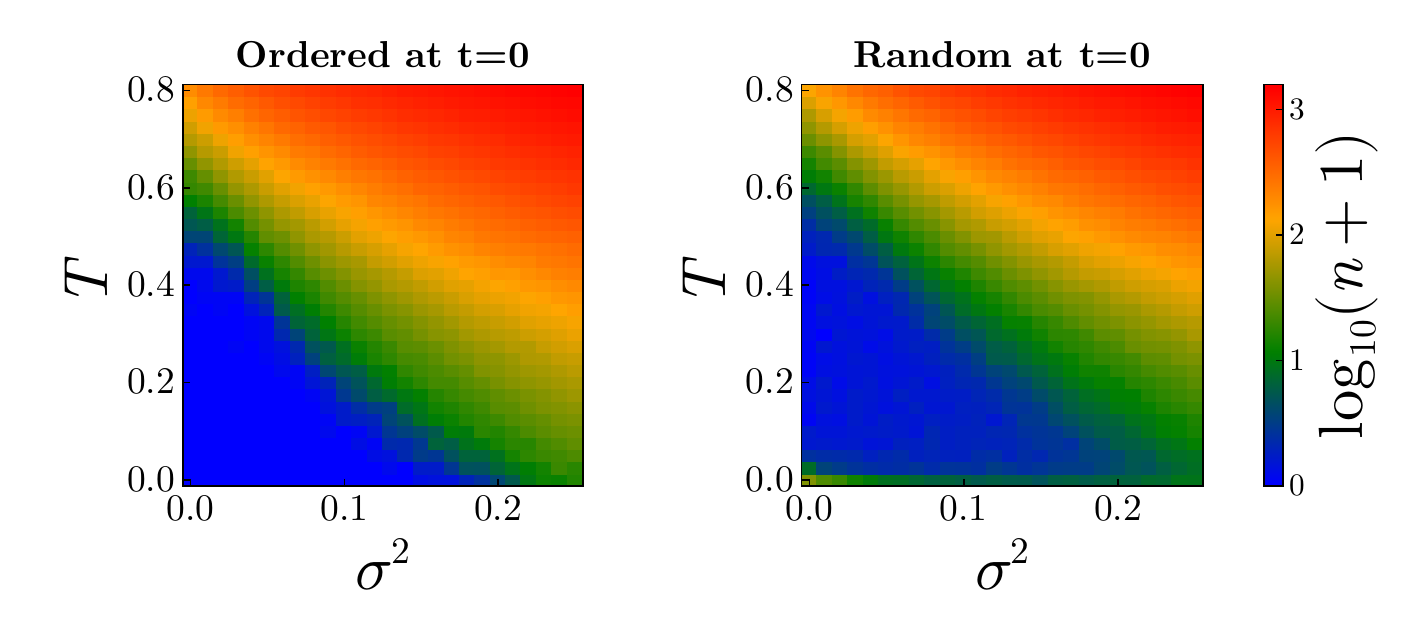}
  \caption{We represent $\log_{10}(1+n)$ over the phase space $\sigma^2-T$, where $n$ is the total number of defects of a $L=200$ system in the steady state for \textbf{(a)} an initially ordered configuration and \textbf{(b)} an initially disordered configuration. 
  }
  \label{fig:kuramoto_phase_space}
\end{figure}

As a useful reference for our analysis, we report a map of the total number of defects in the steady state over the $\sigma^2$-$T$ phase space in Fig.~\ref{fig:kuramoto_phase_space}. 

The number of defects in the steady state increases with the temperature and the self-spinning intensity $\sigma$, as expected.
As $T$ and $\sigma^2$ decrease, the dynamics gets slower and slower, up to the $T = 0$ limit case where the self-spinning alone cannot help annihilating pairs of defects: a system initially disordered remains stuck in a metastable state with a lot more vortices than the number of vortices it would exhibit if relaxed from an initially ordered state, as illustrated in Fig.~\ref{fig:quench_XY_snapshots}(a) for the XY case.

We state once again that the crossover from the defect-rare (blue region) to the defect-rich (red region) regime does not result from an actual phase transition: there is no qualitative change in the correlation length as one moves across the phase space (the system only exhibits short-range order at any $\sigma \neq 0 $ for all $T\geq 0$).
At contrast with the standard literature of the XY model \cite{Kosterlitz1973,Cugliandolo_lecture_notes_phase_transitions}, we do not classify vortices into free and bounded here. Indeed, as discussed in more detail below
, the core mechanism responsible for the BKT phase transition, namely the defects' unbinding at \textit{a finite} temperature, breaks down upon self-spinning.


\section{Boundary formation and defect generation mechanisms}

At this stage, we have studied the coarsening dynamics of the system following a quench from a random initial state to a low temperature regime. In this case, topological defects naturally stem from this initial state. 
However, at this stage, the question of how defects are dynamically generated remains open. This section aims at shedding some light on the underlying mechanisms responsible for the spontaneous creation of domain boundaries and topological defects.

Interested by the emergence of the patterns such as those in the first snapshots (cf. Fig.~\ref{fig:snapshots_kuramoto}(b)), we look at the short time dynamics of the fields $\theta\equiv\theta(x,y)$ and $\dot{\theta}\equiv\dot{\theta}(x,y)$ ($x$, $y$ being the spatial coordinates of a point in the lattice). We do so by following the evolution of the system from an ordered initial condition (all sites with identical phase) in order to isolate the underlying mechanisms at stake and avoid being blurred by the numerous defects inherent to a disordered initial condition. As the time series of $\theta$ is noisy for $T>0$, one has to average in time to obtain a meaningful signal $\dot{\theta}$. We detail the exponential moving average used in that preliminary signal processing step in the Appendix \ref{sec:appendix_exp_moving_avg}. \\

\begin{figure}[h!]
  \centering
  \includegraphics[width=0.99\linewidth]{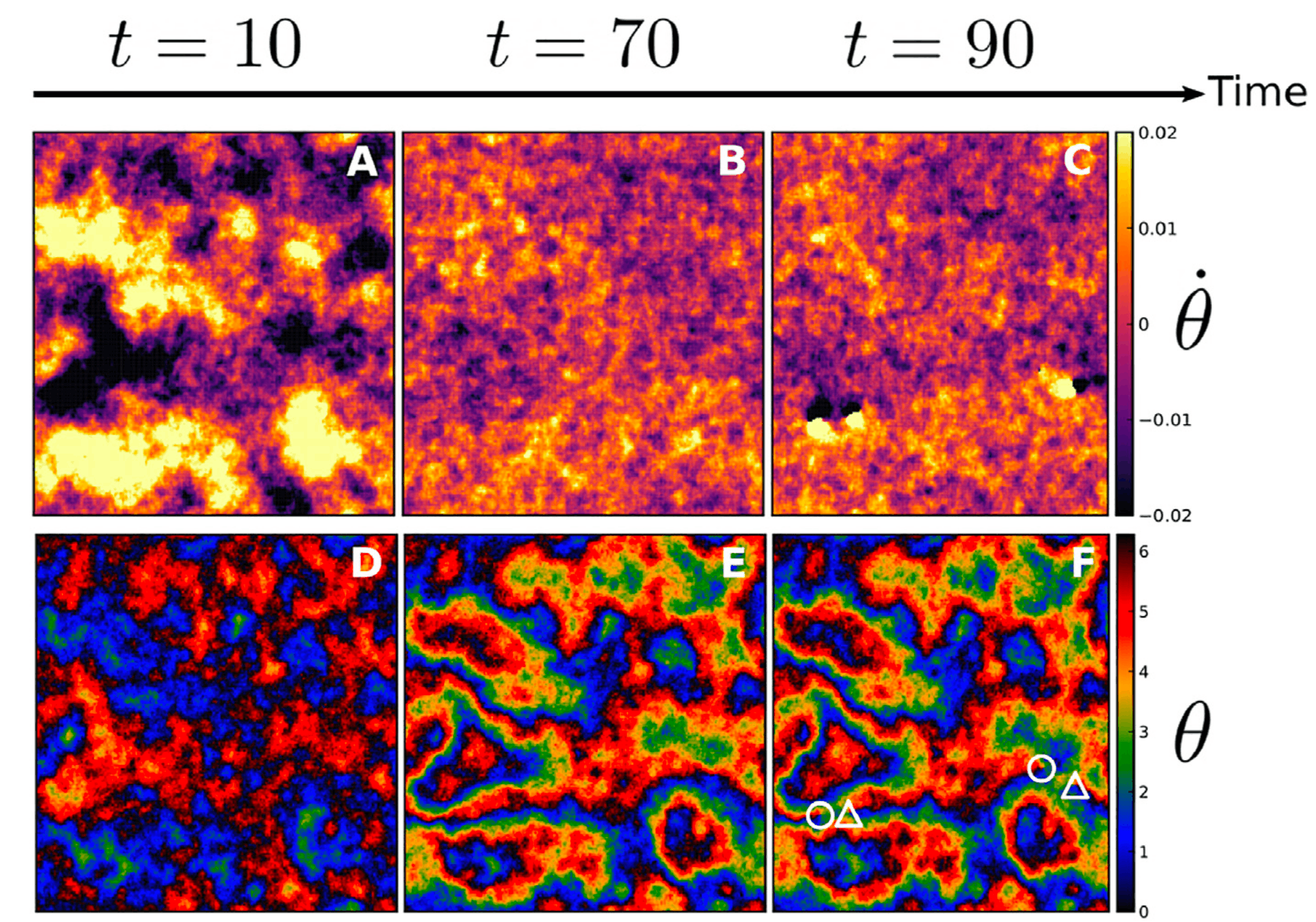}
  \caption{
    Three snapshots over time of the fields $\dot{\theta}(x,y)$ \textbf{(a-c)} and $\theta(x,y)$ \textbf{(d-f)}  for a system initially ordered with $T=0.05$ and $\sigma^2 = 0.2$. In panel (c), defects are clearly identified thanks to the localized and intense amplitude of the instantaneous frequencies. In panel (f), $+1$ defects are highlighted by a circle, $-1$ defects are highlighted by a triangle. 
  }\label{fig:theta_thetadot}
\end{figure}

After very few time steps, a clear picture appears: locally synchronised regions form in the $\dot{\theta}$ field. We recall that since $d=2$ is the lower critical dimension for frequency locking in the noiseless short range Kuramoto model, we do not expect more than local frequency synchronisation. Since the average instantaneous frequency (over the $N$ spins at a given time $t$) scales as $1/\sqrt{N}$ ---as one should expect from a zero-mean distribution of intrinsic frequencies, reciprocal interactions and gaussian white noise---, some regions rotate clockwise and some others counterclockwise; look for instance at the dark and bright regions of Fig.~\ref{fig:theta_thetadot}(a). Naturally, these locally synchronised regions in $\dot{\theta}$ translate  into locally aligned domains in the phase field. This is illustrated in Fig.~\ref{fig:theta_thetadot}: the L-shaped dark region by the left of panel (a) generates a domain of same shape at the same position in panel (d), now in red.


As time goes on, the different domains grow, cf. panel (e), defining a network of  domain boundaries. As they correspond to regions where $|\nabla \theta|$ is large, they can be visually identified by looking for thin elongated regions across which colours change rapidly (though in a smooth way in contrast with, for instance, domain boundaries in the Ising model).
At this stage, the domains are locally ordered, their boundaries concentrate most of the energy of the system and the instantaneous frequencies $\dot{\theta}$ gradually decrease in amplitude (compare panels (a) and (b)). 
\newline

\begin{figure}[h!]
  \centering
  \includegraphics[width=0.99\linewidth]{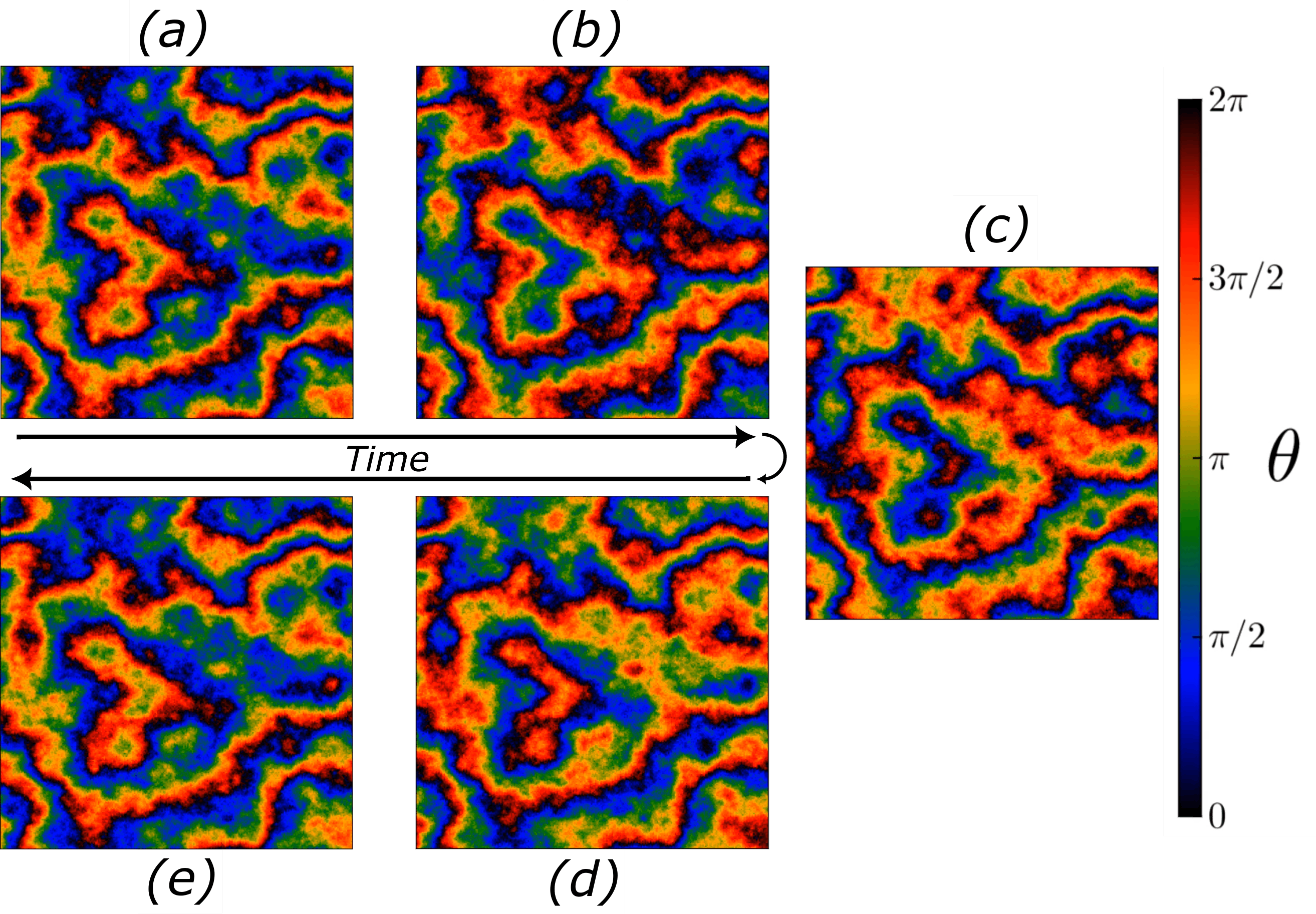}
  \caption{Phase field $\theta$ for a square system of size $L=256$, with $T=0.2$, $\sigma = 0.08$, initially ordered. The differents panels are snapshots at times $t=$
  \textbf{(a)} 27500
  \textbf{(b)} 34700
  \textbf{(c)} 40300
  \textbf{(d)} 46400
  \textbf{(e)} 51400. 
  No defect is present in those configurations. Note the similarity between panels (a) and (e), highlighting the time periodicity of the patterns.
  }
  \label{fig:waves_no_defects}
\end{figure}

{First, let us address the low-$T$, low-$\sigma$ case. In that blue region of the phase space Fig.~\ref{fig:kuramoto_phase_space}, the system remains defectless. It reaches a steady state characterised by synchronised domains, resulting from the competition between the elastic energy and the driving frequencies $\sigma \omega_i$. 
The patterns are visually similar to those observed in presence of defects, and all the phenomenology reported earlier on this model remains true in absence of defects, including the scaling of the correlation length $\xi \sim 1/\sigma$ and the time scaling of Eq.~(\ref{xieq}). 
We show in Fig.~\ref{fig:waves_no_defects} a series of snapshots of the phase field $\theta$ in a system with no defect, for $T=0.2$ and $\sigma=0.08$ (close to the crossover but still below the spontaneous defect creation crossover). 
As seen in the different snapshots over time in Fig.~\ref{fig:waves_no_defects}, these boundaries are dynamic and evolve periodically, as highlighted by the similarity between panels (a) and (e).}

{The dynamics of those boundaries occurs over a time scale much longer than the defect motion that we describe later in this chapter ($\Delta t \approx 15000$ versus $\approx 600$ to displace by $\Delta x\approx50$ for $\sigma = 0.08$). We have not characterised the boundaries thorougly, since the main focus of our study were topological defects. Identifying and tracking those continuous objects is not an easy task, but interesting measurements could be done in the future, such as measuring their wavelength $\lambda$, their frequency $f$ and their speed of propagation $c$. We expect $\lambda$ to scale as $\xi\sim 1/\sigma$ since it is the only lengthscale in the system associated to $\sigma$, $f$ to scale as $\sigma$ since both are frequencies, so that $c=\lambda f$ should be roughly independent of $\sigma$ and depend only on $a\gamma/J$, with $a$ the lattice spacing, $\gamma$ the friction coefficient and $J$ the coupling constant. }

\begin{figure}[h!]
  \centering
  \includegraphics[width=0.99\linewidth]{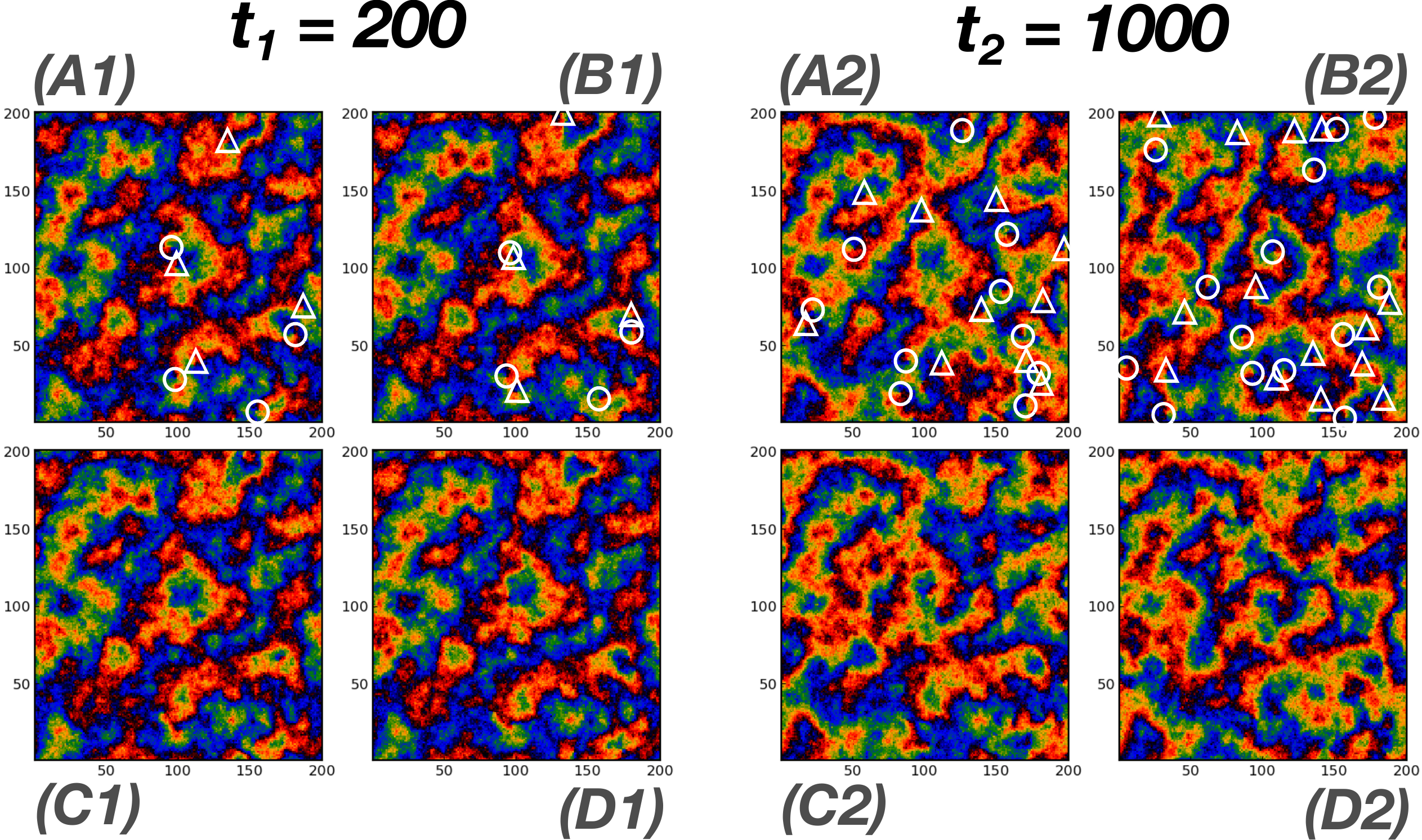}
  \caption{
    Snapshots of four different systems at $T=0.2$ and $\sigma^2 = 0.2$ $\{\omega_i\}$, with the same fixed frequencies : $\omega_{i, A} = \omega_{i, B} = \omega_{i, C} = \omega_{i, D}$ for all spins $i$, at two different times $t_1=200$ and $t_2=1000$. The only difference between the four systems is the realisation of the thermal noise. The initial condition are completely ordered, i.e. $\theta_i(t=0) = 0$ for all $i$.
    The phase of each oscillator is represented by the usual cyclic colour scale.
    For the panels A and B, we highlight the defects by circles for $+1$ defects and triangles for $-1$ defects.
  }\label{fig:snapshots_fixed_frequencies}
\end{figure}

Those patterns are mainly shaped by the realisation of the quenched disorder $\{\omega_i\}$. We show in Fig.~\ref{fig:snapshots_fixed_frequencies} snapshots of four different systems (A,B,C,D) at $T=0.2$ and $\sigma^2 = 0.2$ with the same fixed frequencies $\omega_{i, A} = \omega_{i, B} = \omega_{i, C} = \omega_{i, D}$ for all spins $i$, at two different times $t_1=200$ and $t_2=1000$. The only difference between the four systems is the realisation of the thermal noise. Note how similar the patterns are until $t_1$ (left block, panels A1 to D1). 
This shows that the initial patterns and the location of spontaneous defect generation are mainly governed by the quenched disorder, while the thermal noise only plays a minor role for these processes.

Finally, for sufficiently high $T$ and $\sigma$ (red region), these domain boundaries play a crucial role in the formation of defects. On the boundaries, the important gradients $|\nabla \theta|$ provide most of the energy needed to create and sustain topological defects. The remaining energy contribution eventually comes from thermal fluctuations, explaining why the temperature threshold necessary to spontaneously generate defects decreases as the forcing increases, see Fig.~\ref{fig:kuramoto_phase_space}. 
It also explains why one observes defects creation primarily at the domain boundaries, as illustrated in Fig.~\ref{fig:theta_thetadot}(f) and Fig.~\ref{fig:snapshots_fixed_frequencies}(A1, B1). 

This creation mechanism contrasts with that of the equilibrium case, entirely due to thermal fluctuations and hence homogeneously distributed in space. Yet, the creation mechanism is not the only difference between the active and the passive case: a major difference lies in the motion of these topological defects, as we detail in the last two sections of this chapter.

\section{Defects free motion}  \label{sec:kuramoto_defects_free_motion}
Confirming the intuition from the visual inspection of the system, cf. snapshots in Fig.~\ref{fig:snapshots_kuramoto}, further analysis of the behaviour of a pair of defects shows a clear absence of any relevant defect-defect interacting potential: topological defects are genuinely free in the driven system at \textit{any} temperature $T > 0$. In other words, there exists no finite temperature $T_c$ below which the defects are bounded into pairs. While two XY defects of opposite charge attract each other with a logarithmic potential, we find that the defects, upon self-spinning, generically unbind at any $T \geq 0$ and $\sigma > 0$.

\begin{figure}[h!]
  \centering
    \includegraphics[width=0.7\linewidth]{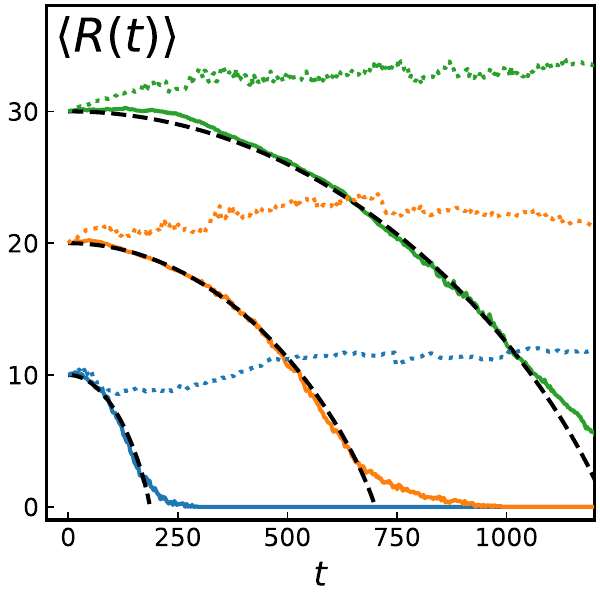}
    \caption{Vortex-antivortex separation $R(t)$, averaged over 120 independent runs, for three different initial vortex-antivortex distances $R_0$ (blue: $R_0 =10$, orange: $R_0 = 20$ and green: $R_0 = 30$). Full coloured lines correspond to $T = 0.1$ and $\sigma^2=0$. Dotted coloured lines correspond to $T = 0.1$ and $\sigma^2=0.1$. Black dashed lines are the XY predictions for an overdamped dynamics with a Coulomb interaction potential, see Chapter \ref{sec:XY_defect_interaction} and \cite{YurkeHuse1993}.}
    \label{fig:kuramoto_Rt}
\end{figure}

We prove it by tracking vortices over time, and considering the average distance between two defects. To do so, we initialize the system in a configuration with a vortex-antivortex pair at a controlled distance $R_0$. We then let the system evolve at low-$T$ and low-$\sigma$: we position ourselves in the blue region of the phase space in Fig.~\ref{fig:kuramoto_phase_space}
such that there is no (or very few) spontaneous creation of vortices to perturb the tracking. We monitor the distance $R(t)$ between our two defects for different initial separations $R_0$ and present the results in Fig.~\ref{fig:kuramoto_Rt}. 

For $\sigma^2=0$ (full lines), the separation $R(t)$ decays with time, as expected in the equilibrium XY model since the defects of opposite charge attract each other until annihilation at contact. The data is consistent with the prediction for an overdamped dynamics with a Coulomb interaction potential, drawn in black dash lines and previously discussed in Chapter \ref{sec:XY_defect_interaction}. \\
\begin{equation}
  \dot R \log R = V'(R) 
\end{equation}
where $V(R) \propto \log R$, a Coulomb 2D potential  \cite{YurkeHuse1993,minnhagen1987coulombgas}. \\

Spinning qualitatively alters this scenario. The analysis of pair trajectories, cf. Fig.~\ref{fig:kuramoto_Rt} reveals that, for $\sigma > 0$, the individual motion of the vortices is uncorrelated: in some cases, they approach and annihilate by chance, but, in many other cases, they move apart. 
As a result, their average distance $R(t)$ remains largely constant over time,
indicating that defects unbind and evolve freely.

One could naively attribute this feature to the non-equilibrium nature of the system. However, deciphering the impact of a non-equilibrium drive on the long-range properties of a system on genereal grounds is a challenging task and one is usually constrained to rely on specific examples where this question has been addressed. 
For instance, it has been shown that the general BKT scenario remains valid (at least not so far from equilibrium) for an XY model with exponentially correlated thermal noise \cite{paoluzzi2018effective} and for the 2D solid-hexatic transition of self-propelled disks \cite{LinoDefects}. In the present model system though, the BKT scenario of the XY model breaks down.

As it shows that long-range influence is lost as soon as the model is made active, the present work conceptually supports the conclusions of Pearce \textit{et al.} \cite{pearce_orientational_2021}, where they report that long-range ordering of defects in active nematics does not take place. 

Following up on the recent work of Pokawanvit \textit{et al.} \cite{pokawanvit2022active}, we underline that our system features an incompressible, dense and immobile ensemble of particles, which, based on their work, will not give rise to any collective motion of topological defects.

\section{Defect superdiffusion}
	
\begin{figure}[h!]
\includegraphics[width=0.98\linewidth]{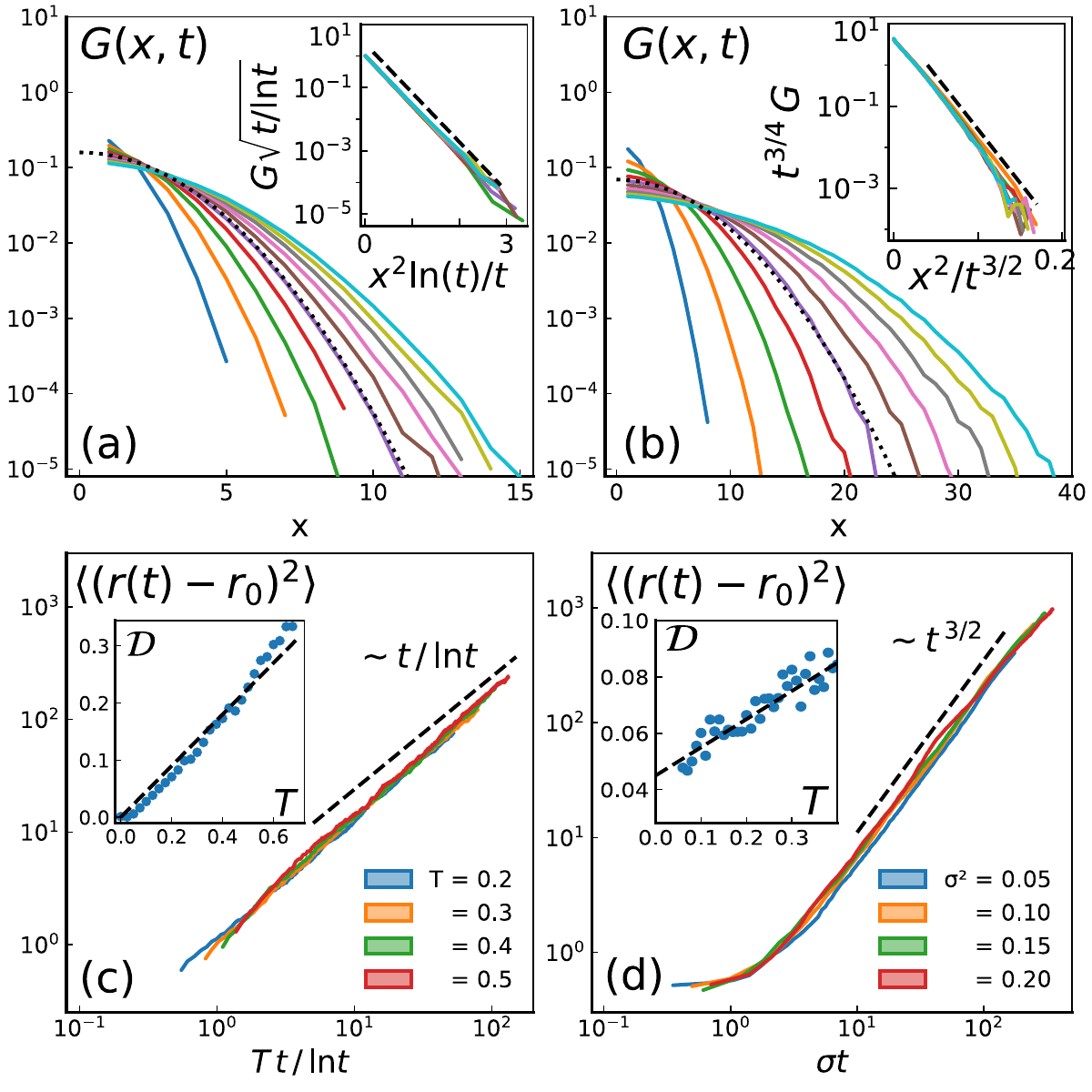}
  \caption{ \textbf{(a)} Distribution of vortex displacements $G(x,t)$ at different times $t=50,100,...,500$ (from left to right) and fixed $\sigma=0$, $T=0.2$. The dotted line shows Eq.~(\ref{eq:GXY}) with $\mathcal{D}=0.069$ and $t=250$.
    \underline{Inset}: these curves collapse when using the scaling Eq.~(\ref{eq:GXY}). The exponential decay of slope $1/(4\mathcal{D})$ is shown in dash. 
    \textbf{(b)} $G(x,t)$ (same times as in (a)) for $\sigma^2=0.025$ and $T=0.2$. The dotted line shows Eq.~(\ref{eq:GNKM}) with $\mathcal{D}=0.066$ and $t=250$. \underline{Inset}: shows an exponential decay of slope $1/(4\mathcal{D}\sigma^{3/2})$. 
    \textbf{(c)} MSD of a single vortex  for $\sigma=0$ and different $T$. 
    The dashed line shows a logarithmic correction to normal diffusion. \underline{Inset}: $\mathcal{D}$ as a function fo $T$, confronted to a linear growth of slope $0.45$. 
    \textbf{(d)} MSD at fixed $T=0.05$ but different $\sigma^2$, as a function of $\sigma t$. The dashed line shows  $\sim t^{3/2}$. \underline{Inset}: $\mathcal{D}$ vs. $T$ for $\sigma^2=0.025$, confronted to a linear growth of slope $0.1$. 
    }
  \label{fig:gmsd}
\end{figure}

Another remarkable property of topological defects upon self-spinning is found in the statistics of their displacement: they feature superdiffusion with exponent $3/2$. Once again we investigate the dynamics of vortices by tracking them. Since we now know that defects are free, we can focus now on the motion a single defect. To do so, we prepare an initial configuration hosting a single $+1$ defect at the center of the simulation box. One can exclusively focus on $+1$ vortices without loss of generality because the equation of motion Eq.~(\ref{eq:model}) is statistically invariant under a change of variable $\theta\to-\theta$ (as long as the distribution of the $\{\omega_i\}$ is centred) and such a transformation transforms a $+1$ vortex into a $-1$ anti-vortex. The dynamics of $\pm 1$ defects are thus statistically identical. We then let the system evolve at low-$T$ and low-$\sigma$ ensuring that new defects are not spontaneously created and that the simulation box is large enough such that boundary effects are negligible. \cite{LoftDeGrand, Olsson1992, Olsson1999, Japo}. The probability distribution function (PDF) of displacements thus obtained, $G(x,t)$, representing the probability that a walker initially at $x=0$ at $t=0$ is at position $x$ at time $t>0$, is shown in Fig.~\ref{fig:gmsd} for different times.

For the 2D XY model, it follows

\begin{equation}\label{eq:GXY}
G(x,t)=\frac{1}{\sqrt{{4\pi \mathcal{D}} \,t/\ln t }}\exp \left[-\frac{x^2}{4\mathcal{D}\, t/\ln t}\right] \ .
\end{equation}
Such form implies anomalous diffusion, in the sense that the ensemble averaged mean square displacement (MSD) behaves, at long times, as
\begin{equation}\label{eq:MSDXY}
\Delta^2(t) = \langle\left[ \bold{r}(t)-\bold{r}(0)\right]^2\rangle=4\mathcal{D}\,t/\ln t\, ,
\end{equation}
showing the expected correction to the normal diffusion scaling, due to the logarithmic dependence of the defect mobility on its size, with a pseudo-diffusion coefficient $\mathcal{D}\propto T$ (see Fig.~\ref{fig:gmsd}(c)) \cite{YurkeHuse1993, Asja, Japo}.

Unlike passive diffusion in equilibrium conditions, upon self-spinning, vortices become superdiffusive, following a Gaussian PDF
\begin{equation}\label{eq:GNKM}
G(x,t)=\frac{1}{\sqrt{{4\pi \mathcal{D}} \,(\sigma t)^{3/2} }}\exp \left[-\frac{x^2}{4\mathcal{D}\, (\sigma t)^{3/2}}\right] \ ,
\end{equation}
which naturally results in
\begin{equation}\label{eq:MSDNKM}
\Delta^2(t) = \langle\left[ \bold{r}(t)-\bold{r}(0)\right]^2\rangle=4\mathcal{D}(\sigma t)^{3/2} \ .
\end{equation}
As shown in Fig.~\ref{fig:gmsd}, these expressions describe our numerical data accurately. The $\sigma$-dependence of the vortex dynamics enters through a rescaling of time: the typical time associated with its motion along a domain is $\propto \xi\sim 1/\sigma$. 
All the $T$-dependance is simply contained in $\mathcal{D}\propto T$. Note that in the Kuramoto limit $T\to0$, vortices are still moving in a lively heterogeneous medium and are thus superdiffusive.

Such behaviour is reminiscent of the anomalous diffusion of tracer particles in quenched random velocity fields \cite{Redner1989superdiff, BouchaudGeorgesPRL1990, PhysA2020} or turbulent flows \cite{BouchaudRev}. However, superdiffusion in this classical context is typically characterised by non-Gaussian statistics and driven by convection rather than an internal drive. 
Recent studies in 2D systems of self-driven units have shown that the dynamics of topological defects is strongly affected by local energy inputs, resulting in spontaneous self-propulsion \cite{Thampi2013,Giomi2013, Giomi2014, Decamp2015, ChateBacteria2019, Shankar2018,shankar2019hydrodynamics, vliegenthart2020filamentous,ChardacBartolo2021}. 
Most of these studies concern active nematics \cite{shankar2019hydrodynamics, giomi2014defect,doostmohammadi2018active,vliegenthart2020filamentous} (see \cite{doostmohammadi2018active} for a review on active nematics), typically described at the coarse-grained level of hydrodynamic theories. Refs. \cite{Decamp2015,ChateBacteria2019, vliegenthart2020filamentous} constitute salient exceptions studying particle-based active nematics and monitoring the motion of defects.
In particular, Vliegenthart \textit{et al.} \cite{vliegenthart2020filamentous}  reports superdiffusion of $+\frac{1}{2}$ disclinations, with $\Delta^2(t)\sim t^{1.8}$, in dense systems of flexible active filaments, and in Bililign \textit{et al.} \cite{Bililign2021} dislocations in dense monolayers of driven colloids show superdiffusion with  $\Delta^2(t)\sim t^{3/2}$.
It is thus remarkable that in our simpler lattice model, we recover similar behaviour for vortices and anti-vortices, which are isotropic defects. In our case, their preferential direction of motion is given by the underlying domain structure.  

\section{$3/2$ exponent explanation}

Since superdiffusion is a frequent phenomenon in active matter, it naturally led to a plethora of random walk models exhibiting transient \cite{caprini2022dynamics,villa-torrealba_run-and-tumble_2020} or long term superdiffusion \cite{fedotov_emergence_2017,han_self-reinforcing_2021,bertoin1996levy}. 
However, we emphasise that as soon as $\sigma\neq 0$, in the time window allowed by our simulations, the MSD display a distinct $3/2$ slope without any sign of a crossover to a longer term diffusive regime. 

\begin{figure}[h!]
	\centering
	\includegraphics[width=0.99\linewidth]{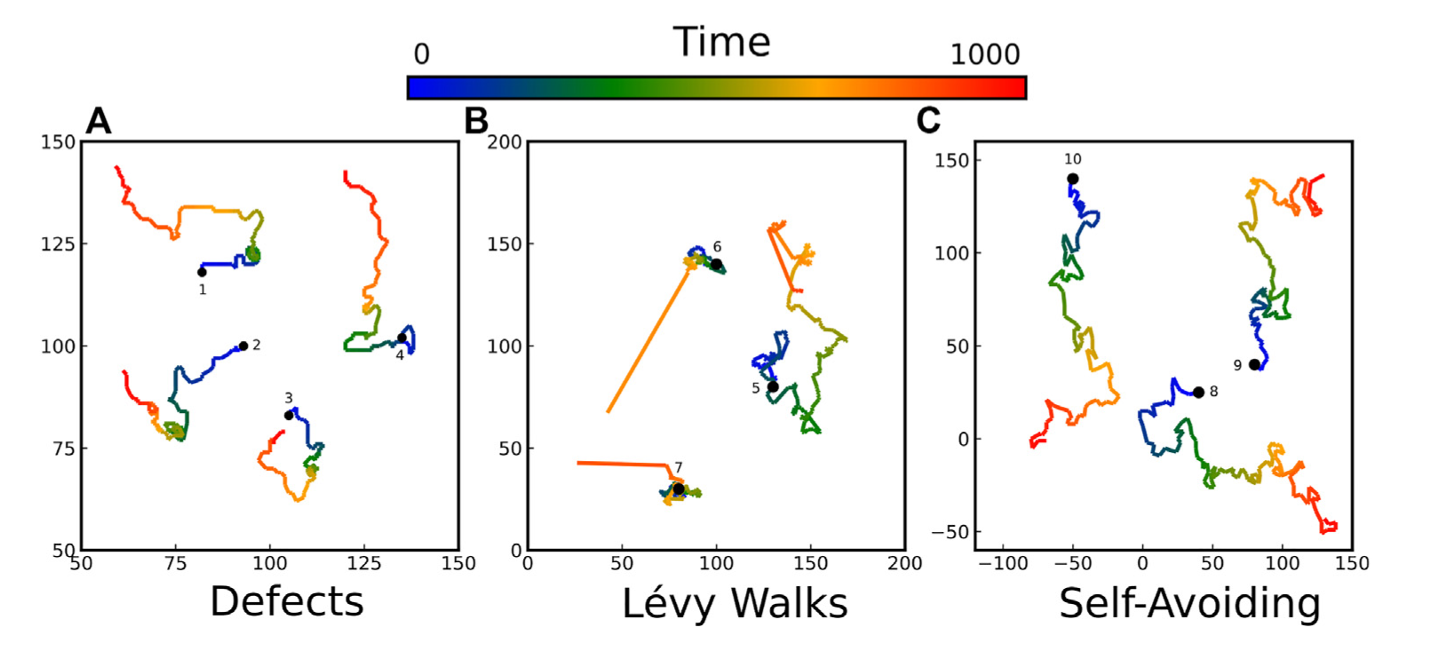}
	\caption{Sample trajectories of \textbf{(a)} topological defects in the actual spin model \textbf{(b)} L\'evy walks with exponent $\gamma= 3/2$ \textbf{(c)} self-avoiding random walks (SAW). 
		Each path departs from a black circle and lasts for $\Delta t= 10^3$, indicated by the colour code.}
	\label{fig:trajectories}
\end{figure}

Among models featuring long-term superdiffusion, the family of L\'evy processes is a popular choice, in particular in biology \cite{ariel2017chaotic, huda2018levy}. In L\'evy walks (respectively L\'evy flights, their discontinuous counterpart), the duration $\tau$ of each walk (respectively the size of each jump) is usually sampled from a fat-tailed distribution $f(\tau) \sim \tau^{-(1+\gamma)}$, where $1<\gamma<2$ is a common choice for superdiffusing L\'evy processes. The resulting MSD follows $\sim t^{3-\gamma}$ (see \cite{levywalks_review} and references therein for a complete review). 
The superdiffusion stems from the infinite variance of this distribution, explaining why the resulting trajectories sometimes feature an enormous jump (as trajectories 6 and 7 depict in Fig.~\ref{fig:trajectories}(b)). Such jumps cannot occur in our system, and thus L\'evy processes do not provide a faithful model to describe the random motion of vortices in the Kuramoto model, see trajectories 1 to 4 in Fig.~\ref{fig:trajectories}(a). In addition, there is no physical argument to support the specific choice of the exponent $\gamma = 3/2$ (the only one to lead to the MSD of our defects).

\begin{figure}[h!]
	\includegraphics[width=0.99\linewidth]{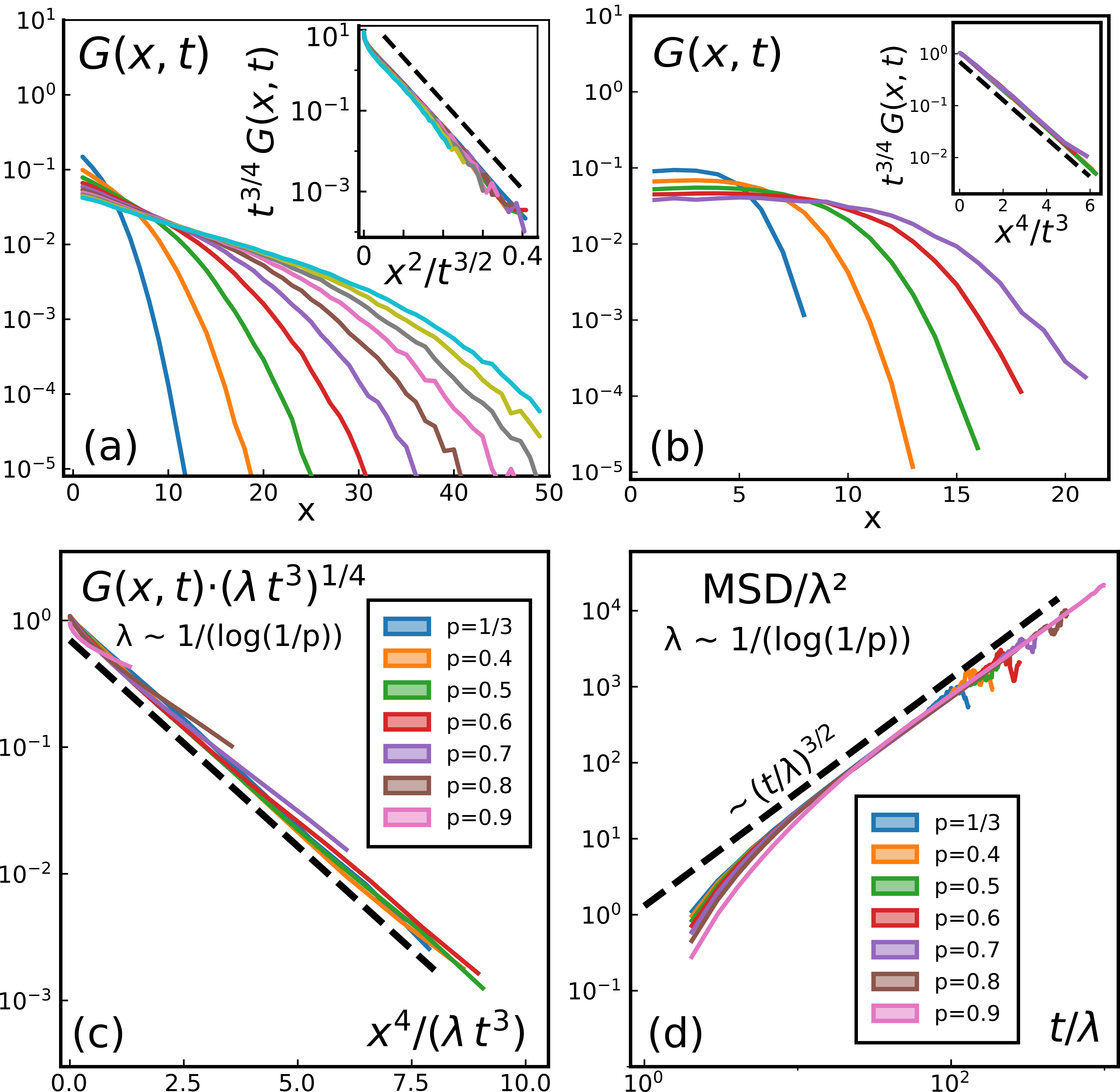}
	\caption{Probability densities of displacement $G(x,t)$ (see main text for definition) for different cases. Both insets represent rescaled data in order to higlight the functional form of $G$. We considered 
		\textbf{(a)} topological defects with $\sigma^2 = 0.025$ and $T = 0.2$, at times (from left to right) $t = 50,100,150,...,500$. Inset: the dashed line follows $f(x) \sim \exp(-60x)$ 
		\textbf{(b)} SAW ($\Delta t = 1$) at times (from left to right) $t = 10,15,20,25,30$. Inset: the dashed line follows $f(x) \sim \exp(-0.85x)$ 
		\textbf{(c)} Persistent SAW (see main text for definition) with different persistence probabilities on rescaled axes to highlight that the functional form is identical for all $1/3\le p<1$. The dashed line follows $f(x) \sim \exp(-3x/4)$.  
		\textbf{(d)} MSD resulting from $G(x,t)$ of panel \textbf{(C)}, for different persistence probabilities $p$. 
	}\label{fig:MSD_4panels}
\end{figure}

An important feature of the system to grasp the dynamics of the vortices is the domain structure of the system. 
Indeed, there is a feedback between the spatio-temporal patterns dynamics and the defects' motion. 
On one hand, defects have a strong preference to move along domain boundaries, where elastic energy is the largest  \cite{Rouzaire_Kuramoto_PRL, ChardacBartolo2021}. 
As a domain boundary separates two neighbouring ordered domains, it is characterised by an excess elastic energy, providing a preferential direction for the motion of the topological defects, in contrast with the isotropic random motion.

\begin{figure}[h!]
\includegraphics[width=0.99\linewidth]{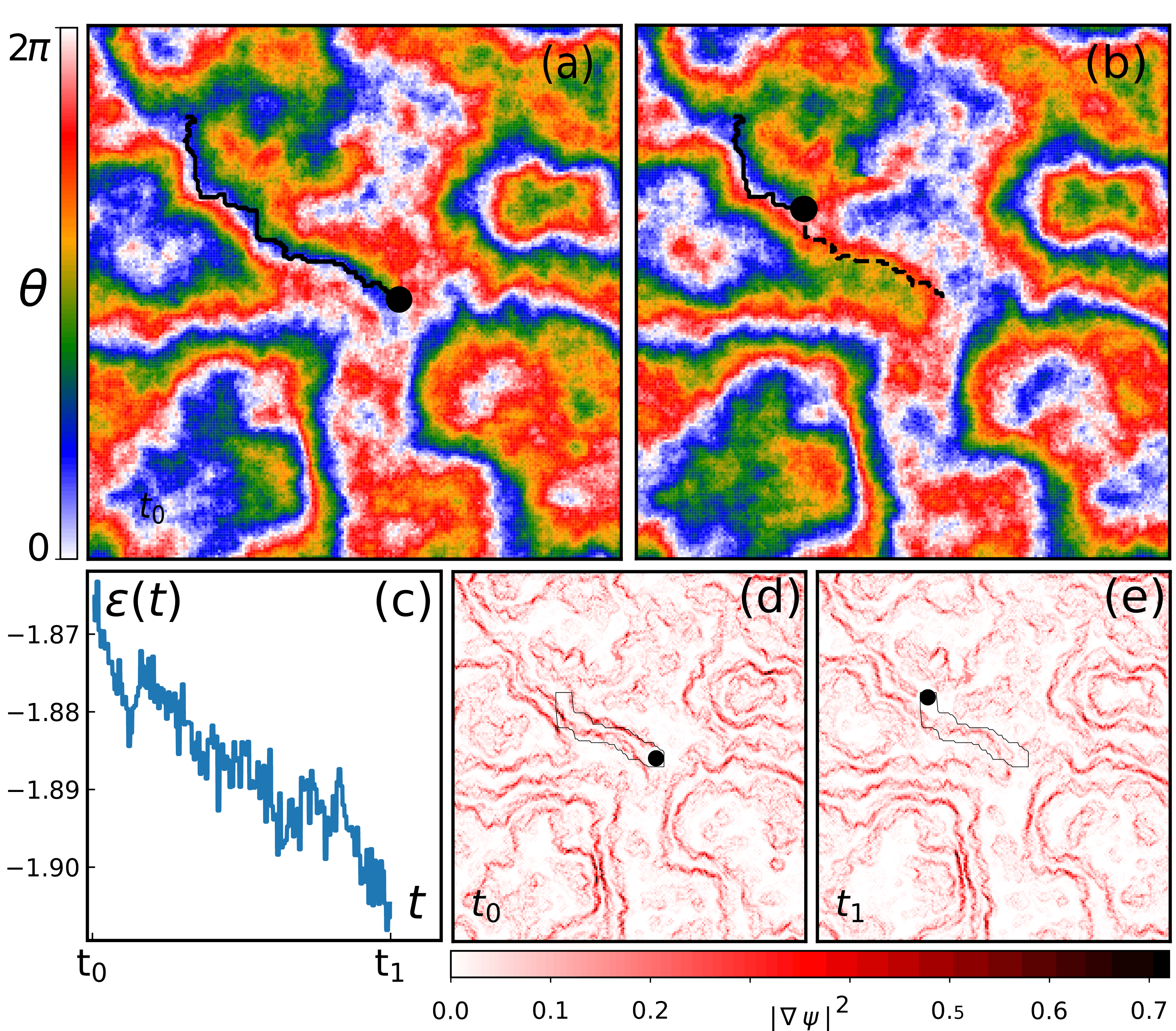}
\caption{
	Typical vortex trajectory at $T = 0.05, \sigma^2=0.1$. \textbf{(a)} Snapshot of the system at $t_0$, showing a vortex with a dot alongside a section of its (upcoming) trajectory and \textbf{(b)}  at $t_1=t_0+500$, showing the path followed by the vortex in broken lines. 
\textbf{(c)}  Decay of the energy per spin of an area around the domain boundary due to the motion of the vortex. Panels \textbf{(d)}  and \textbf{(e)}  display $|\nabla \psi|^2$ at $t_0$ and $t_1$, showing how the vortex removes the boundary it had just ridden, explaining the local decrease in energy.  
}
\label{fig:espilon}
\end{figure}

\begin{figure}[h!]
\includegraphics[width=0.99\linewidth]{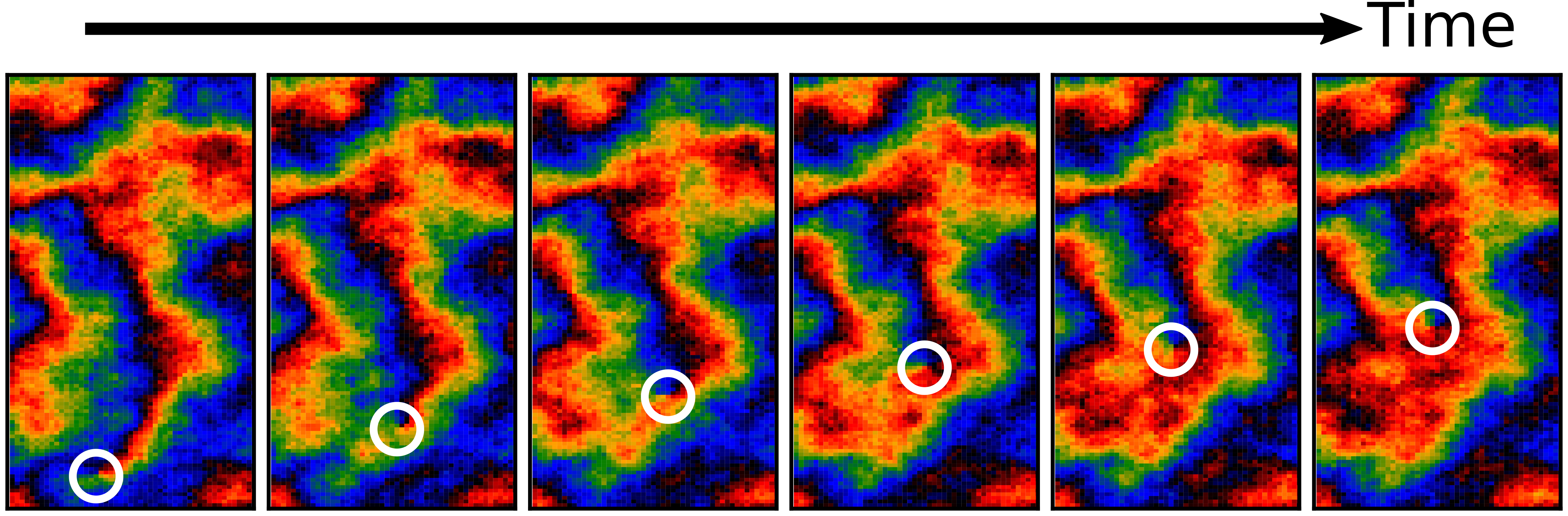}
\caption{
	Successive snapshots of a fixed 50$\times$120 portion of a 200$\times$200 system evolving with $T = 0.1$ and $\sigma^2 = 0.15$. The topological defect of charge $+1$ is highlighted with a white circle. It moves along the domain boundary (thin black line) and erases it in its wake, leaving an aligned region behind.  
}
\label{fig:defects_surf}
\end{figure}

On the other hand, such excess energy is released during the motion of the defects along  domain boundaries, which are erased as vortices move along them.
We illustrate this in Fig.~\ref{fig:espilon} and Fig.~\ref{fig:defects_surf}.

In Fig.~\ref{fig:espilon}, the defect (black dot) is shown at two different times $t_0$ and $t_1=t_0+500$ in panels (a, d) and (b, e), respectively.
In panels (a,b), we plot the orientation field $\theta$, such that it is easy to see that the trajectory of the defect (thick black line) follows the blue domain boundary beneath it.
In panels (d, e), we plot, for the same two times, the squared gradient of the phase field $|\nabla \psi|^2$, which is a measure of the elastic energy density.
We delimitate a region around the domain boundary trajectory (thin black line) and plot the energy per spin in this area as a function of time in panel (c).

The field becomes smoother as the domain boundary is removed by the passage of a defect, dissipating elastic energy: the energy per spin in this area decreases. 

This phenomenon has been observed recently in experimental systems by Kumar \textit{et al.} \cite{kumar2022catapulting} where they report that defects are ‘catapulted’ along these boundaries.

As detailed in the successive snapshots of Fig.~\ref{fig:defects_surf}, defects act like a zip and bringing together the two domains on each side of the boundary (e.g. in first panel Fig.~\ref{fig:defects_surf}(a), these are the two red regions, separated by the thin black line), leaving a synchronised region behind (big red region in last panel Fig.~\ref{fig:defects_surf}(f)). This area is no longer favorable to the future passage of a defect (be it itself or another one), as the elastic energy it previously contained has been dissipated, as reported in Fig.~\ref{fig:espilon}(c).

Altogether, the phase pattern has a strong impact on the motion of the defects and back, defects significantly alter the structure of the system as they move. 
Thus, as vortices remove the domain boundaries as they ride on them, they perform a random walk with memory: it is more likely for a vortex to keep moving along the domain boundary than to retrace its steps. 
The classical framework to treat this kind of motion is the self-avoiding random walk (SAW), which  might be at the origin of the anomalous exponent $3/2$. 
In order to explore whether SAW could provide a useful description of our defects motion, we simulate a few SAW, see Fig.~\ref{fig:trajectories}(c). They qualitatively match the ones we recorded for vortices.

SAW feature a superdiffusion with $\Delta^2(t)\sim t^{\nu}$ with $\nu = 6/(d+2)$ \cite{barat1995statistics}, hence providing an explanation of the $3/2$ exponent we found for the defects' MSD (here $d=2$, the number of spatial dimensions). Yet, the description provided by SAW is not complete, as the full distribution of displacements differ significantly. As shown in Fig.~\ref{fig:MSD_4panels}(a), for our defects, 
\begin{equation}
G(x,t) \sim \exp(-x^2/t^{3/2})
\end{equation} 
while for SAW, one obtains (cf. inset of Fig.~\ref{fig:MSD_4panels}(b)) 
\begin{equation}
G(x,t) \sim \exp(-x^4/t^{3})\,.
\end{equation}

We also investigated the effect of including self-propulsion to mimic the behaviour of vortices on domain boundaries. To do so, we add persistence to the walk, on top of the self-avoiding condition. Instead of continuing straight, turning left or turning right with equal probability 1/3, we set the probability to continue straight to $1/3\le p<1$, and thus turning left or right with probability $(1-p)/2$. This process introduces a typical length $\lambda \sim 1/\log(1/p)$ , as $p^n = e^{-n/\lambda}$. The typical timescale before the trajectory has changed orientation with probability 1/3 also scales as $\lambda$. In other words, a persistent SAW, when expressed in terms of the rescaled space $x/\lambda$ and rescaled time $t/\lambda$, is indistinguishable from a non persistent SAW expressed in terms of $x$ and $t$. This explains why, for all $1/3\le p<1$, the probability density of displacement for a persistent SAW with probability $p$ is given by 
\begin{equation}
	G(x,t) = \frac{1}{2\,\Gamma(5/4)\,(\lambda\,t^3)^{1/4}}\ \exp(-\frac{x^4}{\lambda\,t^3})\ , 
\end{equation}
where $\Gamma(x)$ is the usual Gamma function.
Note that the slope of the curves in Fig.~\ref{fig:MSD_4panels}(c) determines the exact value of the rescaling factor: $\lambda = 4/(3\log(1/p))$. \newline

Overall, a complete, faithful microscopic description of the topological defects' motion in the  short-range Kuramoto model remains to be found. Despite the simplicity of their displacements' functional form $G(x,t) \sim \exp(-x^2/t^{3/2})$, it is likely that collective effects drive the defects' motion in a way simple random walk models cannot capture, calling for a full many-body treatment of the problem.

\section{Appendix A: Signal processing of the instantaneous frequency} \label{sec:appendix_exp_moving_avg}
As the time series of $\theta(t)$ is noisy for $T > 0$ (cf. the blue curves of the top row in Figure~S2), one has to average in time before derivating to obtain a meaningful signal $\dot{\theta}(t)$.

This is done in three steps:

\begin{enumerate}
    \item Perform an exponential moving average on the time series $\theta(t)$.
    \item Differentiate the smoothed signal with a finite difference scheme to obtain $\dot{\theta}(t)$.
    \item Perform an exponential moving average on the time series $\dot{\theta}(t)$.
\end{enumerate}

The exponential moving average $\{y_t\}_1^T$ of a time series $\{x_t\}_1^T$ is defined as follows:

\begin{equation}
y_t = k \sum_{i=0}^{t-1} x_{t-i} (1 - k)^i, \quad \forall t = 2,3,4,\ldots,T \quad \text{and} \quad y_1 = x_1
\end{equation}

with $1/k = 50$ being the window of the moving average.

The backwards finite difference scheme used to differentiate a time series $\{x_t\}_1^T$ is as follows:

\begin{equation}
\dot{x}_t = \frac{x_t - x_{t - \tau}}{\tau},
\end{equation}

with $\tau = 50$ being the window of the derivative operation.

One can check that the meaningful variations of the signal $\theta(t)$ are on a timescale larger than $\Delta t = 50\,dt$ so we are not erasing any major information in the process.

\begin{figure}\centering
  \includegraphics[width = 0.99\linewidth]{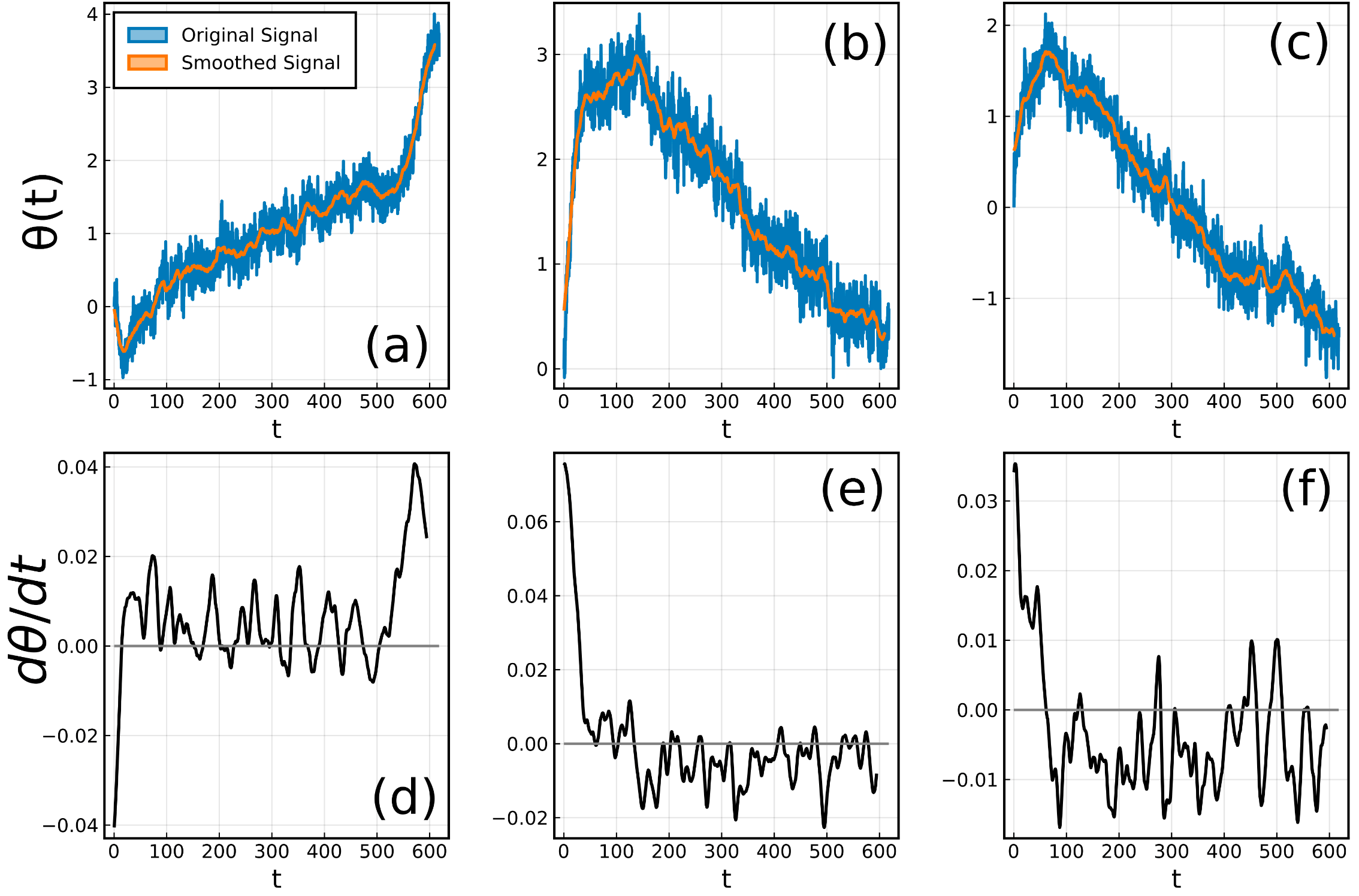}
  \caption{Each column corresponds to a different rotor in the system.
  \textbf{Top row }: original $\theta(t)$ in blue, result of the exponential moving average detailed in the text in orange.
    \textbf{Bottom row }: $\dot\theta(t)$ computed with a simple backwards finite difference scheme detailed in the text, then smoothed with an exponential moving average.
  }  
  \label{fig:appendix_exp_moving_avg}
\end{figure}


\chapter{The short-range off-lattice Kuramoto model} \label{ch:kuramoto_mobile} 
\graphicspath{{figures/Ch6kuramotomobile/}}
\section{Abstract}


Populations of heterogeneous, noisy oscillators on a two-dimensional lattice display short-range order. 
Here, we show that if the oscillators are allowed to actively move in space, the system undergoes instead a Berezenskii-Kosterlitz-Thouless transition 
and exhibits quasi-long-range order. 
This fundamental result connects two paradigmatic models -- XY and Kuramoto model --  and provides insight on the emergence of order in active systems.

This chapter covers the results published in 
\begin{center}
  Rouzaire \textit{et al.}, Physical Review Letters (2025) \cite{rouzaire2025activity}
\end{center}

This is a common work with Parisa Rahmani and Fernando Peruani at the University of Paris-Cergy, in France. This collaboration started from an original idea suggested by Fernando Peruani in a summer school in Nice in 2021, that he co-organised. 
I have lead the project, performed numerical and analytical analysis of the ballistic oscillators (results presented in this chapter). Parisa Rahmani has performed the numerical analysis of the persistent oscillators (qualitatively similar results, presented in the Supplemental Material of \cite{rouzaire2025activity}). 

\section{Introduction}

So far in this thesis, we have focused on two fundamental, strongly related lattice models, lying at the interface between statistical mechanics and dynamical systems: the XY model and the Kuramoto model.

For the classical XY model, the Mermin-Wagner theorem prevents in $d<2$ the emergence of long-range order (LRO) for this system. 
In dimension $d=2$, the system undergoes, below a critical $T_{KT}$, a Berezenskii-Kosterlitz-Thouless (BKT) phase transition 
associated to the unbinding of topological defects that leads, for $T<T_{KT}$, to quasi-long-range order (QLRO) \cite{berezinskii1971destruction,Kosterlitz1973, Kosterlitz1974}. We stress that all these results hold in thermodynamic equilibrium. 

For the $2d$ Kuramoto model, the system can only exhibit short-range order (SRO) and the study of defects in this context have shown that the BKT scenario no longer holds as the defects  unbind and superdiffuse when $\sigma\neq 0$, see Chapter \ref{ch:kuramoto} and references \cite{Rouzaire_Kuramoto_PRL,rouzaire_Frontiers,lee2010vortices}.

Although those paradigmatic models are restricted to a static spatial structure, synchronization patterns can also be found in a large variety of chemical and biological systems where the oscillators move in space. Some examples are fireflies moving in swarms displaying synchronized flashing~\cite{sarfati2021self}, mobile catalytic beads showing coherent oscillatory patterns of the BZ reaction~\cite{taylor2009dynamical},  amoeba exhibiting a synchronized response known as quorum sensing~\cite{gregor2010onset}, motile cells synchronizing their intracellular oscillations in somitogenesis ~\cite{oates2012patterning, uriu2014collective}, or gliding myxobacteria displaying coherent oscillations of the Frz system~\cite{guzzo2018gated}. At a theoretical level, it has been shown that in populations of identical oscillators, spatial diffusion \cite{peruani2010mobility, grossmann2016superdiffusion, LevisPRX}  promotes the emergence of order in finite systems. 

Thus, spatial diffusivity can effectively increase  the range of interactions that, when resulting in an all-to-all coupling, leads to global order~\cite{sevilla2014synchronization, escaff2020flocking}
However, in infinite systems spin waves cannot be suppressed for any finite diffusivity in dimension $d=1$ and thus only SRO is possible~\cite{peruani2010mobility}. While in $d=2$, the emergent order is QLRO, and the system undergoes a BKT-like transition even for diffusive oscillators~\cite{grossmann2016superdiffusion, LevisPRX}. Interestingly, if motion is superdiffusive, LRO can emerge~\cite{grossmann2016superdiffusion}. We stress that these results are restricted to populations of identical oscillators; however, fireflies, amoeba, bacteria, and organisms in general are not identical, but exhibit large inter-individual variability, even within genetically identical populations~\cite{elowitz2002stochastic, raj2008nature}. Thus, it becomes crucial to understand the synchronization of mobile and heterogeneous oscillators, where each of them possesses its own natural frequency.

Here, we address such a fundamental question using a two-dimensional model for actively moving phase oscillators, where each one has its own natural frequency.
Such system can be considered a collection of swarmalators~\cite{okeeffe2017oscillators} where the phase dynamics is coupled with the position but the oscillator position dynamics is independent of the phase.

We show that, while in the absence of motion only SRO can emerge, motility allows the system to achieve QLRO and undergo a BKT transition. The relevance of these results goes beyond the realm of active matter physics.  For instance, while heterogeneous 3d static swarms of fireflies cannot exhibit synchronized flashing, our results indicate that motion can make it possible. Similarly, in somitogenesis, cellular motion might not only enhance synchronization as clearly shown in~\cite{uriu2014collective, uriu2024statistical} using identical phase oscillators, but might be essential to prevent, in heterogeneous populations, the emergence of only SRO; see~\cite{oates2012patterning, uriu2021local, fernandez2022reaction} for a realistic description of somitogenesis.

\section{The model}

We consider $N$ actively moving phase oscillators, at density $\rho$, such that they live in a $L\times L$ plane with periodic boundary conditions, with $L=\sqrt{N/\rho}$. 

The phase $\theta_i$ of oscillator $i$ obeys the Kuramoto Eq.~(\ref{eq:angular_dynamics}) 

\begin{equation}
   \dot{\theta}_i =  \sigma\, \omega_i + \sum\limits_{j\in \partial_i} \sin(\theta_j - \theta_i) + \sqrt{2\,T\,}\eta_i \, ,
    \label{eq:angular_dynamics}
\end{equation}

where $\omega_i$ are drawn from a normal distribution, centered around $0$, of variance $1$, and where the sum over $j$ runs 
over all neighbours of $i$ defined as those oscillators at a distance less than $R_0$. 
The spatial dynamics of the $i$-th oscillator is given by: 

\begin{equation}
\dot{{\bf x}}_i = v_0 \, {\bf e}[\psi_i] =(v_0\,\cos\psi_i\,,\,v_0\,\sin\psi_i)
\label{eq:spatial_dynamics}
\end{equation}

where $v_0> 0$ is the  speed of the oscillator and $\psi_i$ its moving direction. In contrast with the Vicsek model \cite{vicsek1995novel,ginelli2016VicsekRev}, the moving direction $\psi_i$ is completely independent of the phase $\theta_i$.

We analyse two different scenarios: (a) ballistic motion with $\dot{\psi}_i=0$, and (b) persistent random walkers, where

\begin{equation}
  \dot{\psi}_i = \sqrt{2D_{\psi}}\,\nu_i \ ,
\end{equation}

with $D_{\psi}$ constant and $\nu_i$ a $\delta$-correlated noise such that 

\begin{equation}
  \langle \nu_i \rangle=0 
\text{ and  }\langle \nu_i(t)  \nu_j(t')\rangle=\delta_{ij} \delta(t-t') \ .
\end{equation}
Initially, oscillators are placed at random on the $L\times L$ plane, with $\psi_i(t=0)$  randomly selected from the interval $[0, 2\pi)$. 

In the following, we fix  $T=0.1$ and $R_0=1$, and express lengths and times in units of $R_0$ and  $\gamma/J$, {where $\gamma$ is the damping coefficient in Eq.~(\ref{eq:angular_dynamics})}. 
Note that figures shown here correspond to the scenario (a), i.e. ballistic motion, while the same figures, but for the scenario (b), can be found in the Supplemental Material of Ref. \cite{rouzaire2025activity} and have been obtained by Parisa Rahmani. One obtains qualitatively the same results with both dynamics provided $v_0 D_{\psi}^{-1}$ is large enough.

\paragraph{Determination of the timestep $dt$\\}
The timestep $dt$ used varies as a function of the simulation parameters in order to fully resolve the different phenomena at play (we indicate between square brackets which parameters it depends on): 
\begin{enumerate}
    \item The timescale for one particle to cross the interaction radii $[v_0, R_0]$ 
    \item The rotation induced by the highest intrinsic frequency (in absolute value) of the system $[N, \sigma]$.  We use the same bound $\mathbb{E}\max\limits_{i} |\sigma\,\omega_i| < \sigma\sqrt{2\log N}$, see the discussion on the timestep in Chapter \ref{ch:kuramoto}.
    \item The additive alignment force $[J,\rho,R_0]$
    \item  The thermal noise $[T]$
\end{enumerate}
Concretely, 
$$dt = \min \left\{\alpha \frac{2R_0}{v_0}\ ,  \  \frac{\beta}{\sigma\sqrt{2\log N}}  \ ,  \  \frac{\beta}{J\pi\rho R_0^2}\ ,  \  \beta^2 \pi/(4T)  \right\}$$
where $\alpha = 1/10$ and $\beta = \pi/20$ are two arbitrary constants. In the range of parameters explored, the third term typically is the constraining one, leading to timesteps in the range 0.01 - 0.05.

\section{Ordering in finite systems} 

We characterise the emergent order using the (global) polar order parameter, defined as $P=|\sum_{j=1}^N e^{i\theta_j}|/N$. 
In parameter space $\sigma$-$v_0$, we distinguish two clear phases. 

\begin{figure}[h!]
    \centering
    \includegraphics[width=1\linewidth]{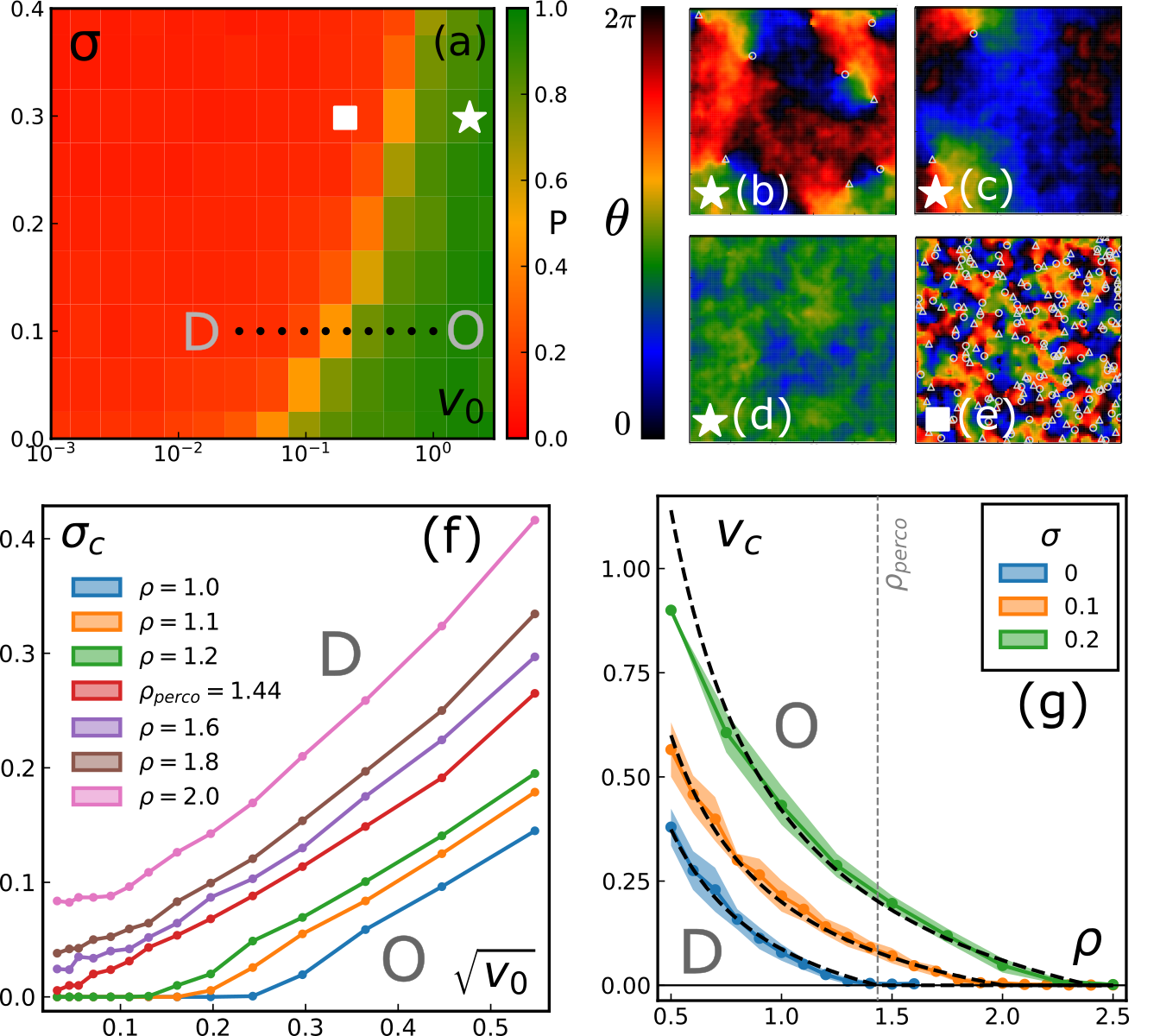}   
    \caption{\textbf{(a)} Steady polarization $P$ map in the $v_0 - \sigma$ plane for $N=10^3$, $\rho=1$.  Symbols correspond to the snapshots shown in  (b)-(e) and the horizontal dotted line to the parameters scanned in Fig.~\ref{fig:2}(a-c) across the disorder (D) to order (O) states. 
    \textbf{(b,c,d)} Coarse-grained $\theta$ field at different times -  $50, 500$ and $1000$ - of a system with $v_0 = 2, \sigma = 0.3$ , $N=10^4, L=100$. {Circles (resp. triangles) are $+1$ (resp. $-1$) topological defects.}
      \textbf{(e)}  Steady-state {coarse-grained configuration} for $v_0 = 0.2, \sigma = 0.3$ , $N=10^4, L=100$.
      \textbf{(f)} Dependence of the threshold $\sigma_c$ on $v_0$,  
      for $N=10^3$ and the different densities shown in the key. 
      \textbf{(g)} Critical velocity $v_c(\rho)= \alpha \big(\rho_c(\sigma)-\rho\big)/\rho$ for $N=10^4$. The black lines follow our theoretical prediction (cf. main text), with the fitting coefficients $\alpha=0.2, 0.2, 0.3$ and $\rho_c = 1.44, 2, 2.4$ for $\sigma=0,0.1,0.2$, respectively. Note that for $\rho_c(\sigma =0) = \rho_p\approx 1.44$, the percolation value, consistent with \cite{LevisPRX}. 
    }
    \label{fig:1}
\end{figure}

\begin{figure}[h!]
    \centering
    \includegraphics[width=1\linewidth]{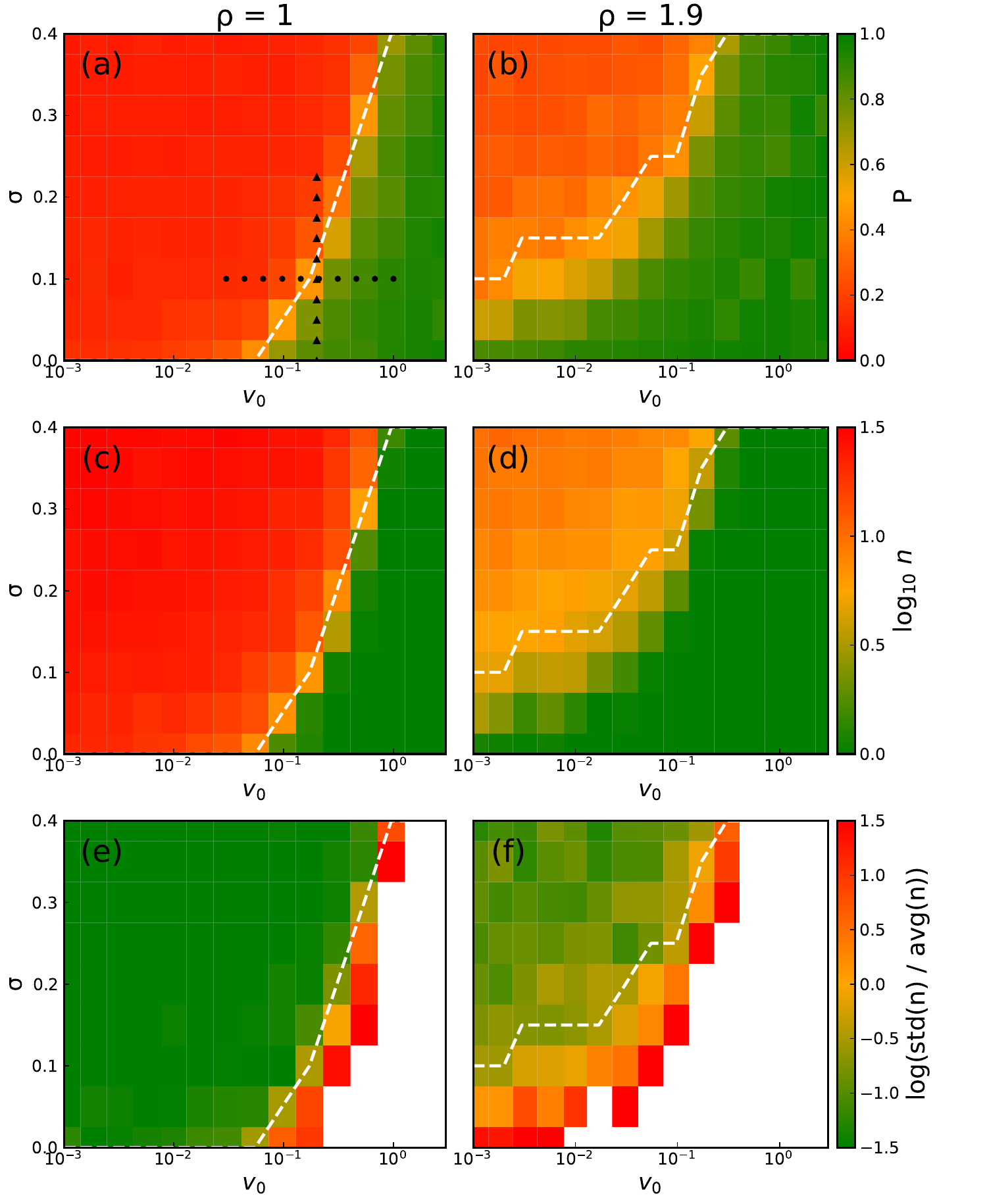}   
    \caption{Phase spaces of \textbf{(a,b)} the polar order $P$, \textbf{(c,d)} the number of defects $n$, and \textbf{(e,f)} the fluctuations of the number of defects (standard deviation divided by the mean), on a logarithmic scale.
\textbf{Left column}: At particle density $\rho=1.0$ and
\textbf{Right column}: At particle density $\rho=1.9$.
Dots mark the horizontal scan through the transition of Fig.~\ref{fig:scan_vertical_horizontal}(a-c). Triangles mark the vertical scan through the transition of Fig.~\ref{fig:scan_vertical_horizontal}(d-f)
}
\label{fig:phase_spaces}
\end{figure}

For low  $v_0$ and high $\sigma$, the system is disordered with $P\ll1$, see Fig.~\ref{fig:1}(a).  
For high $v_0$ and low-enough $\sigma$, our finite-$N$ systems reach an ordered, or synchronized, steady-state, characterised by  $P \sim 1$.  
The order-disorder {transition} in the $\sigma$-$v_0$ plane -- identified, in finite systems, by the condition $P(v_c, \sigma_c) = 1/2$, defining the thresholds $v_c$ and $\sigma_c$ -- can also be equivalently identified by monitoring the number of topological defects $n$. We report in Fig.~\ref{fig:phase_spaces} the corresponding phase space for the number of topological defects $n$ (second row) and the fluctuations in the number of defects (third row). We compute the ratio of the standard deviation over the mean number of defects, where both the standard deviation and the mean are calculated over 40 independent realizations of the thermal noise and initial random configurations. We compare these quantities for $\rho=1$ (below the percolation density $\rho_{perco}\approx 1.44 $, left column) and $\rho=1.9$ (above the percolation density, right column) to show that the transition can be thought either in terms of $P$ or in terms of $n$.
We highlight this shared trend by plotting the $P=1/2$ separation with white dashed lines in the second and third rows. 
In Fig.~\ref{fig:phase_spaces}(e, f), we plot the fluctuations in the number of defects on a logarithmic scale. The white area is the region where the number of defects is identically zero. We see that in the high defect density region (to the left of the white dash line), the fluctuations are small. In contrast, it is at the entrance of the low defect region that the fluctuations are the largest. 

{To identify and track defects in our system of mobile particles, we first coarse-grain the phase field $\theta$ such that each point of an auxiliary square grid is assigned a value, as detailed in the Appendix \ref{sec:appendix_smoothing}. 
A defect is then identified via the standard procedure, as a square plaquette with a winding number $q=\pm1$, defined by the sum of the phase differences along its four bonds $\sumcirclearrowleft\Delta\theta_{i,j}=2\pi q$.}

Starting from a disordered initial state, polar order sets through a typical coarsening process. Topological defects initially span the entire system; they progressively annihilate two by two to eventually completely disappear: the system spontaneously breaks rotational invariance and picks one orientation, hence $P\approx 1$. 
 {We illustrate this dynamics with the coarse-grained phase fields in Fig.~\ref{fig:1}(b-e). }
Occasionally, pairs of defects spontaneously generate but, in contrast to the behaviour of immobile oscillators on a lattice~\cite{Rouzaire_Kuramoto_PRL, rouzaire_Frontiers, lee2010vortices}, they end up annihilating completely and the steady-state typically displays no defects, even for the $\sigma=0$ (XY) axis, cf. Fig.~\ref{fig:1}(d) and Fig.~\ref{fig:phase_spaces}(c,d).

Furthermore, we investigate the impact of the particle density $\rho$ on the transition line (for finite $N$) between order and disorder. For low densities, the system only orders for sufficiently high $v_0$, see Fig.~\ref{fig:1}(f) .  Higher densities promote synchronization because each spin interacts (in an additive fashion) with more neighbours within the fixed radius $R_0$. 
In the limit $\sigma,\,v_0\to0$, the emergence of order -- in a finite system -- requires the density to be above the percolation threshold $\rho_{p} \approx 1.44$ of (penetrable) discs of radius $R_0$, a result consistent with~\cite{LevisPRX}, where the synchronization of moving, single frequency ($\sigma=0$),  oscillators was studied. We observe that the mean coordination number corresponds to $\bar z=\pi\,\rho_{p}\,R_0^2\approx 4.5$.

For $\rho< \rho_{p}$, there exists a threshold in velocity above which the finite system is in the  ordered region. 
To understand the scaling of the threshold speed $v_c$ with the density $\rho$ in Fig.~\ref{fig:1}(g), corresponding to low densities, we estimate the number of encounters (i.e collisions, as a proxy of the number of neighbours), a particle (in the ballistic scenario) experiences in time $\tau_0$ as $z_{\text{eff}}(v_0) \propto \rho R_0 v_0 \tau_0$. Here, $\tau_0\sim J^{-1}$ is the timescale associated to the alignment torque in Eq.~(\ref{eq:angular_dynamics}). 
Requiring that $z_{\text{eff}} \simeq \bar z$, we find that $v_c \propto (\rho_c-\rho)/\rho\,$, where $\rho_c$ is a critical density increasing with $\sigma$ and satisfying $\rho_c(\sigma=0)=\rho_p$ ; see dashed lines in Fig.~\ref{fig:1}(g).

\section{BKT transition}  \label{sec:BKT_transition}
To characterise the nature of the reported phases, we study the coarsening dynamics and compare them with the BKT predictions for the XY model on a $2d$ lattice, see Chapter \ref{ch:XY} and references therein. We focus on the time evolution of the number of topological defects $n(t)$, the polar order parameter $P(t)$, and the spatial correlations $C(r; t)$ here defined as 
\begin{equation}
  C(r; t)=\Big\langle \cos\Big(\theta_i(t)-\theta_j(t)\Big) \Big \rangle_{|{\bf x}_i-{\bf x}_j|=r}
\end{equation}
where $\langle\cdots \rangle$ denotes average over thermal noise, intrinsic frequencies, initial phases, initial locations and direction of travel of individual particles. 


%

We represent the XY predictions as dashed lines in Fig.~\ref{fig:2}(a-c). 

\begin{figure}[h!]
    \centering
    \includegraphics[width=1\linewidth]{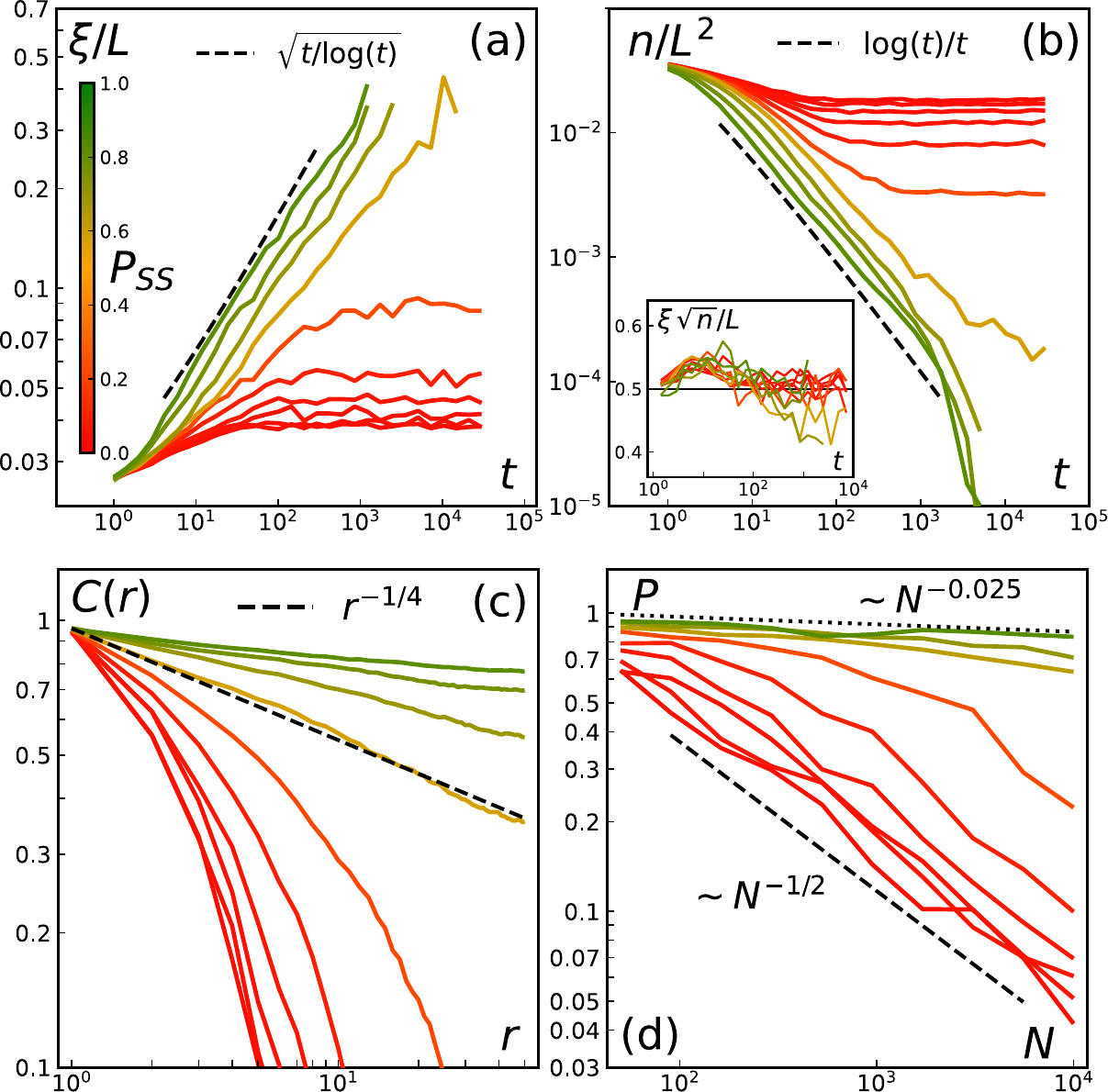}   
    \caption{Time evolution of: \textbf{(a)} the spatial correlation length;  \textbf{(b)} defect density. \underline{Inset }: characteristic length normalised by the average distance between defects.  \textbf{(c)} Spatial decay of the correlation function in steady conditions. \textbf{(d)} Finite size scaling analysis of the polarization, to confirming the different nature of the D and O phases. In all cases: dashed lines are XY predictions, the colour of each curve corresponds to its steady  polarization value,  $N=10^4$, $\sigma = 0.1$ and $0.03 \leq v_0 \leq 1$ . }
    \label{fig:2}
\end{figure}

In systems of moving oscillators, we find that  the defect density $n(t)/L^2$ quickly saturates with time $t$ to a finite value 
when the system is located, on the parameter plane $(v_0, \sigma)$, in the disordered region; 
see red curves in Fig.~\ref{fig:2}. 
In this phase, the spatial correlation function decays geometrically, see Fig.~\ref{fig:2}(c). This is a signature of short-range order (SRO) 
and the correlation length $\xi$ remains always $\xi\ll L$, cf. Fig.~\ref{fig:2}(a). 
In contrast, in the ordered phase, topological defects progressively annihilate and  the density of defects $n(t)/L^2$ decays as 
$n \sim \log t /t$ (see green curves in Fig.~\ref{fig:2}(b)), while the  correlation length $\xi$ increases as $\xi \sim \sqrt{t/\log t\,}$ , see green curves in Fig.~\ref{fig:2}(a). 
As stated in Chatper \ref{ch:XY}, the resemblance of those expressions is not a coincidence: $\xi \sim (n(t)/L^2)^{-1/2}$ implies that the average distance between defects  controls the typical correlation length in the system. 
Moreover, a  distinctive hallmark of the XY model at low temperatures is displayed in the inset of Fig.~\ref{fig:2}(b), where $\xi \sqrt{n}/L\approx 1/2$ for all $t$, 
as observed in the XY model on a lattice \cite{rouzaire_Frontiers}. 
In the ordered phase, the (steady-state) correlation functions are power-laws, as displayed by the green curves in Fig.~\ref{fig:2}(c). 
This is the signature of QLRO. 
Finally, at the critical point, i.e. at the boundary between the SRO and QLRO phase, the  correlation function $C(r)$ scales as $ r^{-1/4}$; see yellow curve in Fig.~\ref{fig:2}(c).  
This is observed when crossing the transition either vertically or horizontally on the $(v_0, \sigma)$ plane, cf. the discussion on Fig.~\ref{fig:scan_vertical_horizontal} in Appendix \ref{sec:appendix_scan_vertical_horizontal}.
In summary, comparing  the dashed (XY predictions) with the green and yellow curves (ordered phase of the mobile Kuramoto model), it becomes evident that  
the analysed out-of-equilibrium system  exhibits all the salient features of the BKT transition.
To further support the QLRO nature of the ordered phase, we performed a finite size analysis (see results Fig.~\ref{fig:2}(d)), finding that polar order decreases with system size as $P\sim N^{-1/2}$ in the disorder phase, while in the order phase, the scaling is $P \sim N^{-\beta}$, with $\beta<1/16$, as would be expected in the  XY model on a $2d$ lattice \cite{nelson1977momentum, frenkel1985evidence}.

\section{Topological defects}

The BKT transition is characterised by the collective behaviour of topological defects.
%
For $T<T_{KT}$, defects are \textit{bounded}. There exists an 
 effective $1/R$ attractive force between defects of opposite topological charges {($q$)} separated a distance $R$~\cite{YurkeHuse1993}. 
 To assess whether the same mechanism is present in the system of actively moving oscillators, we study 
 the temporal evolution of a system whose initial condition consists of {two topological defects, one with $q=+1$ and the other with $q=-1$, } separated by a distance $R(t=0)=L/2$. We set parameters in the ordered phase and monitor the distance $R(t)$ between defects, see illustration in Appendix \ref{sec:appendix_smoothing}. 
We find that $R(t)$ decreases with $t$ (faster for larger $v_0$) as expected in the presence of an attractive force; see Fig.~\ref{fig:3}(a). 

Note that the time $\tau$  for defect annihilation is itself a stochastic variable, cf. inset of Fig.~\ref{fig:3}(a). 
It is useful to study the distance $R$ between defects with respect to the time-to-annihilation $t^*$ (instead of $t$). 
In this equivalent description, $t^*=0$ is the moment of annihilation ($R(t^*\,=\,0) = 0$) and defects move away from each other with increasing $t^*$, see  Fig.~\ref{fig:3}(b). 
The interest of $t^*$ is that now the crucial part of the dynamics, when the $1/R $ force is relevant because $R$ is small, can be analysed with averages that are meaningful and more precise, see \cite{YurkeHuse1993,rouzaire_Frontiers} and Chapter \ref{ch:XY}.
In particular, we observe that the curves for different $v_0$ collapse by rescaling space with $v_0^{-1/2}$ and time by $v_0^{1/2}$; see inset in Fig.~\ref{fig:3}(b).  
Following \cite{YurkeHuse1993}, we solve for the functional form of the interdefect separation and obtain 
\begin{equation}
  R(t^*)  = \exp(W(2\pi\,t^*/\mu)/2)\ , 
\end{equation}
where $W$ is the Lambert function (or Product Log) and $\mu$ is the 
 mobility coefficient, proportional to the damping coefficient $\gamma$ in the Langevin equation Eq.~(\ref{eq:XY_Langevin_adim}).
 We use  $\mu=1/2$, as found in \cite{Rouzaire_Kuramoto_PRL} for the  XY model (cf. Chapter \ref{ch:XY}), to plot the black curve in the inset of Fig.~\ref{fig:3}(b). With no fitting parameter involved, the agreement between the  theoretically predicted $R(t^*)$ and simulations is excellent. 

 \begin{figure}[h!]
    \centering
    \includegraphics[width=1\linewidth]{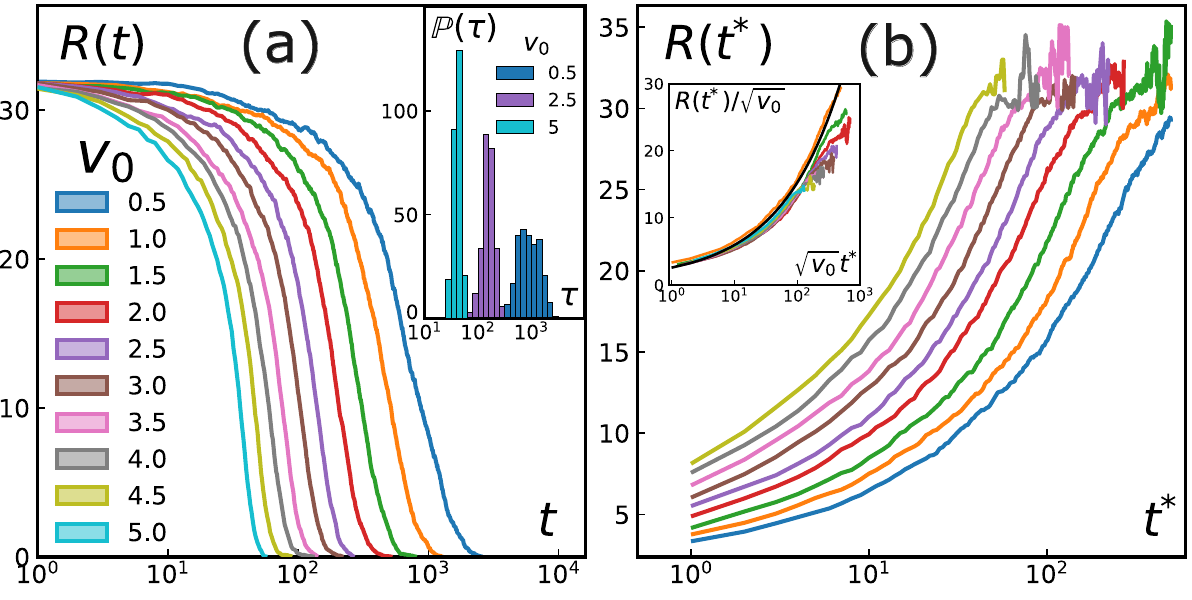}   
    \caption{\textbf{(a)} Inter-defect separation $R(t)$ for different  velocities (in the QLRO phase), $\sigma = 0.1$, averaged over 260 realizations. \underline{Inset}: Histogram of annihilation times $\tau$ for 3 selected velocities.
\textbf{(b)} Same data  plotted against the time to annihilation $t^*$. \underline{Inset}: Rescaling  time by $v_0^{1/2}$  and distance by $v_0^{-1/2}$ makes the curves collapse onto the XY predictions (in black). }
    \label{fig:3}
\end{figure}

\begin{figure}[h!]
    \centering
    \includegraphics[width=1\linewidth]{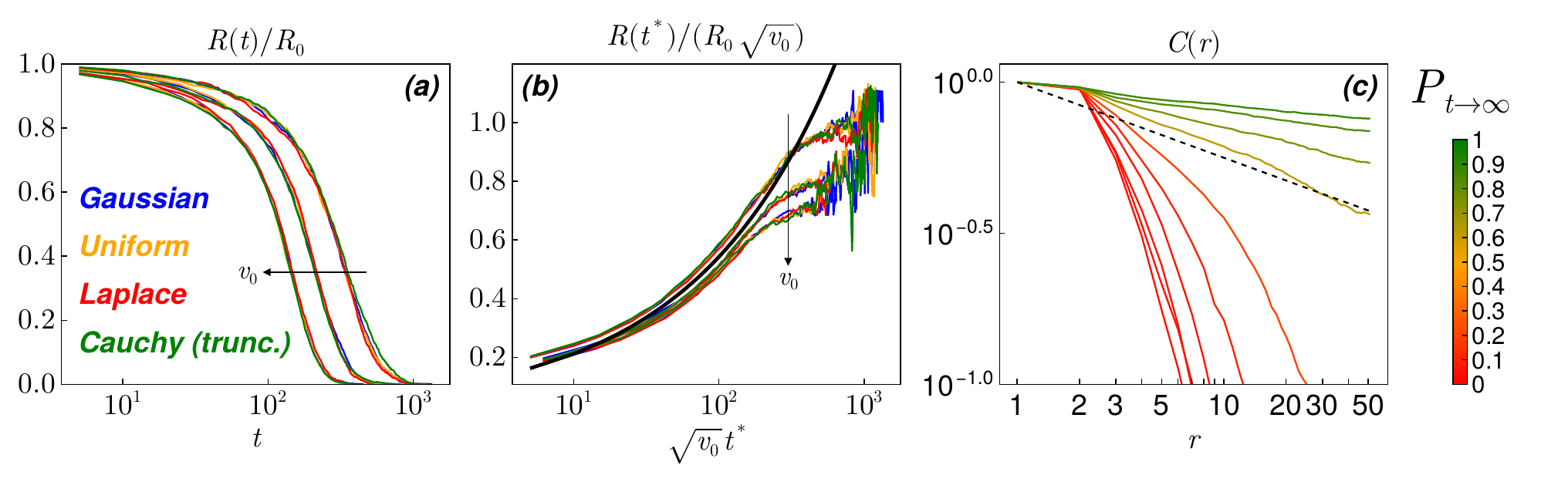}   
    \caption{ {\textbf{(a)} Decay of the distance between two defects manually created at an initial distance $R_0$, for $T=0.1$. Results are averaged over 400 independent realizations. Each group of lines is a different $v_0 = 1, 1.5, 2$ from right to left. Each colour is from a different distribution of the $\omega$. For each distribution, we keep the same standard deviation $\sigma=0.1$ (see text for details). 
    \textbf{(b)} Same data but on rescaled axes. $t^*$ is the time to annihilation and the black line is the XY prediction, computed from the results of \cite{YurkeHuse1993} (both explained in more detail in the main text).  \textbf{(c)} Spatial correlation functions $C(r, t\to \infty)$ for $T=\sigma=0.1$ for the \textit{uniform distribution}. Each curve corresponds to a given velocity $v_0$ (crossing the transition, see circles in Figure.~\ref{fig:phase_spaces}a) and is coloured according to the polarization $P$ in the steady-state (see colourbar).}
    }
    \label{fig:distribution_omegas}
\end{figure}

This shows that  topological defects in our out-of-equilibrium system effectively follow the same interaction mechanism as in the equilibrium XY scenario, which in turn explains why the QLRO nature of the ordered phase is the same in both systems. 
This annihilation mechanism has no reason to depend on the system size, which therefore strengthen our previous results on the existence of a BKT-like transition. 
 {Finally, this key result does not depend on the specific distribution of the intrinsic frequencies. 
We have compared the gaussian case to the uniform, exponential and Cauchy distributions: both the pair annihilation and the coarsening dynamics are similar in all those cases. The results even become \textit{identical} if one imposes the same variance across the different distributions, see Fig.~\ref{fig:distribution_omegas}(a,b). The nature of the ordered phase is QLRO in all cases, see Fig.~\ref{fig:distribution_omegas}(c) for the uniform distribution. 
}

\section{Differences with the XY model}
The system of moving heterogeneous oscillators exhibits some differences with respect to the classical XY model. 

\subsection{Topologically protected patterns}
In particular, we observe frequently the formation of \textit{topologically protected patterns} (TPP), characterised by a phase $\theta(x,y) = 2\pi x/L$; see Fig.~\ref{fig:4}(a).

\begin{figure}[h!]
    \centering
    \includegraphics[width=1\linewidth]{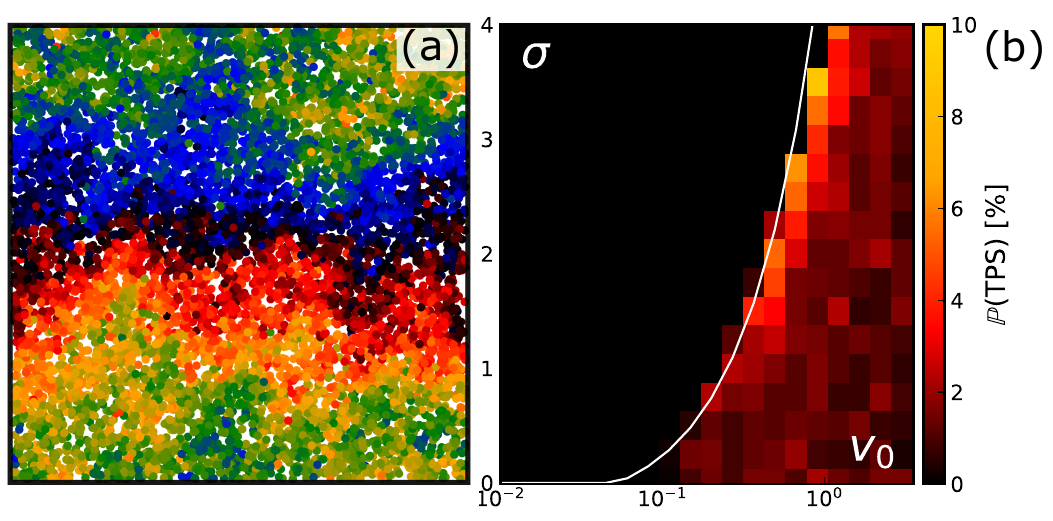}   
    \caption{\textbf{(a)} Representative topologically protected pattern (TPP) pattern, here for a $N=10^4, \rho=1$ system. 
\textbf{(b)}  Probability to observe a TPP against $v_0$ and $\sigma$, for $N=4000,\rho=1$, averaged over 780 realizations. Using disordered initial conditions, a TPP is identified with a defect free configuration with $P<0.5$.  
Such arbitrary threshold does not impact the results as the vast majority of the detected TPP have $P\approx 0.1$. The white line is $\sigma_c =\max(0, v_0 - 0.529) / 2 $, where the numerical factors are extracted from Fig.~\ref{fig:1}(f).}
    \label{fig:4}
\end{figure}
These stable (or metastable) system spanning  structures form when two topological defects annihilate against the attracting force. Imagine for instance the two defects of Fig.~\ref{fig:1}(c) annihilating at the center of the box and not at its borders.

We report in Fig.~\ref{fig:4}(b) the frequency of TPP formation, close to the order-disorder transition, where the probability to observe TPP is the highest. TPP are much less present in the classical $2d$ XY model, and thus constitute a characteristic of systems of moving oscillators. 

\subsection{The $T=0$ case}
\begin{figure}
    \centering
    \includegraphics[width=1\linewidth]{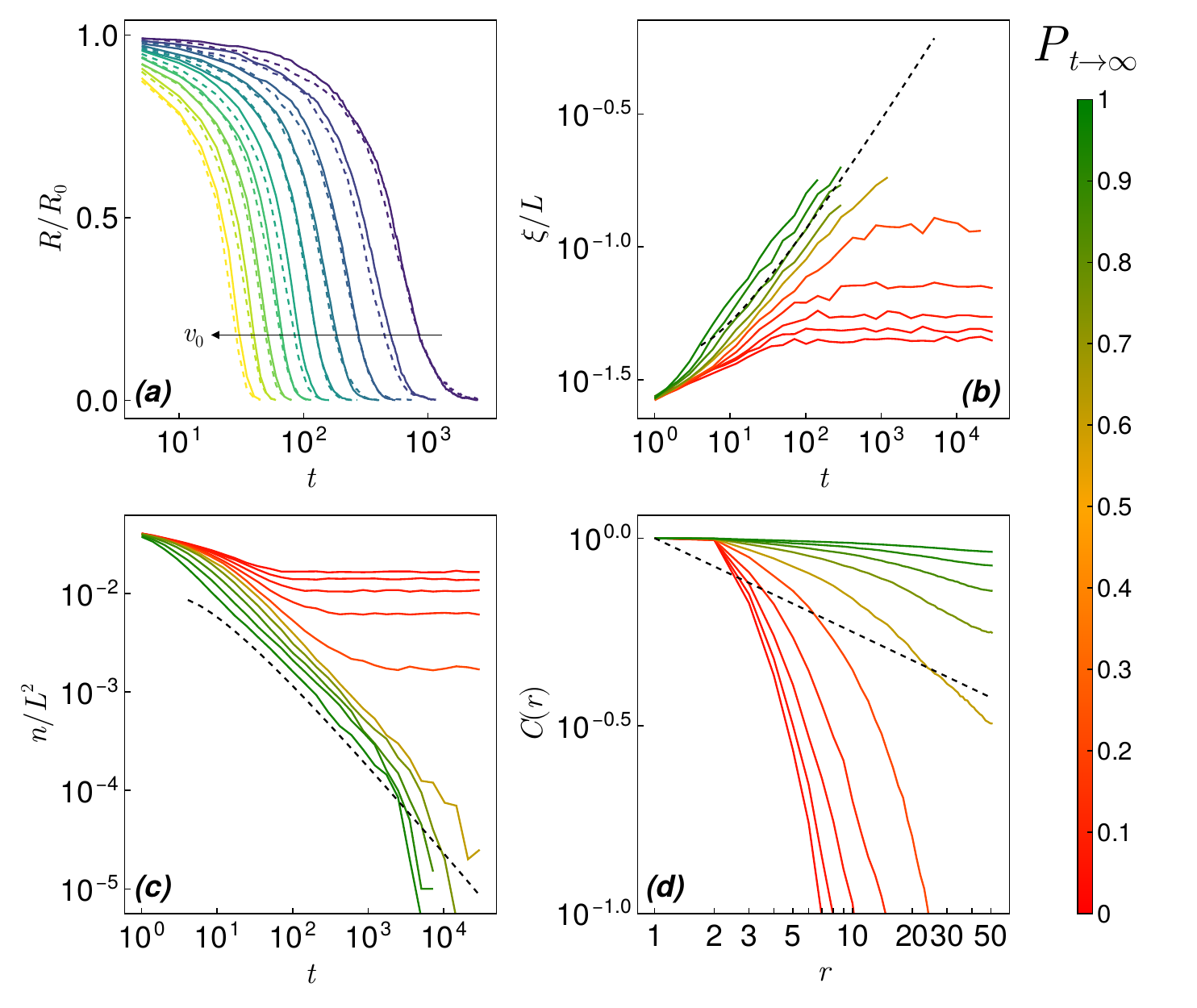}   
    \caption{\textbf{(a)} Distance between 2 defects manually created $R(t)$, for $\sigma=0.1$, different $v_0 = 0.5, 1, 1.5, ..., 5$. In dash, $T=0$, in solid line $T=0.1$. 
    \textbf{(b-d)} Coarsening dynamics for the $T=0$ case, for various parameters $v_0, \sigma$ crossing the transition horizontally (circles in Fig.~\ref{fig:phase_spaces}(a)) . The characteristic lengthscale $\xi(t)/L$ in panel (b), the defect density $n(t)/L^2$ in panel (c) and the spatial correlation functions $C(r)$ in panel (d). 
    }
    \label{fig:comparison_T0}
\end{figure}

Another major difference lies in the the specific case of the absence of thermal noise: $T=0$. 
In the XY model, the $T=0$ case is somewhat singular and the role of noise in the BKT topological phase transition is crucial. In particular, topological defects tend to get trapped in spurious metastable states. In the short-range lattice Kuramoto model of Chapter \ref{ch:kuramoto}, the $T=0$ case is also singular. Indeed, the topological defects are prone to getting trapped in specific regions of space, where the configuration of angles $\theta$ and intrinsic frequencies $\omega$ make them cycle in space roughly periodically, destroying ergodicity. 
Here, the mobility of individual agents implies a constant change in the neighbourhood of each rotor, providing a non-thermal source of fluctuations. This is why we observe qualitatively similar behaviours for $T=0$ and $T>0$. This is true for the defect pair annihilation process and for the coarsening dynamics.
Figure~\ref{fig:comparison_T0}(a) shows that the distance between two manually created defects decreases in the same fashion for $T>0$ (dashed lines) and $T=0$ (solid lines). 
It also shows that the characteristic lengthscale $\xi$ (panel b) and the defect density $n/L^2$ (panel c) both exhibit the same behaviour across the order-disorder transition. In the disordered region of phase space, $\xi$ and $n/L^2$ saturate at a final value. In the ordered region, $n/L^2$ decays to 0 following the same scaling as the one of the XY model (dashed lines), allowing for the typical lengthscale to grow up to the system size. 

The subtlety of the zero temperature case resides in the nature of the order. While for finite temperatures, the spatial correlation function $C(r)$ decays with a powerlow, indicating QLRO (cf. Fig.~\ref{fig:2}(c)), this is no longer the case for the $T=0$ case. When the system is no longer coupled to a thermal bath, the $C(r)$ decays faster than algebraically, indicating true long range order (LRO), see. Fig.~\ref{fig:comparison_T0}(d).

\section{Concluding remarks} 
 {
In~\cite{peruani2010mobility, grossmann2016superdiffusion, LevisPRX} it was shown that non-equilibrium homogeneous systems of identical moving oscillators behave as equilibrium systems of (non-moving) XY spins, while
the synchronization dynamics of heterogeneous systems of moving oscillators with different natural frequencies has so far been uncharted. 
On the other hand, heterogeneous systems of static phase oscillators display only SRO in dimensions less than 5 \cite{Sakaguchi1987, hong2005collective, hong2007entrainment, lee2010vortices,rouzaire_Frontiers,Rouzaire_Kuramoto_PRL}. 
Here, we showed that in $2d$, active motion makes a system of heterogeneous phase oscillators behave as an equilibrium system of static XY spins, exhibiting QLRO and a BKT transition with the same coarsening dynamics. Importantly, these results are independent of the details of the heterogeneity, i.e. whether the distribution of natural frequencies is gaussian, uniform, Laplace, or a truncated Lorentzian. 
This result connects the spatially extended Kuramoto model with the (equilibrium) XY model, and sheds light on how 
order emerges in heterogeneous active systems with quenched disorder. 
Let us recall that in active models such as the Vicsek model~\cite{vicsek1995novel}, a phase $\theta_i$ controls the  active direction of motion \cite{peruani2008mean, chepizhko2010relation, martin2018collective} of the oscillator, and the system exhibits, in $2d$, LRO. 
Furthermore, such a coupling between the phase and active velocity allows heterogeneous active chiral particles to also exhibit LRO~\cite{levis2019activity}. 
Here, we have shown that active motion itself,  even in the absence of a coupling between phase and activity,  promotes the emergence of 
coordinated states that would otherwise be impossible.}

\section{Appendix A: No difference between different crossings through the transition} \label{sec:appendix_scan_vertical_horizontal}
\begin{figure}[h!]
    \centering
    \includegraphics[width=1\linewidth]{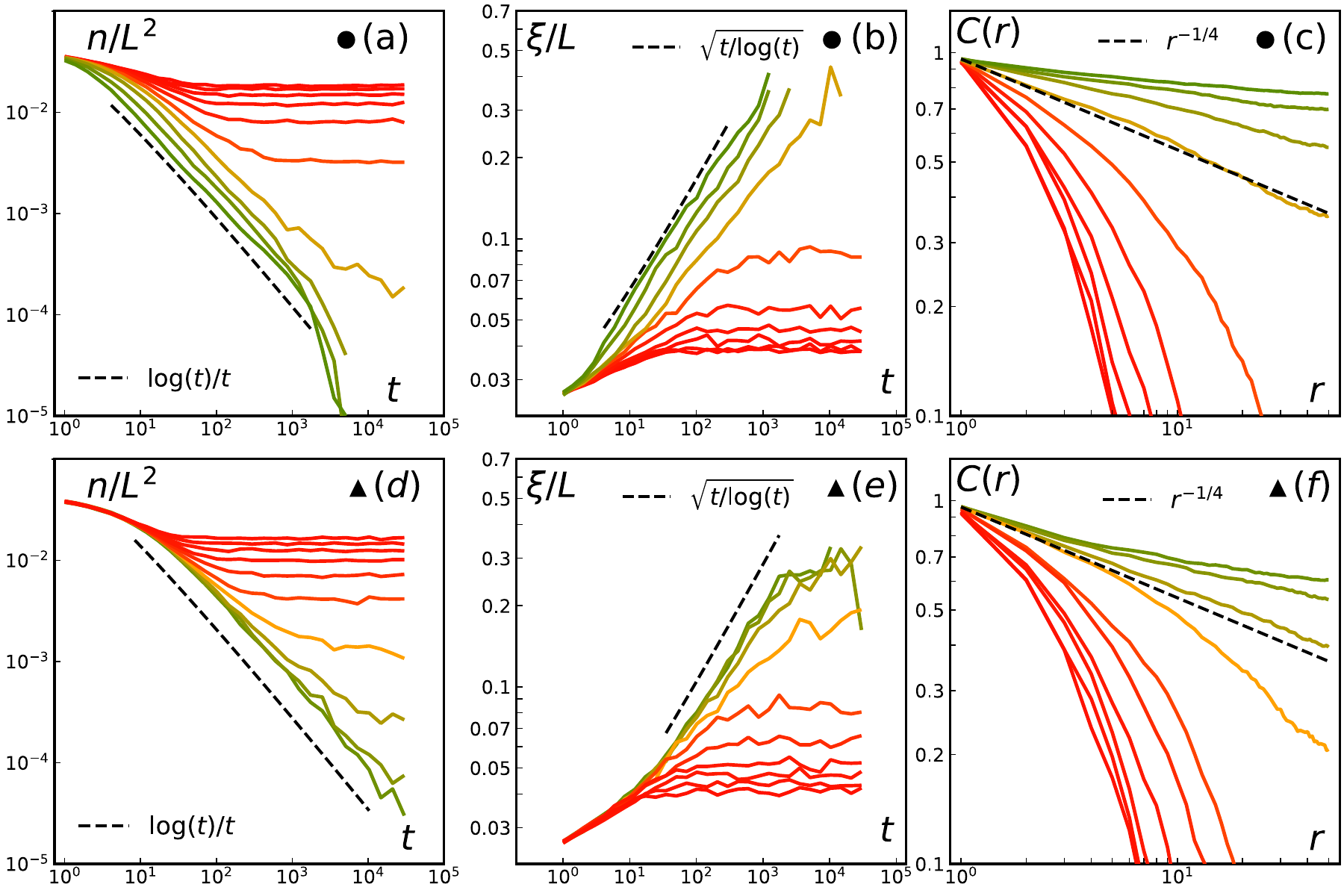}   
    \caption{
    Measurements on a system with $N=10^4$ particles.
Each curve is coloured by its corresponding steady-state polarization.
For the horizontal scan (circles, \textbf{a-c}), $\sigma = 0.1$ and $0.03 \leq v_0 \leq 1$.  For the vertical scan (triangles \textbf{d-f}), $v_0 = 0.2$ and $0.025 \leq \sigma \leq 0.225$.   \textbf{a,d}: defect density over time.\textbf{b,e}: normalised characteristic length over time.\textbf{c,f}: Spatial correlation functions in the steady state. 
\\
}
\label{fig:scan_vertical_horizontal}
\end{figure}

.
In Fig.~\ref{fig:phase_spaces}(a), we added horizontal dots and vertical triangles to highlight the values used in Fig.~\ref{fig:scan_vertical_horizontal} for the horizontal and vertical scans through the transition.  
In Fig.~\ref{fig:2}, we scanned the transition horizontally and measured the defect density, the characteristic length and the spatial correlation functions, observing how they are altered when the transition is crossed. 
In Fig.~\ref{fig:scan_vertical_horizontal}, to show that the two scans are equivalent, we compare the same quantities for both a horizontal scan (first row, panels (a,b,c), corresponding to the $\bullet$ symbols in Fig.~\ref{fig:phase_spaces}(a)), and a vertical scan (second row, panels (d,e,f), corresponding to the $\blacktriangle$ symbols in Fig.~\ref{fig:phase_spaces}(a)). 

\section{Appendix B: Smoothing procedure for visualisation and defect tracking} \label{sec:appendix_smoothing}
The numerical identification of topological defects is complicated when particles do not live on a grid, in particular because the density is not constant neither in time nor in space, even though it remains homogeneous on average.
To overcome this issue, and also for the sake of visualisation, we coarse grain the orientation field $\theta$. 
We decompose the $L \times L$ system into a $2d$ grid with each cell measuring $R_0 \times R_0$. To attribute each cell an effective orientation $\tilde \theta$, we aggregate the contribution of all the particles within a circle $\mathcal{C}$ of arbitrary radius $4R_0$, centered on the considered cell. The contribution of each particle to the considered cell is weighted by an exponentially decaying function of the distance $r_k$ of the particle to the center of the cell: $\tilde \theta = \text{Arg}\left[    \sum \limits_k   \exp{(   i\theta_k - r_k/R_0)}\right]$.

\begin{figure}[h!]
    \centering
    \includegraphics[width=1\linewidth]{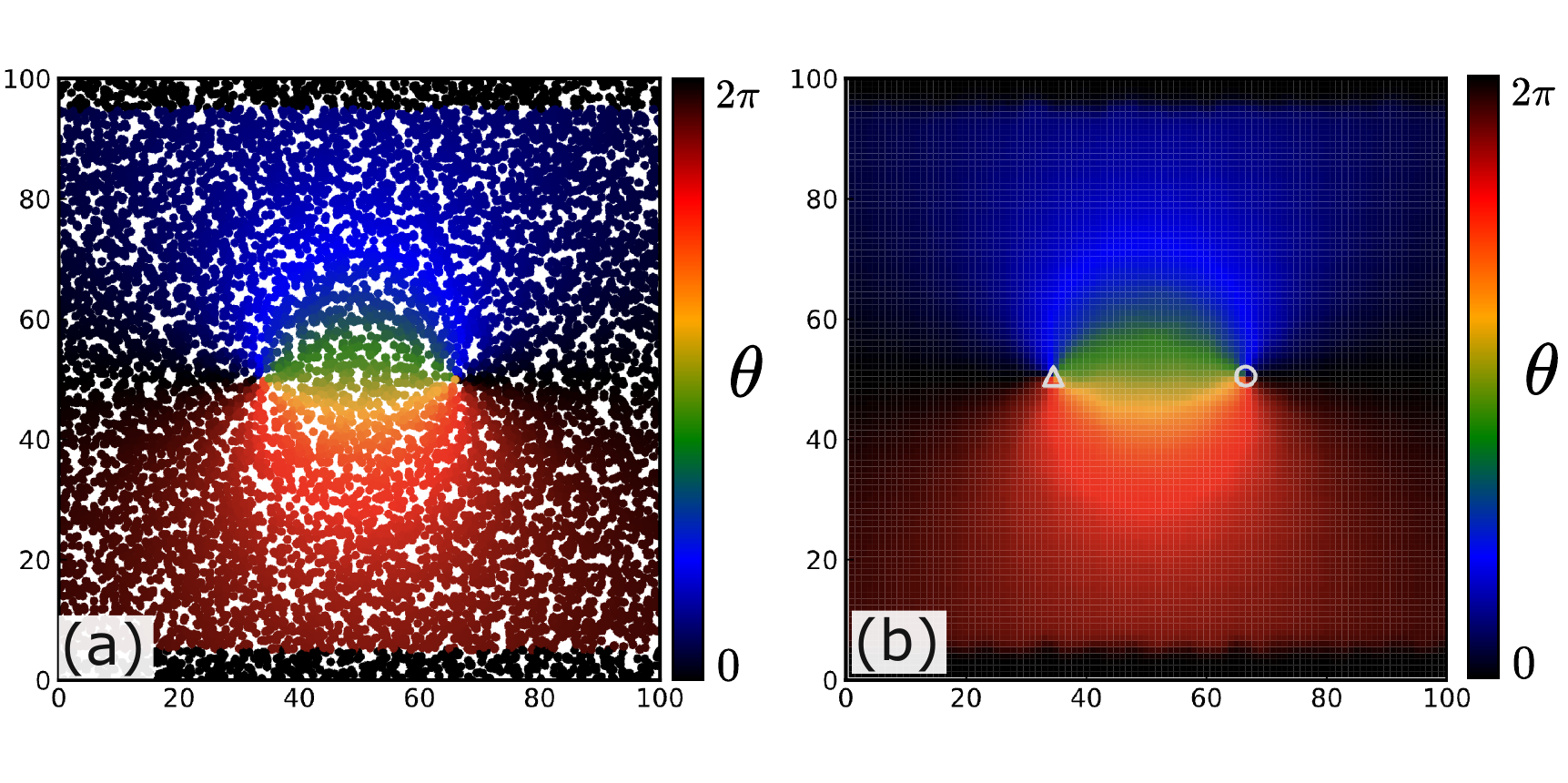}   
    \caption{A manually created pair of defects, with $q=1, N=10^4, \rho =1, L=100$, and $r_0 =32$. \textbf{(a)} Actual scatterplot of the particles.  
\textbf{(b)}  Result of the smoothing procedure. Circles (resp. triangles) are $+1$ (resp. $-1$) topological defects.}
    \label{fig:manually_created_pair}
\end{figure}

\section{Appendix C: Finite size scaling of the relaxation time} \label{sec:appendix_relaxtime}
\begin{figure}[h!]
    \centering
    \includegraphics[width=1\linewidth]{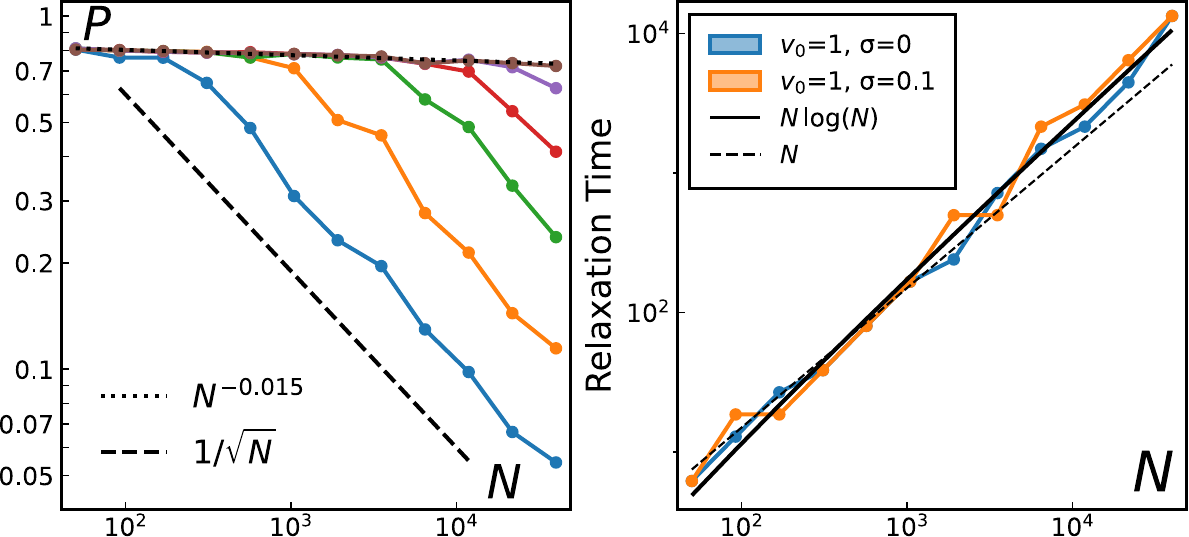}   
    \caption{\textbf{(left)} Finite size scaling of the polarization $P$ against $N$, for $v_0 = 1, \sigma = 0.1$, for different times $t=27, 166, 1035, ..., 40\,000$. (exponentially spaced, from bottom/blue to top/brown).
    \textbf{(right)} Relaxation time, defined as the time necessary to reach 95\% of the final polarization value, for different number of particles $N$. The solid black line is $0.025\,N\log N$. The dashed black line is $0.15\,N$.     }
    \label{fig:relaxation_time}
\end{figure}

In the left panel of Fig.~\ref{fig:relaxation_time}, we plot the polarization $P$ as a function of the number of particles $N = 50, ..., 40000$, averaged over 40 independent realizations, for $v_0 = 1$ and $\sigma = 0.1$ (in the quasi-ordered phase). The different colours represent different times, exponentially spaced, from bottom/blue to top/brown. At $t=4\times 10^4$ (in brown, below the dotted line), the systems up to $N=40000$ have relaxed. We see that smaller systems relax faster than bigger ones. In the right panel of the same Fig.~\ref{fig:relaxation_time}, we report the relaxation time the system needs to relax to its steady state in terms of the polarization. We define this time $t_{relax}$ as the time needed for the system to reach $95\%$ of its steady state polarization:  $P(t_{relax})=0.95\,P(t \to \infty)$. It appears that $t_{relax}$ follows the relationship $t_{relax} \sim N\,\log N$. 

\chapter{The non-reciprocal XY model} \label{ch:nrxy_defects} 
\graphicspath{{NRXY_defects/figures/}}
\graphicspath{{figures/Ch7NRXYdefects/}}
\section{Abstract}


In this chapter, we investigate the dynamics of a non-reciprocal $O(2)$ model, where the interactions between spins break the action-reaction principle.
From the lattice model, where non-reciprocity stems from vision cone like couplings, we derive a continuum description that captures the same physics, in which non-reciprocity translates into a new term depending on the rotational of the orientation field.

We first focus on the impact of non-reciprocity on topological defects. 
We show that in addition to the topological charge $q$, the actual shape of the defects becomes crucial to faithfully describe their dynamics. Non-reciprocal coupling twists the spin field, selecting specific defect shapes, dramatically altering the pair annihilation process. Defect annihilation can either be enhanced or hindered, depending on the shape of the defects concerned and the degree of non-reciprocity in the system, with important consequences on the coarsening dynamics. 

We then describe how this active curl term advects and reshapes patterns,
a generic feature found in many non-reciprocal systems. We characterise the dynamics of $1d$ and $2d$ perturbations, showing that they self-propel, diffuse and acquire an front/back asymmetry. Non-reciprocity also impacts the stability of defectless patterns with non-zero winding numbers and,
above a certain threshold, can even change the topology of the system and allow it to relax to its equilibrium ground state.
\newline

This chapter covers the results published in 
\begin{center}
  Rouzaire \textit{et al.}, Physical Review Letters (2025) \cite{rouzaire2025nonreciprocal} . \\ 
  Rouzaire \textit{et al.}, arXiv preprint (2026) \cite{rouzaire2025emergentdynamicsO2}.
\end{center}

\section{Introduction}

Non-reciprocal (NR) interactions are a common feature to many out-of-equilibrium systems. The recent blossoming of the investigation of this field has revealed a vast panel of rich emergent phenomena. One origin of non-reciprocity is the interaction between several populations, where the goals of the agents and the interactions between them are not necessarily symmetric \cite{saha_scalar_2020,mandal2024robustness, kreienkamp2022clustering}. Another route to non-reciprocity can be found in single species populations, if the (indistinguishable) agents perceive their surroundings through an anisotropic field of view. This is a common feature in nature, where animals can have a limited vision cone  \cite{cavagna_nonsymmetric_2017}. However, in those systems agents typically move in space. Coupling non-reciprocity and mobility makes it hard to disentangle the consequences of each contribution; one can isolate the two components to understand them better. 
The study of reciprocal and motile units in the framework of physics dates back to 1995, when Vicsek and coauthors introduced a minimal agent-based model \cite{vicsek1995novel} to explain the flocking transition in an animal group. Later that year, Toner and Tu \cite{toner1995long} proposed a continuum theory to describe collective behaviours in large groups.
On the other end of the spectrum, recent works have studied non-reciprocal and immobile agent-based systems \cite{dadhichi_journal_2020,loos2023long, rouzaire2025nonreciprocal, popli2025ordering, bandini2025xy} as well as the corresponding continuum models \cite{vafa2022defect, dadhichi_journal_2020, besse2022metastability,  dopierala2025inescapable, huang2024active}. 
While the typical focus of those studies is to establish the nature of the phases, their stability and the transition between them, the literature on the dynamics of such systems is relatively scarce, despite a rich phenomenology.

The study of topological defects in the context of non-reciprocal systems is relatively scarce. In hydrodynamically-driven crystals, dislocations (defects in the crystal structure) split apart and self-propel, leading to a continuous reshaping of grain boundaries and, ultimately, melting of the crystal from its edges. \cite{guillet2025melting}. In continuous polar spins systems, Vafa \cite{vafa2022defect} reported that the shape of positive defects is dynamically selected; Besse and coauthors \cite{besse2022metastability} explained how the spontaneous generation of defect pairs destabilizes the ordered phase. The defect annihilation scenario has first been studied in \cite{rouzaire2025nonreciprocal} and later been expanded by \cite{popli2025ordering}.   
Dynamical patterns have been slightly addressed in \cite{loos2023long} for a non-reciprocal XY (NRXY) lattice model and more in depth in \cite{huang2024active} for a continuum model. Yet, a thorough characterization of pattern dynamics in a non-reciprocal medium is still missing. 
\newline 

In this chapter, we uncover the impact of non-reciprocity on the dynamics of patterns and topological defects in the $O(2)$ model.
First, we present the non-reciprocal XY model (agent-based, on lattice) and how one can derive a continuum description out of it. 
Second, we explain how NR interactions affect the shape of single isolated defects, which in turn impacts their pair annihilation kinetics and the coarsening dynamics of the entire system.
We then describe how a perturbation travels in our non-reciprocal $2d$ medium. 
Finally, we describe the evolution of a topologically protected pattern: the lowest (defectless) energy excitation of the equilibrium $O(2)$ model. We show that for a small non-reciprocity, this configuration gets compressed in a narrow region of space, while the rest of the system orders. Last, we explain how a sufficiently high activity / non-reciprocity destroys this pattern, changes the topology of the system and allows the system to relax to its equilibrium ground state.

\section{The lattice model}
The NR XY lattice model is composed of spins $\hat{\boldsymbol{S}}_i=(\cos\theta_i,\sin\theta_i)$ sitting on the nodes of a triangular lattice of linear size $L$, and evolving accordingly to  
\begin{equation}
\gamma\,\dot{\theta}_i=J\sum_{j \in \partial_i} \,g(\varphi_{i j})\sin \left(\theta_j-\theta_i\right) + \sqrt{2\gamma\, T}\,\eta_i(t) 
 \label{eq:eom}
 \end{equation}
 
where $\gamma$ is the damping coefficient, $J$ the coupling, $\eta$ a Gaussian white noise with zero mean and unit variance and $T$ the temperature of the bath to which the system is coupled (fixing $k_B=1$).  When working with \er{eq:eom}, we set $\gamma=1$ and $J=1$ without loss of generality.  The sum runs over the nearest neighbours $j$ of spin $i$. We numerically integrate \er{eq:eom} using a standard Euler-Maruyama scheme with time step $dt=0.1$.

The kernel $g$ encodes the anisotropic nature of the interactions, thereby introducing NR interactions, and depends on the angle $\varphi_{ij}$ between the orientation of the spin $i$ and the vector $\boldsymbol{u}_{ij} = \boldsymbol{R}_{j} - \boldsymbol{R}_{i}$ joining the position of spin $i$ and the considered neighbour $j$: $\varphi_{ij} = \text{Angle}(\boldsymbol{\hat S}_i\, ,   \boldsymbol{u}_{ij})$.

\begin{figure}[b!]
    \centering
    \includegraphics[width=0.85\linewidth]{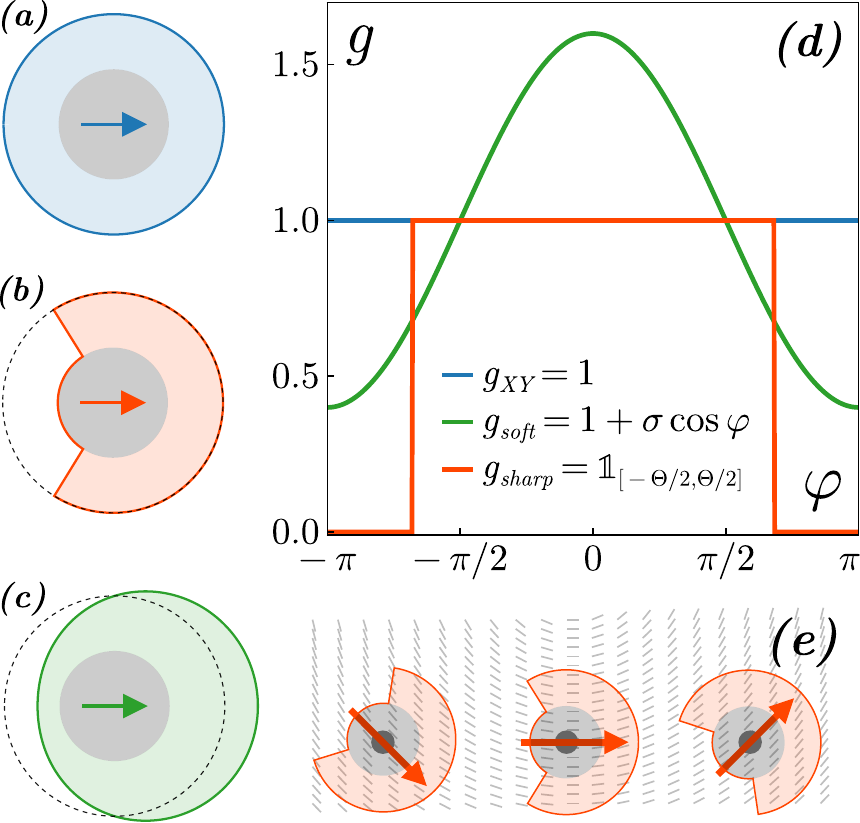}
    \caption{ Three possible kernels describing the angular dependence of the coupling strength: \textbf{(a)} a uniform kernel \textbf{(b)} a sharp vision cone equal to 1 in front and 0 behind. Here, the vision cone aperture is $\Theta=2\pi-2$ \textbf{(c)} a smooth vision cone, smoothly going from $1+\sigma$ (in front) to $1-\sigma$ (behind), with $\sigma = 0.6$.  These three kernels are all centered on the current orientation of the spin, here depicted by the horizontal arrows. The dotted lines in panels (b, c) represent the uniform kernel of (a). \textbf{(d)} Those three kernels represented a function of $\varphi=\text{Angle}(\boldsymbol{\hat S}_i\cdot  \boldsymbol{u}_{ij})$.  
     \textbf{(e)} Sketch of three spins equipped with the kernel in (b), in a field with a positive rotational; see main text for discussion.}
    \label{fig:kernels}
\end{figure}

The reciprocal $2d$ XY model with overdamped (non-conserved) dynamics is recovered for constant $g$, see sketch in Fig.~\ref{fig:kernels}(a).

In panel (b), the kernel is given by a step function $g(\varphi; \Theta) = 1 \text{ if } |\varphi| \le \Theta/2  $ and represents a sharp vision cone. An aperture $\Theta < 2\pi$ creates a blind spot at the rear and makes the interactions generically non-reciprocal, since a configuration where a spin $i$ looks at another spin $j$ but $j$ does not look at $i$ violates the action-reaction principle between $i$ and $j$. This type of vision cone has been used in non-reciprocal agent-based models in \cite{loos2023long, bandini2025xy}, and to construct a hydrodynamic theory in \cite{huang2024active}. 

We consider instead a ``smooth vision cone", see panel (c), based on the Von Mises distribution: $g(\varphi) = \exp(\sigma\,\cos\,\varphi)$ \cite{VonMises} in an effort to attenuate discretisation effects due to the lattice geometry 
, cf. \cite{bandini2025xy, popli2025ordering} and discussion below. 
We used this Von Mises kernel in \cite{rouzaire2025nonreciprocal}. For small non-reciprocity ($\sigma \ll 1$), one has $g(\varphi) \approx 1 + \sigma \cos \varphi$, a kernel used in \cite{popli2025ordering} and in our recent preprint \cite{rouzaire2025emergentdynamicsO2}. All the results presented in this chapter fall in the small $\sigma$ regime and thus can be obtained with both kernels. At high $\sigma$ values, the two kernels could possibly lead to different phenomenology, since $\exp(\sigma \varphi) > 0$ for all $\varphi$, while the $1 + \sigma \cos \varphi$ kernel becomes negative for $\varphi > \pi/2$ if $\sigma > 1$, changing the nature of the interactions from attractive to repulsive. In this chapter, we restrict our view to ferromagnetic interactions ($\sigma \leq 1$). 

The parameter $\sigma$ controls the degree of non-reciprocity, with action-reaction symmetry preserved only when $\sigma = 0$.  Since in the small $\varphi$ regime  $g(\varphi) \sim e^{-\varphi^2 / 2\sigma^{-1}}$,  $\sigma$ plays the role of the inverse variance: higher $\sigma$ values  shrink the vision cone.  Note that $\sigma$ in this chapter has no relation with the $\sigma$ used in the previous chapters on the Kuramoto model. We used the same symbol for historical reasons because in both cases, $\sigma>0$ controls the departure from equilibrium.
\newline

In what follows, we make a series of comments on the model. 
We comment on the symmetries of the system, 
we address the question of the equivalence between sharp and soft vision cones, 
we underline the difference in choosing a Langevin-like dynamics instead of a Glauber one,
and we discuss the crucial interplay between vision cones and the discrete nature of the lattice.

\subsection{Symmetries of the system}
  The equilibrium XY model is invariant under global rotation of the spins $\{ \theta \} \to \{ \theta + \alpha\} $  because the alignment term between two spins $i$ and $j$ only depends on the difference of the spins orientation: 
  \begin{equation}
      \sin ((\theta_j + \alpha) - (\theta_i + \alpha)) =  \sin (\theta_j - \theta_i)
  \end{equation}
  The reciprocal XY model is thus invariant under the addition of a global phase -- we say it is $O(2)$ invariant or, equivalently, $O(2)$ symmetric-- which prevents defects of the same topological charge but of different shapes (different phases) to behave differently.

  In our $2d$ spin system, the parity transformation $\begin{pmatrix}
  x \\ y 
  \end{pmatrix}
  \to 
  \begin{pmatrix}
  -x \\ y 
  \end{pmatrix}$ can be re-expressed in terms of angles with the transformation $\{ \theta \} \to \{ -\theta + \pi\} $. Since the $2dXY$ model is $O(2)$ symmetric, we only have to check that the equation of motion is invariant under the transformation $\{ \theta \} \to \{ -\theta\} $ 
  \begin{equation}
      \begin{aligned}
          \frac{d}{dt}(-\theta_i) = &\  \sin (-\theta_j - (-\theta_i)) \\ 
          -\frac{d}{dt}(\theta_i) = &\  -\sin (\theta_j - \theta_i) \\ 
          \frac{d}{dt}(\theta_i) = &\  \sin (\theta_j - \theta_i) \\
          \dot\theta_i = &\  \sin (\theta_j - \theta_i) 
      \end{aligned}
  \end{equation}
  So the XY model is indeed invariant under parity. Note that the parity transformation transforms a $q=+1$ defect into a $q=-1$ and vice-versa, such that XY defects behave in the same fashion regardless of their charge.

  On the other hand, the non-reciprocal kernel $g$ violates all the previous symmetries. Indeed, while the alignment term depends on the difference of the spins orientation $\theta_j - \theta_i$, $g$ only depends on the orientation of the spin $i$. Therefore, $g(\theta) \neq g(\theta+\alpha)$ and the system is no longer $O(2)$ invariant. This opens the door to different dynamics for defects of the same topological charge but of different shapes (sinks and sources for $+1$  defects for example). 

  In the same spirit, $g(\theta) \neq g(-\theta)$, breaking parity. 
  This allows for the dynamics to be different between $\pm 1$ defects in the non-reciprocal case. 

  Both parity and rotation invariances are true for all kernels that do not depend on $\theta_j - \theta_i$ but only on $\theta_i$. 
  In addition, $\sigma$ breaks time-reversal symmetry \cite{loos_irreversibility_2020, suchanek2023time}. We shall see that those broken symmetries have important consequences on the system dynamics.

  Finally, $\sigma < 0$ flips the modulation of the kernel, putting less weight at the front and more at the back. The system is therefore invariant under the joint transformation $(\theta, \sigma) \to (\theta+\pi, -\sigma)$ and we only consider $\sigma>0$.
  \newline

\subsection{Equivalence between sharp and soft vision cones}
  Sharp and soft vision kernels ($g_\text{sharp}$ and $g_\text{soft}$) lead to similar phenomenology. All the physics described in this chapter can be observed with both types of vision cones. This is not surprising since both kernels break the same symmetries and share the same qualitative front/back asymmetry.

  For soft vision cones, we use the non-reciprocal parameter $\sigma\geq0$. For  sharp vision cones, we use a vision cone $\Theta \leq 2\pi$, centered on the orientation $\theta$ of the spins. \\
  Both kernels have a common property: the total weight in the head-side hemisphere ($\varphi \in [-\pi/2, \pi/2]$) is greater than the total weight of the tail-side hemisphere ($\varphi \in [\pi/2, 3\pi/2]$). We define the ratio $f(\sigma)$ between those hemispheres as 
  \begin{equation}
      f(\sigma) = \int\limits_{-\pi/2}^{\pi/2} d\varphi \, g(\varphi)\bigg/\int\limits_{\pi/2}^{3\pi/2} d\varphi \, g(\varphi)
  \end{equation}
  For a reciprocal (=isotropic) kernel $g(\varphi) = 1$, then $f(\sigma)=1$;  while $f(\sigma)>1$ for non-reciprocal kernels. 
  \newline

  For soft vision cones $g_\text{soft}(\varphi) = \exp(\sigma \,\cos \varphi)$, one obtains
  \begin{equation}
      f_\text{soft}(\sigma) = \frac{\pi(L_0(\sigma) + I_0(\sigma))}{\pi(L_0(\sigma) - I_0(\sigma))}
  \end{equation}
  where {$I_0$} is the modified Bessel function of the first kind and { $L_0$} is the modified Struve function. 
  For small $\sigma$, one can Taylor expand those functions up to second order to obtain
  \begin{equation}
      f_\text{soft}(\sigma) \approx \frac{\pi + 2\sigma + \pi \sigma^2/4}{\pi - 2\sigma + \pi \sigma^2/4} \ .
  \end{equation}
  \newline

  For {sharp} vision cones $$g_\text{{sharp}}(\varphi) = 
  \begin{cases}
  1 \text{ if } |\varphi| \leq \Theta/2 \\
  0 \text{ otherwise}  
  \end{cases}\ ,$$
  one obtains
  \begin{equation}
      f_\text{sharp}(\Theta) = \frac{\pi}{\Theta-\pi} \ .
  \end{equation}

  Finally, equating the head-tail asymmetry ratios ($ f_\text{sharp} =  f_\text{soft}$)  gives an equivalence relation between $\sigma$ and $\Theta$:

  \begin{equation}
      \Theta(\sigma) = 2\pi - \frac{16\pi\sigma}{\pi(4+\sigma^2) + 8\sigma }
      \label{eq:equivalence_short_sharp_vision_cones}
  \end{equation}
  which allows to compare the dynamics of defects found in both versions of the model, as we shall see later in Fig.~\ref{fig:equiv_vision_cones}.  We have not looked in detail whether this equivalence holds for other observables at different scales, such as the coarsening dynamics. 

\subsection{Equation stemming from the XY energy modulated by the kernel}

  Importantly, as $g(\varphi_{ij})$ depends on $\theta_i$ but not $\theta_j$, Eq.~(\ref{eq:eom}) cannot be thought of
  as an equilibrium dynamics driven by a Hamiltonian-like function $\mathcal{H}=-J\sum g(\varphi_{ij}) \,\hat{\boldsymbol{S}}_i\cdot\hat{\boldsymbol{S}}_j$. 
  Generally speaking, the defining feature of a NR system is its dynamics, which here we chose to be Langevin-like \cite{rouzaire2025nonreciprocal, fruchart_non-reciprocal_2021}, in contrast with the Glauber dynamics implemented in \cite{loos2023long,bandini2025xy}. 

  We now compare our Langevin equation of motion to the equation of motion one would obtain from the relaxation driven by an energy function
  \begin{equation}
  E=-\sum_{i, j} g\left(\varphi_{i j}\right) \cos \left(\theta_j-\theta_i\right)
  \end{equation}
  If $g(x) = \exp(\sigma \, \cos x)$, the corresponding equations of motions would be (with $\Delta \theta_{ij} \equiv \theta_j - \theta_i$ and $\varphi_{ij} \equiv \theta_i - u_j$) : 
  \begin{equation}
  \begin{aligned}
  &\dot{\theta}_i  =-\frac{\partial E}{\partial \theta_i} \\
  &=\sum_{j} \frac{\partial}{\partial \theta_i}\left(g\left(\varphi_{i j}\right) \cos \left(\Delta \theta_{ij}\right)\right) \\
  &=\sum_j - g\left(\varphi_{i j}\right) \sin \left(\Delta \theta_{ij}\right) \underbrace{\frac{\partial \Delta \theta_{i j}}{\partial \theta_i}}_{=-1}+\cos \left(\Delta \theta_{ij}\right) \frac{\partial g\left(\varphi_{i j}\right)}{\partial \theta_i} \\
  &=\sum_j g\left(\varphi_{i j}\right) \sin \left(\Delta \theta_{ij}\right)+\cos \left(\Delta \theta_{ij}\right)\left(-\sigma \sin \varphi_{i j}\right) \underbrace{\frac{\partial \varphi_{i j}}{\partial \theta_i}}_{=1} g\left(\varphi_{i j}\right) \\
  &=\sum_j g\left(\varphi_{i j}\right)\left[\underbrace{\sin \left(\Delta \theta_{ij}\right)}_{\text{attraction}}-\underbrace{\sigma \sin \left(\varphi_{i j}\right) \cdot \cos \left(\Delta \theta_{ij}\right)}_{\text{(side) repulsion !}}\right]
  \end{aligned}
  \label{eq:extra_term}
  \end{equation}
  The first term $\sin\left(\theta_j-\theta_i\right)$ corresponds to the alignment term in our Langevin description. 
  The additional second term  in $\cos \left(\theta_j-\theta_i\right)$ favours orthogonal spin configurations but, due to the $\sin \left(\varphi_{i j}\right)$ coefficient, this repulsion mainly takes place on the sides of the spin $i$ where $\varphi_{i j} = \pi/2$ (``ahead" being in the direction of $\theta_i$ where $\varphi_{i j} = 0$). The amplitude of this second term, however, is \emph{a priori} smaller than the first term, since in this work we used $\sigma\leq1$ and since $|\sin \left(\varphi_{i j}\right)|\leq1$.
  Yet, the impact of this additional term, though only to second order, favours the stabilisation of the sources $\mu_+ \approx 0$, a configuration in which spins are indeed perpendicular. 

\subsection{Small vision cones and percolation threshold}
  Most of the results presented in this chapter are for values of $\sigma \leq 0.5$. We pushed up to $\sigma=1$ for Fig.~\ref{fig:3}(e).  
  However, one cannot increase $\sigma$ up to arbitrarily large values. Indeed, the larger $\sigma$ gets, the more peaked the kernel $g(x) = \exp(\sigma \cos x)$ becomes. If almost all the mass of the distribution falls between two neighbours, the spin in fact ``sees" no neighbour at all, and its dynamics is thus only driven by white noise. 
  Here, we work far from this regime. For instance, for $\sigma=0.5$, the deviation from an isotropic kernel is moderate:  $g(0)=e^{0.5} \approx 1.65$ and the smaller coupling value is $g(\pi)=e^{-0.5} \approx 0.6$, far from negligible.
  Such pathological limit $\sigma \to \infty$ is not a specificity of our model. Loos and co-authors \cite{loos2023long, bandini2025xy} report in the same spirit that for sharp vision cones smaller than $\pi/3$ on the triangular lattice ($\pi/2$ on the square lattice), the percolation of the interaction network is no longer guaranteed. 

  Without going to such extreme values of $\sigma$, one can still wonder whether the symmetries of the interactions remain the same for large $\sigma$ values. Indeed, for a kernel $g(\varphi) = 1 + \sigma \cos \varphi$, the interactions are ferromagnetic only for $|\sigma |\le 1$ . Taking $|\sigma |>1$ introduces anti-alignment in the local interactions with neighbors at the back, as the kernel becomes negative in an angular region of aperture $2\pi-2\arccos(-1/\sigma)$. To provide a sense of scale, with $\sigma =1.15$, already 16\% of bonds are antiferromagnetic (implying 1-2 neighbors on a triangular lattice), and this number climbs up to 33\% (2-3 neighbors) for $\sigma= 2$. 

The consequences of such a symmetry change are somehow still unclear. The few references in the literature \cite{dadhichi_nonmutual_2020, besse2022metastability, popli2025ordering} are not fully conclusive regarding the roles of temperature, system size, and mixed ferromagnetic/antiferromagnetic interactions. \\
In an underdamped and noisy system with the same kernel \cite{dadhichi_nonmutual_2020}, it has been shown that the ordered state is unstable to those local anti-aligning interactions upon increasing $\sigma$. The critical value, however, is not simply equal to $J=1$ but scales with the \textit{square root} of an effective coupling constant that also depends on the Gaussian noise intensity $T$, the inertia and  friction coefficients. \\
Instability of the ordered state has also been reported  in an overdamped and noisy continuum model \cite{besse2022metastability}. There, 
a ``foam of aster defects" destabilizing the ordered state is observed for sufficiently large temperature $T$, ``not too small $\sigma$ and large enough systems".  In particular, it has been observed for $\sigma = 0.69$, but at high noise $T = 1.04$ (translated in a framework where the preferred magnitude of the vector field is 1).  A systematic understanding of the role played by  each  parameter in such instability is still lacking. 
Such instability has also been observed in the overdamped $2d$XY lattice model with our choice of kernel $g$ for $\sigma = 2$ and $T=0.4$, while it does not appear for $\sigma \leq 1$ and $T=0.005$~\cite{popli2025ordering}. 
\newline

\subsection{Interplay between vision cones and the discrete nature of the lattice}
Finally, we discuss the interplay between vision cones and a discrete regular lattice. This interaction between the dynamics and the lattice is common to all kinds of vision cones, sharp or soft. One motivation to use a soft vision cone in this work instead of the sharp kernel of Loos \textit{et al.} \cite{loos2023long} is to attenuate the coupling to the discrete symmetry of the underlying lattice.

In \cite{loos2023long}, a spin $j$ is considered to be in the neighbourhood of a spin $i$ if and only if $j$ is in the vision cone of $i$. 
They use a sharp vision cone kernel, as plotted in orange in Fig.~\ref{fig:kernels}, such that $j$ is an interacting  neighbour of $i$ if and only if the angle between the orientation $\theta_i$ of the spin $i$ and the location $u_j$ of the spin $j$ is less than half of the vision cone $\Theta$. 

\begin{figure}[h!]
    \centering
    \includegraphics[width=0.75\linewidth]{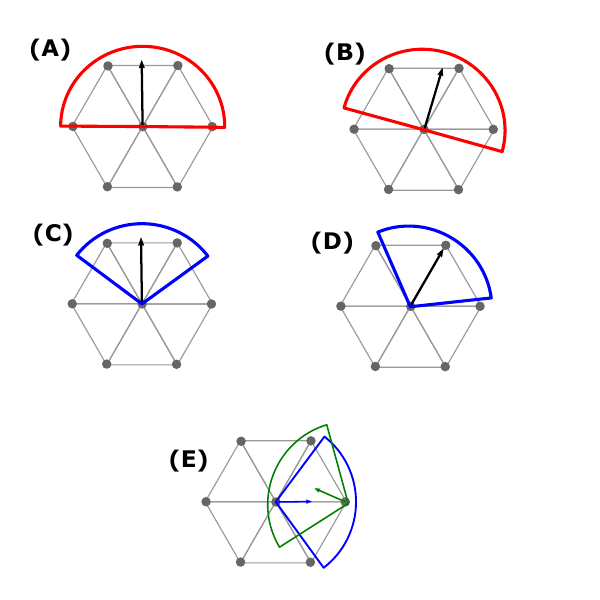}   
    \caption{ 
   \textbf{(a, b, c, d)} A spin can brutally lose or gain one neighbour within its sharp vision cone during the dynamics. \textbf{(e)} The effective temperature felt by two neighbouring spins can be different if their number of neighbours is different.
    }
    \label{fig:issues_vision_cone}
\end{figure}

{The main consequence of the interplay between a sharp kernel and a discrete underlying lattice is the generic {sudden} change in the number of neighbours as the system evolves in time, as we illustrate here on a triangular lattice ($z=6$ is thus the coordination number of the lattice). }
This aspect is discussed in depth in \cite{loos2023long}. The case $\Theta=2\pi/z, \ n=1,...,z$, illustrated in Fig.~\ref{fig:issues_vision_cone}(a,b), is helpful to understand this {feature}. On the triangular lattice, if the orientation of a spin $i$ is $\theta_i = \pi/2$, $i$ has 4 neighbours in its vision cone. If its orientation fluctuates by an infinitesimal amount, it immediately looses one neighbour. The same happens for all values of $\Theta$,  as we sketch in Fig.~\ref{fig:issues_vision_cone}(c,d) for an arbitrary vision cone amplitude. The consequence of this brutal change in the number of neighbours is important. If one works with a Monte-Carlo type of dynamics, since the energy is extensive (as opposed to intensive if normalised by the number of contributing neighbours) there exist $z$ regions energetically favourable
imposed by the symmetry of the underlying lattice. 
This strongly promotes a local alignment with the symmetry of the lattice. 

If instead one works with a Langevin approach based on extensive (i.e. additive) forces, a similar feature remains. We illustrate in Fig.~\ref{fig:issues_vision_cone}(e) a configuration where the blue spin (in the center) has only one {neighbour} within its vision cone, while the green spin (on the right) has two neighbours, among which the one in blue. The Langevin equation of motion looks like $\dot \theta_i = \sum_j \sin(\theta_j-\theta_i) +\sqrt{2T}\eta_i$. For the blue spin, the sum only contains one contribution while for the green spin, it contains 2 terms so it is potentially twice as big. However, the temperature $T$ of the thermal bath is constant. 
The ratio between the thermal angular diffusion and the interacting force thus becomes space and time dependent.

To avoid these consequences, one could imagine normalising the energy (or the force in a Langevin-like approach) by the (time-dependent)  number of neighbours within the vision cone. However, this introduces another source of non-reciprocity, as  the force of green on blue is now twice the force of the blue on green, in absolute value. The introduction of this new source of non-reciprocity has been documented in \cite{chepizhko2021revisiting} and seems of a different nature than the one induced by vision cones. 

{These sharp features are a consequence of using a sharp vision cone kernel on a discrete lattice. 
 However, the interplay with the symmetry of the underlying lattice is a generic feature of such vision cone models (sharp and smooth) on discrete lattices. As such, even if we have not quantified the effective alignment and its impact at larger scales (on defects or at the system scale) in the present work, an interplay with the discrete lattice, providing biased directions for the global order parameter, is also expected; see \cite{bandini2025xy, dopierala2025inescapable,popli2025ordering} for more thorough discussions on that topic. 

Indeed, since the kernel $g(\varphi) = \exp(\sigma\,\cos \varphi)$ is continuous and differentiable, a small change in $\varphi$ can only generate a small change in $g$. Such a feature of vision cone models on regular lattices has been recently discussed in \cite{bandini2025xy,popli2025ordering}.
 }

A possible relevant measure is the sum of the angular weights $\mathcal{G}_i  \equiv \sum_{j=1}^z g\left(\varphi_{i j}\right)$, as it relates to the maximum force a spin can feel. In particular, its dependence on $\theta_i$ is related to the discretisation effects; its dependence on $\sigma$ allows to understand the relative importance of the two terms in the equation of motion: alignment and thermal noise.
We have seen above that, in the sharp vision cone case, both a change in orientation $\theta_i$ or in the vision cone amplitude $\Theta$, can result in a sudden gain of one neighbour, which translates into a sudden increase (+1) of $\mathcal{G}_i$ . 
In the soft vision cone case, we want to study the dependence of $\mathcal{G}_i$ on $\sigma$ and $\theta_i$. To do so, we consider the square lattice for two reasons: the calculations are simpler and discretisation effects in the triangular lattice are less pronounced than for the square lattice; we thus consider the worst-case scenario. The sum over the angular weights gives:
\begin{equation}
\begin{aligned}
\mathcal{G}_i & \equiv \sum_{j=1}^4 g\left(\varphi_{i j}\right) \\
& =\sum_{j=1}^4 e^{\sigma \cos \left(\theta_i-u_{j}\right)} \\
& =\sum_{j=1}^4 e^{\sigma \cos \left(\theta_i-(j-1) \frac{\pi}{2}\right)} \\
& =e^{\sigma \cos \theta_i}+e^{\sigma \sin \theta_i}+e^{-\sigma \cos \theta_i}+e^{-\sigma \sin \theta_i } \\
& =2 \cosh \left(\sigma \cos \theta_i\right)+2 \cosh \left(\sigma \sin \theta_i\right)
\end{aligned}
\end{equation}
Since $|\cos x|\leq1$ and $|\sin x|\leq1$ , for small $\sigma$ one can Taylor expand the cosh:
\begin{equation}
\begin{aligned}
\mathcal{G}_i & = 2+\left(\sigma \cos \theta_i\right)^2+\frac{1}{12}\left(\sigma \cos \theta_i\right)^4+\mathcal{O}\left(\sigma^6\right) \\
& +2+\left(\sigma \sin \theta_i\right)^2+\frac{1}{12}\left(\sigma \cos \theta_i\right)^4+\mathcal{O}\left(\sigma^6\right) \\
& =4+\sigma^2+\sigma^4\frac{\cos ^4 \theta_i+\sin ^4 \theta_i}{12}+\mathcal{O}\left(\sigma^6\right)
\end{aligned}
\end{equation}

We learn that $\mathcal{G}_i $ only varies with the orientation $\theta_i$ to \emph{fourth} order in $\sigma$. 
For $\sigma=0.5$ , it implies that $\mathcal{G}_i(\theta) $ varies at most by 0.06 \% from it maximum value $\mathcal{G}_i(\theta=n\pi/2), \ n=1,...,z$ (to be compared to the 25 \% of the vision cone model when a spin passes from 4 to 3 neighbours). 
Indeed, when $\theta = \pi/4$, the mode of the kernel falls in between 2 neighbours, and

\begin{equation}
\begin{aligned}
\frac{\mathcal{G}_i(\theta=0;\sigma=0.5)-\mathcal{G}_i(\theta=\pi/4;\sigma=0.5)}{\mathcal{G}_i(\theta=0;\sigma=0.5)} & \\
= \frac{4.2552- 4.2526}{4.2552} = 0.0006
\end{aligned}
\end{equation}

Therefore, the soft vision kernel we propose should indeed strongly attenuate the effects stemming from the underlying discrete lattice for the typical range of $\sigma \in [0,1/2]$ used in this work.

\section{The continuum model}

  To provide a continuous description of our model at coarse-grained scales, 
  we first introduce the director field $\boldsymbol{S}(\textbf{r})$: 
 a vector order parameter with two degrees of freedom that can be written as $\boldsymbol{S}=S(\cos\theta,\sin\theta)$. Now,  $\theta=\theta(x,y)$ is  the local phase of the field and the scalar order parameter $S=|\boldsymbol{S}(x,y)|$ its modulus. The scalar order parameter corresponds to the local polarization averaged over some coarse-graining length scale.

 Following \cite{YurkeHuse1993}, the dynamics of $\boldsymbol{S}$ in the equilibrium XY model is governed by 
  \begin{equation}
      \gamma\dot{\boldsymbol{S}} = K\Delta\boldsymbol{S} + \alpha (1-|\boldsymbol{S}|^2)\boldsymbol{S}
      \label{eq:continuousXY}
  \end{equation}
  that can be written as a relaxational dynamics in the presence of dissipation with the following Ginzburg-Landau free energy functional 
  \begin{equation}
      \mathcal{F} = \int d\textbf{r} \left[ \frac{K}{2} (\partial_a\boldsymbol{S}_b)^2 +\frac{\alpha}{4}\left(1-|\boldsymbol{S}|^2\right)^2 \right]
      \label{eq:free_energy}
  \end{equation}
  where repeated indices (due to the square) are summed over following the Einstein convention. 
  The first term of Eq.~(\ref{eq:continuousXY}) penalizes gradients in the director field, $\boldsymbol{S}$, pushing the system towards homogeneous states. The second term penalizes deviations of the scalar order parameter from unity, reflecting the tendency for adjacent spins to align. These two terms are controlled by the parameters $K$ and $\alpha$, and give rise to the length scale $\lambda_K = \sqrt{K/\alpha}$ over which the order parameter can smoothly vary from $0$ to $1$ to compensate for large gradients of the director field. 
  This is particularly important in the neighbourhood of a topological defect, where the gradients of the director field diverge at the core of the defect leading to a decrease of the scalar order parameter. This is why $\lambda_K$ is often referred to as the defect core radius.

  This equilibrium XY scenario is greatly enriched by the introduction of anisotropic couplings, through the non-reciprocal kernel $g$. 
  In \cite{rouzaire2025nonreciprocal}, we have shown that a simple, yet instructive, coarse-graining of the lattice model, built upon identifying lattice differences with derivatives, gives the following field theory
    \begin{equation}
    \begin{aligned}
    \gamma\dot{\boldsymbol{S}}&= - \frac{\delta \mathcal{F}}{\delta \boldsymbol{S}} + \bar\sigma (\nabla \times \boldsymbol{S})\times \boldsymbol{S} \\
    &= K\Delta\boldsymbol{S}
    \!+ \bar\sigma (\nabla \times \boldsymbol{S})\times \boldsymbol{S} + \alpha (1-|\boldsymbol{S}|^2)\boldsymbol{S}
    \end{aligned}
    \label{eq:main_eq}
    \end{equation}
with $\bar\sigma = 2\,\sigma \,L /N_{\text{grid}} $. 
The details of the derivation and the dimensional analysis of the continuum equation are respectively given in the Appendices B and C, at the end of this chapter.
Under the approximation $S\approx 1$, the last (Landau) term becomes negligible and Eq.~(\ref{eq:main_eq}) can be expressed solely in terms of the orientation field $\theta$:
\begin{equation}
\gamma \,\dot \theta = K\Delta \theta + \bar\sigma (\nabla \times \hat{\textbf{S}})_z
= K\Delta \theta + \bar\sigma (\hat{\textbf{S}} \cdot \nabla)\theta 
\equiv f_{\text{EL}} + \bar \sigma f_{\text{NR}}
\ , 
\label{eq:main_eq_1dof}
\end{equation}
where 
\begin{equation}
    \hat{\textbf{S}} = 
\begin{pmatrix}
\cos \theta \\
\sin \theta 
\end{pmatrix} .
\end{equation}
The $z$ index denotes the third component of the rotational (the only one non-zero). 
A further useful simplification of Eq.~(\ref{eq:main_eq_1dof}) is the restriction of the dynamics to the $x$-axis (if $\partial/\partial y=0$), that reads
\begin{equation}
\gamma \,\dot \theta = \ K\partial_{xx} \theta + \bar\sigma \cos \theta \,\frac{\partial \delta}{\partial x}  \ \ \  .
\label{eq:main_eq_1dof_1d}
\end{equation}
  
When working with the continuous model, we numerically simulate Eq.~(\ref{eq:main_eq}) using a square grid of linear size $L$ discretized with $N_\text{grid}$ nodes, with a timestep $dt = 10^{-5}$. 
We always work with periodic boundary conditions. We express times in units of $\gamma/\alpha$, lengths in units of $\sqrt{K/\alpha}$, velocities in units of $\sqrt{K\alpha}/\gamma$ and diffusion coefficients in units of $K/\gamma$, see Appendix C. We choose $L=2\pi$ (it simplifies some computations in which $2\pi/L$ naturally appears, see Appendix I). 
We always work with periodic boundary conditions. We express times in units of $\gamma/\alpha$, lengths in units of $\sqrt{K/\alpha}$, velocities in units of $\sqrt{K\alpha}/\gamma$ and diffusion coefficients in units of $K/\gamma$, see Appendix C. We choose $L=2\pi$ (it simplifies some computations in which $2\pi/L$ naturally appears, see Appendix I)
\newline 

\subsection{Discussion on the non-reciprocal field theory}

\paragraph{Non-reciprocity as an active force \\}

The first equation in our main Equation~(\ref{eq:main_eq}) highlights that non-reciprocity ultimately amounts to an active, state-dependent force. 
Indeed, going back to the lattice model, the missing contribution from a neighbor $j$ lying in the blind spot of spin $i$ can be seen as a virtual force  canceling the symmetric ferromagnetic force of $j$ on $i$. To bridge the microscopic and hydrodynamic descriptions, we represent in Fig.~\ref{fig:kernels}(e) a generic configuration of three spins equipped with vision cones in a field with local positive rotational. 
In the reciprocal XY case, the middle spin would be steady due to the two equal and opposite contributions from its two neighbors, consistent with the rotational invariance of the Laplacian. 
Vision cones break this front/back symmetry and the virtual force canceling the contribution from the neighbor at the rear effectively makes the spins sensitive to the curl of the local orientation field. This justifies the appearance of the curl and its first order derivatives from the coarse-graining procedure. 
\newline

\paragraph{Connection with other field theories of polar active matter\\}

The rotational term drives the system away from equilibrium, has dramatic consequences on the dynamics of topological defects \cite{rouzaire2025nonreciprocal} and allows for the existence of propagative states; cf. \cite{loos2023long} for a short discussion in the agent-based model. In addition, the curl term breaks invariance under global rotation of the spins $\theta \to \theta + \theta_0$.  
Also, since
\begin{equation}
 (\nabla \times \boldsymbol{S})\times \boldsymbol{S}=      (\boldsymbol{S}  \cdot \nabla) \boldsymbol{S} - \frac{1}{2}\,\nabla|\boldsymbol{S} |^2  
 \label{eq:curl_adv}
\end{equation}
the rotational term emerging from the microscopic model bridges immobile, non-reciprocal spins and systems of aligning self-propelled particles. Indeed, Eq.~(\ref{eq:main_eq}) now resembles the Toner-Tu equation for constant-density 
flocks \cite{toner2012birth, besse2022metastability, toner1995long, toner2005hydrodynamics}, that reads
\begin{equation}\label{eq:MtTT}
   \gamma\dot{\boldsymbol{S}}= - \frac{\delta \mathcal{F}}{\delta \boldsymbol{S}}   + \lambda_1(\boldsymbol{S}  \cdot \nabla) \boldsymbol{S} + \lambda_2( \nabla \cdot \boldsymbol{S}) \boldsymbol{S} + \lambda_3\nabla|\boldsymbol{S} |^2   \ .
\end{equation}
A more careful coarse-graining based on Ito calculus, see  \cite{dopierala2025inescapable} and Appendix D, gives the three coefficients $\lambda_{1,2,3}$ in terms of $\bar\sigma$, and after some simple algebra, one can write the resulting continuum equation as 
A more careful coarse-graining based on Ito calculus, see  \cite{dopierala2025inescapable} and Appendix D, gives the three coefficients $\lambda_{1,2,3}$ in terms of $\bar\sigma$, and after some simple algebra, one can write the resulting continuum equation as 
\begin{equation}
   \gamma\dot{\boldsymbol{S}}= - \frac{\delta \mathcal{F}}{\delta \boldsymbol{S}}   + \lambda_1 (\nabla \times \boldsymbol{S})\times \boldsymbol{S} + \lambda_2\left(( \nabla \cdot \boldsymbol{S}) \boldsymbol{S} - \nabla|\boldsymbol{S} |^2  \right) \, . 
   \label{eq:chate_continuum}
\end{equation}
Our Eq.~(\ref{eq:main_eq}), stemming from the microscopic model, thus neglects the divergence $(\lambda_2)$ and part of the self-anchoring $(\lambda_3)$ terms of Eq.~(\ref{eq:chate_continuum}). We acknowledge that these could potentially have a strong impact, for instance on topological defects and, as a result, on the coarsening dynamics of the whole system. Indeed, a $\lambda_2<0$ would stabilize even further sink defects (inward asters), which are now known to be the only stable $+1$ defects and the ones responsible for the slowdown of the annihilation dynamics~\cite{vafa2022defect, besse2022metastability, rouzaire2025nonreciprocal, popli2025ordering}. 
\newline

However, we argue that the curl term captures the essence of non-reciprocity in our framework. First, we have shown in \cite{rouzaire2025nonreciprocal} that Eq.~(\ref{eq:main_eq}) reproduces the phenomenology of the lattice NRXY model, up to a quantitative agreement for the defect annihilation dynamics. 
Second, the two other terms $\propto \lambda_{2,3}$, can be obtained by adding a term $\propto (\nabla \cdot\boldsymbol{S}) |\boldsymbol{S}|^2$ in the free energy $\mathcal{F}$ [see Appendix D for details on Eqs.~(\ref{eq:chate_continuum}) and (\ref{eq:withextraF})] : 
\begin{equation}
\begin{aligned}
       \gamma\dot{\boldsymbol{S}}= &\ - \frac{\delta \mathcal{F}}{\delta \boldsymbol{S}}   + \ \lambda_1 (\nabla\times\boldsymbol{S})\times\boldsymbol{S} \\
   &\ + \frac{\lambda_2 }{2}\left\{\,\frac{\delta}{\delta \boldsymbol{S}}
      \Big[ (\nabla\cdot\boldsymbol{S})|\boldsymbol{S}|^2 \Big]  - \nabla|\boldsymbol{S}|^2\right\}
\end{aligned}
\label{eq:withextraF}
\end{equation}
Equations (\ref{eq:curl_adv}) and (\ref{eq:withextraF}) indicate that in the limit where gradients $\nabla|\boldsymbol{S}|^2$ are small, the rotational term becomes the self-advection one, and that the $\lambda_2$ term can be approximated by an equilibrium term. In this limit, which applies for most of this work, the advective term $(\boldsymbol{S}  \cdot \nabla)\boldsymbol{S}  $ included in the curl is the only genuine out-of-equilibrium term. Thus, in this work, $\lambda_2 = 0$. 

Last, the rotational term not only accounts for non-reciprocity (NR) stemming from \textit{reception} (or vision) \textit{cones} but also for NR stemming from \textit{emission cones}, a concept we briefly explain now. In the model we have presented so far, the reception of the neighbors' information $\theta_j$ is restricted by the vision cone of a test spin $i$. Information about $\theta_i$ is, on the contrary, emitted isotropically: a neighbor $j$ without a vision cone would perceive the same information regardless of the relative position of $j$ and $i$. The kernel $g$ of a reception cone depends on the orientation of the spin $i$ only: $\text{Angle}(\boldsymbol{\hat S}_i\, ,   \boldsymbol{u}_{ij})$. Yet, one could very well imagine a reversed situation where the reception is isotropic but the emission is not. This emission cone case is equally natural if one  considers, for instance,   the directed sound emission in animals. In this scenario, the kernel $g$ now depends on the orientations of the neighbors: $\text{Angle}(\boldsymbol{\hat S}_j\, ,   \boldsymbol{u}_{ij})$. Despite the microscopic models being different, they lead to the same hydrodynamic equation Eq.~(\ref{eq:main_eq}) with a simple sign difference: $\bar \sigma \to -\bar \sigma$, see details in Appendix E.
A $\sigma < 0$ flips the modulation of the kernel, putting less weight at the front and more at the back. 
In other words, at the continuum level, receiving information from ahead and emitting it isotropically is equivalent to emitting information to the back and receiving it isotropically. One can rationalize this in terms of information flux, which in both cases is in the direction opposite to the local polarization. This observation might be useful in cases where it is experimentally simpler to design a setup (of interactions based on light, sound, chemicals, ...) where the emission, rather than the reception, is anisotropic. 
\newline

  This rotational term drives the system away from equilibrium, has dramatic consequences on the dynamics of topological defects \cite{rouzaire2025nonreciprocal} and allows for the existence of propagative states; cf. \cite{loos2023long} for a short discussion in the agent-based model, and the present work for the study in this continuum theory. This curl term also breaks invariance under global rotation of the spins $\theta \to \theta + \theta_0$. 

\section{Defect dynamics}
In this first section, we focus on the dynamics of topological defects and report the new phenomenology induced by non-reciprocity. We first focus on the impact of non-reciprocity on isolated single defects, then on the impact of non-reciprocity on the annihilation of defect pairs, and finally on the impact of non-reciprocity on the coarsening dynamics of a system initially in a disordered state.

We recall that the shape of a $+1$ defect is characterised by the (pseudo) scalar $\mu \in [0, 2\pi]$, giving \emph{sources} ($\mu_+ = 0$), \emph{sinks} ($\mu_+ = \pm\pi$) and \emph{vortices} ($\mu_+ = \pm\pi/2$). The term ``vortex" here refers to this specific divergence-free spin configuration, not to a general non-zero topological charge. 

\subsection{Impact of non-reciprocity on the defect shape}

\begin{figure}[h!]
    \centering
    \includegraphics[width=0.99\linewidth]{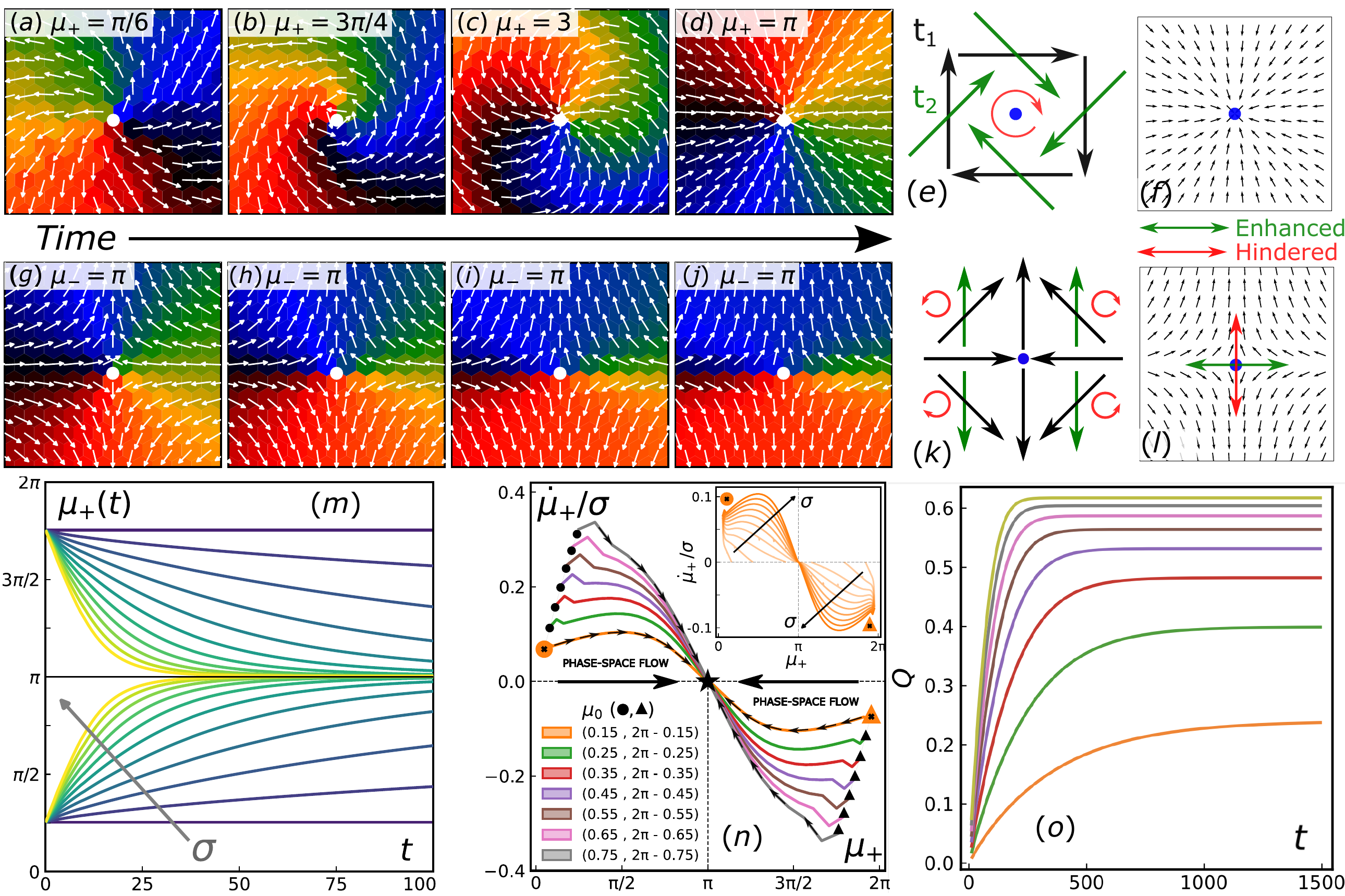} 
    \caption{
    \textbf{(a-d)} Twist of a positive defect towards the sink state ($\mu_+ = \pi$). 
    \textbf{(g-j)} Polarisation of a negative defect, leaving $\mu_-$ unchanged. 
     \textbf{(e,k)} Reshaping mechanism for $+1$  (e) and  $-1$ (k)  defects, at time $t_1$ (black) and $t_2 > t_1$ (green).  The blue circles are the defect cores; the red arrows represent the non-reciprocal torques.  \textbf{(f,l)} Analytical final states computed in Appendix G, for $+1$ (f) and $-1$ (l) defects. For the $-1$ defect, we indicate with a green (resp. red) arrow the direction parallel (resp. perpendicular) to the defect axis, along which the motion of the defect is enhanced (resp. hindered).   
    \textbf{(m)} Time evolution a $q=+1$ defect's shape for different $\sigma$ from 0 (dark blue) to $0.5$ (yellow) by increments of $0.05$, for two initial configurations ($\mu_0=\pi/4,\,7\pi/4$).
     \textbf{(n)} Phase space plot $\mu_+ - \dot{\mu}_+/\sigma$ of a twisting defect at $\sigma = 0.5$. They all flow towards the central star, the stable sink state. \underline{Inset}: same flow plot for two initial shapes $\mu_0 = 0.15$ and $\mu_0 = 2\pi-0.15$, now for different values of $\sigma = 0.05, 0.1, ..., 0.5$. $T=0$ for all panels.
       \textbf{(o)} Nematic order parameter $Q(t)$ of a $q=-1$ defect for different $\sigma$ from 0.05 (bottom, orange) to $0.4$ (top, green) by increments of $0.05$. 
}
    \label{fig:2}
\end{figure}

To understand the effect of NR interactions on $\mu_+$ we analyse an isolated $q=+1$ defect. We observe that at zero temperature, the source and sink are stationary states. All intermediate defect shapes decay to a sink, shown for $\mu_+=\pi/6$ in Fig.~\ref{fig:2}(a-d). By approximating the dynamics of $\theta$ close to a defect, we show that $\mu_+=0$ is an unstable fixed point and $\mu_+=\pi$ is stable, see Appendix G. To highlight the mechanism behind this effect, consider a $q=+1$ vortex defect ($\mu_+ = \pi/2$). The arrows in front of a given spin point slightly towards the defect core, and the arrows behind  slightly away. In a reciprocal system, these influences balance and the defect shape is stable. NR interactions cause an imbalance between these two influences leading to a torque on the spins, reshaping the defect as sketched in Fig.~\ref{fig:2}(e).

To quantify such reshaping, we follow $\mu_{+}(t)$ in time {at $T=0$}. 
As expected, for  $\sigma = 0$, the shape of the defect remains unchanged, see Fig.~\ref{fig:2}(g). For  $\sigma > 0$, $\mu_+$ spontaneously decays to $\pi$ (sink), from any initial value $\mu_0$. 
While $\sigma$ appears to control the decay rate, it does not define a characteristic timescale that can be used to collapse the curves. From Eq.~(\ref{eq:main_eq_1dof}), 
it is clear that $\sigma$ cannot be absorbed in the time unit and that the dynamics are $\mu_+$-invariant around a $q=1$ defect if and only if $\sigma=0$.

The reciprocal part of the kernel is responsible for the elastic force $f_{\text{EL}} = \Delta \theta$ (the $2d$ XY model in the spin wave approximation \cite{chaikin1995principles}) 
while the NR part, by explicitly introducing an asymmetry (here front/back), allows spins to be sensitive to the vorticity of the surrounding spin field.
The resulting twisting torque $f_{\text{NR}}$ powers the twist from the inside of the defect's core, where the amplitude of the rotational is greater. The elastic term distributes the stress isotropically, explaining why the twist radially propagates outwards, {see Fig.~\ref{fig:2}(a-d) and Supplementary Movie 1. Such twisting is also found in continuum theories of constant-density flocks \cite{vafa2022defect}.}
{Last, while $f_{\text{NR}}$ only depends on the current configuration, $f_{\text{EL}}$ retains memory of the original defect configuration $\mu_0$, trying to restore it against $f_{\text{NR}}$.}
As $f_{\text{NR}}=0$ for $\mu_+=0, \pi$, sources and sinks are fixed points. 
{Sinks are attractors of the dynamics, while sources are unstable fixed points. 
For $T>0$, the thermal noise induces perturbations with a non-vanishing vorticity which in turn generates a finite $f_\text{NR}$, driving the defect away from the (unstable) source state and towards the sink state.} We summarize the dynamics of $\mu_+$ in the phase space plot of Fig.~\ref{fig:2}(n), for $\sigma=0.5$ and different initial conditions $\mu_+(t=0)$. All trajectories flow towards the stable sink state. The inset shows the flow for different values of $\sigma = 0.05, 0.1, ..., 0.5$. For small enough $\sigma$, isolated $+1$ defects cannot reach the sink state in a finite time, as $f_{\text{NR}}$ is too weak to overcome the elastic restoring force $f_{\text{EL}}$.
\newline

We now turn our attention on $q=-1$ defects. Figure~\ref{fig:2}(h) shows a stable $q=-1$ defect in the absence of NR interactions { and Fig.~\ref{fig:2}(h-k) shows its time evolution at $T=0, \sigma>0$. }
Under the $f_{\text{NR}}$ active force, an axis of polarisation spontaneously emerges, see the horizontal structure in Fig.~\ref{fig:2}(h,i,j). A simple sketch of the mechanism is given in Fig.~\ref{fig:2}(k), where the 4 red arrows highlight the clear quadrupole like symmetry of the curl of the orientation field. This eventually results in a stable defect shape with a symmetry axis from which the spin vector points predominantly outwards; see Fig.~\ref{fig:2}(j) for a numerical configuration and Fig.~\ref{fig:2}(l) for the final stable shape obtained analytically (see Appendix G).

To quantify the growing dynamics of the polarisation axis, the nematic order parameter 
$Q = \frac{1}{N}\sum_j e^{2i\theta_j}$ 
is an indirect but simple and informative figure of merit. 
We report in Fig.~\ref{fig:2}(o) the time evolution of $Q$ for a system composed of a single negative defect at the origin. The initial condition is a perfectly isotropic defect $\theta(x,y) = -\text{atan}(y/x) + \mu_-$, $Q(t=0) =0$, $P(t=0) =0$. With time, the polarisation axis extends, separating two regions in which the polar order $P$ increases. Yet, those two regions have opposite orientation, so the global $P$ remains zero throughout the process. However, the nematic order parameter $Q$ increases as the contributions from the two regions add up, reflecting the growth of the polarisation axis.
We observe that the growth rate of $Q$ increases with $\sigma$ (from bottom to top), until it reaches a sigma dependent steady length.

The existence and relevance at large scales of such structures has been studied by means of continuum models of constant-density flocks in \cite{besse2022metastability}.  
Importantly, the polarised field around a $q=-1$ defect provides a preferred path for defect motion, {indicated in Fig.~\ref{fig:2}(l) by green and red arrows, respectively for enhanced and hindered motion}. The location of the defect core can slide along the symmetry axis with a small, continuous variation of the spin vector field. On the contrary, it is very unfavourable for the defect to move perpendicular to the axis as it would require a large number ($\sim L$) of spin flips. 

\subsection{Impact of non-reciprocity on the defect annihilation}

\begin{figure}
    \centering
    \includegraphics[width=\linewidth]{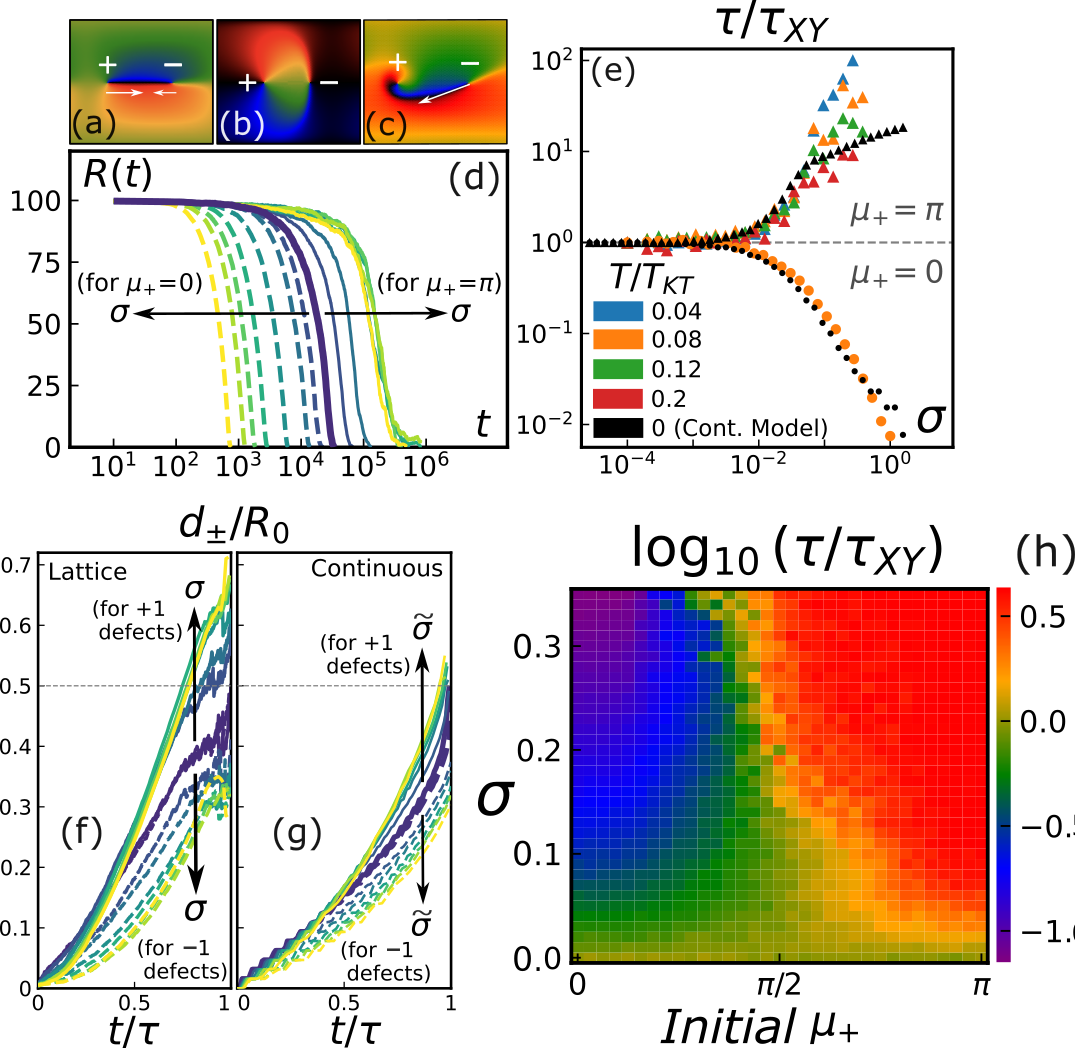}   
    \caption{
     \textbf{(a,b,c)} Configurations of two defects of opposite charge at $T=0, \sigma = 0.35$ with $\mu_+ = 0\,, \pi\,, \pi/2$, respectively. Arrows indicate their velocity. 
     \textbf{(d)} Inter-defect distance $R(t)$ between a  $\mu_+ = 0$ and a $\mu_-=\pi$ (dashed lines) and a $\mu_+ = \pi$ and a $\mu_-=0$ (solid lines) pair, {for $\sigma=0,0.01,0.02, 0.05, 0.1,0.15, 0.2, 0.3$, increasing along the arrows from blue to yellow}. In both cases, $T = 0.08\ T_{KT}, L =2R_0= 200$. 
     \textbf{(e)} Annihilation time $\tau$ rescaled by the one in the $2d$ XY case, for {$\mu_+ = 0$ (lower half) and $\mu_+ = \pi$ (upper half)} at different temperatures ($L=2R_0 = 100$). For the continuous model ($L=256$), we plot the data against $2\tilde\sigma/L$ to compare to $\sigma$. 
     For $\mu_+=0$, we only display the case $T = 0.08\ T_{KT}$ as curves for different temperatures superimpose for all $\sigma$.
     \textbf{(f)} Distance travelled $d_\pm/R_0$ by  $+1$ (solid lines) and $-1$ (dotted lines) defects in the configuration shown in panel (a) ($\mu_+ = 0$, $\mu_-=\pi$) the same values of $\sigma$ as in panel (d). {In the XY model, both defects travel at the same speed : $d_{\pm}(t=\tau) = R_0/2$ (dashed line).}  \textbf{(g)} {Same quantity for the continuous model, for $\tilde\sigma=0, 0.5, 1, 1.5, 2,2.5,3$.  }
     \textbf{(h)} Colour map of the annihilation time in log-scale 
     as a function of initial shape $0\leq\mu_+\leq\pi$ and non-reciprocity $\sigma$ (using the lattice model, $L=64, R_0 = 30$ and $T = 0.08\ T_{KT}$).
     }
    \label{fig:3}
\end{figure}

In the reciprocal $2d$ XY model, defects of opposite charge $\pm q$ at a distance $R$  attract each other with a Coulomb force $F \sim q^2/R$ that drives their annihilation at low  $T<T_{KT}$ \cite{YurkeHuse1993,chaikin1995principles, Rouzaire_Kuramoto_PRL,rouzaire_Frontiers}, see chapter 2 for details.
The XY model is symmetric under parity and thus $\pm q$ defects are equivalent.
The XY model is also invariant under global rotation, meaning that the actual shape of the defects is irrelevant.

As NR interactions break those symmetries and reshape defects, they strongly impact their annihilation dynamics. They thus can \emph{enhance} or \emph{hinder} the annihilation process and induce effective \emph{transverse forces}, depending on the specific shape of the defects involved. 

To explore the annihilation process, we study a pair of defects of charge $q = \pm 1$ and initial shapes $\mu_\pm$  at a distance $R_0$.
{ We create non-twisted configurations by initially imposing $\mu_+ - \mu_- =  \pi$. As explained in Chapter \ref{ch:methods}, this constraint imposes a constant spin orientation along the segment joining the two defect cores.}
We then let the system evolve at $T=0.08\,T_{KT}$, imposing periodic boundary conditions, and track defects over time to obtain their displacement $d_\pm(t)$ and their mutual distance $R(t)$.  
We focus first on two limit situations, involving sources and sinks, that will provide our reference scenarios, to then move to the description of more general annihilation process, as illustrated in Fig.~\ref{fig:3}(a-c). 

A source, $\mu_+ = 0$, can only be initially paired with a  $\mu_- = \pi$ defect that creates a symmetry axis along the core-to-core direction, see Fig.~\ref{fig:3}(a). Such symmetry provides a preferential annihilation pathway. As shown  in Fig.~\ref{fig:3}(d),  $R(t)$ decays faster for larger values of $\sigma$, which translates into a reduced annihilation time $\tau$ (that appears to be independent of $T$, see Fig.~\ref{fig:3}(e)). 
Moreover, $q=\pm 1$ defects are no longer equivalent: $+1$ defects move faster than $-1$ defects in the enhanced annihilation process (Fig.~\ref{fig:3}(f) {for the lattice model and Fig.~\ref{fig:3}(g) for the continuum model}). {This is in contrast with the equilibrium XY model, for which the invariance under the transformation $\theta \to -\theta$ prevents any differentiation, and with the continuum theory studied by Vafa \cite{vafa2022defect}, where the defect interactions appear to be symmetric. We have translated Vafa's results (complex field) in our notations (vector field) in Appendix F for convenience.}

For a sink, $\mu_+ = \pi$ ($\mu_-=0$), the symmetry axis of the $-1$ defect grows perpendicular to the core-to-core direction, as illustrated in Fig.~\ref{fig:3}(b). 
Such structure hinders the motion along the core-to-core direction, explaining why $R(t)$ decays slower (solid lines in Fig.~\ref{fig:3}(d)): the annihilation is dramatically slowed down, up to a factor $10^2$ with respect to the $2d$ XY case, see Fig.~\ref{fig:3}(e).

For any other initial pair, the $+1$ defect experiences a twist due to a non-vanishing $f_{\text{NR}}$ {and its mobility strongly decreases}, while the $-1$ defect grows an oblique symmetry axis that eventually curls to meet the $+1$ defect core, see Fig.~\ref{fig:3}(c). Once the preferential path between the two defects has been created, the $-1$ moves along it until either the defects are close enough to annihilate or the polarisation axis becomes perpendicular to the $+1$/$-1$ direction, {cf. Supplementary Movie 6a}. 
The annihilation time (normalised by $\tau_{XY}, $the corresponding time in the $2d$ XY model) for a pair of defects as a function of $\mu_+-$ and $\sigma$ is shown in Fig.~\ref{fig:3}(h). As $\sigma$ is increased, annihilation times are either increased for $\mu_+\approx \pi$ (red region) or decreased for $\mu_+$ and large $\sigma$ (blue region in Fig.~\ref{fig:3}(h)) with a smooth transition between the two extremes.
\newline

\begin{figure}[h!]
    \centering
    \includegraphics[width=0.9\linewidth]{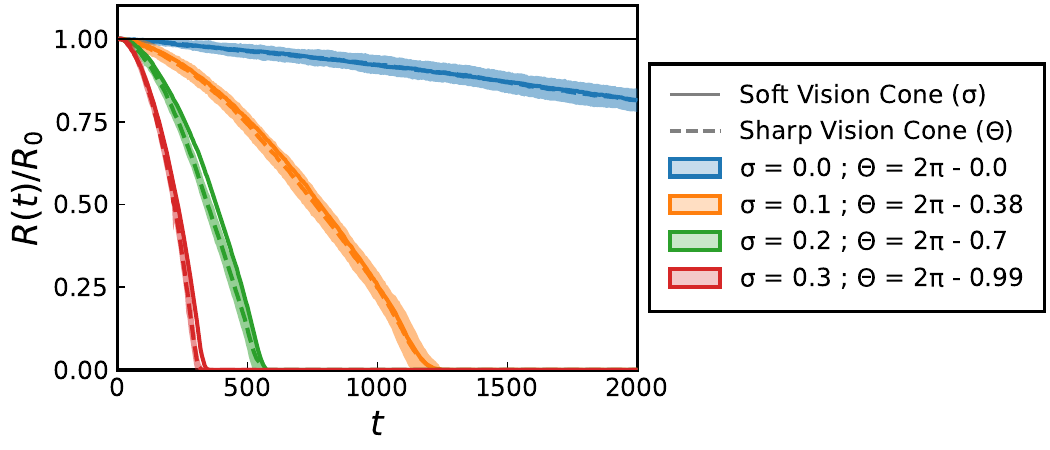}
    \caption{
    Distance separating two defects over time in the enhanced annihilation configuration. The $+1$ defect is a source: $\mu_+=0$ and $\mu_-=\pi$. The annihilation dynamics of both models quantitatively match if the sharp vision cone aperture is taken from Eq.~(\ref{eq:equivalence_short_sharp_vision_cones}). Results for the sharp vision cone are averaged over $R=50$ independent realisations of the thermal noise. The standard deviations are shown with the ribbons.}
    \label{fig:equiv_vision_cones}
\end{figure}

Sharp vision cones give qualitatively similar results, see Fig.~\ref{fig:equiv_vision_cones}. By imposing the same head-tail asymmetry to both sharp and soft kernels (cf. the discussion below the description of the lattice model), we obtain an equivalence relation between $\sigma$ and the sharp vision cone aperture $\Theta$, { which leads to a good quantitative agreement between both models, confirming the robustness and generality of our findings. } 

\subsection{Impact of non-reciprocity on the coarsening dynamics}

\begin{figure}[h!]
    \centering
    \includegraphics[width=1\linewidth]{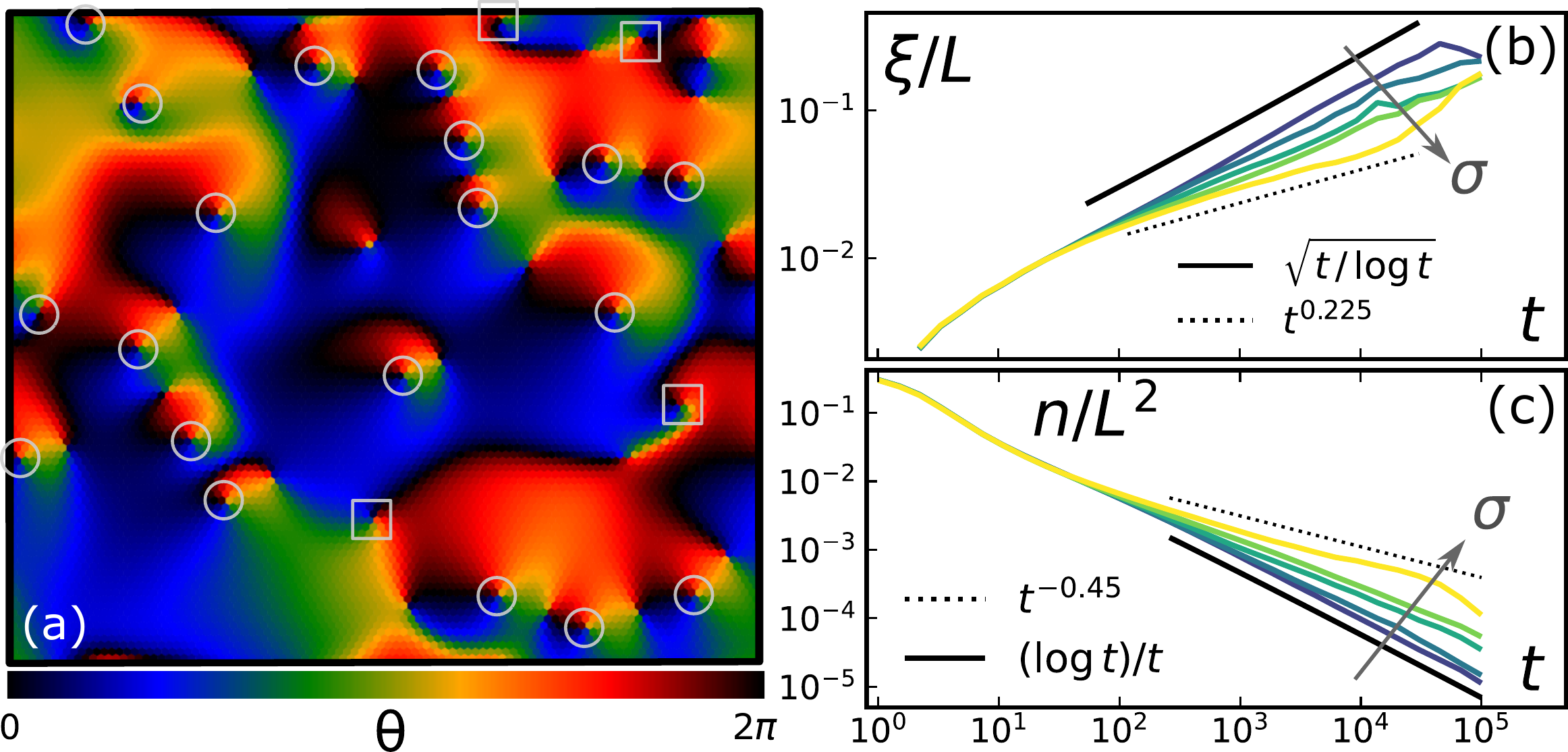}   
    \caption{ 
   \textbf{(a)} Snapshot of a $L=100$ system at $t=200, T=0, \sigma = 0.3$. 
   Sink defects ($q=+1, \mu_+=\pi$) are circled. The other $q=+1$ defects are squared. {Negative defects are not highlighted so the underlying snapshot remains visible, but the total topological charge is 0 due to PBC.  
   \textbf{(b)} Characteristic lengthscale $\xi/L$    
 and \textbf{(c)}  defect density $n/L^2$ for $L=200, T=0.08 \,T_{KT}$ and $\sigma = 0, 0.1, 0.2, 0.3, 0.4$. Black lines are the expected scalings for the equilibrium XY model. The dotted lines are powerlaw fits of the $\sigma =  0.4$ yellow curve.}  }
    \label{fig:4}
\end{figure}

As NR interactions impact significantly the defect annihilation, it is worth exploring its consequences in the coarsening process. We thus analyse the relaxation following a quench from a disordered state, with a large number of defects, to very low temperatures, where only a few bonded defects might persist \cite{rojas1999dynamical, jelic2011quench,rouzaire_Frontiers}. In Fig.~\ref{fig:4}(a), we show a snapshot  of the system during its relaxation at $T=0$.  
{
We define $\xi$ from the spatial correlation function $C(r)= \langle \boldsymbol{S}_i \cdot \boldsymbol{S}_j \rangle 
= \langle \cos (\theta_i - \theta_j) \rangle $ by $C(\xi)=1/e$ and plot its time evolution over the coarsening for different values of $\sigma$ at $T=0.08\,T_{KT}$ in Fig.~\ref{fig:4}(b). At equilibrium ($\sigma=0$, in blue), $\xi \sim (t/\log t)^{1/2}$, recovering the dynamical exponent $z = 2$ up to logarithmic corrections, as expected for the  $2d$ XY model \cite{YurkeHuse1993, jelic2011quench,rouzaire_Frontiers} (recall that $z$ is defined as $\xi\sim t^{1/z}$) . Increasing $\sigma$, the coarsening remarkably slows down and the dynamic exponent also increases up to values above $z=4$, cf. grey dotted lines. This is directly related to the slower decay of the defect density $n/L^2$, cf. Fig.~\ref{fig:4}(c). For $\sigma>0$, all the defects eventually decay to the stable sink shape, giving rise to long lived \emph{chains of} \emph{sinks} (circles in Fig.~\ref{fig:4}(a)) separated by $-1$ defects with symmetry axes that hinder the annihilation with the nearest $+1$ defects. In the $2d$ XY model, those structures are highly unstable and thus never observed. Non-reciprocity stabilises those large-scale structures (cf.  Supplemental Movies in Appendix A), explaining the slowing down of the decay of the defect density. 
} Occasionally, defect pairs spontaneously generate to release local excessive {stress}, but the newly created defects {always} rapidly annihilate (cf. Movie 6b). As such, the steady state is typically defectless at such low temperatures, in agreement with the long-ranged ordered phase reported in a similar lattice model \cite{loos2023long}. We did not find evidence for the metastability of the ordered state reported in \cite{besse2022metastability}, that might be due to either differences between those two approaches or finite-size effects.  
\newline

\section{Patterns dynamics}
We now leave the study of topological defects behind us and address the impact of non-reciprocity on the patterns dynamics of the $O(2)$ model.  We numerically analyse the continuum Eq.~(\ref{eq:main_eq}) using a square grid of linear size $L=2\pi$ discretized with $N_\text{grid}$ nodes,  with a timestep $dt = 10^{-5}$. We always work with periodic boundary conditions. We set $\gamma=1$ and $K=1$ and express times in units of $\gamma$. 
This section is organised as follows. 
First, we describe how a perturbation travels in our non-reciprocal $2d$  medium. To do so, we monitor the time evolution of an initial Gaussian $1d$ perturbation, in particular its unidirectional motion, its flattening, and the development of a front/back asymmetry. We extend those results to the evolution of a $2d$ perturbation. 
Then we describe the evolution of a topologically protected pattern: the lowest (defectless) energy excitation of the equilibrium $O(2)$ model. We show that for a small non-reciprocity, this configuration gets compressed in a narrow region of space, while the rest of the system orders. We explain how a sufficiently high activity / non-reciprocity destroys this pattern, changes the topology of the system and allows the system to relax to its equilibrium ground state. 

\subsection{Advection of $1d$ perturbations in a $2d$ non-reciprocal medium}

Unidirectional pattern propagation is a generic feature in non-reciprocal systems. A true, steady propagative state in a dissipative medium is only possible if the system obeys  conservation laws. This is the case of multi-species mixtures, for which the non-reciprocal extension of the  Cahn-Hilliard model  is a possible hydrodynamic description and can exhibit traveling bands in the steady-state, akin to the chase-and-run long-term dynamics in prey-predator inspired systems  \cite{you2020nonreciprocity, saha_scalar_2020, chiu2023phase, mandal2024robustness, weis2025generalizedNRmultipopulation, pisegna2024emergent}.
In absence of a conserved field (in the latter case, the densities of each species), propagative states can only be  transient. 

This ubiquitous feature is a direct consequence of the (self) advection term that naturally arises from the microscopic details of the dynamics. 
This active term appears in many different models and does not depend on the details of the source of non-reciprocity.
It arises in non-centered spins models where the kernel of interaction, independent of the state of the spin, is spatially translated \cite{seara2023NRIsing}. 
It also follows from vision-cone like interactions, for Ising spins \cite{garces2025phase, rajeev2024ising}, XY spins \cite{rouzaire2025nonreciprocal, bandini2025xy, popli2025ordering, dadhichi_nonmutual_2020}, or even from non-reciprocal steric interactions alone \cite{huang2024active}. 

\begin{figure}[b!]
    \centering
    \includegraphics[width=\linewidth]{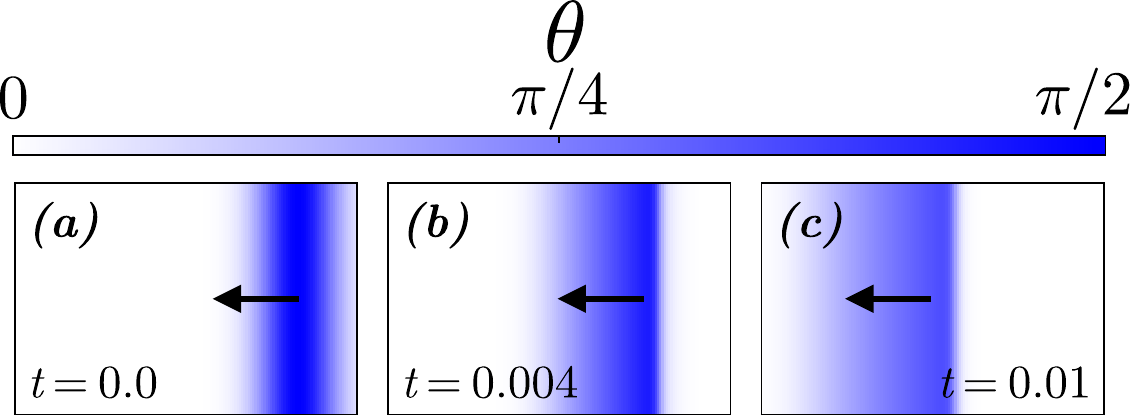}
    \caption{ Evolution of an initial $1d$ gaussian excitation over time. We show a partial rectangular window of a $ N_{\text{grid}}=256$ system. 
    The orientation field $\theta$ is plotted in panels \textbf{(a-c)} for $t = 0, 0.004, 0.01$ respectively. The other parameters are $\alpha =100,\, \bar\sigma = 100,\, \theta_0=0$. The initial profile travels to the left (black arrow) and develops a front/back asymmetry (smoother at the front, sharper at the back). 
}
    \label{fig:snapshot_propagation_1d}
\end{figure}

Here we study an idealized scenario and investigate the propagation of a perturbation $\delta$ along one direction:
\begin{equation}
    \theta(x, y, t) =\theta(x, t) = \theta_0 + \delta(x, t)\, . \\
  \label{eq:perturbation1}
\end{equation}
We take this perturbation to be Gaussian at $t=0$
\begin{equation}
    \delta(x, t=0) = \delta_0 \exp\left(-\frac{(x - x_0)^2}{2w_0^2}\right) 
  \label{eq:perturbation2}
\end{equation}
where  $\delta_0$ is the height of the initial profile, $x_0$ its initial position, $w_0\ll L$ its initial width and $\theta_0$ a global offset. We use $\delta_0 = \pi/2$ and $w_0/L = 0.025$. Since we consider a smooth perturbation, we can work under the approximation $S \approx 1$ and use Eq.~(\ref{eq:main_eq_1dof_1d}). 
Writing the dynamical equation in this form highlights the source of the propagation, as the active term takes the form of an advection term: we thus expect the perturbation $\delta$ to diffuse under the action of the Laplacian and to be advected by the orientation field $\bar\sigma \,\hat{\textbf{S}}$ in the negative $\hat{\textbf{S}}$ direction.  

\begin{figure}[b!]
    \centering
    \includegraphics[width=\linewidth]{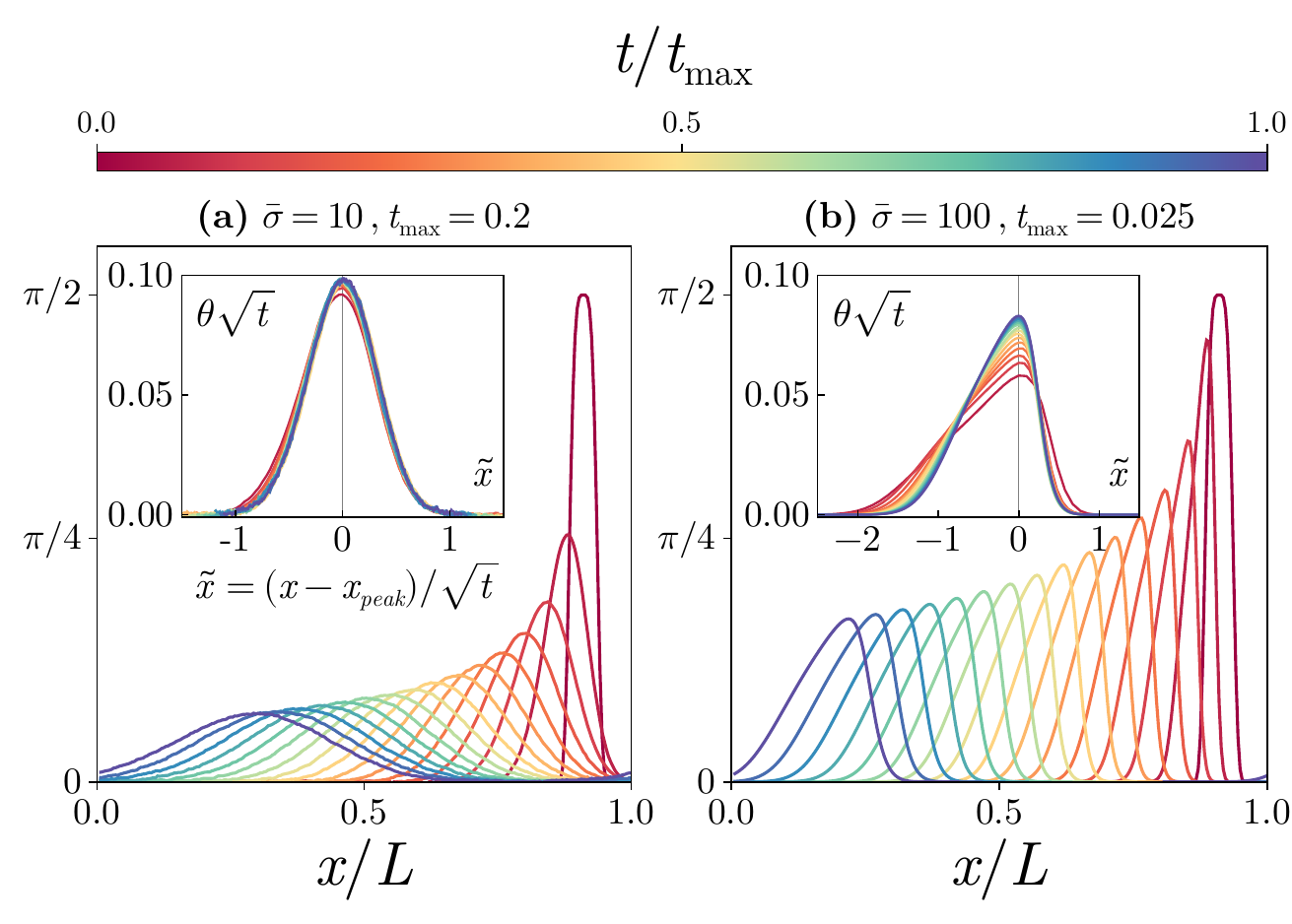}
    \caption{
    \textbf{(a)} Propagation of a perturbation $\theta=\delta(x)$ across the system over time. Parameters: small non-reciprocity $\bar\sigma = 10, t_{\text{max}} = 0.2, N_{\text{grid}} = 256, \alpha = 100, \theta_0=0$. \underline{Inset}: rescaled data $\theta\sqrt{Dt}$ as a function of $\tilde{x} = (x-x_{peak})/L\sqrt{\tilde Dt} $, with the diffusion coefficient $\tilde D = 2.53\,D = 1\cdot K/\gamma$, see Appendix A for details. The curves collapse, indicating that the flattening of the profile follows the diffusion equation behavior. 
    \textbf{(b)} Same but for $\bar\sigma = 100, t_{\text{max}} = 0.025$. \underline{Inset}: same rescaling. For higher non-reciprocity, the profiles are not symmetric anymore: they are stretched in front, and compressed at the back. 
  }
    \label{fig:propagation1}
\end{figure}

We report the time evolution of the $\theta$ profile in Fig.~\ref{fig:propagation1}, from red ($t=0$) to blue ($t=t_{\text{max}}$), for small non-reciprocity ($\bar\sigma = 10$) in panel (a) and high non-reciprocity ($\bar\sigma = 100$) in panel (b). The perturbation spreads, propagates and develops a front/back asymmetry; in what follows, we address those three phenomena. 


\subsubsection{Spreading\ } 
The perturbation decays towards its uniform ground state by spreading over time while moving. For small $\bar\sigma$, the spreading process follows a scale invariant diffusive behavior: if one rescales space and the height of the excitation by the diffusion length $\propto\sqrt{Dt}$, the profiles (centered on their peak) at different times convincingly collapse onto each other, see inset of Fig.~\ref{fig:propagation1}(a).  For larger $\bar\sigma$, the perturbation still spreads over time under the action of the Laplacian, but the $\sqrt{Dt}$ rescaling does not hold anymore because the perturbation develops a strong front/back asymmetry; we shall come back to this in a subsequent paragraph. 

\begin{figure}
    \centering
    \includegraphics[width=0.99\linewidth]{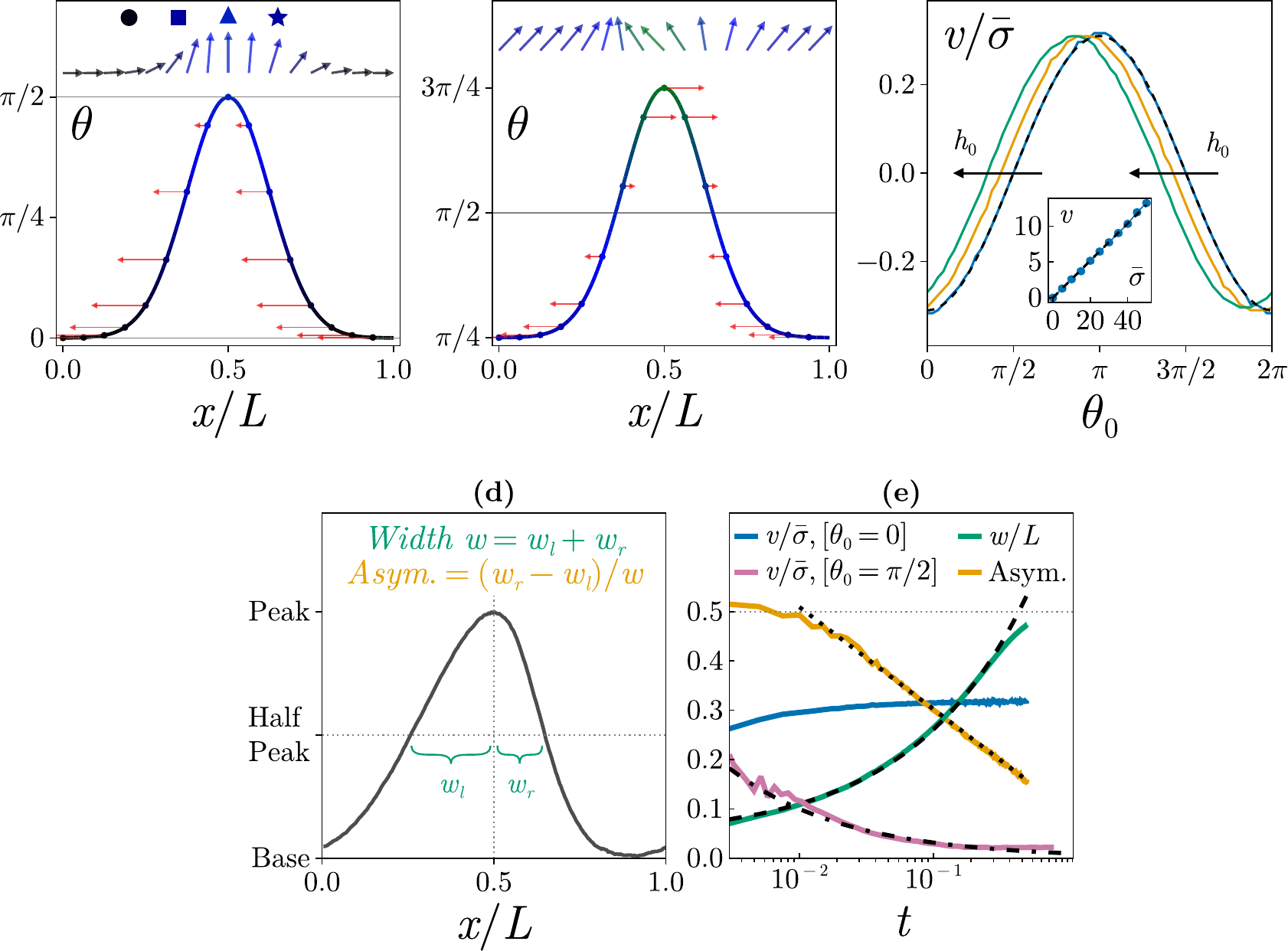}
    \caption{
     \textbf{(a)} Definition of the width $w$ and asymmetry of the profile. 
     \textbf{(b)} Time evolution of $w$ and asymmetry, up to $t_{\text{max}} = 0.5$, for $\bar\sigma = 100$. While the other results are obtained at $T=0$, this simulation is performed with a small temperature $T=0.02$,  
     and averaged over $32$ independent realizations. 
  The dash line fitting the width is $ w_0/L+ 0.8\sqrt{t}$, with $w_0/L = 0.025$. The dotted line fitting the asymmetry is $ 0.095- 0.09\log(t)$. 
  \textbf{(c)} A gaussian profile of height $\delta_0 = \pi/2$ and background orientation $\theta_0 = 0$, used as a support of the explanation in the main text. 
  Colors indicate the orientation $\theta$, see colormap. The red arrows represent $-\cos\theta$, which physically corresponds to the local convective velocity, see main text for details. The symbols at the top of the boxes are used in the main text to guide the reader. 
    \textbf{(d)} Similar to panel (c), but with a background orientation $\theta_0 = \pi/4$. 
     \textbf{(e)} Velocity of the peak for $\theta_0=0$ (blue)  and $\theta_0=\pi/2$ (pink)  at $T = 0.02$.
          The dash line corresponds to $1.25(1 - c^2/2t)$ and the dotted one to  $c/\sqrt{t}$ , with $c = 0.0314$ (see Appendix A). 
     \underline{Inset}: steady-state peak velocity showing  a linear growth $ v_{\text{max}} = 1.25\,\bar\sigma$.     
     \textbf{(f)} The peak velocity as a function of $\theta_0$.  The dashed line is $-1.25 \cos \theta_0$, compatible with the expected modulation. The curves shift to the left for increasing initial amplitude of the perturbation $\delta_0 = \pi/20,5\pi/20, 9\pi/20. $ }
    \label{fig:propagation2}
\end{figure}

Given a generically asymmetric profile, like the one sketched in Fig.~\ref{fig:propagation2}(a), we define the width $w$ as the sum of the left and right widths at mid-height: $ w = w_l + w_r$. In the same spirit, we define the asymmetry as the difference between the two: $(w_l - w_r) / w$ . We then report in Fig.~\ref{fig:propagation2}(b) the rescaled width $w(t)/L$  for $\bar \sigma = 100$. At short times, its  evolution is perfectly fitted by $w_0/L + \text{cst}\cdot\sqrt{t}$~. When $w\approx L/2$, the perturbation starts to self-interact through periodic boundary conditions and the spreading departs from the diffusive scaling. 
{By that time, the perturbation has traveled over a distance of the order of $10\,L$, impacting multiple time each point in the system. 
Interestingly, the same result is obtained for an agent-based model with a sharp vision cone with Glauber dynamics \cite{loos2023long}. This suggests that the excitation dynamics just described is general and robust and calls for a more thorough investigation of the propagation phenomena. 

}

\subsubsection{Propagation \ }
The propagation of the Gaussian excitation can be rationalized in different ways. The vision cone perspective sheds light on this problem and provides a first intuitive explanation of the observed dynamics. To guide the reader through the following paragraph, we have drawn in Fig.~\ref{fig:propagation2}(c), in addition to the profile $\theta(x)$, the corresponding arrows and four symbols (circle $\bullet$, square $\blacksquare$, triangle $\blacktriangle$ and star $\star$). 
Let us first consider the spin $\blacksquare$. It points slightly to the right, so its vision cone makes it more sensitive to the spin $\blacktriangle$ than to the spin $\bullet$. It will preferentially align with $\blacktriangle$ and thus rotate counterclockwise. On the other hand, the spin $\star$ looks away from the perturbation and will therefore rotate clockwise to align with the horizontal background. One can also rationalize this phenomenon thanks to the field vorticity: the torque $(\nabla \times \boldsymbol{S})\times \boldsymbol{S}$ is zero at the peak, clockwise on its left and counterclockwise on its right.
Altogether, the peak of the profile displaces to the left over time, see Fig.~\ref{fig:propagation1}, as first reported in \cite{loos2023long}.  

The study of the advective term in $1d$, see Eq.~(\ref{eq:main_eq_1dof_1d}), provides a better understanding of the speed of the propagation. 
The convective velocity, $\cos \theta\ \bar\sigma/\gamma$, is local and time dependent since $\theta = \theta_0 + \delta(x,t)$. 
We track the peak of the perturbation over time and extract its speed $v_{\text{peak}}=\dot x_{peak}$. Since the perturbation is advected following Eq.~(\ref{eq:main_eq_1dof_1d}), we expect it to move at a velocity related to $\cos \theta\ \bar\sigma/\gamma$. However, the displacement of the peak of the perturbation does not depend on the advection only. It results from the competition between the advection, the diffusion and, as we will see in the next paragraph, the tendency of the perturbation to develop an asymmetry. Therefore, we do not expect $v_{\text{peak}}$ to be strictly equal to $\cos \theta_{\text{peak}}\ \bar\sigma/\gamma$ but rather to be proportional to it: we write $v_{\text{peak}}=v_{\text{max}} \, \cos \theta_{\text{peak}}$. In fact, $v_{\text{max}}$ is the maximum velocity of the perturbation, realized when the background medium is parallel to the propagation ($\theta_0 = 0, \pi$) and at long times
(keeping in mind that, strictly speaking, at $t\to \infty $ there is no perturbation anymore). The inset of Fig.~\ref{fig:propagation2}(e) confirms that $v_{\text{max}}  = 1.25\, \bar \sigma$, cf. Appendix A for details. 

Since the rotational term in Eq.~(\ref{eq:main_eq}) is \textit{not} invariant under a global rotation of the spins, changing the orientation of the background medium $\theta_0$ will affect the excitation dynamics. 
We have seen that the amplitude of the perturbation diffuses out, such that the peak amplitude $\delta_{\text{peak}} \sim c/\sqrt{t}$, with $c$ depending only weakly on $t$, see insets of Fig.~\ref{fig:propagation1} and Appendix A. At some point, $\delta_{\text{peak}}$ becomes small and one can Taylor expand the cosine as follows:
\begin{equation}
\begin{aligned}
    v_{\text{peak}} = & \ v_{\text{max}}\,\cos \theta_{\text{peak}} \, \\
    = & \ v_{\text{max}}\, \cos (\theta_0 + \delta_{\text{peak}}) \\
    = & \ v_{\text{max}}\,\Big[ \cos \theta_0 \cos \delta_{\text{peak}} -\sin \delta_{\text{peak}}  \sin \theta_0  \Big] \\
     \approx & \ v_{\text{max}}\,\Big[ \cos \theta_0 (1 - \delta_{\text{peak}}^2/2) -\delta_{\text{peak}} \sin \theta_0  \Big] \\
      \sim & \ v_{\text{max}}\,\left[ \cos \theta_0 \left(1 - \frac{c^2}{2t}\right) - \frac{c}{\sqrt{t}}\,\sin \theta_0   \right] \\
    \end{aligned}
    \label{sigmacos}
\end{equation}
If $\cos \theta_0 \neq 0$ , the first term dominates. The short time behavior is well fitted by the last equation in Eqs.~(\ref{sigmacos}), see the blue saturating curve in Fig.~\ref{fig:propagation2}(e). 
The perturbation propagates at a steady velocity ($v_{\text{max}}$ in this case because $\theta_0=0$) to the left because the background medium transmits information in the $-\hat{\mathbf{S}}$ direction. 
For all $\theta_0\neq\pm\,\pi/2$, the orientation of the background medium has a non-zero component along $x$ and can therefore sustain a transient but long-lived propagation of the excitation. We report in Fig.~\ref{fig:propagation2}(f) the $\cos \theta_0$ modulation of the steady velocity of its peak. We report in Appendix H various plots of the motion of the perturbation for many different $0\le \theta_0 < 2\pi$ , see Fig.~\ref{fig:profiles_different_theta0}.

If instead $\cos \theta_0 = 0$, the background medium points perpendicular to the perturbation motion: there is no net information flux anymore, the first term in Eq. (\ref{sigmacos})  vanishes  and no long-lived  propagation can be sustained. This limit case is singular. It will nevertheless prove useful in the $2d$ perturbation analysis to explain transverse motion. 
Since the first term $\propto \cos\theta_0$ in Eq.~(\ref{sigmacos}) is identically zero, the second term, now negative, becomes relevant, and the perturbation propagates to the right before quickly vanishing: see the pink decaying curve in Fig.~\ref{fig:propagation2}(e). 
One could thus wonder why there is any propagation at all. To explain this, we need to turn to the last generic consequence of non-reciprocity on perturbations: the tendency to acquire a front/back asymmetry.


\subsubsection{Asymmetry}  
In this paragraph, we explain why the front end of a typical perturbation gets stretched while its trailing end gets compressed, see the $\theta$ profile in Fig.~\ref{fig:propagation1}(b) or the sketch in Fig.~\ref{fig:propagation2}(a).  This asymmetry is a direct consequence of the $\cos \theta(x)$ modulation of the \textit{local} speed of the perturbation, see Eq.~(\ref{eq:main_eq_1dof_1d}). The $x$-dependence is crucial here to account for this phenomenon, while one can forget about the $t$-dependence of the perturbation.
In the sketch of Fig.~\ref{fig:propagation2}(c), the modulus of the red arrows corresponds to the magnitude of $\cos \theta(x)$. Here $\theta_0=0$ and the height of the perturbation is $\delta_0=\pi/2$. The local advection speed is maximal at the base of the perturbation ($\cos \theta = \cos \theta_0 = 1$) and minimal at the peak ($\cos \theta = \cos \delta_0 = 0$). This creates velocity shear: the base wants to move faster that the peak, hence the stretch on the left and the compression on the right, skewing the initially symmetric profile.  The situation is even clearer if the perturbation crosses the  $\theta = \pi/2$ line, implying that the base and the tip of the profile point in opposite directions. We illustrate such a profile in Fig.~\ref{fig:propagation2}(d), with $\theta_0=\pi/4$ and $\delta_0=\pi/2$. 
There, the background medium   points to the right ($\cos \theta > 0$) while the central part of the profile points to the left ($\cos \theta < 0$). This is a special and short-lived case where the tip of the perturbation travels in a direction opposite to that of its base. 

We can now better understand the peculiar propulsion mechanism in the $\cos \theta_0 = 0$ case. There, the Laplacian  plays a role in the unidirectional propagation, in contrast with the $\cos \theta_0 \neq 0$ case where one does not need to appeal to elastic forces to explain the motion of the profile. 
Under the action of the velocity shear, the perturbation gets skewed and the peak displaces to the right. This creates higher local gradients $\nabla \delta$, that the elastic Laplacian dissipates isotropically. As a result, the whole perturbation moves to the right: it is powered from the inner non-reciprocal active force, and kept in a coherent shape by the reciprocal elastic force.

In all cases, the asymmetry decays over time, because the diffusive spreading lowers the amplitude of the perturbation, hence decreasing the difference between $\cos \theta(x)$ at the base and at the top. We plot the asymmetry (defined in Fig.~\ref{fig:propagation2}(a)) in yellow in Fig.~\ref{fig:propagation2}(b). Its time dependence seems to follows a  $\sim\log(1/t)$ scaling for which we don't have an explanation. 

We have thus characterized and rationalized the dynamics of the first 3 moments of the profile: the propagation (displacement of the mean), the flattening (increase of the standard deviation) and the asymmetry (non-zero skewness). We mention here that higher moments do not appear to play any role. This is consistent with the fact that the dynamics is dominated by low-order derivatives. One can reproduce the exact same results with an initial quartic profile $\propto\exp(-x^4)$: after a very short transient time, the profile is indeed indistinguishable from the initial gaussian case. 

\subsection{Propagation of a $2d$ perturbation}

\begin{figure}[b!]
    \centering
    \includegraphics[width=\linewidth]{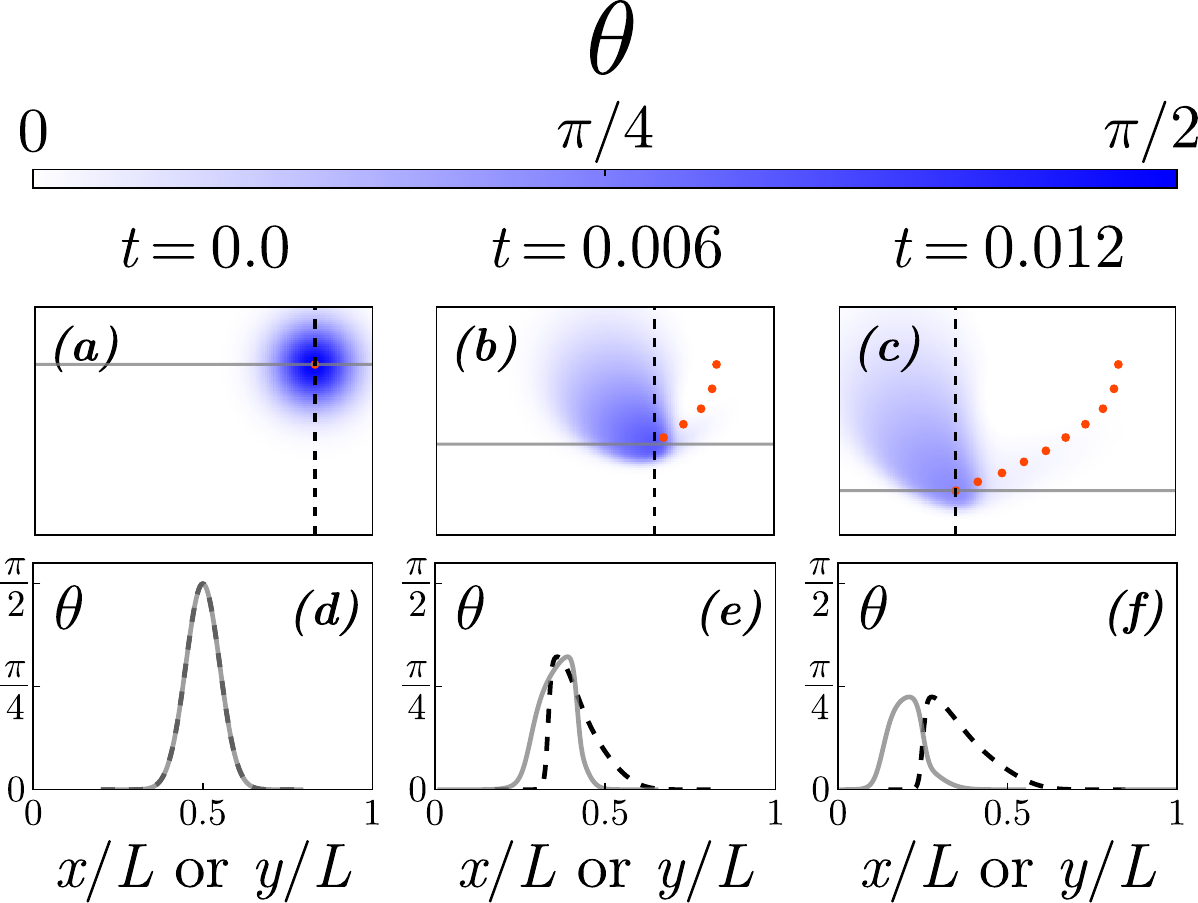}
    \caption{ Evolution of an initial isotropic $2d$ gaussian perturbation over time, with $\theta_0 = 0, \delta_0 =\pi/2$. We show a partial rectangular window of a $ N_{\text{grid}} =256$ system. 
    The orientation field $\theta$ is plotted in panels \textbf{(a-c)} for $t = 0, 0.006, 0.012$ respectively. The dots track the past positions of the peak. Other parameters: $\alpha =100, \bar\sigma = 100$. 
    \textbf{(d-f)}  cross-sections of the orientational field $\theta$ along a vertical (dash, against $y/L$) and horizontal (solid, against $x/L$) lines passing though the peak of the perturbation in panels (a-c). 
  }
    \label{fig:snapshot_propagation_2d}
\end{figure}

\begin{figure*}
    \centering
    \includegraphics[width=\linewidth]{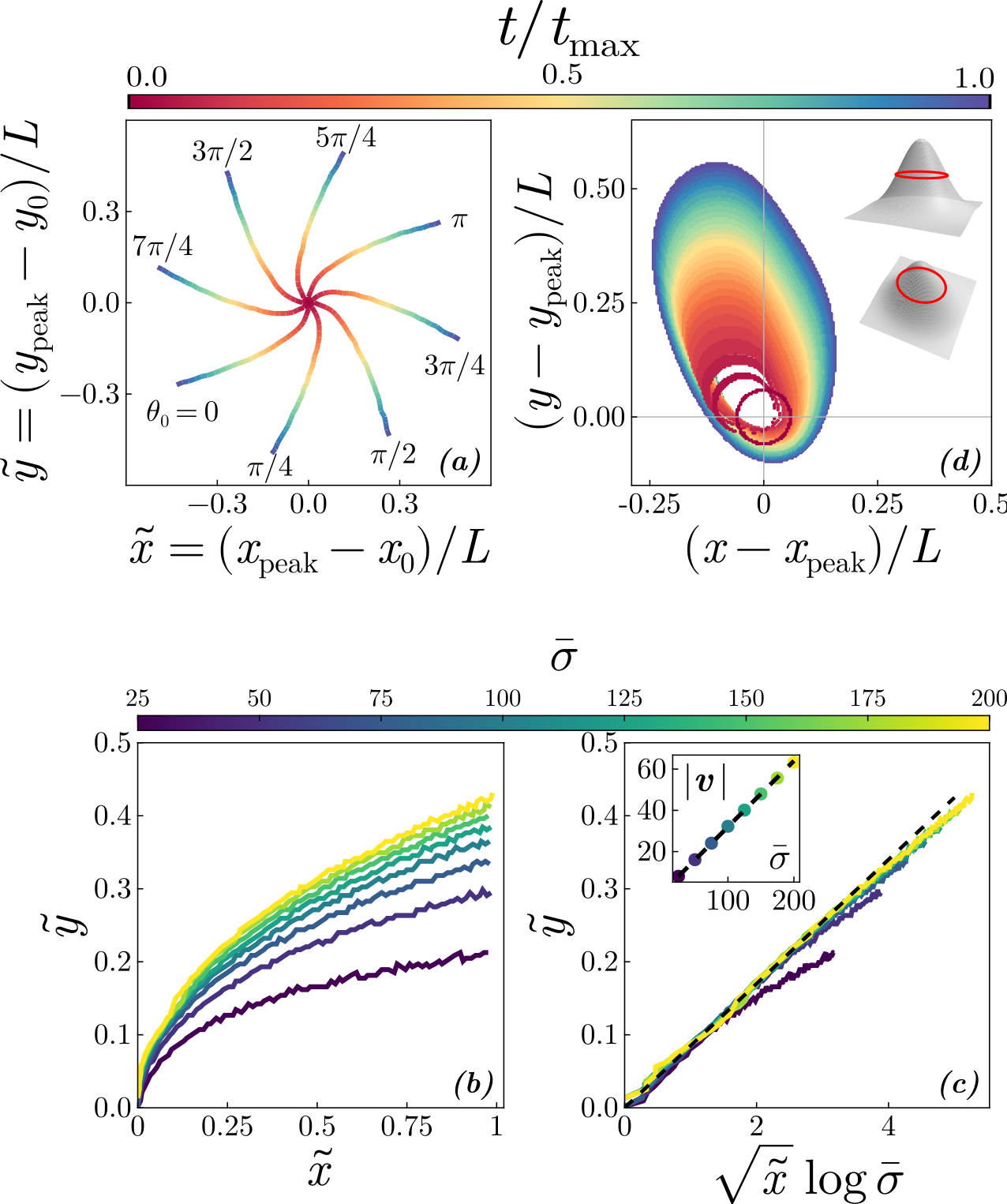}
    \caption{
    \textbf{(a)} Trajectory of the peak, from its initial location $(x_0, y_0)$. $\tilde{x}, \tilde{y}$ are defined as $(x_\text{peak} - x_0)/L$ and $(y_\text{peak} - y_0)/L$. The colorbar indicates time, from red ($t=0$) to blue ($t_\text{max} = 1.8$). The different branches of the star are the trajectories for different background orientations $\theta_0$, indicated at the end of each branch. For this panel, $\bar\sigma =100$.  For all 4 panels: $\delta_0 = \pi/2 \text{ (the initial height)}, \alpha = 100, N_{\text{grid}}= 256$. 
    \textbf{(b)} Variation of the trajectory $\tilde y(\tilde x)$ (here $\theta_0 = \pi$) with different $\bar\sigma$, corresponding to different colors. The simulation time $t_\text{max}$ is adapted for each $\bar\sigma$ such that the distance traveled along the $x$-axis is the same for all the curves: $t_\text{max}(\bar\sigma) =  t_\text{max, ref} \cdot \bar\sigma_\text{ref} / \bar\sigma = 1.8 \cdot 200 / \bar\sigma $ 
    \textbf{(c)} Same data, but on rescaled axes. The curves for different $\bar\sigma$ collapse, indicating that $\tilde y \sim \sqrt{\tilde x} \, \log \bar\sigma$. The dashed line is a linear function of slope 0.085. \underline{Inset}: the norm of the velocity of the peak $|\boldsymbol{v}_{\text{peak}}| = 1.25\, \bar\sigma/\gamma$ (dash lines) 
    \textbf{(d)} Flattening and asymmetry of the profile over time, for $\bar\sigma =100$. For different times [in different colors, from red ($t=0$) to blue ($t_\text{max} = 5$)], we plot the mid-height manifold centered on the peak, as depicted in red in the sketches. 
    }
    \label{fig:propagation2d}
\end{figure*}
We now expand the discussion to a $2d$ isotropic gaussian perturbation
\begin{equation}
    \delta( x,y,t=0) = \delta_0 \exp\left(-\frac{(x - x_0)^2 + (y - y_0)^2}{2\,w_0^2}\right) 
\end{equation}
where $\delta_0$ is the height of the initial profile, $(x_0, y_0)$ the initial position of its peak, $w_0\ll L$ its initial width (the same along $x$ and $y$). We use $\delta_0 = \pi/2$ and $w_0/L = 0.05$, as for the $1d$ perturbation. We show snapshots of the perturbation over time in Fig.~\ref{fig:snapshot_propagation_2d}(a-c), with the trajectory of the peak plotted with red dots. 
{The perturbation takes a curved path. 
This comes from the sign of $\delta_0$ which indicates whether the perturbation features a clockwise or anti-clockwise rotation of the background.}
In the bottom row, we plot the cross-sections of the orientational field $\theta$ along a vertical (dash) and horizontal (solid) lines passing through the peak of the perturbation. 
Even though the profiles are different from the $1d$ excitation case, the three main features remain: its motion,  flattening and shearing (giving rise to an asymmetric profile). 

We first investigate the motion of the perturbation. We monitor the displacement of the peak over time and present the results in Fig.~\ref{fig:propagation2d}(a). The colors represent the time, from red to blue. 
Each branch of the star corresponds to a different background orientation $\theta_0$ (multiples of $\pi/4$, indicated at the end of each trajectory), and all the branches are rotationally symmetric. 
{This is because a change in $\theta_0$ can be mapped to a global rotation of the system now that the perturbation at $t=0$ is a rotationally symmetric gaussian.}

We thus select one branch (the one for $\theta_0=\pi$), vary $\bar\sigma$ (we have checked that varying $\alpha$ does not change the results, consistent with the fact that $S\approx 1$ everywhere) and present the results in Fig.~\ref{fig:propagation2d}(b). Recall that our dynamics is noiseless: the roughness of the curves is only due to the discretisation of the position of the peak on a $N_{\text{grid}}=256$ grid. The trajectory seems to follow a square root, confirmed by Fig.~\ref{fig:propagation2d}(c), where the same data plotted against $\log \bar\sigma\,\sqrt{(x_\text{peak} - x_0)/L}$ becomes linear, while the curves for different $\bar\sigma$ collapse. 
{From our consideration of the $1d$ perturbation, we expect the perturbation to quickly reach its steady state velocity along the $(-\cos \theta_0, -\sin \theta_0) $ direction. Since we have chosen to focus on $\theta_0 = \pi$, this implies a linear relationship $\tilde{x}\sim-t$, where $\tilde x = (x_{\text{peak}}-x_0)/L$ is the normalized displacement of the peak from its initial position. As we discussed in the $1d$ perturbation, we expect the motion perpendicular to the background (the $\theta_0 = \pm \pi/2$ direction) to give a speed $d{\tilde{y}}/dt\sim 1/\sqrt{t}$, such that $\tilde{y}\sim\sqrt{t}$. Combining these leads to the $\tilde{y}\sim\sqrt{\tilde{x}}$ that we observe in Fig.~\ref{fig:propagation2d}(c).}

We also computed the vectorial velocity $\boldsymbol{v}_{\text{peak}}$ of the peak; its norm $|\boldsymbol{v}_{\text{peak}}|$ is constant over time and, in the long time limit, it obeys the exact same scaling as in the $1d$ perturbation case $|\boldsymbol{v}_{\text{max}}| = 1.25\, \bar\sigma/\gamma$ , see inset of Fig.~\ref{fig:propagation2d}(c). This indicates that the (scalar) speed of a perturbation does not depend on its actual shape, nor on its dimensionality. 

Finally, we look at the deformation of the perturbation over time. For a $1d$ profile, we defined the width at mid-height [see sketches in Fig.~\ref{fig:propagation2d}(d)]; in the same spirit, for a $2d$ perturbation we extract the mid-height manifold and present the results in Fig.~\ref{fig:propagation2d}(d) for different times (different colors). More precisely, we center the data such that the peak of the perturbation is at the origin $(0,0)$ at all times and we plot all the spins $(x,y)$ such that $|\theta(x,y) - \frac{1}{2}\,\underset{x,y}{\max} \ \theta(x,y,t)| \le  0.04$ (within a small, constant and arbitrary margin 0.04). 
The initially isotropic gaussian flattens and evolves into a complex, fully asymmetric shape that is hard to rationalize.    

In this section, we have thus shown that a simple perturbation propagates in the direction opposite to the medium it travels in, at a speed proportional to the non-reciprocity parameter and to the component of the background medium parallel to its motion. Over time, the perturbation flattens under the action of the Laplacian, and develops a front/back asymmetry.  

\section{Extended topologically protected excitations}

\begin{figure}
    \centering
    \includegraphics[width=\linewidth]{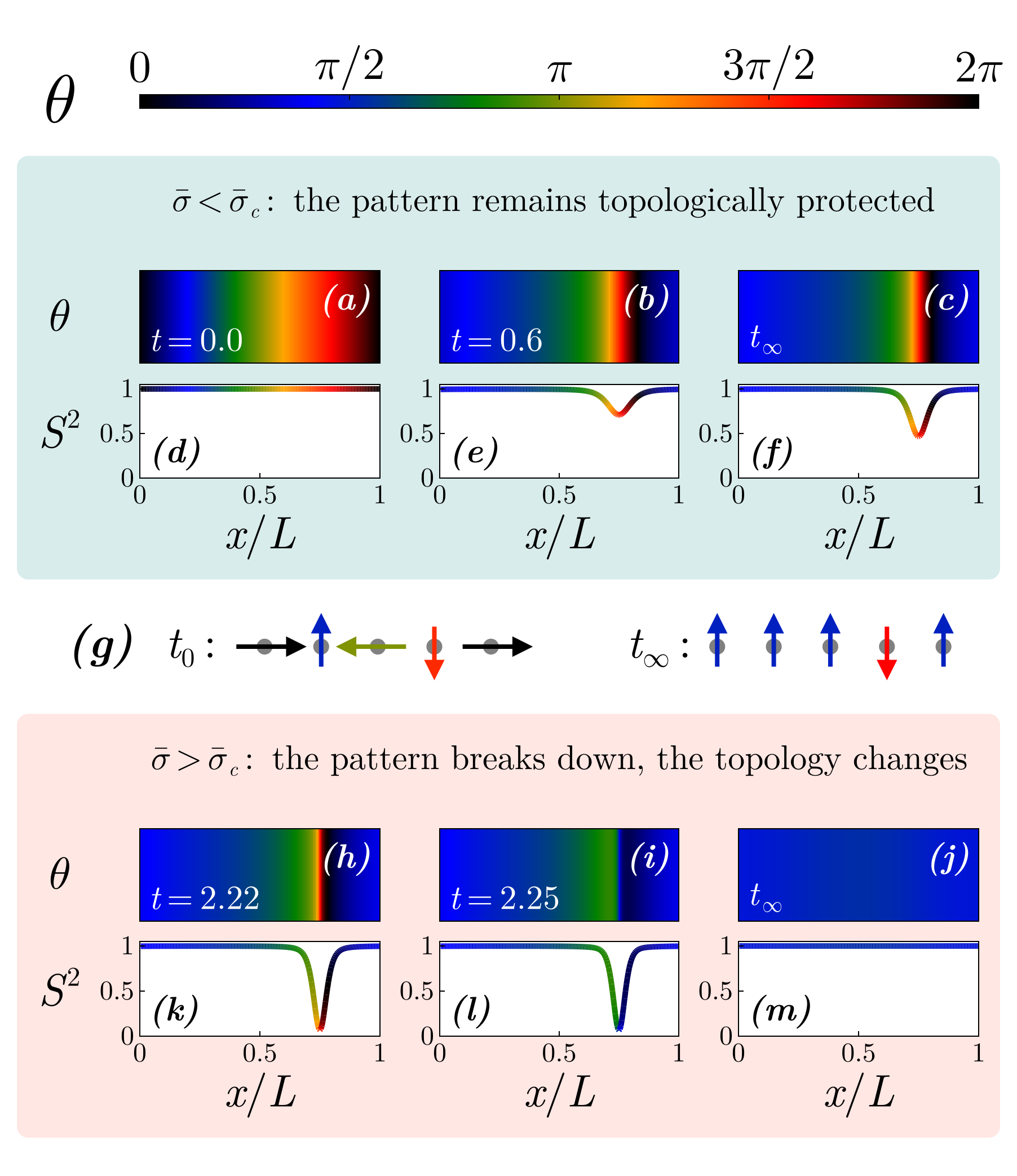}
    \caption{ Evolution of a topologically protected state over time, for 2 different values of $\bar\sigma$. In panels (a-f), $\bar\sigma = 1.95$ is below the critical value $ \bar\sigma_c$ (see explanation in the main text). 
    The orientation field $\theta$ is plotted in panels \textbf{(a-c)} for $t = 0, 0.6, 0.9$ (already in the steady state), respectively. The corresponding magnitude field $S$ is plotted in panels \textbf{(d-f)} for the same times. 
    In panels (h-m), $\bar\sigma = 2$ is above the critical value. $\theta$ is plotted in panels \textbf{(h-j)} for $t = 2.22, 2.25, 2.30$ (already in the steady state). $S$ is plotted in panels \textbf{(k-m)} for the same times. 
    We show a partial rectangular window of a $ N_{\text{grid}} =512$ system, for $\alpha =100$. 
    In panel \textbf{(g)}, we sketch the mechanism that reshapes the spinwave.
  }
    \label{fig:snapshots_concentration}
\end{figure}

We now turn to a different kind of excitation that is not localized but spans the whole system: a modulation of the orientation field with a winding number equal to one (ie. the orientation field $\theta$ wraps around $2\pi$); it is therefore topologically protected and reads 
\begin{equation}
    \theta(x,y) = 2\pi\,x/L \text{ and } S(x,y)=1 \ .
    \label{eq:spinwave}
\end{equation}
Such a configuration is apolar: $P = L^{-2} |\sum_k \exp(i\theta_k)|=0$. We represent it in Fig.~\ref{fig:snapshots_concentration}(a). In the limit $L \to \infty$, the elastic cost to sustain a small angular difference $\varepsilon = 2\pi/L$ between two neighbors is $\sim \varepsilon^2 \sim L^{-2}$ so the energy of this pattern is finite in the thermodynamic limit.

\begin{figure*}
    \centering
    \includegraphics[width=0.99\linewidth]{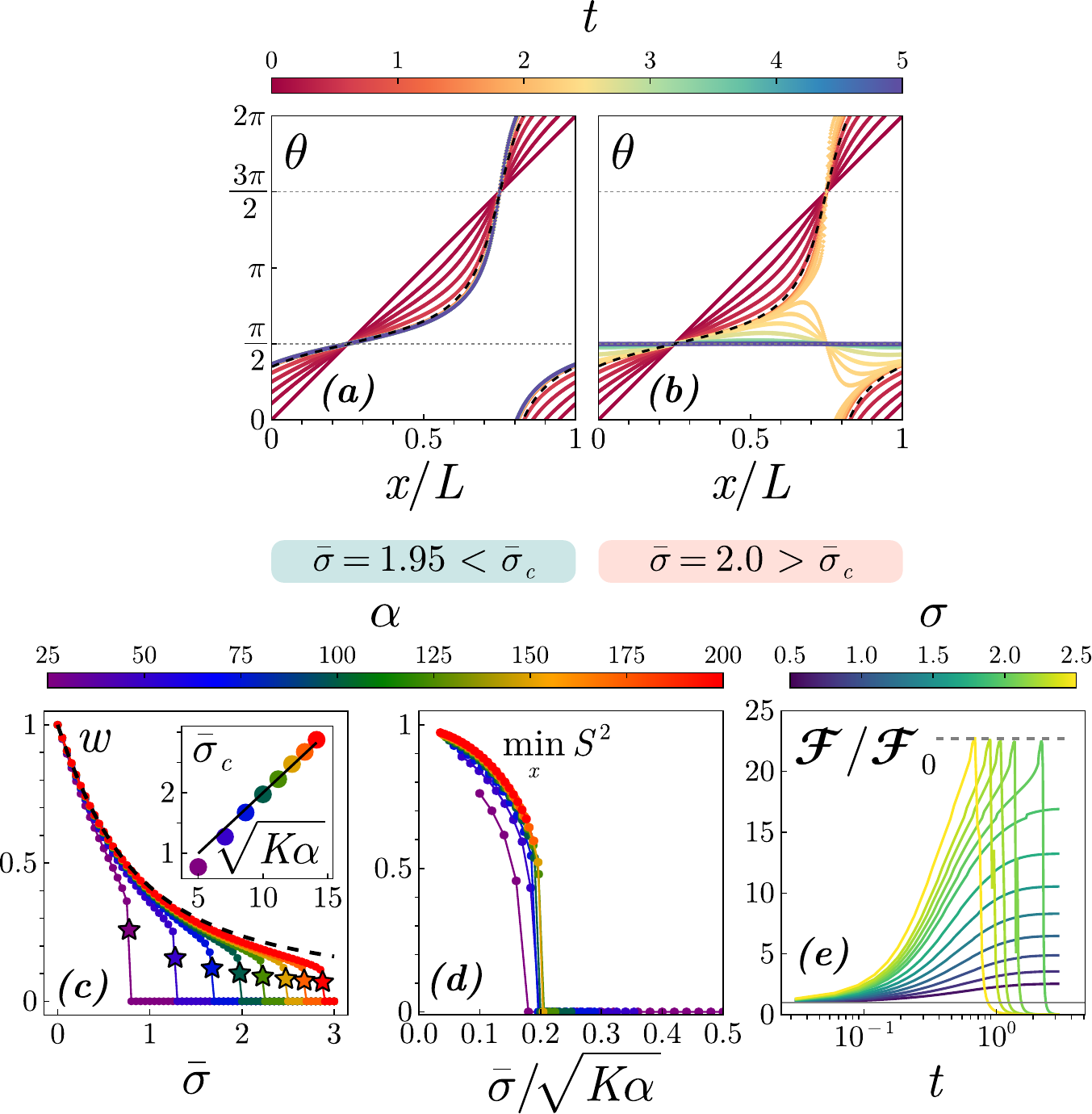}
    \caption{
    \textbf{(a)} Time evolution of the profile $\theta(x)$ for $\alpha=100, \bar\sigma =1.95, N_{\text{grid}} = 512$. The different lines and colors correspond to different times, from red ($t=0$, $\theta(x) = 2\pi\,x/L$) to blue (steady state). The dash line corresponds to Eq.~(\ref{eq:profile_derivation}).
    \textbf{(b)} Same as in panel (a) but for $\bar\sigma =2$. 
    The topologically protected state breaks down at $t=2.24$ . 
    \textbf{(c)} The  width $w$ as a function of $\bar\sigma$, for different $\alpha$, in different colors, and $ N_{\text{grid}} = 512$. The stars are located at $(\bar\sigma_{c}, 0.2\,K/(w_0\,\bar\sigma_{c})$, where $\bar\sigma_{c}=\bar\sigma_{c}(\alpha)$ is the highest value of $\bar\sigma$ for which the topologically protected state does not break down. \underline{Inset}: $\bar\sigma_{c}$ against $\sqrt{\alpha}$. The black line is  $\bar\sigma_{c}=0.2\sqrt{K\alpha}$ . 
    \textbf{(d)} Value of the minimum of $S^2$ over the profile, against $\bar\sigma / \sqrt{K\alpha}$ , for different $\alpha$ (see colorbar). A lower minimal value corresponds to a more compressed state. The breakdown occurs at $\bar\sigma / \sqrt{K\alpha} = 0.2$ for all $\alpha$.
    \textbf{(e)}  Time evolution of the free energy $\mathcal{F}$, normalized by its initial value $\mathcal{F}_0$, for $\alpha = 100$ and different $\bar\sigma$ (different colors). The thin grey line is $\mathcal{F}/\mathcal{F}_0 = 1$ and the dashed grey line at the top is to guide the eye to the common maximum of the curves.
  }
    \label{fig:concentration1}
\end{figure*}

In this section, we study the impact non-reciprocity has on such a pattern.  Since it is topologically protected, it cannot smoothly vanish as the localized $1d$/$2d$ gaussian profiles of the last section did. Equation~(\ref{eq:main_eq}) is not invariant under a global rotation of the spins, so the transformation $\theta\to \theta + \theta_0$ might affect the dynamics. However, for this specific profile combined with periodic boundary conditions, it is equivalent to a global translation along $x$,  so we can generically set $\theta_0=0$ and consider the profile of Eq.~(\ref{eq:spinwave}).
While Eq.~(\ref{eq:spinwave}) is a steady state solution of Eq.~(\ref{eq:continuousXY}), it is no longer the case when $\bar\sigma > 0$. Indeed, the active curl contains $\cos \theta$ , see Eq.~(\ref{eq:main_eq_1dof_1d}),
which is generally non-zero.
Figure~\ref{fig:snapshots_concentration} illustrates the two possible scenarios for the evolution of the initial configuration, given by Eq.~(\ref{eq:spinwave}) and illustrated in Fig.~\ref{fig:snapshots_concentration}(a). Both have in common that the linear profile is deformed: the phase field gets compressed around $\theta = 3\pi/2$ and stretched around $\theta = \pi/2$, the two stable points the $\cos \theta$ in Eq.~({\ref{eq:main_eq_1dof_1d}}), see Fig.~\ref{fig:concentration1}(a). In the compressed regions, the system reacts to the non-reciprocal drive as a passive system ruled by Eq.~(\ref{eq:continuousXY}) would do: the magnitude $S$ drops to compensate for the increased energetic cost of the gradients in $\theta$, see Fig.~\ref{fig:snapshots_concentration}(d-f). 
This evolution of the profile is easily understood from the agent-based perspective of the non-reciprocal term $\bar\sigma\,(\nabla \times \boldsymbol{S})\times \boldsymbol{S} $. 
Let us sketch the initial profile at $t_0 =0 $ as in Fig.~\ref{fig:concentration1}(g, left). The leftmost spin ($\rightarrow$)  tends to preferentially align  with the spin it looks at, and thus tends to align vertically. In such an effective $1d$ profile, a vertical spin has no preferential influence, and will thus remain unaltered. The profile therefore tends to the one in Fig.~\ref{fig:concentration1}(g, right) as $t\to t_\infty$ and finally reaches a steady state where the gradients powered by the curl are compensated by the restoring elastic force $K\Delta \boldsymbol{S}$.

For very small $\bar \sigma$, one can perturb the initial profile as:
\begin{equation}
    \theta(x,t) = \frac{2\pi x}{L} + \bar \sigma \,\delta(x,t)\ .
\end{equation}
We inject this in Eq.~(\ref{eq:main_eq}), we Taylor expand the cosine and we crudely retain only the terms of order $\mathcal{O}(\sigma^0)$, thus neglecting $\partial\delta/\partial x$. In the steady state, it gives 
\begin{equation}
    0 = \Delta \delta(x) + \cos\left(\frac{2\pi x}{L}\right)\frac{2\pi}{L}\ .
\end{equation}
The cosine is an eigenmode of the Laplacian and appears in the forcing term. So we make the ansatz $\delta(x) = A\,\cos\left(2\pi x/L\right)$  and   find $A = L/2\pi$. One thus has 
\begin{equation}
    \theta(x,t) = \frac{2\pi x}{L} + \bar \sigma \,\frac{L}{2\pi}\,\cos\left(\frac{2\pi x}{L}\right) \ ,
\end{equation}
which fits well the data for $\bar \sigma \lessapprox 0.6$ (not shown). 
To reproduce the data for larger $\bar \sigma $, one needs to go beyond the small gradient expansion used here. 
We provide a more refined derivation and discussion in Appendix I. 
Assuming only one degree of freedom, $\theta$, and for the specific value $L=2\pi$, one obtains the steady state profile
\begin{equation}
    \theta(x) = 2 \arctan \left\{\frac{1}{C}\, \Big[ \bar\sigma +  \tan \Big( \pi (x-x_0)\Big)\Big]\right\} \ ,
    \label{eq:profile_derivation}
\end{equation}
with $C^2 = K^2+\bar \sigma^2 $ and $x_0 = 0.17$ a constant of integration fitted to the data. 
We plot Eq.~(\ref{eq:profile_derivation}) in dash in Fig.~\ref{fig:concentration1}(a) and we see that they reproduce almost perfectly the steady state profile, even at $\bar\sigma = 1.95 \gg 0.6$. The only region where the prediction deviates from the actual data is the region $0.6<x/L<0.8$, where the order parameter $S$ drops, see Fig.~\ref{fig:snapshots_concentration}(e,f). 

To follow the deformation of the spinwave, we define the steady state width $w$ of the compressed region as 
\begin{equation}
    w=\frac{2\pi}{ \partial_x\theta(\frac{x}{L}=\frac{3}{4})} \ ,
\end{equation}
evaluated at the compression (stable) point $x/L = 3/4$, or, equivalently, $\theta = 3\pi/2$. 
For the initial profile Eq.~(\ref{eq:spinwave}), one obtains $w=1$. 
The long-term width is the result of the competition between the non-reciprocal force and the elastic force.
In first approximation, these two effects balance at the length scale $\lambda_\sigma = K/\bar\sigma$, such that $w \sim \lambda_\sigma $~. 
As reported in Fig.~\ref{fig:concentration1}(c) for different $\alpha$, $w$ decreases for increasing $\bar\sigma$. 
Using the expression Eq.~(\ref{eq:profile_derivation}), one can derive (see Appendix I)
\begin{equation}
    w  = \frac{K}{\sqrt{\bar{\sigma}^2 + K^2}+ \bar \sigma} \ ,
\end{equation}
which nicely fits the data, see dashed line in Fig.~\ref{fig:concentration1}(c). It also confirms a posteriori the scaling $w \sim\lambda_\sigma $ for large $\bar \sigma$. 
\newline

So far, we have seen that non-reciprocity reshapes the $\theta$ profile but the winding number remains unchanged and the state topologically protected.
In contrast, if $\bar\sigma$ exceeds a certain threshold,
non-reciprocity is found to change the topology of the initial configuration. 
When the width of the profile, $\lambda_{\bar\sigma}\sim K/\bar\sigma$, becomes comparable to the length scale at which distortions drive a reduction in order parameter, $\lambda_\alpha\sim\sqrt{K/\alpha}$, we expect $S$ to vanish locally, see Fig.~\ref{fig:snapshots_concentration}(k).  
In the region where $S \approx 0$, the equation of motion Eq.~(\ref{eq:main_eq}) becomes largely independent of the orientation field $\theta$, thus allowing its drastic reorientation, cf. the change from red to blue ($\theta = 3\pi/2 \to \theta = \pi/2$) between Fig.~\ref{fig:snapshots_concentration}(h) and \ref{fig:snapshots_concentration}(i). The corresponding $\theta$ profiles are plotted in Fig.~\ref{fig:concentration1}(b).  Such apparent change in topology is possible because of the relaxation of the spherical constraint: the amplitude of the field can locally vanish to allow its orientation to relax to its ground state.   
This is why the derivation of Appendix I, giving Eq.~(\ref{eq:profile_derivation}), cannot account for it, see dashes in Fig.~\ref{fig:concentration1}(b).
This is why the derivation of Appendix I, giving Eq.~(\ref{eq:profile_derivation}), cannot account for it, see dashes in Fig.~\ref{fig:concentration1}(b).


We rationalize such a breakdown of the spin-wave excitation, occurring for $\bar\sigma > \bar\sigma_{c}(\alpha)$, by a simple scaling argument. Imposing $\lambda_{\bar\sigma} \sim \lambda_\alpha$ leads to $\bar\sigma_{c}\sim\sqrt{K\alpha}$, confirmed by the inset of Fig.~\ref{fig:concentration1}(c). 
The ratio $\lambda_\alpha/\lambda_{\bar\sigma }= \bar\sigma/\sqrt{K\alpha}$  describes almost all of the variation in the order parameter through this process. Figure~\ref{fig:concentration1}(d) shows that $S$ vanishes at a value of $\bar\sigma_\textrm{c}/\sqrt{K\alpha} \approx 0.2$, in good agreement with the fitting parameter used in Fig.~\ref{fig:concentration1}(c) (inset). Additionally, the stars in Fig.~\ref{fig:concentration1}(c) correspond to the predicted breakdown width $ w =  K/\bar\sigma_{c}$.

After the spin-wave breaks down, the winding number is zero and the profile flattens to the steady state value $\theta = \pi/2$, as indicated in blue in Fig.~\ref{fig:concentration1}(b) or illustrated in Fig.~\ref{fig:snapshots_concentration}(j). At this point, the field $\theta$ is uniform, and $S$ tends to 1 everywhere, see Fig.~\ref{fig:snapshots_concentration}(m), thereby reaching the ground state of the equilibrium system. 
The time evolution of the free energy $\mathcal{F}$, defined in Eq.~(\ref{eq:free_energy})  and normalized by its initial value $\mathcal{F}_0$, is plotted in Fig.~\ref{fig:concentration1}(e), for different $\bar\sigma$, increasing from blue to yellow. Active forces~$\sim\bar \sigma$ inject energy into the system. Interestingly, above $\bar\sigma_{c}$, this system provides a good example where non-reciprocity helps a system escaping a local minimum of the elastic energy $\mathcal{F}$ and reaching its equilibrium ground state $\mathcal{F}=0$.


We have therefore shown that the topologically protected pattern Eq.~(\ref{eq:spinwave}) is unstable to non-reciprocal forces. Small values of non-reciprocity slightly deform the pattern, moderate values compress it to a narrow region of space and larger activity can even change the topology of the configuration and help the system to recover the ground-state of the equilibrium $\mathcal{O}(2)$ model, a scenario impossible without any active force. 
 \newline

\section{Concluding remarks}

To conclude, we have shown how NR interactions in the vision-cone spirit impact topological defects in a paradigmatic continuous spin system. Non-reciprocity breaks the parity and rotational symmetry of the $2d$XY model and drives the system out of equilibrium. For slightly anisotropic vision cones, non-reciprocity translates into a novel term in the continuum dynamical equation of the spin field, proportional to its vorticity. Notably, this new term gives (and constrains) two of the three active terms in the constant density Toner-Tu equations \cite{toner1995long}, a field theory developed for motile agents. As such, our approach conceptually relates non-reciprocal systems of immobile agents and active flocks.  In the same spirit, a recent work \cite{huang2024active} also derives a hydrodynamic theory that encompasses Active Model B+ \cite{nardini2017entropy, tjhung2018cluster},  
from a microscopic model of non-motile particles interacting in a non-reciprocal fashion. 
The impact of such rotational term is to twist the phase field and to discriminate defects according to their specific shape (on top of their charge), leading to a rich phenomenology for pair annihilation: the motion of $\pm1$ defects is asymmetric, they can be speeded up or slowed down without altering the shape of the stable configuration. This is in contrast with the dynamics of liquid crystals, ruled by the relaxation of a functional, for which the relative values of bend and splay rigidity play a crucial role in the dynamics of topological defects, but cannot directly tune the relative velocity of $\pm 1$ defects without altering the stability of different shapes \cite{stannarius2016defect, missaoui2020annihilation, harth2020topological, huang2023structures}. Our intricate scenario goes well beyond the isotropic annihilation of defects in the $2d$XY model and appears to be robust and general, 
 as it remains valid at the hydrodynamic level. It neither depends  on the specific shape of the interaction kernel, as we have found an equivalence relation between sharp and soft vision cones based on its anisotropy. 
\newline 

We have also studied how non-reciprocity advects and reshapes patterns, a ubiquitous feature found in many non-reciprocal systems, independently of the symmetry and the nature of the spins themselves. We have focused
on localized perturbations and on extended topologically protected patterns. A simple localized perturbation propagates opposite to the background orientation, at a speed proportional to the non-reciprocity parameter and modulated by the component of the background medium parallel to its motion. This modulation explains the development of an asymmetry in the perturbation, as well as the curved trajectory in case of a $2d$ perturbation.
These phenomena are generic and should be observable in systems where the non-reciprocity stems from anisotropy in the interaction in the spirit of vision cones. 
Finally, we have shown that non-reciprocity can destabilize a topologically protected spin-wave excitation. Above a critical value of non-reciprocity, the winding number of the pattern changes and the system reaches the ground state of the equilibrium $\mathcal{O}(2)$ model, a scenario impossible without any active force.

\section{Appendix A: Supplemental Movies}
We provide short movies to visualize the dynamics at play at the single defect level, at the pair level and at the system's level (chains of sinks). \\
All videos can be found on \href{https://www.youtube.com/playlist?list=PLDLpeAHw9Khf5NMAXeNH-mesFoR8pAe1B}{this Youtube channel}.
\\

It helps sketching the configuration to understand which spins are mainly influenced by which spins. Here is a correspondence that might help :\\
Black : $\theta = 0 \ (\rightarrow)$ \\
Blue : $\theta = \pi/2 \ (\uparrow)$ \\
Green/yellow : $\theta = \pi \ (\leftarrow)$ \\
Red/orange : $\theta = 3\pi/2 = -\pi/2 \ (\downarrow)$ \\

\paragraph{Single defects } \  \\
Both movies in Appendix A are at $T=0, \sigma = 0.3$. 
Both movies in Appendix A are at $T=0, \sigma = 0.3$. 
\begin{itemize}
    \item ``1- Twist" : \\ movie of the twist of a positive defect with an initial shape $\mu_+(t=0) = 1$. The final state is a sink, easily recognised by the black colour being on the left of the defect. 
    \item ``2-Polarisation" : \\ movie of the twist of a negative defect with an initial shape $\mu_-(t=0) = 1$.  
\end{itemize}

\paragraph{Pairs of defects}  \  \\
For these videos, the initial separating distance $R_0 = 48$, {the} linear size of the system $L=100$ lattice spacings. $T=0$. Hereafter, {the shapes indicated in the titles are the initial shapes at $t=0$}. 
\begin{itemize}
    \item ``3-XY annihilation" : \\Annihilation of a pair. XY equilibrium case : $\sigma=0$. Note how the $\theta$ field conserves its symmetry with respect to the defecs cores {over} the process.
    \item ``4a-enhanced annihilation mu0" : \\Enhanced annihilation of a pair $\mu_+ = 0, \mu_- = \pi$ for $\sigma=0.3$ . Note how the $\theta$ field looses its equilibrium symmetry {over} the process. The polarisation axis progressively grows in the direction of the positive defects. Until now, none of the defects move in space. Then the axis reaches the $+1$ defect. At from this point, both defects run along the axis that now connects them until annihilation. 
    \item ``4b-enhanced annihilation mu0.2" : \\Enhanced annihilation of a pair $\mu_+ = 0.2, \mu_- = \pi-0.2$ for $\sigma=0.3$ . Here the initial shape of the positive defect $\mu_+ = 0.2\neq 0$ so the positive defect should, if it were isolated, decay to the sink state. It does indeed start to twist in the first instants of the movie but the negative defect rapidly polarises its surroundings: the polarisation axis reaches the positive defect before it has time to complete its twist. The two defects are now connected by the axis and eventually annihilate following the scenario described above for a pure source $\mu_+ = 0$.  
    \item     ``5a-hindered annihilation mu pi" : \\ Hindered annihilation of a pair $\mu_+ = \pi, \mu_- = 0$ for $\sigma=0.3$ . Typical configuration where the axis of the negative defect grows perpendicularly, hindering the annihilation process.
    \item ``5b-hindered annihilation mu pi-0.5" : \\ Hindered annihilation of a pair $\mu_+ = \pi-0.5, \mu_- = 2\pi - 0.5$ for $\sigma=0.3$. Since $\mu_- \neq 0,\pi$ the polarisation axis (initially)  grows obliquely. Yet, complex dynamics of the surrounding $\theta$ field impose an effective force on the axis, such that it becomes perpendicular to the core-to-core direction. Thus, it is hard for the $-1$ defect to move: the XY annihilation process is hindered.
    \item ``6a-transverse motion mu pi2" : \\ Transverse motion, for  $\mu_+ = \pi/2, \mu_- = 3\pi/2$ and $\sigma=0.3$ . The polarisation axis grows obliquely while the positive defect decays to a sink. The negative defect then follows the polarised path just created. It eventually places itself below the positive defect, because in this configuration the axis is now perpendicular to the core-to-core direction. We are now in a configuration that leads to the hindering of the annihilation process. 
    \item ``6b-transverse motion creation defects mu pi2" : \\ Same initial configuration, but now for  $\sigma=0.5$ . Note how the polarisation axis, separating two domains of opposite orientation becomes so thin that the natural elasticity of the $\theta$ field cannot support the large gradients $|\nabla \theta|$ at this location. It therefore breaks to relieve the local tension, creating a new pair of defects: the new $+1$ is on the right and meets/{annihilates} with the old $-1$. The new $-1$ follows the rest of the polarised axis and places itself below the $+1$, for the same reasons as explained above.
    \item     ``6c-transverse motion annihilation mu pi2" : \\ Same initial configuration, but now for  $\sigma=0.2, T=0.12\,T_{KT}$ . The negative defect first follows the path traced by the axis, then stabilises below the positive defect for the same reasons as explained above. The difference with the two previous videos is that the present one is at finite temperature, such that the thermal fluctuations help the defects to annihilate.
\end{itemize}
 
\paragraph{Coarsening dynamics of the system}  \  \\
We provide 3 movies in Appendix A of a $L=200$ triangular lattice, at $T/T_{KT} =0.08$. 
We provide 3 movies in Appendix A of a $L=200$ triangular lattice, at $T/T_{KT} =0.08$. 
\begin{itemize}
    \item ``7a-XY system sigma 0" : \\ XY case: $\sigma =0$ The field around defects usually is isotropic and non-twisted. Defect annihilate {pairwise} following the scenario shown for isolated pairs of defects.
    \item ``7b-NRXY system sigma 0.15" : \\ Non-reciprocal case: $\sigma =0.15$ . Note how positive defects twist and how negative defects {grow} their polarisation axes, creating straight lines in the system.  The behaviours described at the level of a single pair of defect are still valid in this many-body system of defects.
    \item ``7c-NRXY system sigma 0.3" : \\Non-reciprocal case: $\sigma =0.3$ . Same {phenomenology}, but the effects are intensified.
\end{itemize}

\paragraph{Miscellaneous}  \  \\
\begin{itemize}
    \item ``8a-NRXY propagation sigma 0.5" : \\ Unidirectional propagation of gradients, as already observed by \cite{loos2023long} .
    \item  ``9-XY-NRXY chain of sinks" : \\ A chain of sinks $+1$ defects (and their corresponding $-1$ defects with $\mu_-=\pi$ ) is unstable in the equilibrium XY model (left) and stable for the NR XY model ($\sigma = 0.3$). The video is at $T=0$ for visualisation purposes but the phenomenon exists at finite $0<T<T_{KT}$. 
\end{itemize}

\section{Appendix B: derivation of continuum equation}
In this section, we detail the steps to write the equation of motion 
\begin{equation*}
  \dot{\theta}_i= \sum_j \sin \left(\theta_j-\theta_i\right)\ g(\varphi_{ij})
\end{equation*}
in a continuous form, based on derivatives of the  field $\theta = \theta(x,y)$, which emphasizes the reciprocal and non-reciprocal contributions of the kernel. 

Consider a spin $i$ on the square lattice: it has 4 nearest neighbouring spins $j=1,2,3,4$ (right, top, left, bottom), the orientations of which we note $\theta_{1,2,3,4}$ . Those 4 neighbours can be thought to lie on a unit circle centred on the spin $i$. We call $u_{j}$ the angle formed by horizontal axis and the location 
of spin $j$, see Fig.1a of the main text. In a square lattice, $u_j = \frac{\pi}{2}(j-1), \ j=1,...,4$. In a triangular lattice, $u_j = \frac{\pi}{3}(j-1), \ j=1,...,6$ . Note that $u_j$ and the orientation $\theta_j$ are independent.

We start by explaining how \emph{not} to proceed. One could be tempted to identify, from the beginning, $\theta_j-\theta_i$ with the first order derivative $\frac{\partial \theta}{\partial k}$ ($k=x,y$). The issue is, if one takes the simplest case of the XY model, where the kernel $g=1$, one would obtain 
\begin{equation}
    \begin{aligned}
\dot{\theta}_i= & \sum_j \sin \left(\theta_j-\theta_i\right) \\ 
= & \  \sin\left(\frac{\partial \theta}{\partial x}\right) +  \sin\left(\frac{\partial \theta}{\partial y}\right)+ \sin\left(-\frac{\partial \theta}{\partial x}\right)+  \sin\left(-\frac{\partial \theta}{\partial y}\right)\\
= & \  0 \
\end{aligned}
\end{equation}
for all fields $\theta(x,y)$ without any assumption. The reason why all terms cancel is that implicitly, one here uses the non-centred numerical definition of derivatives, therefore saying that $\frac{\partial \theta}{\partial x} = \theta_1-\theta_i = \theta_i-\theta_3$, where the last equality is incorrect in general and especially around a topological defect. 

To avoid this problem, one has to go to second order (not surprising if one wants to account for an elastic term) and consider both the spins on the right-left (resp. top-bottom) of the spin $i$ of interest.  

{In the following derivation, we work under the assumptions of \textit{small anisotropy/non-reciprocity }$\sigma \ll 1$ and \textit{small spatial gradients} of the field $\theta$.
The limit $\sigma \ll 1$ allows to linearise the  coupling kernel to first order in $\sigma$ ($e^{\sigma \cos x} \approx 1+\sigma \cos x$), such that :}
\begin{equation}
    \begin{aligned}
\frac{\gamma}{J} \dot{\theta}_i= & \sum_j \sin \left(\theta_j-\theta_i\right)\ g(\varphi_{ij}) \\
= & \sum_j \sin \left(\theta_j-\theta_i\right)\left[1+\sigma \cos \left(\varphi_{ij}\right)\right]\\
= & \sum_j \sin \left(\theta_j-\theta_i\right)\left[1+\sigma \cos \left(\theta_i-u_{ j}\right)\right]\\
= & \sum_j \sin \left(\theta_j-\theta_i\right)\left[1+\sigma \cos \left(\theta_i-\frac{\pi}{2}(j-1)\right)\right] \\
= & \sin \left(\theta_{1}-\theta_i\right)\left(1+\sigma \cos \theta_i\right) \\
& +\sin \left(\theta_2-\theta_j\right)\left[1+\sigma \cos \left(\theta_i-\frac{\pi}{2}\right)\right] \\
& +\sin \left(\theta_3-\theta_i\right)\left[1+\sigma \cos \left(\theta_i-\pi\right)\right] \\
& +\sin \left(\theta_4-\theta_i\right)\left[1+\sigma \cos \left(\theta_i-\frac{3 \pi}{2}\right)\right] \\
= & \sin \left(\theta_1-\theta_i\right)\left(1+\sigma \cos \theta_i\right) \\
& +\sin \left(\theta_2-\theta_i\right)\left(1+\sigma \sin \theta_i\right) \\
& +\sin \left(\theta_3-\theta_i\right)\left(1-\sigma \cos \theta_i\right) \\
& +\sin \left(\theta_4-\theta_i\right)\left(1-\sigma \sin \theta_i\right) \\
= & \sin \left(\theta_1-\theta_i\right) + \sin \left(\theta_3-\theta_i\right) + \sin \left(\theta_2-\theta_i\right) + \sin \left(\theta_4-\theta_i\right) \\
& + \sigma  \cos(\theta_i)\big(\sin \left(\theta_1-\theta_i\right)-\sin \left(\theta_3-\theta_i\right))\\
& + \sigma \sin(\theta_i)\big(\sin \left(\theta_2-\theta_i\right)-\sin \left(\theta_4-\theta_i\right))\\
\end{aligned}
\end{equation}
One has (the same goes for the vertical direction with $\theta_2$ and $\theta_4$) :
\begin{equation}
     \begin{aligned}
   &\sin \left(\theta_1-\theta_i\right) + \sin \left(\theta_3-\theta_i\right)\\
    &=2\sin \left[\frac{1}{2}\left(\theta_1-2 \theta_i+ \theta_3\right)\right] \cos \left[\frac{1}{2}\left(\theta_1-\theta_3\right)\right]
    \end{aligned}
\end{equation}
\begin{equation}
         \begin{aligned}
&\sin \left(\theta_1-\theta_1\right)-\sin \left(\theta_3-\theta_1\right)\\
&=2 \cos \left[\frac{1}{2}\left(\theta_1-2 \theta_i+\theta_3\right)\right]\sin \left[\frac{1}{2}\left(\theta_1-\theta_3\right)\right]
     \end{aligned}
\end{equation}

Now, if one identifies the differences in the continuum as
\begin{equation}
    \begin{aligned}
& a\frac{\partial \theta}{\partial x}=\frac{\theta_1-\theta_3}{2} \\
& a\frac{\partial \theta}{\partial y}=\frac{\theta_2-\theta_4}{2} \\
& a^2\frac{\partial^2 \theta}{\partial x^2}=\theta_1-2 \theta_i+\theta_3 \\
& a^2\frac{\partial^2 \theta}{\partial y^2}=\theta_2-2 \theta_i+\theta_4
\end{aligned}
\end{equation}
one obtains 
\begin{equation}
\begin{aligned}
\frac{\gamma}{J} \frac{\dot{\theta}}{2} & = \cos (a\,\theta_x) \sin (\frac{a^2}{2}\,\theta_{x x})+\cos (a\,\theta_y) \sin  (\frac{a^2}{2}\,\theta_{y y}) \\
& +\sigma\left(\cos \theta \sin (a\,\theta_x) \cos  (\frac{a^2}{2}\theta_{x x})+\sin \theta \sin (a\,\theta_y) \cos  (\frac{a^2}{2}\theta_{y y})\right)
\end{aligned}
    \label{eq:SM_theta_cos}
\end{equation}
where $\theta =\theta(x,y)$, $\theta_k=\frac{\partial \theta}{\partial k}$ and $\theta_{kk}=\frac{\partial^2 \theta}{\partial k^2}, k=x,y$.

 From this expression, it is now clear that (i) $\sigma$ cannot be absorbed in the time unit {(ii) the dynamics is $\mu$-invariant (ie. it only depends on derivatives of the field $\theta$) if and only if $\sigma=0$}. 

{Now using the small gradients approximation, and} $\cos \alpha \approx 1$ and $\sin \alpha \approx \alpha$,  
one obtains 
\begin{equation}
    \frac{\gamma}{J} \dot{\theta} =  a^2(\theta_{x x}+\theta_{y y}) + 2a\sigma\, (\cos (\theta) \cdot \theta_x +\sin (\theta)\cdot \theta_y)
\end{equation}

The first two terms give the Laplacian operator $\Delta$ that physically represents the elasticity of the $\theta$ field.
The last two terms can be rewritten as follows: 
\begin{equation}
\cos (\theta) \cdot \theta_x +\sin (\theta)\cdot \theta_y  = \partial_x (\sin \theta) - \partial_y(\cos \theta) = (\nabla \times \boldsymbol{\hat{S}})_z
\end{equation}
where $\boldsymbol{\hat{S}}=(\cos\theta,\sin\theta)$ is the (unit length) spin vector field, thus leading to:
\begin{equation}
    \frac{\gamma}{J} \dot{\theta} =  a^2\,\Delta \theta + 2a\sigma\, (\nabla \times \boldsymbol{\hat{S}})_z
    \label{eq:rotational}
\end{equation}
Note that here we have kept $\gamma, J$ and $a$ , even though $\gamma=J=a=1$ in the adimensionalized main text version, to highlight the unitless nature of $\sigma$. 

\ \newline 
Inspired by the formulation in Eq.~(\ref{eq:rotational}) in terms of the field
\begin{equation}
\begin{aligned}
\boldsymbol{\hat S} & =\left(\begin{array}{c}
\hat{S}_x \\
\hat{S}_y
\end{array}\right)=\left(\begin{array}{c}
\cos \theta \\
\sin \theta
\end{array}\right)\ , 
\end{aligned}
\end{equation}
we look for an equivalent formulation solely in terms of the vector field $\hat{S}$.
Importantly, for now the magnitude of $\boldsymbol{\hat{S}}$ still remains strictly fixed to $|\boldsymbol{\hat{S}}|=1$, such that $d{\boldsymbol{\hat{S}}}/dt \perp \boldsymbol{\hat{S}}$. 
Indeed, one has: 
\begin{equation}
\begin{aligned}
d{\boldsymbol{\hat{S}}}/dt & =\dot{\theta}\left(\begin{array}{c}
-\sin \theta \\
+\cos \theta
\end{array}\right)=\dot{\theta}\left(\begin{array}{l}
-\hat{S}_y \\
+\hat{S}_x
\end{array}\right)
\end{aligned}
\end{equation}

We start by the elasticity term $\Delta \boldsymbol{\hat{S}}=\left(\begin{array}{c}
\Delta \hat{S}_x \\
\Delta \hat{S}_y
\end{array}\right)$: 
\begin{equation}
    \begin{aligned}
& \frac{\partial^2}{\partial x^2} \hat{S}_x=\frac{\partial^2}{\partial x^2}(\cos \theta) \\
& =\frac{\partial}{\partial x}\left(-\frac{\partial \theta}{\partial x} \sin \theta\right) \\
& =-\left(\frac{\partial^2 \theta}{\partial x^2} \sin \theta+\left(\frac{\partial \theta}{\partial x}\right)^2 \cos \theta\right) \\
& =-\left(\frac{\partial^2 \theta}{\partial x^2}\hat{S}_y+\left(\frac{\partial \theta}{\partial x}\right)^2 \hat{S}_x\right) \\
& =-\frac{\partial^2 \theta}{\partial x^2} \hat{S}_y \\
&
\end{aligned}
\end{equation}
Where the last equality results from the approximation that we only consider first order gradients.
Also,
\begin{equation}
    \begin{aligned}
\frac{\partial^2}{\partial x^2} \hat{S}_y & =\frac{\partial^2}{\partial x^2}(\sin \theta) \\
& =\frac{\partial}{\partial x}\left(\frac{\partial \theta}{\partial x} \cos \theta\right) \\
& =\frac{\partial^2 \theta}{\partial x^2} \cos \theta-\left(\frac{\partial \theta}{\partial x}\right)^2 \sin \theta \\
& =\frac{\partial^2 \theta}{\partial x^2} \hat{S}_x-\left(\frac{\partial \theta}{\partial x}\right)^2 \hat{S}_y \\
& = \frac{\partial^2 \theta}{\partial x^2} \hat{S}_x
\end{aligned}
\end{equation}

Similarly, $\frac{\partial^2}{\partial y^2} \hat{S}_x =-\frac{\partial^2 \theta}{\partial y^2} \hat{S}_y $ and $\frac{\partial^2}{\partial y^2} \hat{S}_y =+\frac{\partial^2 \theta}{\partial y^2} \hat{S}_x $, such that

\begin{equation}
    \Delta \boldsymbol{\hat{S}} =\left(\frac{\partial^2 \theta}{\partial x^2}+\frac{\partial^2 \theta}{\partial y^2}\right)\left(\begin{array}{l}-\hat{S}_y \\ +\hat{S}_x\end{array}\right)=\Delta \theta\left(\begin{array}{c}-\hat{S}_y \\ +\hat{S}_x\end{array}\right)
\end{equation}

On the other hand, one has :
\begin{equation}
\begin{aligned}
& (\nabla \times \boldsymbol{\hat{S}}) \times \boldsymbol{\hat{S}}=(\nabla \times \boldsymbol{\hat{S}})_z\left(\begin{array}{l}
-\hat{S}_y \\
+\hat{S}_x
\end{array}\right) \\
\end{aligned}
\end{equation}

Combining both parts leads to
\begin{equation}
\begin{aligned}
a^2 \Delta \boldsymbol{\hat{S}}+2\sigma a\,(\nabla \times \boldsymbol{\hat{S}}) \times \boldsymbol{\hat{S}}
&=\left[a^2 \Delta \theta+2\sigma a\,(\nabla \times \boldsymbol{\hat{S}})_z\right]\left(\begin{array}{c}
-\hat{S}_y \\
+\hat{S}_x
\end{array}\right) \\
& =\frac{\gamma}{J} \dot{\theta}\left(\begin{array}{l}
-\hat{S}_y \\
+\hat{S}_x
\end{array}\right) \\
& =\frac{\gamma}{J} \, d{\boldsymbol{\hat{S}}}/dt \\
&
\end{aligned}
\end{equation}

Defining $\bar \sigma = 2\,\sigma\,a$, one obtains 

\begin{equation}
    \frac{\gamma}{J}\, d{\boldsymbol{\hat{S}}}/dt = a^2 \Delta \boldsymbol{\hat{S}}+\bar \sigma \,(\nabla \times \boldsymbol{\hat{S}}) \times \boldsymbol{\hat{S}}
\end{equation}
Finally, one can relax the hard constraint $|\boldsymbol{\hat{S}}|=1$ and only softly enforce a preferred magnitude (here unity without loss of generality) via a Lagrange multiplier term $\alpha\,(1-\boldsymbol{S}^2)\boldsymbol{S}$:

\begin{equation}
    \frac{\gamma}{J}\dot{\boldsymbol{S}} = a^2 \Delta \boldsymbol{S}+\bar \sigma \,(\nabla \times \boldsymbol{S}) \times \boldsymbol{S} + \alpha\,(1-\boldsymbol{S}^2)\boldsymbol{S}
\end{equation}
where now $\boldsymbol{S}$ is a vector field of arbitrary magnitude $S$ : 
\begin{equation}
\begin{aligned}
\boldsymbol{S} & =\left(\begin{array}{c}
S_x \\
S_y
\end{array}\right)=S\left(\begin{array}{c}
\cos \theta \\
\sin \theta
\end{array}\right)\ .
\end{aligned}
\end{equation}
As $\sqrt{J/\alpha}$ physically sets the defects core size, it has to be larger than the simulation mesh spacing. 
In the simulations of the continuum model, we use $\alpha = 100$, $J=1$, and typically $N_{\text{grid}}=256$. The mesh spacing is $1/256 \approx 0.004$ while $\sqrt{J/\alpha}=0.1$. 

\section{Appendix C: dimension analysis of the continuum equation}
In this section, we extract the natural scales from Eq.~(\ref{eq:main_eq}) 
\begin{equation}
\gamma\dot{\boldsymbol{S}} = K\Delta\boldsymbol{S}
\!+ \bar\sigma (\nabla \times \boldsymbol{S})\times \boldsymbol{S} + \alpha (1-|\boldsymbol{S}|^2)\boldsymbol{S}
\end{equation}
Because $\boldsymbol{S}$ is a vector field, it has no units, so Eq.~(\ref{eq:main_eq}) is dimensionally equivalent to 
\begin{equation}
\frac{\gamma}{t}= \frac{K}{x^2} + \frac{\bar \sigma}{x} + \alpha
\end{equation}
where $t$ means time and $x$ means space. Each term has to have the same unit. 
From there, one deduces the only time scale of our problem: 
\begin{equation}
    \gamma / \alpha
\end{equation}
One can construct 3 distinct length scales: 
\begin{equation}
    \sqrt{\frac{K}{\alpha}}\ , \ \frac{\bar \sigma}{\alpha} \text{ and } \frac{K}{\bar \sigma}
\end{equation}
Following, one can construct 3 speed scales: 
\begin{equation}
    \frac{\sqrt{K\alpha}}{\gamma}\ , \ \frac{\bar \sigma}{\gamma} \text{ and } \frac{K\alpha}{\bar \sigma \,\gamma}
\end{equation}
Because we want scales that do not change when $\bar\sigma$ is varied, we will work in terms of the length scale  $\sqrt{K/{\alpha}}$ and velocity scale ${\sqrt{K\alpha}}/{\gamma}$. Diffusion coefficients are in units of $K/\gamma$. 
\newline 

We now explain how we implemented the equation in the code. First, we divide Eq.~(\ref{eq:main_eq})  by $K$.
Then we redefine the parameters as follows 
\begin{equation}
    \begin{aligned}
        \tilde t &= t \cdot \frac{K}{\gamma} \\ 
         \tilde \sigma &= \bar \sigma / K  \\ 
        \tilde \alpha &=  \alpha / K  \\ 
    \end{aligned}
\end{equation}
Recall that we set $L=2\pi$, in units of $\sqrt{K/{\alpha}}$ . 
In the code, the lengths $\tilde x = i/N_\text{grid} \ \in [0,2\pi]$ are unitless and the speeds $\tilde v = d\tilde x /d \tilde t$ are therefore in units of $\alpha / K$. The diffusion coefficients are also in units of $\alpha / K$. 

The complete correspondence between the quantities of the code (with tilde) and the quantities stemming from Eq.~(\ref{eq:main_eq}) (without tilde) is: 
\begin{equation}
    \begin{aligned}
        \tilde t &= t \cdot \frac{K}{\gamma} \\ 
         \tilde \sigma &= \bar \sigma / K  \\ 
        \tilde \alpha &=  \alpha / K  \\ 
         \tilde x &=  x / 2\pi \quad \to\quad \frac{\partial}{\partial \tilde x} = 2\pi\,\frac{\partial}{\partial x}\\ 
         \tilde v &=  v\,\frac{\gamma\sqrt{\alpha}}{2\pi\,K^{3/2}}  \\ 
         \tilde D &= D\,\frac{\gamma\,{\alpha}}{(2\pi\,K)^2} 
    \end{aligned}
        \label{eq:SM_units}
\end{equation}

Now, recall that we have set $\gamma = K = 1$ and we usually work with $\alpha = 100$. Focusing on the numerical conversion for these values, one has
\begin{equation}
    \begin{aligned}
        \tilde t &= t \cdot  \\ 
         \tilde \sigma &= \bar \sigma   \\ 
        \tilde \alpha &=  \alpha  \\ 
         \tilde x &=  x / 2\pi \quad \to\quad \frac{\partial}{\partial \tilde x} = 2\pi\,\frac{\partial}{\partial x}\\ 
         \tilde v &=  v\,\frac{10}{2\pi} = 1.59\,v \\ 
         \tilde D &= D\,\frac{100}{(2\pi)^2}  = 2.53\,D \equiv \bar D
    \end{aligned}
    \label{eq:SM_units_value}
\end{equation}
Applied to the diffusion length, it gives $\sqrt{\tilde D \tilde t\,}= \sqrt{\bar D t\,} = \sqrt{2.53\, D t} = 1.59\,\sqrt{D t} $. 
\newline

\begin{figure}[h!]
    \centering
    \includegraphics[width=0.9\linewidth]{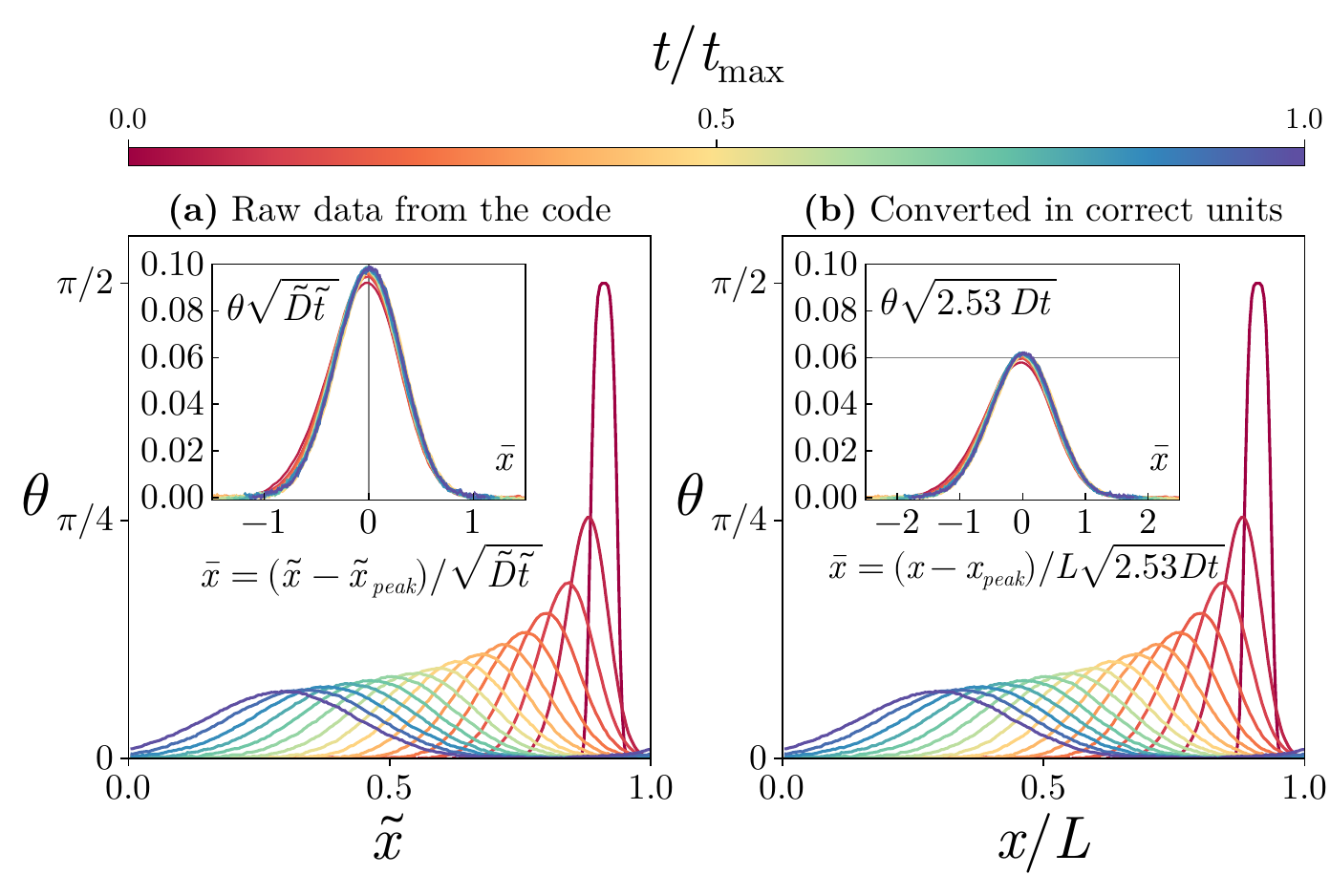}
    \caption{
     Propagation of a perturbation $\theta=\delta(x)$ across the system over time. Same parameters for both panels: small non-reciprocity $\bar\sigma = 10, t_{\text{max}} = 0.2, N_{\text{grid}} = 256, \alpha = 100, \theta_0=0$. 
     \textbf{(a)} In units of the code (with tilde). The length $\tilde x$ is unitless. \underline{Inset}: rescaled data $\theta\sqrt{\tilde Dt}$ as a function of $\bar{x} = (\tilde x-\tilde x_{peak})\sqrt{\tilde D\tilde t} $, with $\tilde D = 2.53\,D$. 
    \textbf{(b)} In units compatible with the main Equation~(\ref{eq:main_eq}) (without tilde). The length $x$ has units of length, hence the normalization by the length of the simulation box $L$. \underline{Inset}: same rescaling, $\theta\sqrt{2.53\,Dt}$ as a function of  $\bar{x} = ( x- x_{peak})/L\sqrt{ 2.53D t} $.  
  }
    \label{fig:propagation1_tilde}
\end{figure}

\paragraph{Numerical value of the variable $c$ \\}
In Figure~\ref{fig:propagation1_tilde}, we determine the numerical value of $c$, from the simulation units (left panel, variables with tilde), to its numerical value in units of the main Equation~(\ref{eq:main_eq}) (right panel, without tilde). Here, because $\theta_0=0$, $\theta = \delta$. 
Recall that we defined $c$ in the main text as 
\begin{equation}
    \delta = \frac{c}{\sqrt{t}} 
\end{equation}
with $c$ a coefficient that depends on $\bar \sigma$, on $\alpha$ but only weakly on $t$. We have shown that the perturbation decay follows the usual diffusion behavior, since the curves collapse when rescaling both axes by the diffusion length (=diffusion coefficient $\cdot$ time). In simulation units, this diffusion length is equal to $\sqrt{\tilde D \tilde t}$, see panel (a). When converting to units of Equation~(\ref{eq:main_eq}), the diffusion length is equal to $\sqrt{2.53\,D  t}$,  see panel (b). Recall that $D = K/\gamma$, so it has a numerical value of 1.
The height of the rescaled curves in panel (b) is 0.06. Thus, in this case ($\bar \sigma = 10, \alpha = 100)$ one has $ \theta \sqrt{2.53\,D  t} = 0.06$, so $c = 0.06 / \sqrt{2.53} = 0.0377$. 

If one follows the same reasoning for $\bar \sigma = 100, \alpha = 100$, one gets $ \theta \sqrt{2.53\,D  t} = 0.05$, so $c = 0.05 / \sqrt{2.53} = 0.0314$ ,see inset of Figure~\ref{fig:propagation1}(b) in the main text. This is the value we will use to fit the velocity curves for $\bar \sigma = 100, \alpha = 100$ of Figure~\ref{fig:propagation2}(e), as we explain now. 
\newline 

\paragraph{Numerical value of the variable $v_{\text{max}}$ \\}
Recall that in $1d$, the advective term reads $\bar \sigma \, \cos \theta \, \frac{\partial \delta}{\partial x}$ . 
The raw data from the simulation gives the following steady state velocity of the peak of the perturbation: $v_{\text{max}} / \bar \sigma = 0.316$. 

Because $v_{\text{max}} $ is a velocity, one could think, based on Equations.~(\ref{eq:SM_units}) and (\ref{eq:SM_units_value}) that a simple rescaling $v_{\text{max}}\to v_{\text{max}}\,L/\sqrt{\alpha} =0.628\,v_{\text{max}}$ will give the correct numerical value. For the special case of the convective velocity, it is incorrect. Indeed,  the last gradient term is also involved and should be in the correct units. Because in the simulation, lengths $\tilde x$ are unitless, one also has to convert $\tilde x = x/L$ to obtain
\begin{equation}
    \tilde{ \sigma} \, \cos \theta \, \frac{\partial \delta}{\partial \tilde x} = \bar \sigma \, \cos \theta \, \frac{\partial \delta}{\partial (x/L)} = L \,\bar \sigma \, \cos \theta \, \frac{\partial \delta}{\partial x} \ , 
\end{equation}
making an extra $L$ factor appear in front of the gradient. So, with $L=2\pi$, the correct rescaling for the convective velocity is $v_{\text{max}}\to v_{\text{max}}\,L^2\sqrt{\alpha} =3.948\,v_{\text{max}}$ , such that the numerical value of $v_{\text{max}} / \bar \sigma $ in the correct units is $0.316 \cdot 3.948 = 1.25$. 
\newline

Using the values $c = 0.0314$ and $v_{\text{max}} / \bar \sigma = 1.25$, we perfectly fit the velocity of the peak, see Figure~\ref{fig:propagation2}(e) of the main text.

\section{Appendix D: Rewriting the results of Ref. [217] }\label{sec:rewriting_chate}

In this section, we explicit the simple algebraic rewriting of the Equation (2) of 
\textit{Inescapable anisotropy of nonreciprocal XY models}, Phys. Rev. Lett.,  (2025)
\cite{dopierala2025inescapable}, in our notation (their $J_0$ is our $J$, their $J_1$ is our $\sigma$). We show that the Toner-Tu like equations, with the coefficient derived in \cite{dopierala2025inescapable}, can almost be rewritten in terms of a curl term and a functional derivative term.   
\begin{align}
\lambda_1 &= \tfrac{\sigma}{2}\left(-3 + \tfrac{J}{T}\right), \\[6pt]
\lambda_2 &= \tfrac{\sigma}{2}\left(1 - \tfrac{J}{T}\right), \\[6pt]
\lambda_3 &= \tfrac{\sigma}{4}\left(1 + \tfrac{J}{T}\right).
\end{align}

One can show that 
\[
\lambda_3 = -\tfrac{1}{2}\lambda_1 - \lambda_2,
\]
so
\begin{align}
& \ \lambda_1 (\boldsymbol{S}\cdot\nabla)\boldsymbol{S} 
  + \lambda_2 (\nabla\cdot\boldsymbol{S})\boldsymbol{S} 
  + \lambda_3 \nabla |\boldsymbol{S}|^2 \\[6pt]
= &\ \lambda_1 \Big[ (\boldsymbol{S}\cdot\nabla)\boldsymbol{S} 
     - \tfrac{1}{2}\nabla|\boldsymbol{S}|^2 \Big] 
   + \lambda_2 \Big[ (\nabla\cdot\boldsymbol{S})\boldsymbol{S} 
     - \nabla|\boldsymbol{S}|^2 \Big] \\[6pt]
= &\ \lambda_1 (\nabla\times\boldsymbol{S})\times\boldsymbol{S} 
   + \lambda_2 \Big[ (\nabla\cdot\boldsymbol{S})\boldsymbol{S} 
     - \nabla|\boldsymbol{S}|^2 \Big] \\[6pt]
=& \ \lambda_1 (\nabla\times\boldsymbol{S})\times\boldsymbol{S} 
   + \frac{\lambda_2 }{2}\left\{\,\frac{\delta}{\delta \boldsymbol{S}}
      \Big[ (\nabla\cdot\boldsymbol{S})|\boldsymbol{S}|^2 \Big]  - \nabla|\boldsymbol{S}|^2\right\}
\end{align}

In \cite{dopierala2025inescapable}, the Lagrange term is 
\begin{equation}
    (\alpha - \gamma|\boldsymbol{S}|^2)\boldsymbol{S}
\end{equation}
To compare with our framework $(1 - |\boldsymbol{S}|^2)\boldsymbol{S}$,  which penalizes magnitude deviations from \textit{unity}, one simply applies the rescaling 
\begin{equation}
    \boldsymbol{S} \to \boldsymbol{S} \, \sqrt{\alpha/\gamma}
\end{equation}
Note that this rescales the $\lambda$ coefficients: 
\begin{equation}
    \lambda \to \lambda \sqrt{\alpha/\gamma}
\end{equation}

\section{Appendix E: Emission cones and reception cones }

In a fully reciprocal system of agents, each particle emits and receives information isotropically, independently of the orientation of the emitter or of the receiver. 
In a system of agents with \textit{reception cones}, each particle emits information isotropically but the reception is weighted by a kernel depending on the orientation of the receiver. You will receive a higher signal if you look at your neighbor. An intuitive realization of such a system is precisely the well-known \textit{vision cone} in the animal kingdom. 
However, one can think of the reversed case, ie. \textit{emission cones}, where each agent perceives information isotropically but emits it in a preferential direction, as in the case of the voice. There, the emission is angularly weighted by the orientation of the emitter. You will receive a higher signal if your neighbor looks at you. 

This motivates the study of the following equation of motion, which describes a system of agents with emission cones (note the red index $j$ instead of the index $i$ for the reception cones, and the change of the sign of $\sigma$)
\begin{equation}
    \dot{\theta}_i  =\sum_j e^{-\sigma \cos \left(\theta_{\color{red}{\textbf{j}}}-u_{i j}\right)} \sin \left(\theta_j-\theta_i\right) + \sqrt{2T} \eta_i(t)
\end{equation}
Working on a square lattice at $T=0$, one can explicitly write the sum on the 4 nearest neighbours to obtain
\begin{equation}
\begin{aligned}
\dot{\theta}_i & =\left(1-\sigma \cos \theta_1\right) \sin \left(\theta_1-\theta_i\right) \\
& +\left(1-\sigma \cos \left(\theta_2-\frac{\pi}{2}\right)\right) \sin \left(\theta_2-\theta_i\right) \\
& +\left(1-\sigma \cos \left(\theta_3-\pi\right)\right) \sin \left(\theta_3-\theta_i\right) \\
& \left.+1-\sigma \cos \left(\theta_4+\pi / 2\right)\right) \sin \left(\theta_4-\theta_i\right) \\
\dot{\theta}_i & =\left(1-\sigma \cos \theta_1\right) \sin \left(\theta_1-\theta_i\right) \\
& +\left(1-\sigma \sin \theta_2\right) \sin \left(\theta_2-\theta_i\right) \\
& +\left(1+\sigma \cos \theta_3\right) \sin \left(\theta_3-\theta_i\right) \\
& +\left(1+\sigma \sin \theta_4\right) \sin \left(\theta_4-\theta_i\right)
\end{aligned}\\
\end{equation}
\begin{equation}
\begin{aligned}
& \dot{\theta}_i=\sum_j \sin \left(\theta_j-\theta_i\right) \\
& -\sigma\left\{\cos \theta_1 \sin \left(\theta_1-\theta_i\right)-\cos \theta_3 \sin \left(\theta_3-\theta_i\right) +\sin \theta_2 \sin \left(\theta_2-\theta_i\right)-\sin \theta_4 \sin \left(\theta_4-\theta_i\right)\right\}
\end{aligned}
\end{equation}

For the non-reciprocal part in $\sigma$, one can approximate, to first order in the Taylor expansion ($a$ is the lattice spacing)
\begin{equation*}
\begin{aligned}
& \theta_1=\theta_i+a \frac{\partial \theta}{\partial x} \equiv \theta_i+\theta_x \\
& \theta_3=\theta_i-\theta_x \\
& \theta_2=\theta_i+a \frac{\partial \theta}{\partial y} \equiv \theta_i+\theta_y \\
& \theta_4=\theta_i-a \frac{\partial \theta_y}{\partial y}=\theta_i-\theta_y
\end{aligned}
\end{equation*}
one thus obtains
\begin{equation*}
\begin{aligned}
& \quad \cos \left(\theta+\theta_x\right) \sin \theta_x+\cos \left(\theta-\theta_x\right) \sin \theta_x \\
& +\sin \left(\theta+\theta_y\right) \sin \theta_y-\cos \left(\theta-\theta_y\right) \sin \theta_y
\end{aligned}
\end{equation*}

If one assumes small gradients, one can approximate $\cos \alpha \approx 1$ and $\sin \alpha \approx \alpha$. One thus obtains
\begin{equation*}
2\left(\theta_x \cdot \cos \theta+\theta_y \sin \theta\right)=2(\nabla \times \textbf{S})_z
\end{equation*}

For the reciprocal part, one needs to go up to the second order in the Taylor expansion because the first orders are equal and opposite, leaving us with no reciprocal part and $\dot\theta = \mathcal{O}(\sigma)$ 
\begin{equation*}
\begin{aligned}
& \theta_1=\theta_i+a \frac{\partial \theta}{\partial x}+\frac{1}{2} a^2 \frac{\partial^2 \theta}{\partial x^2} \equiv \theta_i+\theta_x+\theta_{x x} \\
& \theta_3=\theta_i-a \frac{\partial \theta}{\partial x}+\frac{1}{2} a^2 \frac{\partial^2 \theta}{\partial x^2} \equiv \theta_i-\theta_x+\theta_{x x} \\
\end{aligned}
\end{equation*}
Which gives
\begin{equation*}
\sin \left(\theta_1-\theta_i\right)+\sin \left(\theta_3-\theta_i\right) = \sin \left(\theta_x+\theta_{x x}\right)+\sin \left(-\theta_x+\theta_{x x}\right) = 2 \sin \theta_{x x} \cos \theta_x
\end{equation*}

which, in the small gradients limit, is simply equal to $\frac{\partial^2 \theta}{\partial x^2}$
Following the same procedure for the $y$ axis ( with $\theta_2$ and $\theta_4$ ), one obtains the Laplacian $\Delta \theta=\frac{\partial^2 \theta}{\partial x^2}+ \frac{\partial^2 \theta}{\partial y^2}$
Finally, one obtains
\begin{equation*}
\dot \theta=a^2 \Delta \theta-2 a\, \sigma (\nabla \times S)_z
\end{equation*}
Differing from the reception cones (=vision cones) only by a minus sign, which can be interpreted as: a system with vision cones (greater reception in front) is equivalent to a system with emission cone (greater emission behind). 
This equivalence might be useful if it is easier to build an experimental setup where the emission, rather than the reception, is anisotropic. 

In fact, to second order in $\sigma$, emission and reception cones have equal sign contributions. This implies that a combination of emission and reception cones with the same value of $\sigma$ gives 
\begin{equation}
    \dot{\theta} / a^2  = \Delta \theta+\sigma^2 \cos (2 \theta)\left[\left(\frac{\partial \theta}{\partial x}\right)^2+\left(\frac{\partial \theta}{\partial y}\right)^2\right]
\end{equation}

\section{Appendix F: rewriting Ref. [215] in terms of vector calculus operators}
In their work \textit{Defect dynamics in active polar fluids vs. active nematics}, Soft
Matter (2022)\cite{vafa2022defect}, Vafa considered the following equation: 

\begin{equation}
\begin{aligned}
   \partial_t p=\mathcal{I}(p)&=-\frac{\delta \mathcal{F}(\{p\})}{\delta \bar{p}}+\lambda \mathcal{I}_\lambda(p) \\
&=4 \partial \bar{\partial} p+2 \epsilon^{-2}\left(1-|p|^2\right) + \lambda(p \partial+\bar{p} \bar{\partial})p
\end{aligned}
\end{equation}

They defined

\begin{equation*}
\begin{aligned}
p & =p_x+i p_y \\
\bar{p} & =p_x-i p_y \\
\partial & =\frac{1}{2}\left(\partial_x-i \partial_y\right) \\
\bar{\partial} & =\frac{1}{2}\left(\partial_x+i \partial_y\right) \\
\end{aligned}
\end{equation*}

Therefore

\begin{equation*}
\begin{aligned}
4 \partial \bar{\partial} & =\left(\partial_x-i \partial_y\right)\left(\partial_x+i \partial_y\right)=\partial_x^2+\partial_y^2=\Delta \\
2 p {\partial} & =\left(p_x+i p_y\right)\left(\partial x-i \partial_y\right) \\
& =p_x \partial_x+p_y \partial_y+i\left(p_y \partial_x-p_x \partial_y\right) \\
2 \overline{p} \bar{\partial} & =\left(p_x-i p_y\right)\left(\partial_x+i \partial_y\right) \\
& =p_x \partial_x+p_y \partial_y-i\left(p_y \partial_x-p_x \partial_y\right)
\end{aligned}
\end{equation*}

So $(p \partial+\bar{p} \bar{\partial})p=(\vec{P} \cdot \nabla) \vec{P} \quad$ where $\quad \vec{P}=
\begin{pmatrix}
           p_x \\
        p_y 
         \end{pmatrix}$ .

Finally, one can rewrite Vafa's model as:
\begin{equation}
\partial_t \vec{P}=\Delta \vec{P}+\epsilon^{-2}\left(1-|\vec{P}|^2\right)\vec{P}+\lambda(\vec{P} \cdot \nabla) \vec{P}
\end{equation}
, highlighting the convective nonlinearity of the model and its similarity with the model of Besse \textit{et al.}, PRL, 2022 \cite{besse2022metastability}.

\section{Appendix G: analytical defect shapes}
To support the intuitive argument sketched in Fig.~\ref{fig:2}(e, k) of the main text, we analytically compute the dynamics of the field around ideal isolated topological defects.
We start from the  equation of motion
	\begin{equation}
		\gamma\,\dot{\theta}_i=J\sum_{j \in \partial_i} \,g_{\sigma}(\varphi_{i j})\sin \left(\theta_j-\theta_i\right) + \sqrt{2\gamma\, T}\,\eta_i(t) 
	\end{equation}

	We first study this dynamics in the continuum limit, where the orientation field is a function of a position $\underline{x}$, and consider a neighbourhood around this position. We also ignore noise and set $J/\gamma = 1$ for simplicity.
	\begin{equation}
		\dot{\theta}(\underline{x}) = \int_{-\pi}^{\pi}g_{\sigma}(\phi_1 - \theta(\underline{x}))\sin \left(\theta(\underline{x} + \delta \hat{\underline{\phi}}_1)-\theta(\underline{x})\right) d\phi_1
	\end{equation}
	
	Where $\phi_1$ is the polar angle around position $\underline{x}$ and $\delta$ is a small finite distance which could be thought of as the interdefect spacing and $\hat{\underline{\phi}}_1$ is a unit vector with orientation $\phi_1$. {Approximating $\theta(\underline{x} + \delta \hat{\underline{\phi}}_1)$ by its first order Taylor expansion, one gets}:
	
	\begin{align}
		\dot{\theta}(\underline{x}) &= \int_{-\pi}^{\pi}g_{\sigma}(\phi_1 - \theta(\underline{x}))\sin \left(\theta(\underline{x}) + \delta \hat{\underline{\phi}}_1\nabla\theta(\underline{x})-\theta(\underline{x})\right) d\phi_1\\
		&= \int_{-\pi}^{\pi}g_{\sigma}(\phi_1 - \theta(\underline{x}))\sin \left(\delta \hat{\underline{\phi}}_1\nabla\theta(\underline{x})\right) d\phi_1		
	\end{align}

	We now convert our coordinate system to polar coordinates around the center of a defect, which we denote $(\phi,r)$ with corresponding orthonormal basis vectors $[\underline{\hat{r}},\underline{\hat{\phi}}]$ shown in the figure below. 
	\begin{figure}[h!]
		\centering
		\includegraphics[width=0.5\columnwidth]{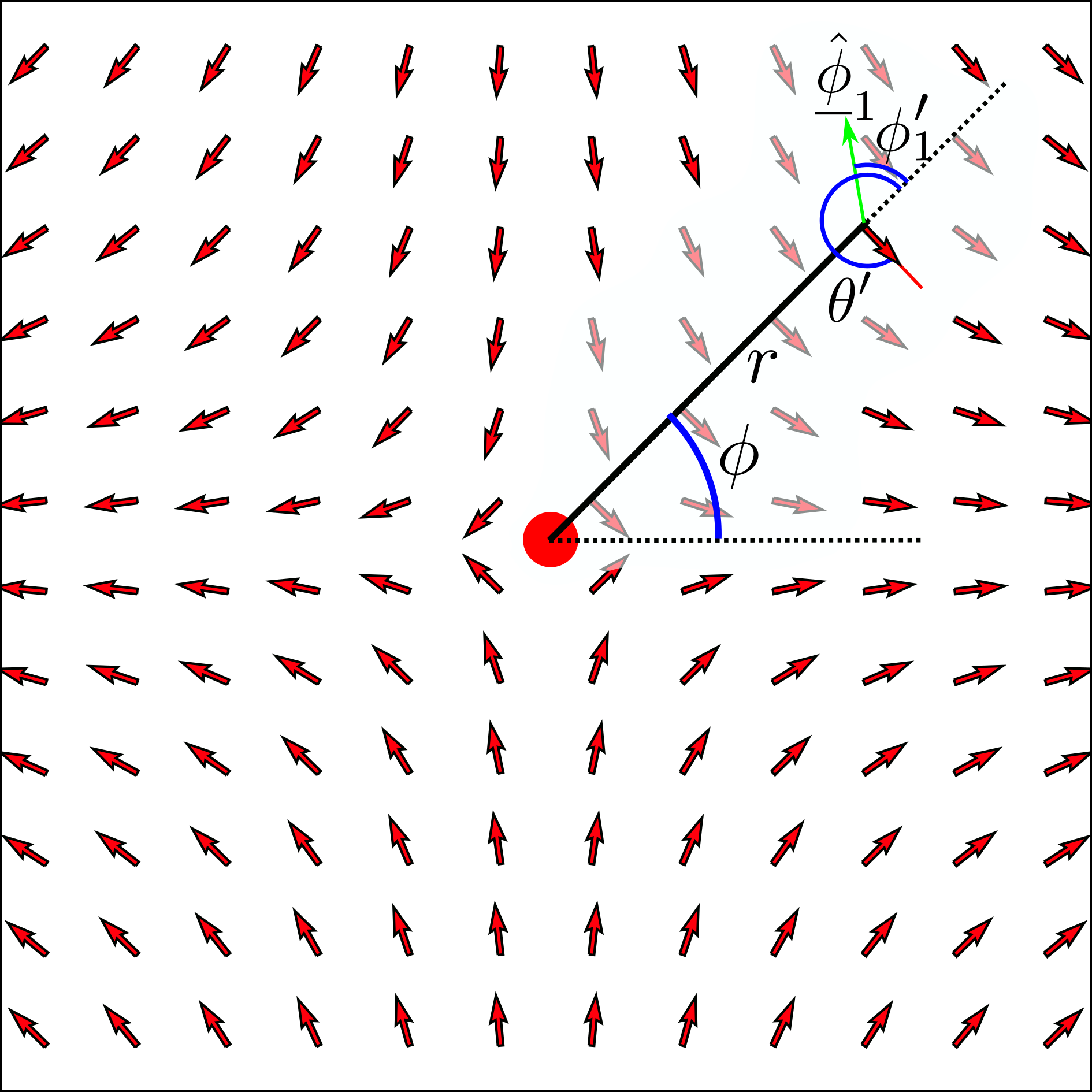}\caption{Coordinate system for the calculation.\label{fig:Coords}}
	\end{figure}

	In this coordinate system, the new director field can now be written relative to the $\underline{\hat{r}}$ direction as:
	\begin{equation}
		\theta'(r,\phi) = \theta(\underline{x}) - \phi \implies \dot{\theta}' = \dot{\theta}
	\end{equation}
	 We must also convert the coordinate system over which we are performing the integration, thus:
	 \begin{align}
	 	\phi_1' &= \phi_1 - \phi\\
	 	\hat{\underline{\phi}}_1 &= \cos(\phi_1')\underline{\hat{r}} +  \sin(\phi_1')\underline{\hat{\phi}}
	 \end{align}
	 
	This leaves us at:
	\begin{equation}
		\dot{\theta}' = \int_{-\pi}^{\pi}g_{\sigma}(\phi_1' - \theta')\sin \left(\delta \hat{\underline{\phi}}_1\nabla\theta(\underline{x})\right) d\phi_1'
	\end{equation}
	Note, we have not changed the limits of the integration as all functions are periodic over the interval $[-\pi,\pi]$ so shifting by a constant does not change the integral.

	Finally, we assume that $\theta$ does not vary radially around a topological defect, thus $\partial_r\theta' = 0$. This is true for an isolated defect in the reciprocal XY model and allows us to write the gradient of $\theta$ in the polar basis:	 
	\begin{equation}
 		\nabla\theta = \frac{(\partial_\phi\theta' + 1)}{r}\underline{\hat{\phi}}
	\end{equation}

	Which gives us
	\begin{equation}
		\nabla\theta.\underline{\hat{\phi}}_1 = \frac{(\partial_\phi\theta' + 1)}{r}\sin(\phi_1)
	\end{equation}

	Using these identities we can convert our equation to the new coordinate system to arrive at:
	\begin{equation}
		\dot{\theta} = \int_{-\pi}^{\pi}g_{\sigma}(\phi_1 - \theta)\sin \left(\delta (\partial_\phi\theta + 1)\frac{\sin(\phi_1)}{r} \right) d\phi_1		
		\label{eq:E}
	\end{equation}
	Where we have dropped all $'$. 
	
	We consider the non-reciprocal kernel as $g_{\sigma}(\varphi) = \exp(\sigma\cos(\varphi)) \approx 1 + \sigma\cos(\varphi)$ for small $\sigma$. 
	\begin{align}
		\dot{\theta} &= \int_{-\pi}^{\pi}[1+\sigma\cos(\phi_1 - \theta)]\sin \left(\delta (\partial_\phi\theta + 1)\frac{\sin(\phi_1)}{r} \right) d\phi_1	\\
		&= \int_{-\pi}^{\pi}\sin \left(\delta (\partial_\phi\theta + 1)\frac{\sin(\phi_1)}{r} \right) d\phi_1 + \sigma\int_{-\pi}^{\pi}\cos(\phi_1 - \theta)\sin \left(\delta (\partial_\phi\theta + 1)\frac{\sin(\phi_1)}{r} \right) d\phi_1	
	\end{align}

	\paragraph{Stability of a +1 defect\\}
	
	For a $+1$ defect, $\partial_\phi\theta' = 0$ and $\theta' = \mu$. With $\mu = 0$ signifying a source and $\mu = \pi$ being a sink. In this scenario there is no variance on the polar angle around the defect and we can write a single update equation for the shape of the defect $\mu$.
	
	\begin{equation}
	\dot{\mu} = \cancelto{0}{\int_{-\pi}^{\pi}\sin \left(\delta \frac{\sin(\phi_1)}{r} \right) d\phi_1} + \sigma\int_{-\pi}^{\pi}\cos(\phi_1 - \mu)\sin \left(\delta\frac{\sin(\phi_1)}{r} \right) d\phi_1	
	\end{equation}

	Since $\sin \left(\delta\frac{\sin(\phi_1)}{r} \right)$ is anti-symmetric about $\phi_1=0$, the second integral will only give zero in the case where $\cos(\phi_1 - \mu)$ is symmetric about zero. This is only true for $\mu = 0, \pi$, thus the only fixed points of this equation are these values. 
	
	We now linearise this equation around $\mu$ using angle addition formula.
	\begin{align}
		\dot{\Delta\mu} &= \sigma\int_{-\pi}^{\pi}\cos(\phi_1 - \mu - \Delta\mu)\sin \left(\delta\frac{\sin(\phi_1)}{r} \right) d\phi_1 - \dot{\mu}\\
		& = \sigma\int_{-\pi}^{\pi}[\cos(\phi_1-\mu)\cos(\Delta\mu) + \sin(\phi_1-\mu)\sin(\Delta\mu)]\sin \left(\delta\frac{\sin(\phi_1)}{r} \right) d\phi_1 - \dot{\mu}\\
		&= \cancel{\sigma\int_{-\pi}^{\pi}\cos(\phi_1-\mu)\sin \left(\delta\frac{\sin(\phi_1)}{r} \right) d\phi_1} + \Delta\mu\sigma\int_{-\pi}^{\pi}\sin(\phi_1-\mu)\sin \left(\delta\frac{\sin(\phi_1)}{r} \right) d\phi_1 - \cancel{\dot{\mu}}\\
		&= \Delta\mu\sigma\int_{-\pi}^{\pi}\sin(\phi_1-\mu)\sin \left(\delta\frac{\sin(\phi_1)}{r} \right) d\phi_1
	\end{align}

	This is solved trivially by the equation $\Delta\mu = A\exp[\omega t]$ with 
	\begin{equation}
		\omega = \sigma\int_{-\pi}^{\pi}\sin(\phi_1-\mu)\sin \left(\delta\frac{\sin(\phi_1)}{r} \right) d\phi_1
	\end{equation}
	Which indicates stability for $\mu = \pi$ (sink) and instability for $\mu = 0$ (source). The magnitude of $\omega$ is greatest around $r/\delta = 0.5$, which means the evolution speed of $\mu$ decreases as $r$ increases (above the lattice spacing).

	\paragraph{Shape of a -1 defect\\}

	For a $-1$ defect, $\theta = -2\phi$ minimizes the energy in the reciprocal XY model. This is a defect in which the director points out along the $x$ axis like the one shown in Fig.~\ref{fig:Coords}. This gives 
	\begin{equation}
		\dot{\theta}(\phi) = \cancelto{0}{\int_{-\pi}^{\pi}\sin \left(-\delta \frac{\sin(\phi_1)}{r} \right) d\phi_1} + \sigma\int_{-\pi}^{\pi}\cos(\phi_1 + 2\phi)\sin \left(-\delta \frac{\sin(\phi_1)}{r} \right) d\phi_1	
	\end{equation}
	Once again the reciprocal terms vanish by symmetry. We can expand the second term using angle addition formula. 
	
	\begin{align}
		\dot{\theta}(\phi) &= \sigma\int_{-\pi}^{\pi}\cos(\phi_1 + 2\phi)\sin \left(-\delta \frac{\sin(\phi_1)}{r} \right) d\phi_1	\\
		&= \sigma\int_{-\pi}^{\pi}[\cos(\phi_1)\cos(2\phi) - \sin(\phi_1)\sin(2\phi)]\sin \left(-\delta \frac{\sin(\phi_1)}{r} \right) d\phi_1	\\
		&= \cos(2\phi)\sigma\cancelto{0}{\int_{-\pi}^{\pi}\cos(\phi_1)\sin \left(-\delta \frac{\sin(\phi_1)}{r} \right) d\phi_1} - \sin(2\phi)\sigma\int_{-\pi}^{\pi}\sin(\phi_1)\sin \left(-\delta \frac{\sin(\phi_1)}{r} \right) d\phi_1	\\
		&= \sin(-2\phi)\sigma\int_{-\pi}^{\pi}\sin(\phi_1)\sin \left(-\delta \frac{\sin(\phi_1)}{r} \right) d\phi_1
	\end{align}
	
	This has quadrupole-like symmetry such that $\theta$ will grow in the first and third quadrant, and shrink in the second and fourth, leading to a nematic symmetry of the -1 defects.
	It is not easy to perform a similar linear stability analysis for the $-1$ defect as the solution for $\dot{\theta} = 0$ is non-trivial.
	
	\paragraph{Numerical results\\}
	
	We can estimate the stable solutions of Eq.~(\ref{eq:E}) for different winding numbers using finite different methods. The winding number is fixed by the initial conditions and $\theta$ is evolved using Eq.~(\ref{eq:E}) until a stable solution is reached. The results are shown in Fig.~\ref{fig:Defs} below.

	\begin{figure}[h]
		\centering
		\includegraphics[width=0.75\columnwidth]{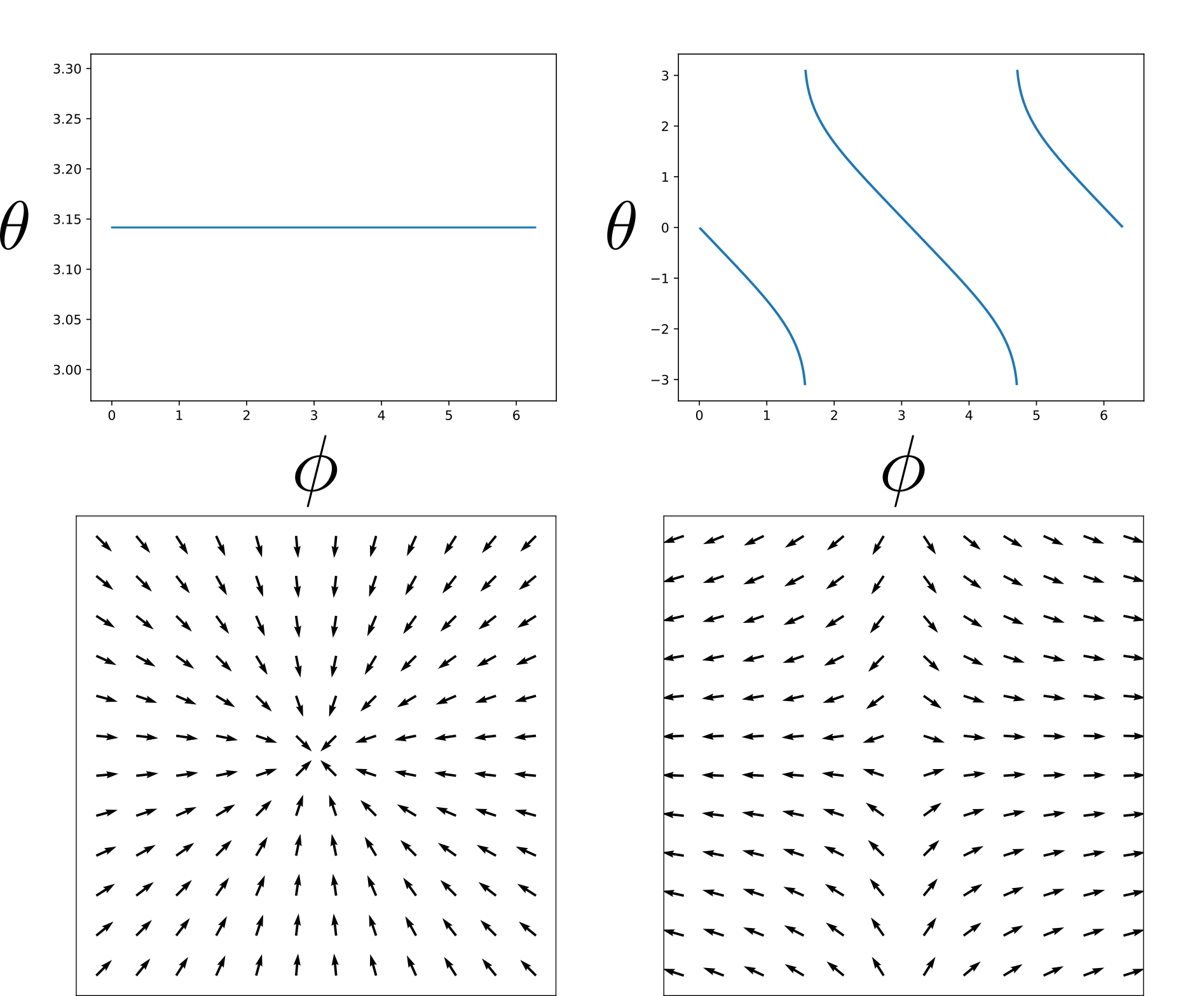}\caption{Stable configurations for +1 (left) and -1 (right) defect. Both solutions obtained by numerically solving Eq.~(\ref{eq:E}). \label{fig:Defs}}
	\end{figure}

\section{Appendix H: $1d$ propagation} \label{sec:profiles_different_theta0}
In the spirit of the Figure~\ref{fig:propagation1}, we plot the profiles $\theta(x)$ of a $1d$ perturbation over time, from $t=0$ in red to $t_\text{max} = 0.2$ in blue, for different background orientations $\theta_0 = n\pi/8 \text{, with } n = 0, ..., 15$, as indicated above each panel. 

\begin{figure*}
    \centering
    \includegraphics[width=\linewidth]{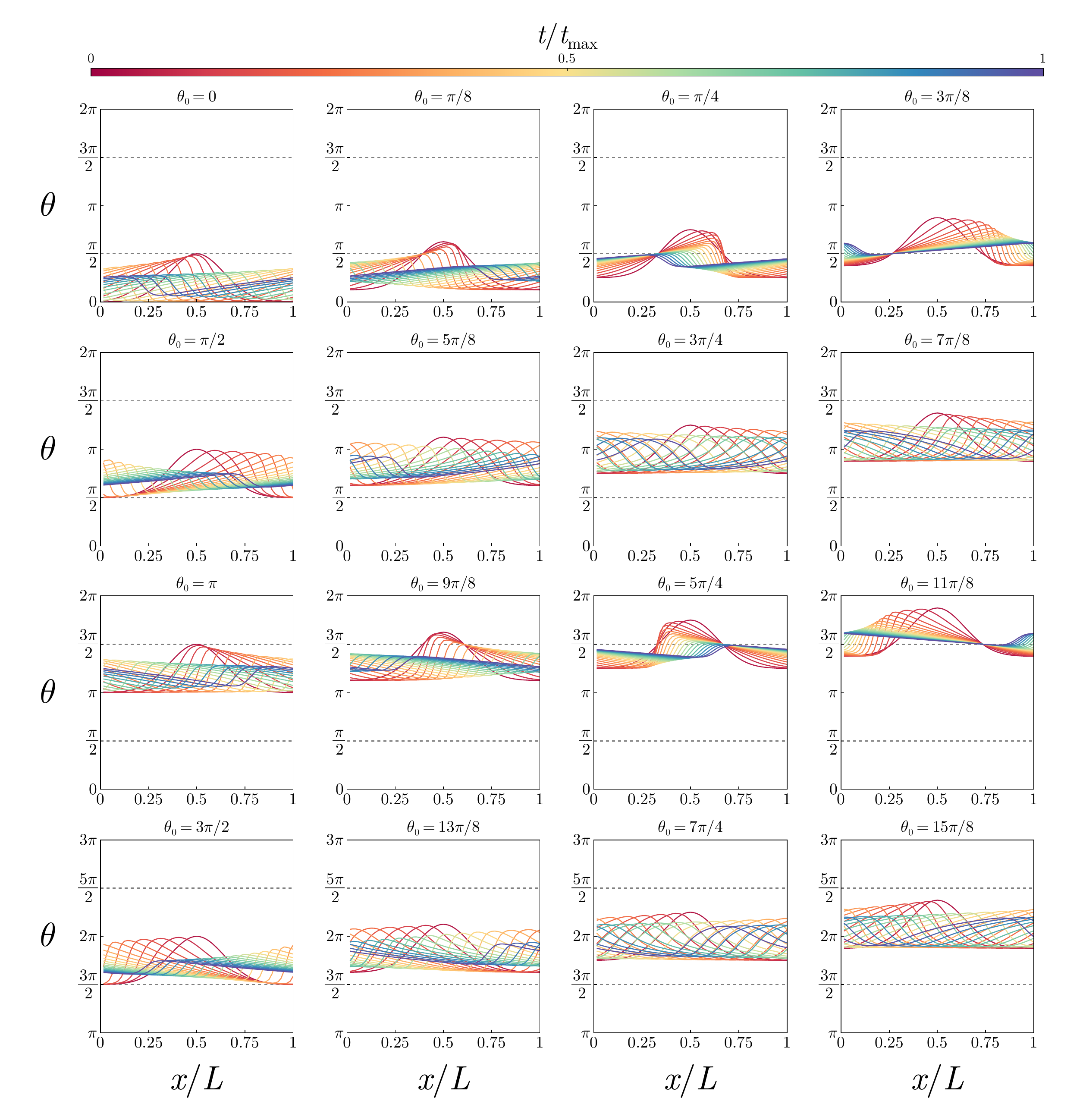}    
    \caption{
    Profiles $\theta(x)$ of a 1d perturbation over time, from $t=0$ in red to $t_\text{max} = 0.2$ in blue, for different background orientations $\theta_0 = n\pi/8 \text{, with } n = 0, ..., 15$, as indicated above each panel. \\ We plot horizontal dashed lines to mark the $\theta = \pi/2$ and $\theta = 3\pi/2$ lines, where the information flux changes sign and orientation. For $\pi/2<\theta<3\pi/2$, the profiles are pushed to the right, for $\theta<\pi/2$ and $\theta>3\pi/2$, the profiles are pushed to the left. 
Parameters : $L=64, \bar\sigma = 30, t_\text{max}=0.2, \alpha = 100 $}
    \label{fig:profiles_different_theta0}
\end{figure*}
\newpage

\section{Appendix I: Derivation of the stationary solution of the TPP} \label{sec:derivation}

For simplicity, let us work with a reduced space variable $x/L \to x$. Thus, $0\leq x \leq 1$. We will finish by inverting it back at the end with $x \to Lx$.

We start from 
\begin{equation}
\partial_t \theta
= \partial_{xx} \theta + \bar{\sigma}\, \cos(\theta)\,\partial_x \theta 
\label{eq:SM_first_eq_derivation}
\end{equation}
A stationary profile $\theta_s(x)$ satisfies
\begin{equation}
0 = \partial_{xx} \theta_s(x) + \bar{\sigma}\, \cos\!\big(\theta_s(x)\big)\,\partial_x \theta_s(x)
\end{equation}
Thanks to the change of variable $p(x)=\partial_x \theta_s(x)$, we will reduce the order of our equation. Note that the chain rule implies
\begin{equation}
    \partial_{xx} \theta_s = \partial_xp = \frac{\partial p}{\partial\theta}\frac{\partial\theta}{\partial x}\,= \frac{\partial p}{\partial\theta}\,p
\end{equation}
Assuming a monotone profile ($p= \partial_x \theta_s\neq 0$), we can divide by $p$ and integrate: 
\begin{equation}
\frac{dp}{d\theta} = -\,\bar{\sigma}\, \cos \theta
\quad\Rightarrow\quad
p(\theta) = C - \bar{\sigma}\, \sin \theta
\end{equation}
One thus has
\begin{equation}
\partial_x \theta_s = C - \bar{\sigma}\, \sin \theta_s
\label{eq:SM_dtheta}
\end{equation}
where $C$ is a constant of integration that we will specify below. For readability, we drop the index $s$ because we only work under the steady state assumption: $\theta_s \to \theta$. We now want to integrate this expression once more: we will separate the variables $\theta_s$ and $x$:
\begin{equation}
dx = \frac{d\theta}{C-\bar{\sigma}\, \sin \theta}
\qquad \text{such that } \qquad
x-x_0 = \int \frac{d\theta}{C-\bar{\sigma}\, \sin \theta}
\label{eq:SMeq1}
\end{equation}
where $x_0$ is a constant of integration.  To perform the integration on the right hand side of Eq.~(\ref{eq:SMeq1}), we resort to the change of variable 
\begin{equation}
    t=\tan\left(\frac{\theta}{2}\right)
\end{equation}
which implies $\theta = 2\arctan t$ . Recall that 
    \begin{align}
        &\sin(2x)=2\sin x \, \cos x \\
        &\cos \arctan x = 1 / \sqrt{1+x^2} \\
        &\sin \arctan x = x / \sqrt{1+x^2} \\
        &d\arctan(x)/dx = 1 / (1+x^2)
    \end{align}
Following, one has 
\begin{equation}
    \sin \theta = \frac{2t}{1+t^2} \qquad \text{and} \qquad d\theta = \frac{2\,dt}{1+t^2}
\end{equation}
Thus
\begin{equation}
\int \frac{d\theta}{C-\bar{\sigma}\, \sin \theta}
= \int \frac{2\,dt}{\,C(1+t^2)-2\bar{\sigma}\, t\,}.
\end{equation}
Let us focus on the denominator and complete the square to make the next integration easier: 
\begin{equation}
    C(1+t^2)-2\bar{\sigma}\,t
= C\Big[(t-\tfrac{\bar{\sigma}}{C})^2 + (\tfrac{\Delta}{C})^2\Big]
\end{equation}
where we defined $\Delta=\sqrt{C^2-\bar{\sigma}^2}$ .  Rewriting, one has
\begin{equation}
\frac{2}{C}\int \frac{dt}{(t-\tfrac{\bar{\sigma}}{C})^2 + (\tfrac{\Delta}{C})^2} = \frac{2}{C}\int \frac{d(t - \bar\sigma / C)}{(t-\tfrac{\bar{\sigma}}{C})^2 + (\tfrac{\Delta}{C})^2}
\end{equation}
We now use the formula 
\begin{equation}
\int \frac{d u}{u^2+a^2}=\frac{1}{a} \arctan \frac{u}{a}
\end{equation}
to obtain 
\begin{equation}
I \equiv \int
\frac{d\theta}{C-\bar{\sigma}\, \sin \theta}
= \frac{2}{\Delta}\,
\arctan\!\Big(\frac{C\,t-\bar{\sigma}}{\Delta}\Big)+c_1
\end{equation} 
When $\theta$ runs from 0 to $2\pi$, $t=\tan(\theta/2)$ runs from $-\infty$ to $+\infty$ and the arctan changes by $\pi$. 
One thus has
\begin{equation}
   \left.
   \frac{2}{\Delta}\,
\arctan\!\Big(\frac{C\,t-\bar{\sigma}}{\Delta}\Big) 
\right|_{\theta=0}^{\theta=2\pi} = \frac{2}{\Delta}\pi = c_1
\end{equation}
Now we can determine the value of the constant of integration $C$. Since $I = \int\limits_0^1 dx = 1$ (see Eq.~(\ref{eq:SMeq1})) ,  
\begin{equation}
\begin{aligned}
    & \ \Delta = 2\pi \\ 
    \Rightarrow   &\ \Delta^2 = 4\pi^2 \\ 
    \Rightarrow &\ C^2-\bar{\sigma}^2 = 4\pi^2 \\
    \Rightarrow &\ C = \sqrt{\bar{\sigma}^2 + 4\pi^2}
    \end{aligned}    
\end{equation}
Recall that if one works with $x/L$ instead of $x$, one has $I = \int\limits_0^L dx = L$ and thus 
\begin{equation}
\begin{aligned}
    & \ \Delta = \frac{2\pi}{L}\\ 
    \Rightarrow &\ C = \sqrt{\bar{\sigma}^2 +\left( \frac{2\pi}{L}\right)^2}
    \end{aligned}    
\end{equation}

Finally,  we isolate $t$. Recall that we need to perform the transformation $x\to Lx$, to obtain
\begin{equation}
t = \frac{1}{C}\, \left[ \bar\sigma + \Delta \tan \left( \frac{L \Delta}{2}(x-x_0)\right)\right]
\end{equation}
And $\theta = 2\arctan t$. In order to fit the data of Fig.~\ref{fig:concentration1}(a,b), we use $x_0 = 0.17$. 
If one has $L=2\pi$, the expression simplifies considerably. One has $C^2 =  + \bar \sigma^2 $ and $\Delta = $, so 
\begin{equation}
t = \frac{1}{C}\, \Big[ \bar\sigma + \tan \Big( \pi(x-x_0)\Big)\Big]
\end{equation}
and $\theta = 2\arctan t$. 
\newline 

Finally, one can derive the width $w$ defined as 
\begin{equation}
    w=\frac{2\pi}{ \partial_x\theta(x=3L/4 )} \ .
\end{equation}
One could derive the expression obtained for $\theta$, but it is much simpler to use the expression Eq.~(\ref{eq:SM_dtheta}). Since $x=3L/4 $ is a fixed point of the profile,  $\theta(x=3L/4 ) = 3\pi/2$ at all times, for all $\bar \sigma$. Therefore, $\sin \theta(x=3L/4 )=-1$ and $\partial_x \theta = C + \bar \sigma$. Therefore, 
\begin{equation}
    w = \frac{2\pi}{(C + \bar \sigma)} = \frac{2\pi}{ \sqrt{\bar{\sigma}^2 + 4\pi^2}+ \bar \sigma}
\end{equation}
In the simplified case $L=2\pi$, recall that one needs to transform back $x\to Lx$, including in the derivative $\partial/\partial x \to \partial/\partial (x/L) = L \cdot \partial/\partial x$, thus giving
\begin{equation}
    w = \frac{2\pi}{L \cdot (C + \bar \sigma)} = \frac{1}{\sqrt{\bar{\sigma}^2 + 1}+ \bar \sigma}
\end{equation}
For simplicity, we have derived all these equations omitting the elastic constant $K=1$. If one wants to reintroduce it, looking at Eq.~(\ref{eq:SM_first_eq_derivation}), one need to change $\bar\sigma\to\bar\sigma/K$. In particular, this gives $w = K/(\sqrt{\bar{\sigma}^2 + K^2}+ \bar \sigma)$ for $L=2\pi$.




	\chapter{Conclusion}
	\fancyhead[L]{Conclusion}
        \renewcommand{\bibsection}{\subsubsection*{References}}		
\graphicspath{{conclusion/figures/}}

\section{Summary}

In this thesis, we have extended the XY model, a paradigmatic model of phase transitions and topological defects, to non-equilibrium situations relevant in particular to biological systems. From synchronisation phenomena in coupled oscillators to non-reciprocal systems, we have explored how intrinsic disorder, mobility, and broken action-reaction symmetry affect the large scale properties of model systems through a modification of the topological defects behaviour in these systems.
\newline

We first provided a comprehensive review of the equilibrium XY model, the  topological defects and their role in the Berezinskii-Kosterlitz-Thouless transition and the coarsening dynamics. We introduced an efficient numerical method to simulate systems on a triangular lattice while using the same data structure as for square lattices. We also detailed the numerical method to compute the spatial correlation function using Fourier transforms, which is much more efficient than the direct space computation. 
Both approaches were particularly useful to port our code to GPUs to speed up the simulations. 
\newline

We then considered the steep XY model introduced by Domany \textit{et al.} \cite{Domany}. This equilibrium extension of the XY model, where the steepness of the potential is controlled by a parameter $p$, is known to exhibit a first order phase transition for large enough $p$. We argued that the estimated critical value $p_c \approx 15$ is in agreement with a converging body of observations and analogies with other models. We numerically gave the critical temperature $T_c(p)$.   
We established that there are two kinds of topological defects in this model, namely split XY defects and domain walls, and introduced an algorithm to discriminate between them. We showed that the pointlike, unit-charge defects of the XY model split in two half-integer defects separated by a line with non-zero tension of length $\ell \sim p$.  
\newline 

We then turned to the Kuramoto model, a paradigmatic model for synchronisation phenomena where each spin has its own intrinsic frequency. The Kuramoto model is traditionally widely studied in the dynamical systems community, which focuses in particular on chaos and bifurcation theory, and the emergence of synchrony. However, its connections to statistical physics and topological defects have been largely overlooked even though the simple coupling of the Kuramoto model to a thermal bath further highlights its relevance to statistical mechanics.
As such, we briged the gap between these two communities by studying the Kuramoto model from a statistical mechanics perspective, focusing on phase transition, topological defects and ordering dynamics.

We showed that the presence of quenched disorder destroys the quasi-long-range order of the XY model and generates unbound, superdiffusive defects \cite{Rouzaire_Kuramoto_PRL}.
The analogy of our defects' trajectories with self-avoiding random walks not only gives the correct superdiffusion exponent $3/2$, but also provides a simple and generic argument to explain the superdiffusive behaviour in other systems \cite{rouzaire_Frontiers}. 
We have also argued that the conclusions of our work are susceptible to apply to a wide range of systems \cite{kumar2022catapulting, Bililign2021}, and suggest that the simple lattice Kuramoto model could be a minimal model for superdiffusive defects in various quenched disorder systems.
\newline

We then extended the Kuramoto model to $2d$ mobile oscillators, and demonstrated that sufficient mobility spectacularly restores XY-like ordering and the standard defect annihilation law \cite{rouzaire2025activity}. This out-of-equilibrium journey also sheds new light on the equilibrium XY model itself, revealing the robustness of its quasi-long-range order and defect dynamics against substential perturbations. The smoothing procedure introduced in this context is general and can be applied to coarse-grain off-lattice particle systems~\cite{koehler2026flow}. 
\newline

Finally, we analysed a non-reciprocal XY model where directional interactions break the action-reaction principle. We introduced a continuous and non-negative weighting kernel to tune the non-reciprocity of the interactions in a smooth manner. This allowed us to come up with a very simple derivation to obtain a continuum theory from the microscopic XY model. We highlighted the emergence of an active curl term that has no equilibrium counterpart, thus clarifying the active nature of the non-reciprocal XY model and its connections to the celebrated Toner-Tu theory of flocking \cite{toner1995long}. 

Our derivation and our results enriched the literature on single species non-reciprocal systems, in particular regarding the current debate on what is a good continuum model for such systems~\cite{popli2025ordering,dopierala2025inescapable} or on the nature of the long-range order \cite{loos2023long, bandini2025xy,besse2022metastability,popli2025ordering}.

In addition, the rotational term explains the rich phenomenology of the non-reciprocal XY model: the active curl reshapes defects, which in turn opens the way to new defect annihilation scenarios unattainable in equilibrium \cite{rouzaire2025nonreciprocal}. 
This active term also advects patterns, leading to unidirectional propagation phenomena or even powering changes in the topology of the system \cite{rouzaire2025emergentdynamicsO2}. We believe that those results, potentially valid beyond the specific case of the $\mathcal{O}(2)$ model, will stimulate further research on general non-reciprocal systems, especially in the context of topological defects and pattern propagation.
\newline


\section{Perspectives}

I then wish to conclude this thesis by giving perspectives of possible extensions of my works.  
\newline

\paragraph{Non-reciprocal systems\\}
In the continuity of my research on non-reciprocal systems, one could study the non-reciprocal XY model on curved surfaces. 
On the torus for instance, one expects a competition between curvature-induced and non-reciprocity-induced forces on defects.
Curvature induces a geometric potential that acts on defects like an external field, attracting positive defects to regions of positive Gaussian curvature and negative defects to regions of negative Gaussian curvature. 
The addition of non-reciprocal forces might lead to new stable configurations unattainable in equilibrium, such as positive defects stabilised in regions of negative Gaussian curvature and vice-versa. It could also help the system to relax faster to its equilibrium ground state: in the case of a torus, the total topological charge is zero; yet, for a "fat enough" torus (with a small hole in the middle), the positive and negative defects are trapped on the inner and outer sides of the torus respectively, preventing their annihilation. Non-reciprocal forces might help the system to overcome this barrier and reach its equilibrium ground state,  in a spirit similar to the topological breakdown described in Chapter~\ref{ch:nrxy_defects}. In the case of a sphere, where the total topological charge is $+2$, one expects non-reciprocal forces to lead to new stable configurations, possibly with higher total topological charge, in the spirit of the stabilization of chains of defects described in Chapter~\ref{ch:nrxy_defects}. 
\newline

Another possible perspective is to address the current lack of a consensus on the correct continuum theory for the non-reciprocal XY model. In particular, a correct continuum theory should not reflect nor be impacted by the discrete symmetries of the underlying microscopic lattice. 
In particular, the order parameter of a system should remain $\mathcal{O(2)}$ symmetric, and not reflect the $Z_4$ ($Z_6$) symmetry of the square (triangular) lattice. In the same spirit, one should not observe a pinning of the defects polarisation axes along the lattice directions. This is a current debate in the community, with recent articles \cite{popli2025ordering,dopierala2025inescapable} discussing the relevance and the implications of different models, including the one we developed. Because the discrete nature of the regular lattice is the source of this problem, one could try to study the non-reciprocal XY model on random/disordered lattices, or even on off-lattice particles. However, even the reciprocal XY model on random lattices is not fully understood, as the particle density spatial fluctuations create effective quenched disorder that does affect topological defects properties such as the mobility. The defects tend to be trapped in low-density regions, where the effective noise is higher. This hinders their mobility and their ability to annihilate. Since the BKT scenario relies on the collective behaviour of defects, this in turn affects the long-range order of the system. One could also work with a regular lattice and look for a clever weighting kernel that would not propagate this lattice anisotropy to the continuum description. Instead, a hybrid approach is promising: working on a regular lattice, which is numerically efficient, while introducing some randomness in the interactions to smooth out the lattice effects. For instance, one could introduce a small amount of (dynamic) noise in the kernel:
\begin{equation}
    g(\varphi) = 1 + \sigma \cos (\theta_i - u_{ij} + \nu(t))
\end{equation}
where $\nu(t)$ is a random variable sampled from a uniform distribution in $[-\kappa, \kappa]$ at each timestep. 
The limit case $\kappa =0$ corresponds to the standard non-reciprocal XY model, while $\kappa = \pi$ effectively blurs the actual location of the spins, and should thus, on average, give the same physics as the equilibrium XY model. By tuning $\kappa$, one could be able, while keeping the physical impact of non-reciprocity, to smooth out the lattice effects, subtlety delocalising the neighbours as if the spins were submitted to thermal diffusion (but neglecting couplings between spins, which would give rise to phonons).
\newline

Third, building on the model of Juliane Klamser (CNRS, Montpellier) \cite{klamser2025directed}, one could study non-reciprocal steric forces. In this thesis, non-reciprocity was implemented at the level of social-like alignement forces. However, non-reciprocal interactions can be taken into account for other physical forces, such as steric repulsion, ie. excluded volume. An intuitive example is that of a dense crowd of pedestrians in a festival, where individuals tend to avoid more strongly being compressed from ahead than from their back. This leads to non-reciprocal steric forces, which can be implemented in a minimal model of self-propelled hard disks with usual repulsive potential, weighted by a non-reciprocal kernel like the one used in Chapter \ref{ch:nrxy_defects}. Following the approach described in that chapter, one could decompose the kernel in a reciprocal component, which one could express as the gradient of a potential, and a non-reciprocal component, which would give rise to an active force. Studying the collective behaviour and synchronisation of these active forces could account for the emergence of a global drift, as observed in \cite{klamser2025directed}. The elegant model of \cite{grossmann2020particle} might also be a good angle of attack study, as it would allow to systematically study the impact of the agents' aspect ratio (modelled as ellipses) on the collective dynamics under non-reciprocal steric forces.
This could also lead to new insights on the physics of dense crowds, in particular impacting the wave propagation in such crowds as we investigated in Chapter \ref{ch:nrxy_defects} and in \cite{rouzaire2025emergentdynamicsO2}. An interesting, wider perspective would be to compare these models to the literature on actual human crowds, to see whether non-reciprocal steric forces could help understand some aspects of crowd dynamics phenomena. 
\newline

\paragraph{Mixed symmetry systems\\}
Another interesting perspective, seen as a minimal, lattice model of self-propelled rods \cite{ginelli2010large, peruani2006nonequilibrium, bar2020self}, would be to study an active extension of the nematic XY model, where spins, living on a lattice, are equipped with a polar propulsion mechanism. One could use the Von Mises distribution to implement a preference to self-propel in the direction of spins' orientation, in the spirit of our kernel for non-reciprocal interactions. As both would be implemented through similar mechanisms, it would make the comparison between the role of activity and non-reciprocity easier.
The consequences of the mixed symmetry of nematic interactions and of polar activity are not clear, in particular regarding topological defects. 
In such a model, activity breaks the nematic symmetry of the interactions, since the particles have a well-defined head and tail. In this case, it is not clear why and when topological defects could be of charge $\pm 1/2$ (from nematic symmetry) or $\pm 1$ (from polar symmetry). One might expect a coexistence of both types of defects, or even new defect types with mixed symmetries, akin to the split defects observed in the equilibrium steep XY model \cite{Domany} studied in Chapter \ref{ch:steepXY}. 
In active nematics, where the activity has a nematic symmetry \cite{giomi2014defect}, the $+1/2$ defects are known to self-propel, while the $-1/2$ defects do not due to their rotationally symmetric 3-fold structure. 
The impact of a polar propulsion mechanism and of the particle density (vacancies) on their dynamics is not clear. Very preliminary simulation results indicate a nuanced scenario, where some $+1/2$ defects could self-propel, while others would not, depending on their shapes and on the local density of vacancies. 
The role of $+1$ and $-1/2$ defects in this system and their relation with the vacancies is also an open question.
Overall, the interplay between the nematic symmetry of the interactions and the polar symmetry of the activity could lead to new defect types, new isolated defects dynamics and new defect-defect interactions.
\newline

\paragraph{Machine learning\\}
Machine learning has revolutionized many fields of research, from image recognition to natural language processing. Its impact on condensed matter physics has also grown rapidly in the last decade, from simulation acceleration \cite{wu2024noise, tzivrailis2025uncertainty, mukhamadiarov2025controlling}, to data-driven identification of governing partial differential equations \cite{brunton2016discovering, colen2024interpreting} to discovery of order parameter \cite{du2025discovering}, to phase transition detection \cite{zhang2019machine}. 
In the context of topological defects, Deep Learning has already shown promising results in the identification and characterisation of defects in various systems \cite{suzuki2025machine, braghetto2025variational}. 

However, the architectures of the neural networks used so far in condensed matter are relatively simple (fully connected networks or convolutional neural networks), and do not leverage the powerful capacities of attention mechanisms \cite{vaswani2017attention}. 
Yet, in the equilibrium XY model, it seems that most of the information of the orientation field ($N$ spins, typically of the order of $10^4-10^5$) is contained in very few variables related to $n$ topological defects (position $x,y$, charge $q$ and shape $\mu$), typically of the order of $4n\sim 10^2$ at the end of the coarsening process. To leverage this condensed information effectively, one could use an attention based neural network, to self-adapt to the topological configuration of the system, without any need for systematic prior detection of the topological defects.
A good proof of concept would be to train an autoencoder network to generate a full configuration from the condensed information ($x,y,q,\mu$). It might be possible that this technique allows for a substential speed-up of the simulations by cleverly mixing machine learning and classical simulations, in the spirit of \cite{wu2024noise}. 
Once the equilibrium case is understood, one could turn to the non-reciprocal XY model and try to obtain the governing equations for the dynamics of a pair of defects, for which, as discussed in \cite{rouzaire2025nonreciprocal}, is a priori shape and charge dependent and highly non-trivial.

I am very interested in this topic. This is why, upon my own initiative, I gave a series of mini-lectures ($5 \times 2$ hours) in my department on Deep Learning based on the two very good courses of Francois Fleuret (University of Geneva) on Deep Learning (\href{https://fleuret.org/dlc/}{link}) and of Steve Brunton (University of Washington, famous for the SINDy pipeline \cite{brunton2016discovering}) on the interface between physics and Deep Learning (\href{https://www.youtube.com/@Eigensteve}{link}). 
The goals of this lecture was to explain from scratch the functioning principles of deep neural networks, to give a broad overview of main concepts and tools, to develop intuition on those complex machineries and showcase state-of-the-art tools that I know of: GPT, Gemini, MidJourney, Copilot. I plan to give those mini-lectures once more in the upcoming months to students and interested researchers. \\
\newline

\paragraph{Linguistics\\}
Finally, I would like to mention a more unconventional perspective, on the application of established statistical physics tools and of the methods developed throughout this thesis to the concept of complexity in linguistics. Dan Dediu and Alejandro Garcia Matarredona, researchers in linguistics at the University of Barcelona, study the complexity of texts using theoretical tools from statistical physics and agent-based simulations. An interesting perspective could be to model the interaction between individuals in a population by the famous $\mathcal{O}(n)$ model of statistical physics, where the language of each individual is represented by a spin in an $n$-dimensional feature space, and the interactions between individuals are represented by couplings between spins. A first idea is to develop an agent-based model to replicate the language acquisition by children, where non-reciprocal interactions between adults and children have been shown to play a crucial role \cite{macwhinney2017first}. 
A second idea would be to introduce foreign agents learning the language from natives, and to study what shapes the complexity of the acquired language.
Altogether, this model could help us understand possible mechanisms that describe how adults, children or foreigners learn a language from their environment, and how the structure of the language they acquire is influenced by the property of the interactions within the base population.


	\clearpage
	\thispagestyle{empty}


\renewcommand{\bibsection}{\chapter{Bibliography}}		
\fancyhead[L]{Bibliography}

\makeatother
\bibliographystyle{apsrev4-2}
\bibliography{These_cleaned}

 \label{LastMainPage} 

\clearpage
\thispagestyle{empty}

\appendixtitleon 
\appendixtitletocon
\begin{appendices}

 \titleformat{\chapter}[display]{\Huge\bfseries}{Appendix\ \thechapter: }{40pt}{\Huge} 
\titlespacing*{\chapter}{40pt}{100pt}{100pt} 

\newcounter{AppendixEquation} 
\makeatletter
\@addtoreset{equation}{AppendixEquation}
\makeatother

\newcounter{AppendixFigure}
\makeatletter
\@addtoreset{figure}{AppendixFigure}
\makeatother
 \makeatletter \renewcommand\@biblabel[1]{[#1]}
\makeatother
\pagenumbering{arabic} 
\fancyhead[L]{Appendix \thechapter}
\fancyhead[R]{\thepage / \pageref{LastPageapp}}
\renewcommand*{\thepage}{A-\arabic{page}} 
 \renewcommand{\bibsection}{\section*{Appendix references}}	
 \includefrom{appendix/}{appendix}
\end{appendices}

\cleardoublepage
\thispagestyle{empty}

\makeatletter
\newcommand{\clearevenpage}{\clearpage\if@twoside \ifodd\c@page
    \hbox{}\newpage\if@twocolumn\hbox{}\newpage\fi\fi\fi}
\makeatother
\thispagestyle{empty}

\clearevenpage
\thispagestyle{empty}

\dernierepage

\end{document}